**Mohammad Khalil, MSc.**

# Learning Analytics in Massive Open Online Courses

**DOCTORAL THESIS**

to achieve the university degree of

Doktor der technischen Wissenschaften

submitted to

**Graz University of Technology**

Supervisor/First reviewer:

Priv.-Doz. Dipl.-Ing. Dr.techn. Martin Ebner

Graz University of Technology

Second reviewer:

Prof. Dr. Carlos Delgado Kloos

Universidad Carlos III de Madrid

## AFFIDAVIT

I declare that I have authored this thesis independently, that I have not used other than the declared sources/resources, and that I have explicitly indicated all material which has been quoted either literally or by content from the sources used. The text document uploaded to TUGRAZonline is identical to the present doctoral thesis.

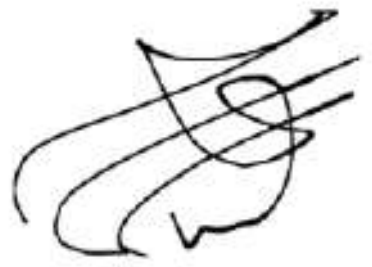

| 01.04.2017 | |
|---|---|
| Date | Signature |

# ABSTRACT

Educational technology has obtained great importance over the last fifteen years. At present, the umbrella of educational technology incorporates multitudes of engaging online environments and fields. Learning analytics and Massive Open Online Courses (MOOCs) are two of the most relevant emerging topics in this domain. Since they are open to everyone at no cost, MOOCs excel in attracting numerous participants that can reach hundreds and hundreds of thousands. Experts from different disciplines have shown significant interest in MOOCs as the phenomenon has rapidly grown. In fact, MOOCs have been proven to scale education in disparate areas. Their benefits are crystallized in the improvement of educational outcomes, reduction of costs and accessibility expansion. Due to their unusual massiveness, the large datasets of MOOC platforms require advanced tools and methodologies for further examination. The key importance of learning analytics is reflected here. MOOCs offer diverse challenges and practices for learning analytics to tackle. In view of that, this thesis combines both fields in order to investigate further steps in the learning analytics capabilities in MOOCs. The primary research of this dissertation focuses on the integration of learning analytics in MOOCs, and thereafter looks into examining students' behavior on one side and bridging MOOC issues on the other side. The research was done on the Austrian iMooX xMOOC platform. We followed the prototyping and case studies research methodology to carry out the research questions of this dissertation. The main contributions incorporate designing a general learning analytics framework, learning analytics prototype, records of students' behavior in nearly every MOOC's variables (discussion forums, interactions in videos, self-assessment quizzes, login frequency), a cluster of student engagement, and a conceptualization and implementation of a student motivational model. In the end, this dissertation presents a set of security and ethical challenges faced by learning analytics and contributes a conceptualization and implementation of a de-identification approach that offers a good solution to the ethical implications of learning analytics in MOOCs.

# Kurzfassung

Lehr- und Lerntechnologien haben in den letzten fünfzehn Jahren viel an Bedeutung gewonnen. Derzeit findet man unter dem Begriff der Lehr- und Lerntechnologien verschiedene online Lernumgebungen und -bereiche. Learning-Analytics und Massive-Open-Online-Courses (MOOCs) sind zwei Bereiche, die derzeit höchste Aktualität in dieser Domäne genießen. MOOCs sind auf reges Interesse bei den Lernenden gestoßen, da sie offen und meistens auch frei zugänglich sind. Deshalb kann die Anzahl der Teilnehmerinnen und Teilnehmer bei einem MOOC, je nach Themengebiet, auf mehrere Hunderttausend ansteigen. Diese rasante Entwicklung hat das Interesse zahlreicher Expertinnen und Experten von unterschiedlichen Disziplinen auf sich gezogen. MOOCs haben gezeigt, wie das Bildungswesen in den unterschiedlichen Gebieten bereichert werden kann. Verbesserte Lern- und Lehrergebnisse, Senkung der Kosten und Ausbau der Verfügbarkeit sind einige Vorteile die sich aufgezeigt haben. Die starke Nutzung von MOOCs-Plattformen erzeugte enorme Datenmengen, welche mit fortgeschrittenen Techniken und Technologien analysiert werden können. Hier kommt Learning-Analytics in Spiel. MOOCs bieten eine Vielzahl von Herausforderungen sowie Methoden für Learning-Analytics, die es zu bewältigen gilt. In Hinblick darauf, verbindet diese Doktorarbeit die beiden oben genannten Felder und untersucht die weiteren Einsatzmöglichkeiten von Learning-Analytics in MOOCs.

Der wichtigste Forschungsbereich dieser Dissertation befasst sich mit der Integration von Learning-Analytics in MOOCs und der anschließenden Untersuchung des Lernverhaltens der Lernenden, sowie den forschungsrelevanten MOOCs-Themen im Speziellen. Die Untersuchung wurde auf der österreichischen xMOOC-Plattform „iMooX“ durchgeführt. Um auf die Forschungsfragen dieser Doktorarbeit einzugehen wurden Prototypen entwickelt und Case-Studies durchgeführt. Der Hauptteil dieser Arbeit umfasst den Entwurf eines allgemeinen Learning-Analytics-Frameworks, einen Learning-Analytics-Prototypen, das Lernverhaltensprotokoll der Lernenden nach nahezu allen MOOCs Variablen (Diskussionsforen, Interaktionen in den Videos, Self-Assessment-Tests, Login-Häufigkeit), ein Cluster von Engagement der Lernenden und die Konzeptualisierung bzw. Implementation eines Motivationsmodels der Lernenden. Abschließend werden etliche sicherheitsbezogene und ethische Herausforderungen die Learning-Analytics gegenüberstehen präsentiert, welche durch die Konzeptualisierung und Implementation einer De-Identifikationsmethode zu einer guten Lösung der ethischen Probleme von Learning-Analytics in MOOCs beitragen.

# ACKNOWLEDGMENTS

This dissertation is dedicated to the memory of my great father. I will never forget him. He was a big support to me during my childhood and youth. Without his efforts, achieving this important step of my life would have never happened.

I would also like to express my deep appreciation and thanks to my great supervisor who has recently been promoted to Adjunct professor, Dr. Martin Ebner, who supported me through my career. Dr. Martin taught and helped me. He was not just an advisor, but a friend. I would also thank the Erasmus Mundus – Avempace III scholarship coordinators for giving me the opportunity to continue my PhD career at the Graz University of Technology.

A special thanks to my family as well. I am deeply thankful to the endeavors of my wife. She paid great patience and effort to see me through holding the doctoral degree. She gave me love, care, encouragement and supported me in all my pursuits. Words cannot express my gratefulness to my mother as well. Without her love, prayer and wisdom, I would never have reached this level. I dedicate the dissertation to my son, too. He sparked my incentive towards working hard.

I gratefully acknowledge my brothers Firas and Hamzeh and all my sisters for standing with me in all the hard and good times. My thanks also goes to my father-in-law and mother-in-law for their help and assistance.

In the meanwhile, I just lost my beloved grandmother. She always prayed for me to success and waited the moment that I graduate. I sincerely dedicate my dissertation to her memory. You will never be forgotten.

Lastly, I appreciate the help of all my colleagues, especially Behnam Taraghi, Markus Ebner, Walther Nagler and Alexei Scerbakov. Thank you very much!

Mohammad Khalil

March 2017

# CONTENTS

# List of Figures

# List of Tables

# List of Abbreviations

| Abbreviation | Meaning |
|---|---|
| ACM | Association for Computing Machinery |
| AECT | Association for Educational Communications and Technology |
| ASCILITE | Australasian Society for Computers in Learning in Tertiary Education |
| CAI | Computer-Assisted Instruction |
| CAPTCHA | Completely Automated Public Turing Test to Tell Computers and Humans Apart |
| CC | Creative Commons |
| CIA | Confidentiality, Integrity and Availability |
| cMOOCs | connectivism Massive Open Online Courses |
| CSCL | Computer Supported Collaborative Learning |
| CSV | Comma-Separated Values |
| DPD | Data Protection Directive |
| EC-TEL | European Conference on Technology and Enhanced Learning |
| ECTS | European Credit Transfer and Accumulation System |
| EDM | Educational Data Mining |
| EDs | Editors |
| EDUCON | IEEE Global Engineering Education Conference |
| EFA | Education For All |
| eLAT | Learning Analytics Toolkit for Teachers |
| e-Learning | Educational Technology |
| FERPA | Family Educational Rights and Privacy Act |
| GUI | Graphical User Interface |
| HE | Higher Education |
| HIPPA | Health Insurance Portability and Accountability Act |
| ICALT | International Conference on Advanced Learning Technologies |
| ICCE | International Conference on Computer in Education |
| ICDE | International Council for Open Distance Education |
| ICMI | International Conference on Multimodal Interaction |
| ICT | Information and Communication Technology |
| iLAP | iMooX Learning Analytics Prototype |
| ILE | Interactive Learning Environments |
| IT | Information Technology |
| ITS | Intelligent Tutoring System |
| JLA | Journal of Learning Analytics |
| L@S | Learning at Scale Conference |
| LAK | Learning Analytics and Knowledge |
| LASyM | Learning Analytics System for MOOCs |
| LDA | Latent Dirichlet Allocation |
| LIVE | Live Interaction in Virtual Learning Environments |
| LMS | Learning Management System |
| LOCO | Learning Object Context Ontology |
| MIT | Massachusetts Institute of Technology |
| MOLAC | MOOC Learning Analytics Innovation Cycle |
| MOOCs | Massive Open Online Courses |
| NIST | National Institute of Standards and Technology |
| OCLOS | Open e-Learning Content Observatory Services |
| OCW | Open Courseware |
| OER | Open Educational Resources |
| OU | Open University of England |
| PLE | Personal Learning Environments |
| RQ | Research Question |
| SLA | Social Learning Analytics |
| SNA | Social Network Analysis |
| SoLAR | Society of Learning Analytics Research |
| SR | Success Rate |
| STEM | Science Technology, Engineering, and Mathematics |
| TEL | Technology Enhanced Learning |
| TF-IDF | Term Frequency–Inverse Document Frequency |
| UNED | Universidad Nacional de Educación a Distancia |
| UNESCO | United Nations Educational, Scientific and Culture Organization |
| USSR | Union of Soviet Socialist Republic |
| VLE | Virtual Learning Environment |
| VPN | Virtual Private Network |
| xMOOCs | extended Massive Open Online Courses |

# 1 INTRODUCTION

The accessibility of the distance learning movement has gained much impetus over the last few years. Some time ago, Open CourseWare was introduced, and this enabled the promotion and growth of open and free online learning as we now know it. Following the debut of Open CourseWare, the educational technology community witnessed the onset of new courses with massive student numbers which are open for all and available online. These types of courses are called “Massive Open Online Courses,” more commonly known as “MOOCs” (McAuley et al., 2010).

The growth of MOOCs in the modernistic era of online learning is bolstered by the millions of participants from all over the world who choose to enroll in these massive courses. MOOCs bring revolutionary innovation to elementary education as well as to higher education (Khalil & Ebner, 2015a). In fact, the number of available MOOCs has exploded in recent years with over 4,500 courses provided by renowned universities across the USA and Europe.

The first rendition of the MOOC movement was developed by George Siemens and Stephan Downes. These courses are commonly referred to as cMOOCs (Holland & Tirthali, 2014). Their original MOOC was based on the connectivism theory of networking information over social channels. Thereafter, other versions of MOOCs emerged, such as xMOOCs or extended MOOCs, and xMOOCs became more popular. Due to their newfound popularity, xMOOCs were adopted broadly across MOOC providers. One of the most prominent and successful xMOOCs was offered by Stanford University professor Sebastian Thrun in 2011. His group launched an online course called “Introduction to Artificial Intelligence” that attracted over 160,000 students (Yuan & Powell, 2013). The introduction of this first xMOOC proved that providing free learning sessions that are taught by experts from prominent universities in a ubiquitous context drives large numbers of learners from heterogeneous backgrounds to join MOOCs (Alario-Hoyos et al., 2013).

MOOCs have the potential to advance education in many different fields and subjects. A report by Online Course Report (2016) showed that computer science and programming represented the largest percentage of offered MOOCs. Substantial growth of MOOCs has also been seen in Science, Technology, Engineering, and Mathematics (STEM) fields. The anticipated success of MOOCs vary between business purposes such as saving costs and scenarios of improvement involving the pedagogical and educational concepts of online learning (Alario-Hoyos et al., 2013; Khalil & Ebner, 2016e). In addition, their benefits include the improvement of educational outcomes and the extension of accessibility and reach. Another advantage of MOOCs is their long-term ability to contribute to lifelong learning as well as Technology Enhanced Learning (TEL) contexts (Ebner et al., 2014).

Although MOOCs have created a revolution in online education, they continue to suffer from high attrition rates among registered users. Dropout rates can go up to 95% (Daniel, 2012); because of this, MOOC attrition is a considerable area of concern. Reasons behind the profound risk of dropout can be explained by the free nature of MOOCs (Alario-Hoyos et al., 2013), student ability to self-regulate their learning (Zimmerman, 2000), or personal reasons. In addition, MOOCs face the universal educational challenge of keeping students engaged and motivated (XU & YANG, 2016). For instance, it has been found that student motivation decreases significantly following the first weeks of a MOOC (Lackner, Ebner, & Khalil, 2015). This factor has created an abundance of research questions with respect to patterns of engagement and categorization of students in MOOCs (Kizilcec, Piech, & Schneider, 2013; Alario-Hoyos et al., 2016; Ferguson & Clow, 2015; Khalil & Ebner, 2015a). Furthermore, the lack of interaction between learners and instructor(s), and the controversial argument regarding MOOCs' pedagogical approach act as roadblocks to the advancement of MOOCs. As a consequence of these challenges, research studies in MOOCs from surveys to case studies have heavily increased in the last five years (Liyanagunawardena, Adams, & Williams, 2013; Yousef et al., 2014; Kloos et al., 2014). In fact, the large data sets generated by student interactions in MOOCs provide an excellent opportunity to expand the research capabilities in and around MOOCs.

Because of the large number of MOOC enrollees, rich information content is stored in the databases of MOOCs servers. The accumulation of such data leads to so-called "Big Data." This term, which has become quite familiar in recent years, refers to "datasets whose size is beyond the ability of typical database software tools to capture, store, manage, and analyze" (Manyika et al., 2011). However, and of note, when noisy, unstructured and steep complex data are filtered, examination and analysis become conceivable. In 2011, the field of learning analytics emerged from the growing need to understand behavior and attitudes of learners in online learning platforms and from the need for advice in the domain of learning (Siemens, 2010). Learning analytics is firmly related to other fields such as web analytics, educational data mining, academic analytics and business intelligence (Elias, 2011). One of the key uses of learning analytics is in the identification of trends, discovery of patterns, and evaluation of learning environments such as MOOCs. Ubiquitous technologies have made tracking students online easier than ever before. Some significant factors that drive the evolution and development of learning analytics are: a) technology in educational categories, b) the Big Data available from learning environments, and c) the availability of analytical tools.

## 1.1 Motivation

In the current literature, much of the learning analytics research focuses on building predictive models of performance and students-at-risk in online environments and learning management systems (Dawson et al., 2014). However, the research of learning analytics in MOOCs carries with it the inherent possibility to examine more esoteric factors that might influence learning (Reich, 2015). Learning analytics faces particularly challenging demands and dilemmas when applied to MOOCs (Clow, 2013), but as Knox (2014) discussed, learning analytics carries with it a strong potential for discovery when it is applied to MOOC datasets. At the time of inception of this dissertation study in September 2014, there were very few research studies that combine learning analytics and MOOCs. In fact, there still remain very few research studies, confirmed by the absence of learning analytics practices and research in MOOCs (Moissa, Gasparini & Kemczinski, 2015; Vogelsang & Ruppertz, 2015; Kloos et al., 2016; Ruipérez-Valiente et al., 2016).

The application of learning analytics in MOOCs is critical to unveil hidden information and patterns contained in large educational data sets. Additionally, the demand for learning analytics in MOOCs has materialized to provide decision support, to find relevant solutions, to optimize learning, and to engage students to achieve more commitment and a higher level of success. According to Clow (2013) and because of the relative novelty of both the fields of MOOCs and learning analytics, research studies that target the combination of learning analytics and MOOCs have not yet been extensively carried out. When we reviewed the literature in 2014, it was obvious that some researchers shed light on the shortage of research on learning analytics in MOOCs (Clow, 2013; Kizilcec, Piech, & Schneider, 2013; Knox, 2014; Chatti et al., 2014; Neuböck, Kopp, & Ebner, 2015). The integration of learning analytics and MOOCs together is an essential thesis of discussion which contains a number of challenges to overcome. Among these challenges are questions about the design and implementation of learning analytics in MOOCs (Khalil & Ebner, 2015b; Khalil & Ebner, 2016e), MOOC completion rates and student motivation (Khalil & Ebner, 2017; Khalil & Ebner, 2016c), MOOC variables analyses, and last but not least, student behavior, profiles and categories in MOOCs (Khalil & Ebner, 2016a; Khalil, Kastl, & Ebner, 2016). The motivation of this dissertation comes from our belief that the discovery potential of learning analytics will provide us with answers to our research questions.

## 1.2 Research Framework

The research in this dissertation was carried out and established using the first Austrian xMOOC platform, known as the iMooX. iMooX is an online learning stage founded in 2013 as a project between the University of Graz and the Graz University of Technology. iMooX has over 14,000 enrollees and more than 30 offered MOOCs. This has created an engaging environment for testing, questionnaires, surveys and research studies. Since 2014, more than 80 publications of bachelor's, master's, and Ph.D. research studies and articles have been done on the iMooX-MOOC platform. This dissertation is an examination of some open questions that need broad investigation and a kernel for new challenges that we tackle using the latest research outcomes of the eLearning field.

## 1.3 Research Questions

The overall purpose of this thesis is to investigate the potential of learning analytics in Massive Open Online Courses (MOOCs). Our research focuses on the integration of learning analytics in MOOCs and thereafter examines what learning analytics can tell us about learning in MOOC platforms on one side, student behavior on the second side, and bridging MOOC issues from yet another side. The analysis enclosed provides answers to several research questions related to the implementation of learning analytics in MOOCs and its challenges, student patterns, and profiling as well as motivating student engagement with MOOC variables. Hence, our present dissertation answers the following research questions:

- How should we design a learning analytics framework in xMOOCs that can help us implement a learning analytics tool?

- How can we implement learning analytics prototypes in xMOOCs for research and administrative use?

The following questions are related to the generated educational data from MOOCs with the help of learning analytics:

- How is it possible to build activity profiles in MOOCs for students when the MOOC is dedicated to school children but also open to the public? And what student behavior can we identify through self-assessment quizzes from such MOOCs?
- What is the capability of our learning analytics tool in monitoring student patterns of MOOC videos? And is there a difference between certified students and non-certified students in dealing with interactive videos?
- What participant types can be classified (clustered) in MOOCs based on their MOOC engagement level? And is there a difference between MOOCs' credited higher education students (undergraduates) and external students?
- Given the results from the categorization of MOOC students, how can we distribute the student categories on the Cryer's scheme of Elton (1996)?
- Is there a specific point in xMOOCs where learners tend to drop the course or become lurkers?
- What student behavior exists in the discussion forum of the iMooX-MOOC platform? And how can Gilly Salmon's Five Stage Model foster student contributions in xMOOCs discussion forums?
- How can we motivate MOOC students in iMooX and make them more active which therefore reduces the lurking ratio among learners?
- What are the security constraints of learning analytics in general and learning analytics of MOOCs in particular? And what can be proposed to overcome inadvertent disclosure of learners' private information?

The field of learning analytics uses different tools and techniques that can lead to the improvement of online environments from different perspectives. The key answers to previous questions will clarify the promises of learning analytics in MOOCs.

## 1.4 Thesis Outline

The thesis at hand is established and organized into seven chapters as the following:

- Chapter 2 describes the followed research methodology that is used to carry out the research studies and questions in this thesis.
- Chapter 3 presents the related work of this dissertation. This chapter includes a large literature review that covers educational technology, online learning, MOOCs, learning analytics, and learning analytics of MOOCs.
- Chapter 4 gives an overview of the iMooX platform as well as describes the learning analytics framework. In addition, the chapter describes the development of the learning analytics prototype and its evaluation.
- Chapter 5 describes detailed analyses of various case studies. The chapter introduces several case studies that mainly aim at describing the potential of learning analytics in every different learning scenario from iMooX. This chapter includes six cases studies: Chapter 5.1 titled as “Activity Profile and Self-Assessment Quizzes,” Chapter 5.2 entitled “Tracking Videos Activity in MOOCs,” Chapter 5.3 has a title of “Dropout Investigation,” Chapter 5.4 is the case study of “Clustering Patterns of Engagement to Reveal Student Categories,” Chapter 5.5 is presented with “Case Study: Fostering Forum Discussions in MOOCs,” and Chapter 5.6 has a title of “Fostering Student Motivation in MOOCs.”
- Chapter 6 points out the learning analytics constraints and presents the De-Identification model as well as the Anonymizer.
- Finally, Chapter 7 summarizes every aspect of this thesis. This chapter includes conclusions and future research directions.

# 2 RESEARCH METHODOLOGY

In order to conduct the research in this work, this thesis followed two main steps: (1) Prototyping, and (2) Case studies research methodology.

## 2.1 Prototyping

Prototyping first came to discussion in the late 1970's, with the emergence of high-level languages (Budde et al., 1992). Prototyping has improved to become a convenient methodology in software development. It is effective when changes are required (Carr & Verner, 1997). In fact, prototyping allows researchers to develop parts of the solution to prove functionality in a limited period of time. Budde et al. (1992) presented a broad range of guidelines to perform prototyping similar to our research studies. Budde et al. (1992) set preliminary characterizations of prototyping as the following:

- Prototyping is a scheme based on an evolutionary view of software development, and it has an impact on the whole development process as a whole.
- Prototyping comprises production of early working versions (or prototypes) of the future application.
- Prototyping offers a great way of opening a communication channel with groups involved in the development process.
- And finally, prototyping allows for more experiments and therefore is a convenient model to do research and case studies.

The prototyping model that we used in this dissertation followed, to some extent, what is shown in Figure 1. It starts with the first version of the software based on specific requirements, then we test the pilot version, and after that, if the software meets our needs, we proceed to case studies. Otherwise, we revise the prototype until satisfaction.

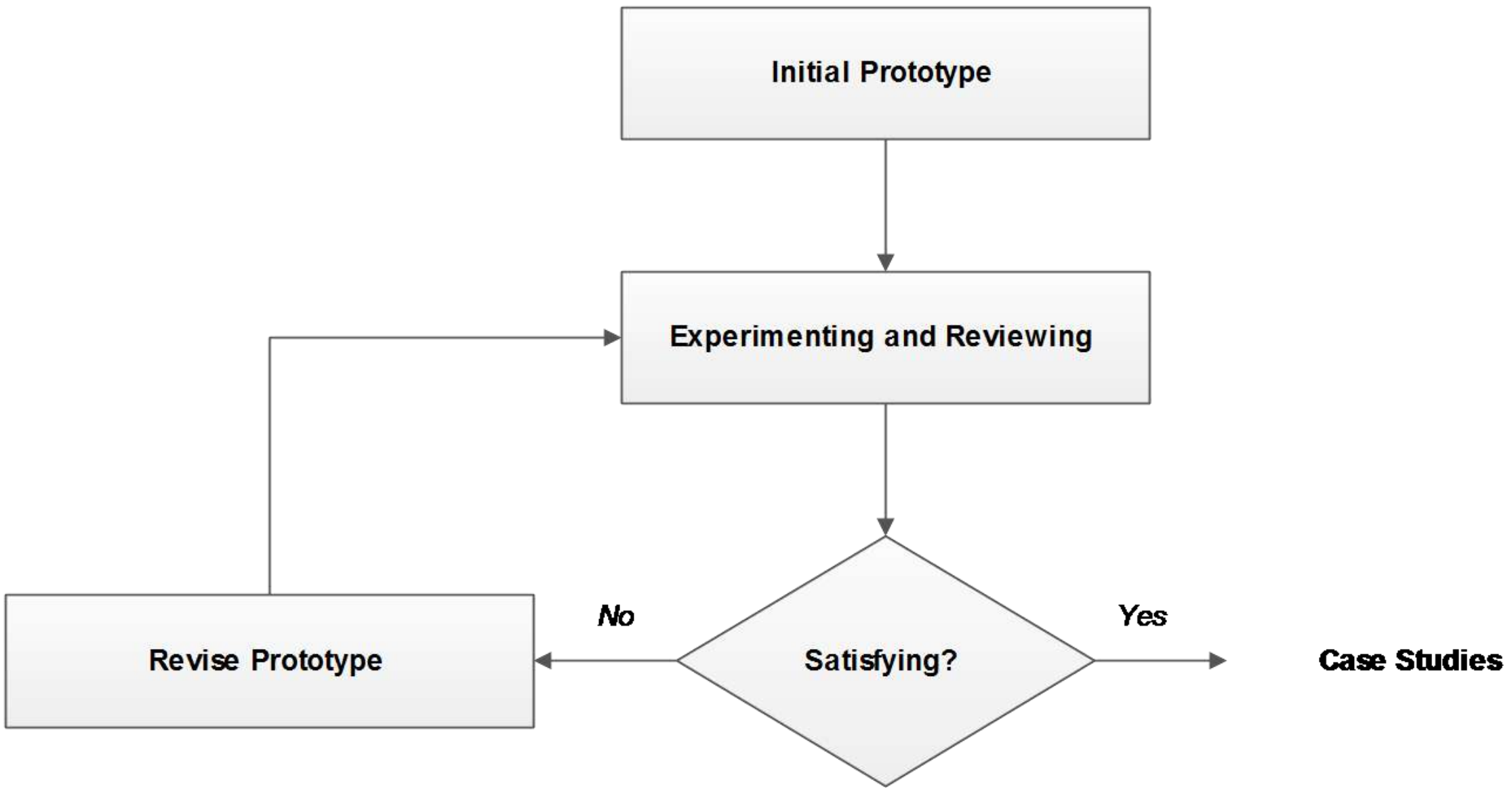

**Figure 1.** The prototyping methodology (Budde et al., 1992; Carr & Verner, 1997)

## 2.2 Case Studies

The next step after the prototype software is the creation of case studies. This dissertation has extensively employed case study research methodology adopted from Yin (2003). The essence of "a case study… is that it tries to illuminate a decision or set of decisions, why they were taken, how they were implemented, and with what results" (Schramm, 1971, P. 6). Yin (2003) said that a research design of case studies is "an action plan for getting from here to there, where here may be defined as the initial set of questions to be answered, and there is some set of conclusions (answers) about these questions" (p. 19). Similar to what we did in this dissertation, the distance between "here" and "there" included data collecting, content analysis, and extracting results. The content analyses of case studies followed the Neuendorf (2002) method by which units of analysis (MOOC indicators in this thesis) get measured and benchmarked based on qualitative decisions such as comprehensive decisions and surveys.

Thus, to summarize the methodology of our thesis, we first did a broad literature review on Massive Open Online Courses and learning analytics. Furthermore, in each of the following chapters, we present a short related work/background of each case study. After analyzing the accumulated literature, we present our developed learning analytics prototype. With the collected data from the learning analytics prototype, we develop case studies, interpret the data, report findings, and make conclusions (Figure 2).

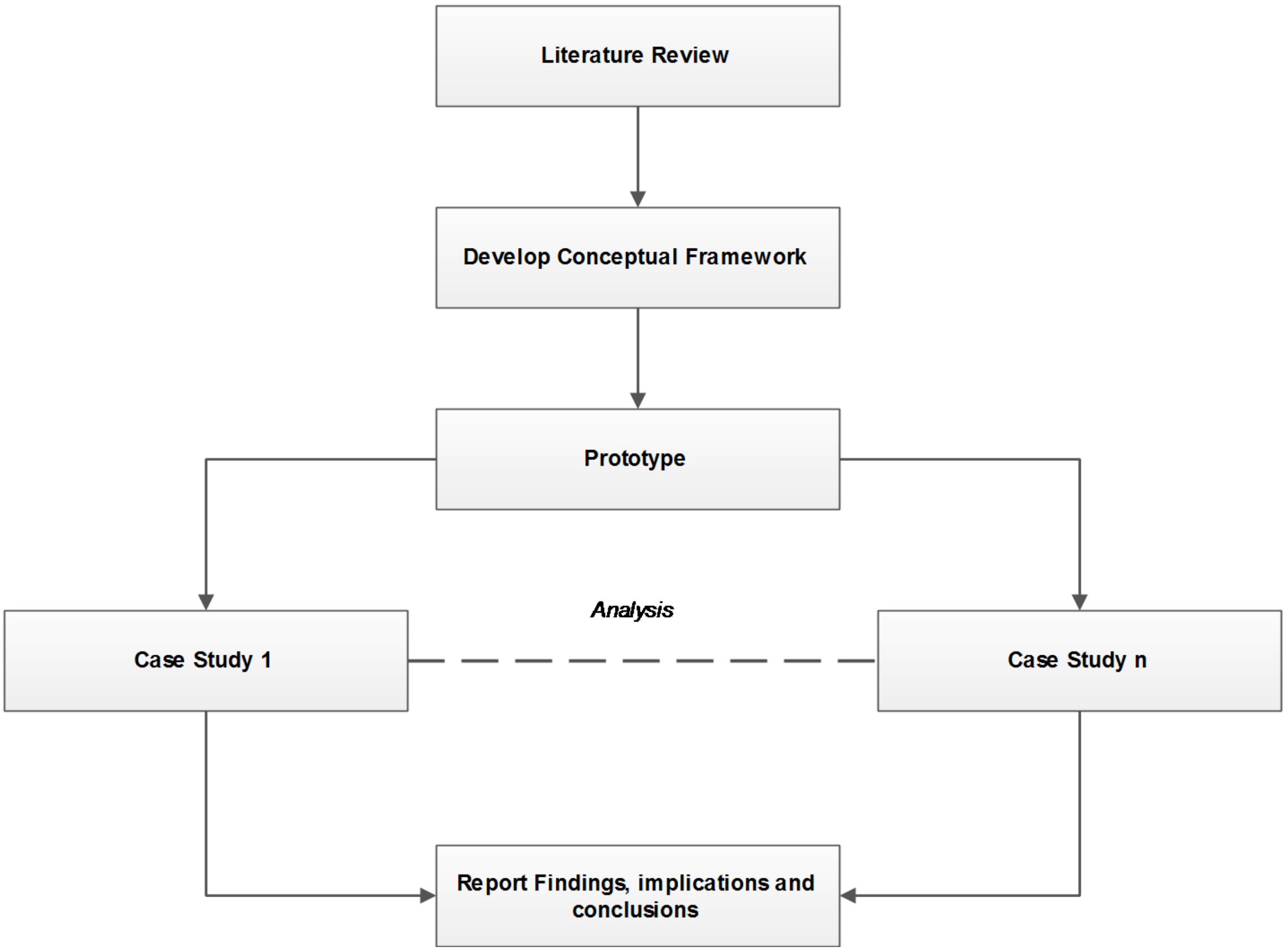

**Figure 2.** The research methodology process followed in this thesis. Methodology trails studies of (Budde et al., 1992) and (Yin, 2003)

# 3 RELATED WORK[1]

In the last few years, technology and the Internet have evolved rapidly and changed the world of communication and information. The rhythm of modernization has been further applied in the fields of mathematics, engineering, and healthcare. Similar to other areas, education has entered the revolution of change. Several fields evolved within the education scope and created one of the groundbreaking developments called educational technology or web-based education (e-Learning). Since then, e-Learning has altered the classical learning, training, and teaching landscape. Despite the fact that e-Learning as a term was frequently used in the late '90s and the 21st century, it has a significant historical background in the first decade of the 20th century. Sidney Pressey, a professor at Ohio State University, developed the first mechanical teaching machine in 1924 (Hilgard & Bower, 1975) which examined students based on multiple-choice questions (see Figure 3). Pressey's machine is considered one of the first e-Learning devices.

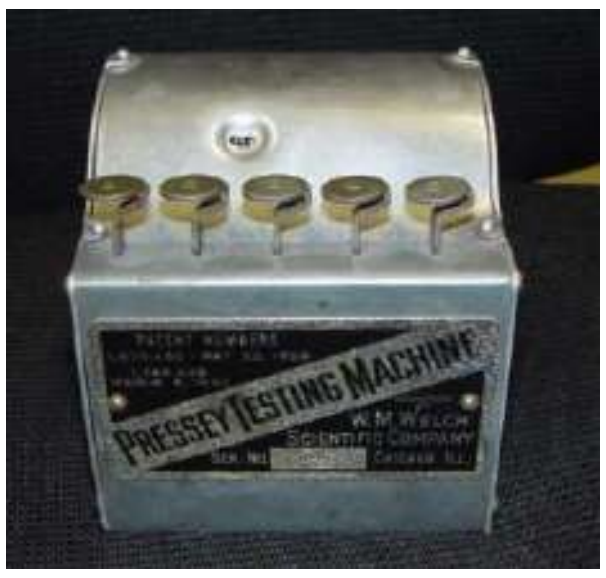

**Figure 3.** Sidney Pressey's multiple-choice machine is considered one of the first e-learning applications (Image available at: http://www.nwlink.com/~donclark/hrd/history/pressey.jpg, last accessed 01.November.2016)

---

[1] Parts of this chapter have been published in:

Khalil, M., Brunner, H., & Ebner, M. (2015). Evaluation Grid for xMOOCs. *International Journal of Emerging Technologies in Learning*, 10(4).

Khalil, M., & Ebner, M. (2016d). What is Learning Analytics about? A Survey of Different Methods Used in 2013-2015. *Proceedings of Smart Learning Conference, Dubai*, UAE, 7-9 March, Dubai: HBMSU Publishing House, 294-304.

Khalil, M., & Ebner, M. (2016f). When Learning Analytics Meets MOOCs – a Review on iMooX Case Studies. In *Proceedings of the 16th International Conference on Innovations for Community Services (I4CS)*, 2016. Vienna, Austria, pp. 3-19. doi: 10.1007/978-3-319-49466-1_1

Leitner, P., Khalil, M., & Ebner, M. (2017). Learning Analytics in Higher Education: A Literature Review. *Learning analytics: Fundaments, applications, and trends: A view of the current state of the art*. Springer International Publishing. http://dx.doi.org/10.1007/978-3-319-52977-6_1

After the Union of Soviet Socialist Republics (USSR) launched the first satellite to space in 1957, the United States planned to widen the higher education level to include a large majority of people. This plan increased the needs for more educational supplements and resources. Therefore, more calls were made to computerize learning. A series of machines with televised display like PLATO (see Figure 4) were then assembled to assist automated learning and help students to evaluate and revise materials (Bitzer, Braunfeld & Lichtenberger, 1961). Years later, institutions around the world began to utilize technology to deliver education to a larger population where physical presence became harder, using video conferencing, distance learning courses and the World Wide Web.

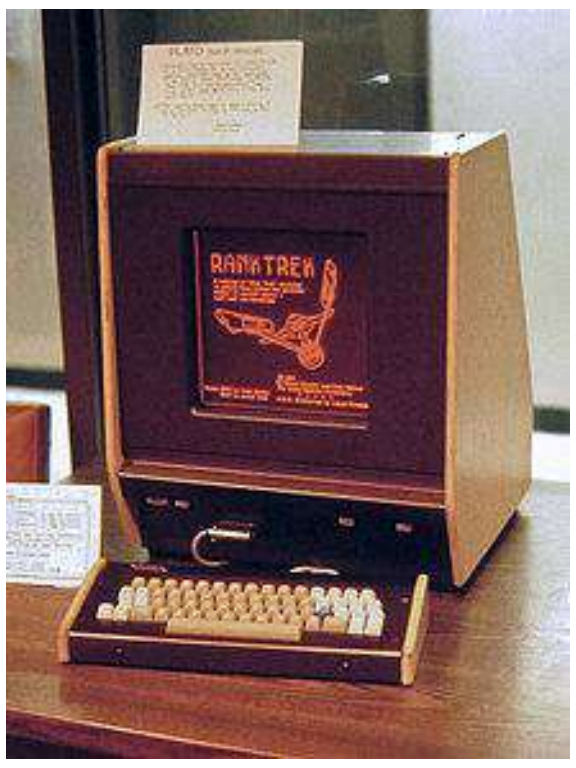

**Figure 4.** Plato V is a televised e-Learning device (Image available at: https://upload.wikimedia.org/wikipedia/commons/thumb/e/ed/Platovterm1981.jpg/220px-Platovterm1981.jpg, last accessed 01.November.2016)

Researchers from disciplines like psychology and pedagogy have identified that e-Learning models are mainly based on technology prototypes (Ajzen & Fishbein, 1977; Davis, Bagozzi & Warshaw, 1989). Although e-Learning is considered the main term when technology is ascribed to education, there are many other related terms that people in the community use under the umbrella of e-Learning like: Technology-Enhanced Learning (TEL), technology based learning, seamless learning, Computer Supported Collaborative Learning (CSCL), distance learning, online education, online learning, Computer-Assisted Instruction (CAI), Information and Communication Technology (ICT) in education, mobile learning and digital education (Ebner & Schiefner, 2008; Sharples et al., 2009; Goodyear and Retalis, 2010).

E-Learning has recorded quite a high demand in the business sector as well as schools and institutes of higher education. According to (Sun et al., 2008), the e-Learning market has shown a strong growth rate of over 35% in the last couple of years. Also, a recent report by Docebo on e-Learning market trends forecasted that revenues in this area will reach the $50 billion mark by 2016 (Trends, 2014). Such market predictions confirm the high value of e-Learning and its significant role in influencing the current ICT domain.

Practitioners and researchers have attempted to define e-Learning according to their background and experience. Welsh and co-authors defined it as “the use of computer network technology, primarily over an intranet or through the Internet, to deliver information and instruction to individuals” (Welsh et al., 2003, p. 246). Guri-Rosenblit (2005) has described e-Learning as “the use of electronic media for a variety of learning purposes that range from add-on functions in conventional classrooms to full substitution for the face-to-face meetings by online encounters” (p.469). The Association for Educational Communications and Technology (AECT, 2008) defined it as “the study and ethical practice of facilitating learning and improving performance by creating, using, and managing appropriate technological processes and resources.” The great benefits of e-Learning incorporate the digital interactions between learners and teachers from one side, and learners and learners from the other side; however, these interactions happen in the outlying space (like internet or networks) and without time limitations.

Educationalists and practitioners modeled e-Learning frameworks across time. A research study by Sun and his colleagues (Sun et al., 2008) have summarized six dimensions of e-Learning from learners’ point of view that would result in definite e-learner satisfaction. These dimensions are: learners, instructors, courses, technology, design, and the environment. Under each dimension, the authors gathered factors that yield to fulfillment of online learner satisfaction. Figure 5 depicts the e-Learning six dimensions, and the factors that leave a high impact on each dimension.

In the early ‘90s, e-Learning was about presenting the traditional learning content with the help of multimedia interactions, simulations, and animations (Ebner, 2007). Nevertheless, due to the development of the World Wide Web and the emergence of social networks and blogs, Stephen Downes, an educationalist and specialist in online learning technologies, argued that online learning is coming to the edge of extending the previously discussed e-Learning to a more advanced version called e-Learning 2.0 (Downes, 2005). E-Learning 2.0 theories are about engaging students more in the learning process. The introduction of Web 2.0 technologies has completely changed Internet social media and users’ interactions on the digital web (O'reilly, 2007). Web 2.0 has brought tagging, news syndication and blogging ideology to life. This revolution has changed the way the Internet behaves. As a consequence, podcasts, weblogs, wikis and web sharing applications induced e-Learning 2.0 to emerge (Ebner, 2007). Downes (2005) expressed the idea of e-Learning 2.0 in a body of questions:

> What happens when online learning ceases to be like a medium, and becomes more like a platform? What happens when online learning software ceases to be a type of content-consumption tool, where learning is 'delivered,' and becomes more like a content-authoring tool, where learning is created? The model of e-learning as being a type of content, produced by publishers, organized and structured into courses, and consumed by students, is turned on its head.

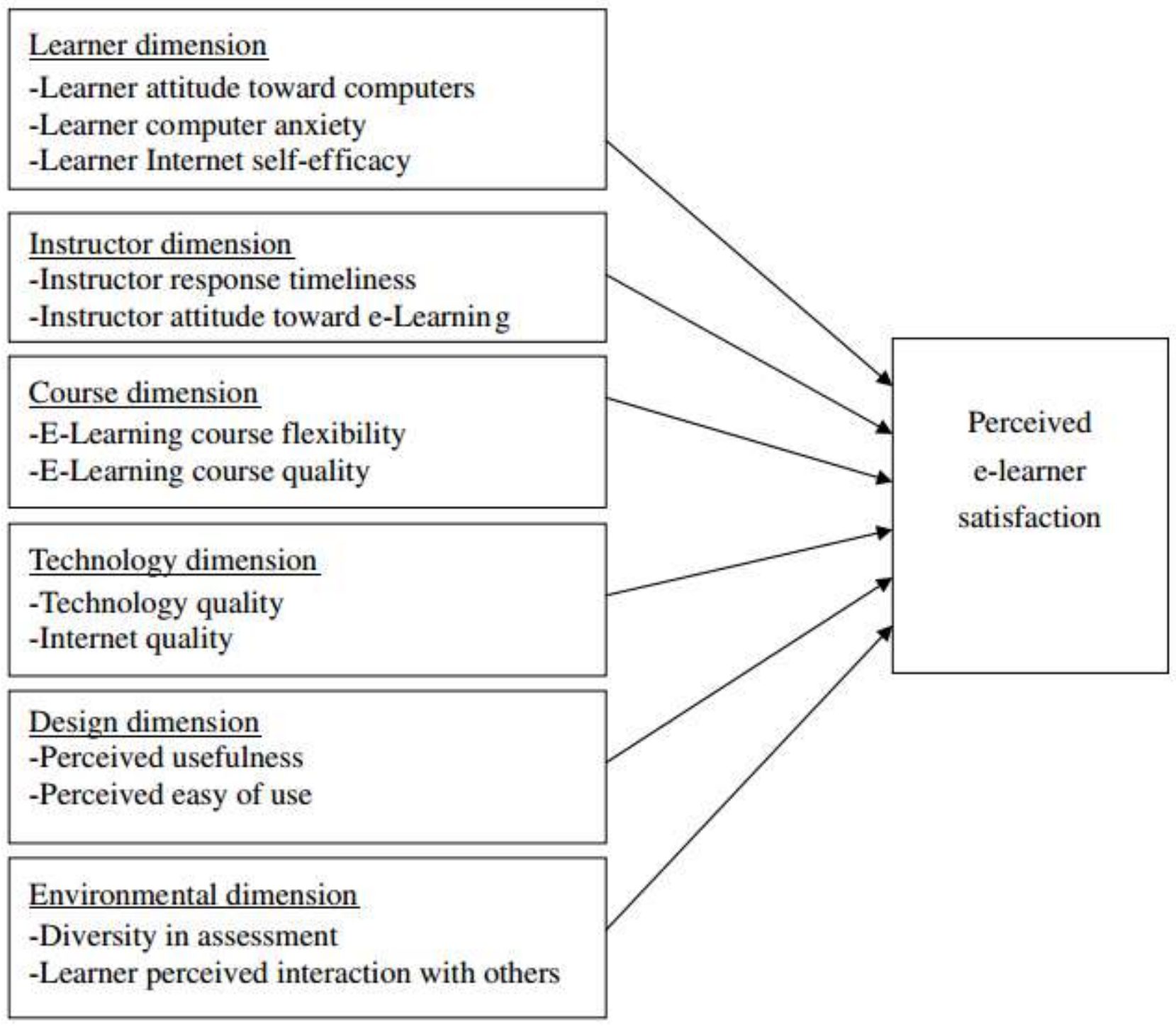

**Figure 5.** E-Learning six-dimensions and their factors that affect learners perceived satisfaction (Sun et al., 2008)

Today's e-Learning is split into two main classes, synchronous and asynchronous learning. The synchronous e-Learning requires students to be 'on-site' in front of their computers (Welsh et al., 2003) or their mobile phones (Sharples et al., 2009) at a specific time. This type of online learning is less common than the other type because of the time restriction and the confined learner freedom. The asynchronous e-Learning is more popular and is entitled to have ‘recorded’ learning lectures (Rosenberg, 2001). This type has been more accepted across online education because it offers students better time and location flexibility.

## 3.1 Distance Learning

The first practical use of distance education returns to an advertisement by Caleb Philipps in 1728 in the *Boston Gazette* newspaper who claimed to send written lessons by mail for interested people residing outside Boston (Holmberg, 2005). Later, academies started looking for alternative methods to deliver education to unreachable students who do not reside within their area. Many higher education institutes thenceforth adapted mail learning as a solution, such as the University of London in 1858 and the Australian University of Queensland in 1911 (Store & Chick, 1984). Afterwards, with the development of open education, universities like the Spanish Universidad Nacional de Educación a Distancia (UNED) in 1972 (National University of Distance Education, n.d.), and Germany's FernUniversität in Hagen in 1974 (University of Hagen, n.d.) began supporting distance education.

The ‘Distance Learning’ term was first used in the 1970s and became more popular in 1982 after the International Council for Correspondence Education changed its name to the International Council for Distance Education (Holmberg, 2005). The organization is currently known as the International Council for Open Distance Education (ICDE) and promotes open, flexible, distance and online education. The 21st Century witnessed substantial changes and key prominence of distance learning in international forums. Definitions of distance education vary depending on the time during which it is defined. Michael Moore (1973), for instance, defined it as “The family of instructional methods in which the teaching behaviors are executed apart from the learning behaviors, includes those that in contiguous situations would be performed in the learner’s presence, so that communication between teacher and the learner must be facilitated by print, electronics, mechanical or other devices.” (p. 664). However, we notice that a recent definition of distance learning by Kaplan and Haenlein (2016, p. 441) which states, “… providing education to students who are separated by distance and in which the pedagogical material is planned and prepared by educational institutions" agrees with using technology to offer education to students at distance.

Distance education emerged from the high demands of delivering education to those who cannot engage in face-to-face courses. Experts in the field have related Information Technology (IT) to the emergence of distance education (Niper, 1989; Guri-Rosenblit, 2005). Distance education is also often adopted with open learning, whilst others use distance learning as an alternative term to e-Learning. However, Sangrà and his colleagues in their paper "Building an inclusive definition of e-learning: An approach to the conceptual framework" stated that both e-Learning and distance learning contribute to the use of ICT for educational purposes, but neither one equals the other (Sangrà, Vlachopoulos & Cabrera, 2012). Sarah Guri-Rosenblit (2005) has also mentioned that scholars use both terms "interchangeably as synonyms" which creates misunderstood thoughts. She further added that "distance education in most higher education systems is not delivered through new electronic media and vice versa." Her conclusive claims depended on a 2002 study by the US. Department of Education in which they found that 85% of students use e-learning forms, while only 7.6% of undergraduates take some distance learning courses (Guri-Rosenblit, 2005). Based on previous literature, distance education and e-Learning overlap in use and definition. Nevertheless, through this doctoral dissertation, we believe distance education and e-Learning (including subcategories like TEL, CSCL, etc.) would hardly emerge without the help of technology and the new educational paradigms. Later, this thesis will employ "online learning" as an essence of the distance learning concept.

## 3.2 Open Educational Resources (OER)

While distance learning was developing and shifting firmly in higher education institutes, it suffered from the limited number of served scholars because of production, reproduction, communication and distribution costs (Caswell et al., 2008). Production of Courseware requires funding capital allocated for teachers, materials, copyrights and distribution to finally reach students at colleges and universities. However, ICT, Internet, telecommunications, and technology have reduced costs of reproduction and sharing of educational materials.

Nowadays, there are many non-profit and international organizations that call for the reaches of education to extend to everyone, including both genders and children. The United Nations

Educational, Scientific and Cultural Organization (UNESCO) launched the Education for All (EFA) movement that demands governments provide quality basic education to adults, children, and youth (UNESCO, 2000). Following these examples, Open Educational Resources (OER) were firstly coined at the Forum on Impact of Open Courseware for Higher Education in Developing Countries by UNESCO (2002) in which the OER movement calls for expanding education through openness and the reuse of educational resources. UNESCO has defined it as:

> Teaching, learning, and research materials in any medium, digital or otherwise, that reside in the public domain or have been released under an open license that permits no-cost access, use, adaptation and redistribution by others with no or limited restrictions. (UNESCO, 2002)

Besides that, other organizations have provided more OER definitions; however, most of these definitions agree on the general right to access, adapt and share. Table 1 shows what is permissible and what is unallowable according to each organization's rules.

**Table 1.** OER popular organization definitions allocated rights (Creative Commons, 2016b)

| | Open copyright license required | Right of access, adaptation, and republication | Non-discriminatory (rights given to everyone, everywhere) | Does not include non-commercial limitations |
|---|---|---|---|---|
| **Hewlett Foundation** | ✓ | ✓ | ✓ | ✓ |
| **OECD** | | ✓ | | |
| **UNESCO** | ✓ | ✓ | ✓ | ✓ |
| **Cape Town Declaration** | ✓ | ✓ | ✓ | |
| **Wikieducator OER** | | ✓ | ✓ | ✓ |
| **OER Commons** | | ✓ | ✓ | ✓ |

Guntram Geser (2007) from the Salzburg research center summarized OER potential benefits to the community from the viewpoint of teachers and students. The list was in the context of the first European funded project on OER called the Open e-Learning Content Observatory Services (OCLOS) and included:

- OER offers a broader range of subjects to select from, and allows for extra flexibility in choosing material for learning and teaching.
- OER saves time and efforts through reusing of resources.
- OER allows for employing teachers in leveraging the educational value of resources through their own assessments and personal feedback.
- OER promotes collaborative learning environments through the Web 2.0 features like blogs and Wikis.
- And finally, OER enhances lifelong learning and open education.

With the emergent technologies and tools in distance learning and the rising momentum of the OER movement across higher education institutes in the early 2000s, the philosophy of openness materialized like never before. As a result, initiatives like Open CourseWare (OCW) acted as enablers to the achievement of universal and open education (Caswell et al., 2008). Unlike classical distance learning programs which charge students and are very limited to a certain number of participants, OCW offers free education to everyone through online courses. The United States Massachusetts Institute of Technology announced a program called MIT OCW targeting free available content by 2012 over the Internet (MIT, 2001). The program aims to encourage institutions around the world to directly use MIT OCW materials for curriculum development, provides learners with supplementary materials and serves as a big database of information and multidisciplinary ventures. Following MIT's pioneering action, other universities like Delft University of Technology launched their own Open CourseWare targeted to publish high-quality educational content online at no cost (Hennis, 2008). Many OER and Open CourseWare are currently governed under organization initiatives of creative work like Creative Commons.

### 3.2.1 Creative Commons

While OER offers great benefits as previously described, some countries show particularly slow growth in relation to openness and production of rights. For instance, the OER movement in the German-speaking countries (Austria, Germany, and Switzerland) faces very strict laws of copyrights and privacy. Arnold and her co-authors described a case in Germany where using photos from an older textbook with a projector in universities is considered a breach of rights (Arnold et al., 2013). The authors discussed some solutions like using the Creative Commons (https://creativecommons.org/, last accessed: 03.November.2016) licenses which gives a clear legal description and provides an alternative answer to legal challenges.

Creative Commons defines licensing of materials use (including educational resources) based on a spectrum of possibilities between completely free, all rights reserved, full copyright and no rights reserved. Figure 6 depicts the Creative Commons license ranks scale. Usually these ranks are represented as symbols placed in images, videos, and websites. For instance, CC0 indicates full open and no rights reserved, which enables users to freely use, modify and share without any restrictions. CC BY-NC-ND only gives limited usage of sharing, but without remixing or building upon material.

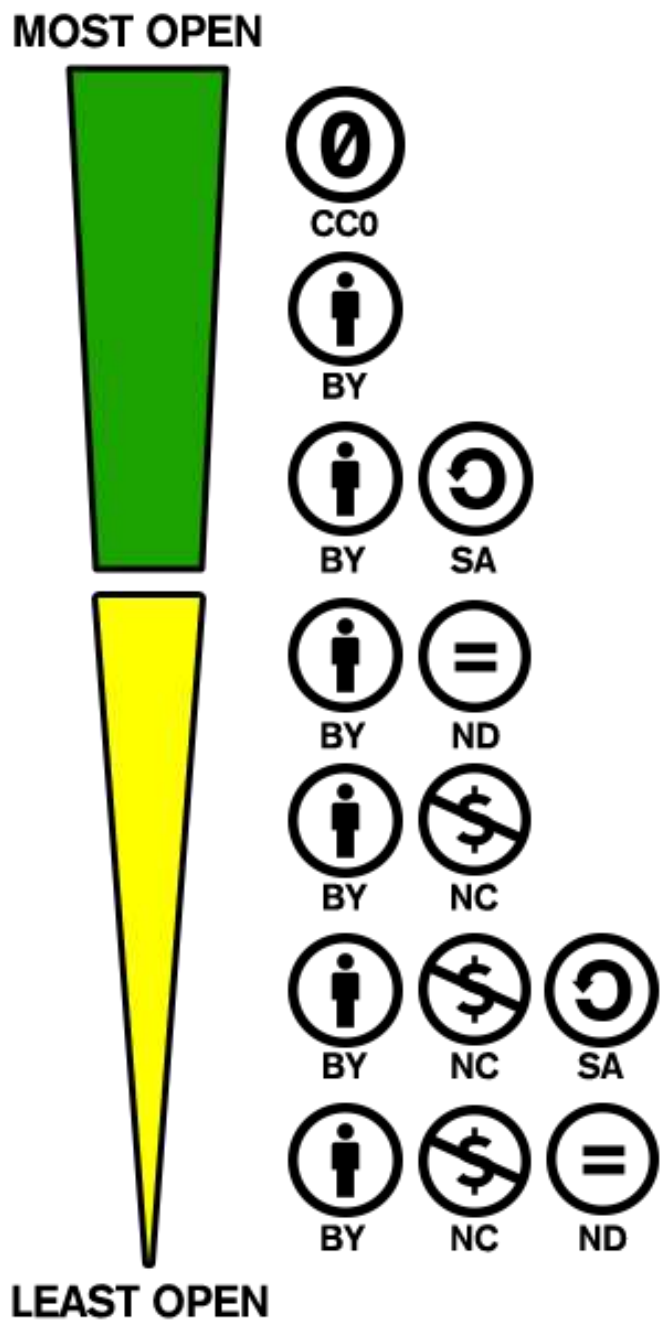

**Figure 6.** Creative Commons licenses ranked from the most open to the least open (Creative Commons, 2016a)

Distance learning educational materials in open platforms like OCW, or later on MOOCs, are not only available online for free but are also possible to share, remix and even reuse under licenses like Creative Commons.

## 3.3 Massive Open Online Courses (MOOCs)

The openness of the distance learning movement has gained impetus over the last couple of years. Programs like Open CourseWare in some respect have pushed this movement forward. A short while after the UNESCO’s adoption of OER, the educational technology community witnessed the start of a new era in courses which are massive in terms of student numbers, open for all and are available online. These types of courses are called “Massive Open Online Courses” or more commonly by the abbreviation “MOOCs.” The term MOOC was first coined in 2008 by David Cormier and George Siemens, describing a twelve-week course on connectivism and connected knowledge at the University of Manitoba, Canada (Cormier & Siemens, 2010). The course material was designed by Downes and Siemens and had 25 enrollees who registered and paid for university credit points. Later, the course was opened to the public and attracted over 2,300 participants. Such a large number of enrollees created a significant course experience and garnered large-scale attention. MOOC is an acronym for:

- **Massive**: Refers to large in scale, amount or degree, in which the number of participants exceeds the so-called Dunbar's number (Wedekind, 2013). The Dunbar number describes the cognitive limit to the number of social relationships with other people (Dunbar, 2010). Massive indicates that enrollees are much larger than regular classes where the number of participants exceeds hundreds to thousands of participants.
- **Open**: The openness of MOOCs usually refers to the free access to online courses and learning materials. The course’s curriculum, assessment, and information should be open (Rodriguez, 2012). Learners can participate in a course without fulfillment of other formal requirements or other additional restrictions. Everyone can enroll without prerequisites. Thus, learners can access the courses and the education materials whenever and wherever they like.
- **Online**: The management, the information system, as well as the course itself are exclusively online. Communication between the course participants and the learning contents takes place via a specially accredited course that is available online and introduced on a web page, for instance (Wedekind, 2013). Likewise, physical attendance is nonexistent, and all classes are dealt remotely.

- **Course**: The course can be summarized as a collection of learning materials that are introduced by teachers in the form of a program. These courses usually have a predetermined start date and end date. Courses can be taught by more than one teacher according to the content itself and the online course provider (Wedekind, 2013).

McAuley, Stewart, Siemens, & Cormier (2010) clearly defined MOOCs as:

> An online course with the option of free and open registration, a publicly shared curriculum, and open-ended outcomes. MOOCs integrate social networking, accessible online resources, and are facilitated by leading practitioners in the field of study. Most significantly, MOOCs build on the engagement of learners who self-organize their participation according to learning goals, prior knowledge and skills, and common interests.

The results of MOOCs depend on different perspectives. For example, in higher education, institutions were looking forward to improving pedagogical and educational concepts by providing high-quality teaching principles and saving costs of university level education. This can happen when an instructor has thousands of students who attend a hypothetical class instead of a real room which cannot handle groups of more than a hundred learners. On the other hand, education reformers see a glimmer of hope in the Internet-based models, like MOOCs, which help more students to earn college degrees or certificates at a lower cost to themselves, their families, and the government (Quinton, 2013). The MOOCs' objectives thus varied between saving costs and increasing revenues, improving educational outcomes, extending reachability as well as accessibility of learning material to everyone (Hollands & Tirthali, 2014) and also providing support for the Open Educational Resources (Arnold et al., 2015).

### 3.3.1 MOOC Types

On the Web, there are a variety of MOOC types available from different providers. Siemens distinguishes, for example between cMOOCs, xMOOCs and quasi-MOOCs (Siemens, 2013a). The cMOOCs concept was developed by George Siemens and Stephan Downes based on the philosophy of connectivism. The idea of cMOOCs concerns itself with knowledge and knowledge construction by self-organized networks (Wedekind, 2013). The "c" in the cMOOCs comes from the roots of the underlying learning theory of connectivism (Siemens, 2006). cMOOCs are based on phases of an iterative process "Aggregate, Remix, Repurpose & Feed Forward" (Mackness et al., 2013). Through this process, the learners in cMOOCs produce and reflect their content and share their new knowledge. Moreover, the learning process is generated with the help of learners themselves (Robes, 2012).

In contrast, xMOOC is an online mass course with a strongly predetermined learning path, communication tools and assignments (Wedekind, 2013). The prefix "x" finds its origin afforded by the famous universities such as Harvard and Stanford and serves as the abbreviation of “extended” (Downes, 2013). Online platform providers started to distribute additional information, learning resources and activities to lectures, which made these courses open and easily accessible by general users (O’Toole, 2013). Unlike cMOOCs, which focus on distributing information on networks, xMOOCs are based on the traditional instruction-driven principle. Information is made available via an online learning platform for a large group of students (Rodriguez, 2012). The study by Langer & Thillosen (2013) reveals that the main tool for distributing information in xMOOCs is done by video sequences. These often follow the model of traditional lectures. Moreover, xMOOCs offer multiple-choice questions, asynchronous discussion forums and work with essays (Langer & Thillosen, 2013).

Stacey (2014) argued that MOOC pedagogy is boring and not interactive unless the online pedagogies are open, connections between the elements of MOOCs which are learners, instructors and context are open on the web, and online learning happens when students are involved in blogs, discussion forums, and group assignments. xMOOC providers propose badges or certificates to students who successfully complete courses as a type of encouragement or extrinsic motivation. The first real massive open online course (xMOOC) by Sebastian Thrun and his colleagues attracted over 160,000 participants from all continents (Yuan & Powell, 2013), and the story of magnetizing more participants continues with the ongoing MOOC providers. As an example, a team from Harvard and Massachusetts Institute of Technology released their research study on the Harvardx and MITx MOOC platform (edX) in which they examined 1.1 billion logged events of 1.7 million students (Ho et al., 2015). It is a logical development for each MOOC platform to seek influence, achieve popularity and also to attract as many participants as possible (Khalil, Brunner, & Ebner, 2015).

On the other hand, quasi-MOOCs are a loose collection of web-based tutorials or Open Educational Resources (OER) elements. These have neither an interaction as in cMOOCs, nor an instruction-driven curriculum as in xMOOCs (Siemens, 2013a). There are obvious common areas among the three types of MOOCs. Figure 7 shows a scheme covering intersection points between the listed types of Massive Open Online Courses: a) xMOOCs, b) cMOOCs and c) the Quasi-MOOCs.

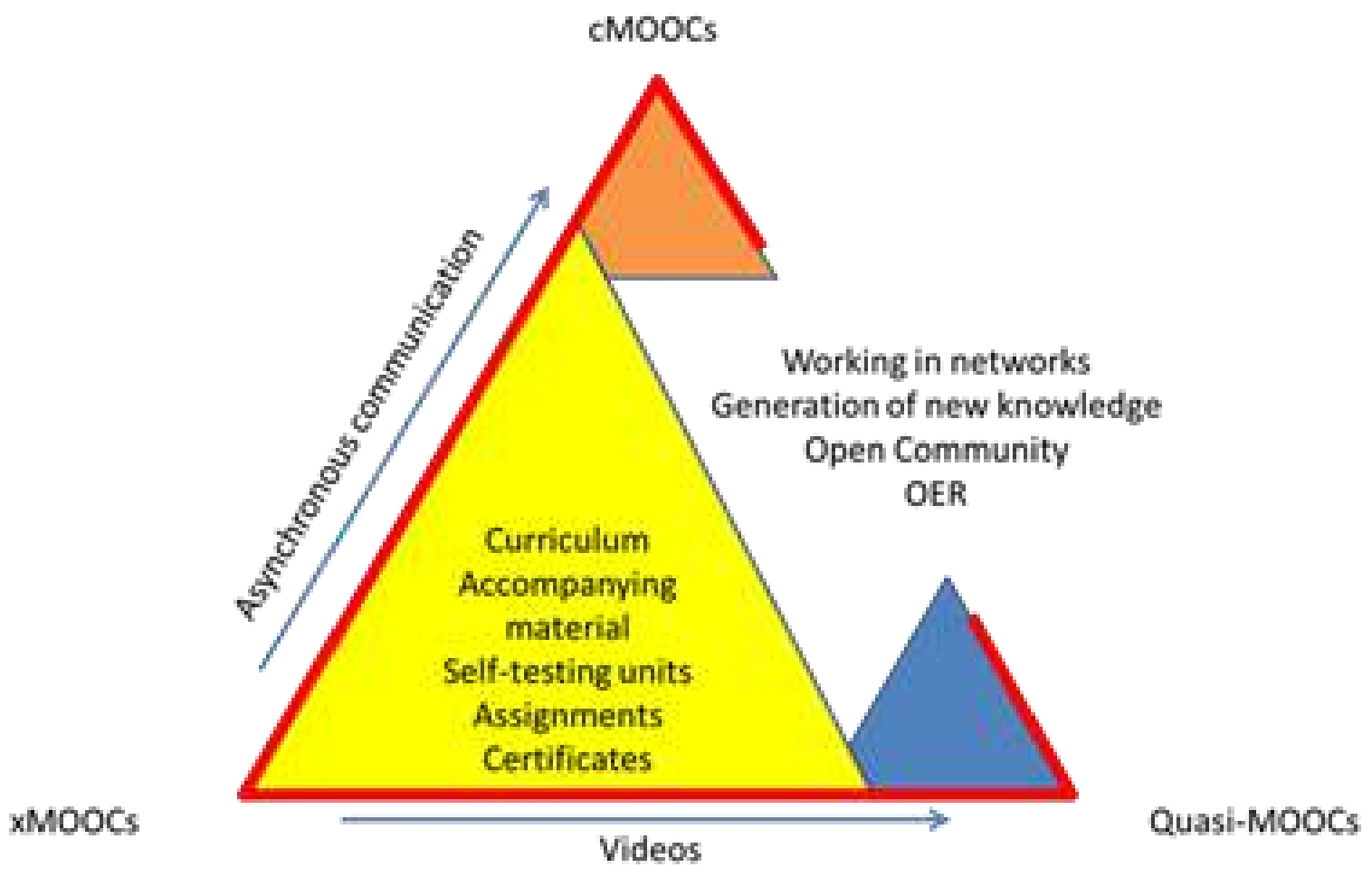

**Figure 7.** Intersection points between cMOOCs, xMOOCs and Quasi-MOOCs (Brunner, 2014)

The criteria "asynchronous communications" can be achieved in cMOOCs according to its definition, which encourages the learners to share and reflect their learning content (Baker, 2012). In accordance with the studies by (Robes, 2012; Wedekind, 2013), cMOOCs can have a high score of interaction criteria. Quasi-MOOCs are courses which are authored by non-certified authors (Siemens, 2013a). Therefore, Quasi-MOOCs lack asynchronous communication and interaction.

The pedagogical approach in MOOCs mainly consists of learning and teaching exchange with the combination of watching videos, downloading course materials, completing quizzes, and turning in assignments as well as participating in the social discussion forums between the learners themselves and the learners with the course instructor(s). Taking a deeper look at the pedagogical approaches of MOOCs, Anderson & Dron (2011) explained that distance learning pedagogical models are classified as connectivist, cognitive-behaviorist and social-constructivist. Rodriguez (2012) postulated that cMOOCs belong to connectivism, which depends on building networks of information, and xMOOCs belong to the cognitive-behaviorist model wherein guided learning and providing feedback are acquired. However, it is summarized that each MOOC depends on (1) learners, and those who register in a MOOC platform and then enroll in one of the courses, (2) instructors and those who appear in video lectures, explaining the materials to the learners and giving assignments, and (3) Context, which includes topics, videos, documents, etc.

### 3.3.2 Evaluation Grid for xMOOCs

The number of MOOCs has steadily increased worldwide since Sebastian Thrun offered his first xMOOC for more than 160,000 students (Yuan & Powell, 2013). Nowadays, decision makers and students, as well as lecturers are inquiring as to the quality of such courses. In this section, we describe our results of a live experiment we did on 15 randomly chosen MOOCs. The experiment was developed through a master thesis research project (Brunner, 2014). At the end of this section, we present an evaluation grid for xMOOCs that helped us in:

- Providing a new grounded categorization of the evaluation criteria of MOOCs
- Evaluating 15 xMOOCs from 12 MOOC providers

#### 3.3.2.1 Fundamental Elements of xMOOCs

In our study, we strongly concentrated on investigating xMOOCs rather than cMOOCs. Therefore, we considered an appropriate observation on the 15 xMOOCs based on various references and extracted the following crucial elements for further research in details: curriculum, videos, self-testing units, accompanying material, asynchronous communication, assignments, certificates, and technical implementation. The following shows a short description of each element:

**Curriculum**

Most of xMOOCs are offered as multi-week courses. The typical duration is from 6 to 12 weeks (Sharples et al., 2012). The curriculum is mainly introduced in weekly intervals (Khalil & Ebner, 2013; Wedekind, 2013). Within a boundary-timing curriculum, the concentration among participants increases rapidly (Sharples et al., 2012).

**Videos**

The most common way of transmitting information to students is through lecture videos. In addition to videos, short trailers play the part of marketing the courses. This can be seen across different MOOCs providers.

**Self-testing Units**

Fundamental components of xMOOCs are quizzes and multiple-choice tests (Lipson, 2013). These elements are referred to as self-test units (Kerres, & Preußler, 2013). Some courses tend to provide frequent quizzes after a predetermined set of information units or after completing the whole course. To enhance the social element in xMOOCs, some courses offer exchanging the quiz information and discussing the quiz material among the discussion forums (Bremer & Thillosen, 2013).

**Accompanying Material**

In addition to the video lectures, MOOC organizers offer supplementary and accompanying material to achieve voluntary deepening purposes (Wedekind, 2013). Accompanying material can be formed as simple texts, lecture notes, case studies or simple hyperlinks that lead to external resources. Kerres & Preußler confirmed that the additional materials in xMOOC play a critical role and give individuals better support for their learning activities (Kerres & Preußler, 2013).

**Asynchronous Communication**

Information and factual knowledge are well communicated through asynchronous communication (Eppler, 2005). The social structures among MOOC providers are usually similar between each other. For instance, the communication between learners and/or teachers happens in discussion forums (Wedekind, 2013). These discussions are used to clarify questions regarding the content of a MOOC (Khalil & Ebner, 2013). Learners feel the positive effect when they experience the cooperation between the tutor and themselves.

**Assignments**

There are different methods to assess the performance of participants in xMOOCs. The learners process tasks on a weekly basis, which are commonly referred to as assignments (Wedekind, 2013). Different types of assessments are available for xMOOCs: a) Automatic assessment: this is an automated process of evaluating quizzes provided by items such as multiple-choice tests, b) Self-assessments: here, the students evaluate themselves and assess each other as to whether they achieved the course goals, c) Peer-assessment: here, students evaluate each other in small groups and provide feedback on their experience (Kerres & Preußler, 2013).

**Certificates**

After achieving the minimum number of points required to pass a course, students can pay to get certificates. A certificate is an extrinsic motivator for many course participants. Unlike cMOOCs, where the participants are motivated to extend the collective capabilities of the course's network, participants of xMOOCs are eager to achieve a good score to be able to pursue a badge or a certificate (O'Toole, 2013). Badges can be used as a proof of performance. These were firstly introduced in order to satisfy the demand for certificates in cMOOCs. Badges can be used as a feedback instrument too (Hodges, 2004). However, in this evaluation, we only considered certificates because the research study in this section was conducted in the period of 2013-2014 and most MOOC platforms did not offer badges at that time.

**Technical Implementation**

There are some requirements that xMOOCs must meet with respect to the technical implementation of xMOOCs, such as: quiz functionality, navigation based on the weekly courses principle, powerful search function in the discussion forums, availability of social media components, assurance of video accessibility during peak hours, as well as the representation of learning progress and the generation of certificates (Meinel & Willems, 2013). In general, these technical principles should be fully working in any xMOOCs environment.

#### 3.3.2.2 The Research Design of the Evaluation

In this study, the fundamental elements of xMOOCs are refined. Through our active observations of different xMOOCs from multiple providers, characteristics of the basic instructional elements of xMOOCs are determined. We divided them into subcategories and presented criteria in order to clarify the main categories which exist in xMOOCs. Different types of methodologies were used to reveal and evaluate the upcoming results. These methods are qualitative content analysis, personal observations, document analysis, and experts' opinion. In a period of three months, an observation of 15 courses from 12 xMOOCs providers was carried out. We applied different types of criteria for examination purposes. Finally, each criterion was scaled from 1 to 5 according to the didactic dimension-model by Baumgartner (Baumgartner, 2009): (a) Grade "1": Very clear,

(b) Grade “2”: Clear, (c) Grade “3”: Sufficient, (d) Grade “4”: Unclear, (e) Grade “5”: Does not exist.

**Document Analysis**

We followed Punch’s (1998) methodology of qualitative document analysis. We examined the relevant contents of xMOOCs. xMOOC tutors provide instructions to the learners in different ways such as emails with different instructions, motivational information, and reports before, during and after the course kicks off. During courses, assignments and general discussions are exchanged using documents, forums, and video comments. This data sheds light on entries made by learners about the course subject and the learning environment in general. Contents of forum threads and the main difficulties learners post in discussions were tracked. Furthermore, analyses on the forum’s topics and their frequency have been performed.

**Observations**

As we acted as test learners, a total of 15 xMOOCs were observed. Through a brief look at the learning contents, participation in forum discussions, as well as resolved assignments and tasks, all steps and difficulties faced during the observation were documented. Additionally, we assessed the interactions and discussions between teachers and students in the learning environment atmosphere. Passive participation and observations of similar courses were compared together to enhance the results akin to the study by (Nonnecke & Preece, 2001).

**Data Collection**

Data collection was based on the role of a student in 15 xMOOCs. We watched video lectures, observed discussion forums and resolved quizzes. Within the personal participation, we compared the workload and the provided learning content. We posted in forums and recorded the response time from students, teachers and teacher’s assistants. Additionally, we looked into visualizations and progress in the courses. The time and efforts needed for quizzes, and how hard they were to solve, have been all documented. We distinguished between courses in respect to time boundaries and participation rate. Table 2 shows the evaluated xMOOCs of this study.

**Table 2.** The analyzed xMOOCs and their providers in the evaluation grid study (Brunner, 2014)

| Course Provider | Course Name |
|---|---|
| Moodle | Learn Moodle |
| Coursera | Learn to Program: The Fundamentals |
| Coursera | Foundations of Virtual Instruction |
| OpenCourseWorld | Learn how to lead |
| Udacitiy | Introduction to Computer Science |
| edX | Introduction to Biology - The Secret of Life |
| Canvas Network | College Foundations: Reading, Writing, and Math |
| Canvas Network | Exploring Engineering |
| openHPI | In-Memory Data Management 2013 |
| NovoEd | Technology Entrepreneurship Part 1 |
| Open2Study | Concepts in Game Development |
| Standford OpenEdX | SciWrite: Writing in the Sciences |
| Standford OpenEdX | Solar: Solar Cells, Fuel Cells and Batteries |
| Waikato University | Data Mining with Weka |
| University of Amsterdam | Introduction to Communication Science |

#### 3.3.2.3 Findings

During our research studies on xMOOCs, we became familiar with the fundamental elements of the provided courses. This study comes up with different criteria to differentiate between courses. We considered the common issues of xMOOCs and cMOOCs to classify our criteria into categories and subcategories (Schulmeister, 2013).

### Categories

In Table 3, we list the main categories of evaluating xMOOCs, subcategories, and the criteria. We categorized xMOOCs into three main parts according to literature study and our observations. It is noteworthy to mention that learner satisfaction and dropout rate are not listed in our categories.

**Table 3.** Categories, subcategories and criteria of xMOOCs (Brunner, 2014)

| Category | Subcategory | Criteria |
|---|---|---|
| System | General | Course content, conditions of participations, certificates |
| | Information | Requirements, target group, learning objective, workload |
| | User Interface | Course clarity, availability and durability forums searching feature |
| Interaction | Nature of Information | Emails prior to the course, emails during the course |
| | Interactivity | Interactive elements, forums activity, replies intensity, course activity, motivations |
| | Asynchronous communication | Average response time, teachers' reply and assistance, invitations |
| Contents | Media elements | Video duration, scripts and documents, download feature |
| | Evaluation | Self-study plans, self-assessment, quiz level, transparency, assignments level, learning strategy, learning experience integration |

### Knockout Criteria

Through our survey on several xMOOCs, we discovered that sometimes a criterion was specified as more important than other criteria. For instance, on a course web page, the definition of the target group is essential for each xMOOC. Each xMOOC has to identify the target students who will take part in the course (Wedekind, 2013). Other factors like a clear learning objective (Bremer & Thillosen, 2013) and a well-planned workload were considered as serious criteria. Likewise, we recognized the factors of course clarity (Meinel & Willems, 2013), assistance to students who need help (Bremer & Thillosen, 2013; Langer & Thillosen, 2013; Schulmeister, 2013), and quick feedback (Wedekind, 2013). Finally, a criterion of providing an environment inclusive of self-study medium such as self-assessment (O'Toole, 2013) and making teaching materials like scripts and documents available anytime are very relevant when doing a MOOC evaluation (Kerres & Preußler, 2013). Figure 8 depicts an overview of the knockout criteria of xMOOCs evaluation.

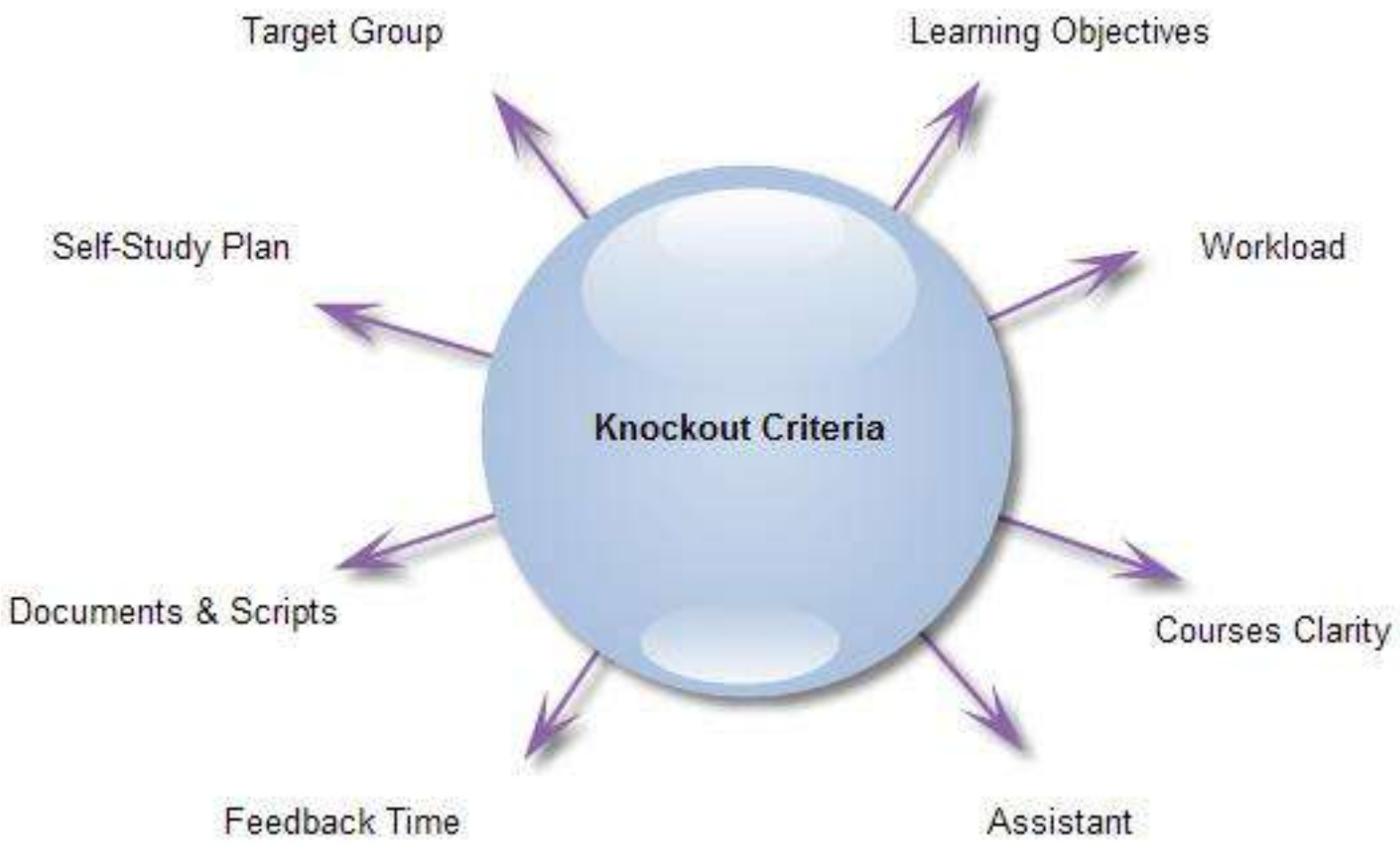

**Figure 8.** An overview of the knockout criteria of xMOOCs evaluation

### The Evaluation Grid

Based on the criteria we extracted through our observations, we present an evaluation grid that summarizes every MOOC criterion weight (see Table 4). This evaluation grid supported us in navigating xMOOCs, creating a background of xMOOCs' most important principles, and identifying strengths and weaknesses of xMOOCs.

In the end, the outcome list is still not immutable to human judgment errors, but it can be seen as a proposed structure to evaluate MOOCs in a non-systematic way.

**Table 4.** The xMOOCs evaluation grid (Brunner, 2014)

| Criterion | MD_Learn | CO_FUPRO | CO_FOVIR | OC_LLEAD | UD_INIBI | edX_INTBI | CN_COLRD | CN_EXENG | OH_DATMG | NE_TECENT | O2_SCIWR | OX_SOLCE | OX_SOLCE | WA_DAMIN | UA_INCSI |
|---|---|---|---|---|---|---|---|---|---|---|---|---|---|---|---|
| Course content | 1 | 1 | 2 | 1 | 2 | 1 | 1 | 1 | 1 | 1 | 1 | 1 | 1 | 1 | 1 |
| Condition of participation | 1 | 1 | 1 | 2 | 1 | 1 | 1 | 1 | 1 | 1 | 1 | 1 | 1 | 1 | 1 |
| Certificates | 3 | 5 | 2 | 2 | 3 | 2 | 2 | 4 | 2 | 5 | 2 | 2 | 5 | 5 | 2 |
| Requirements | 3 | 1 | 2 | 1 | 1 | 2 | 1 | 1 | 2 | 1 | 2 | 2 | 1 | 2 | 2 |
| Target group | 2 | 2 | 1 | 3 | 1 | 2 | 1 | 1 | 4 | 1 | 2 | 2 | 4 | 3 | 3 |
| Learning objective | 1 | 1 | 1 | 2 | 2 | 1 | 1 | 2 | 3 | 2 | 1 | 2 | 3 | 1 | 2 |
| Workload | 3 | 2 | 5 | 1 | 3 | 2 | 3 | 3 | 5 | 2 | 2 | 2 | 5 | 5 | 1 |
| Course clarity | 3 | 1 | 2 | 2 | 1 | 1 | 2 | 1 | 1 | 3 | 2 | 2 | 2 | 2 | 2 |
| Availability | 2 | 1 | 2 | 4 | 2 | 1 | 2 | 1 | 2 | 1 | 2 | 2 | 2 | 2 | 2 |
| Forums search | 4 | 1 | 1 | 5 | 2 | 2 | 1 | 1 | 1 | 1 | 3 | 3 | 3 | 1 | 4 |
| Emails prior to course | 1 | 2 | 3 | 4 | 4 | 1 | 4 | 2 | 4 | 3 | 2 | 3 | 4 | 4 | 4 |
| Email during course | 1 | 1 | 2 | 5 | 4 | 4 | 5 | 1 | 2 | 3 | 3 | 1 | 4 | 2 | 4 |
| Interactive elements | 1 | 1 | 5 | 5 | 1 | 1 | 3 | 1 | 5 | 2 | 5 | 3 | 3 | 5 | 5 |
| Forum activity | 1 | 1 | 1 | 5 | 4 | 1 | 3 | 1 | 1 | 1 | 3 | 3 | 3 | 1 | 3 |
| Teachers assistance | 1 | 1 | 2 | 5 | 2 | 1 | 2 | 1 | 1 | 2 | 1 | 3 | 3 | 1 | 1 |
| Average response time | 1 | 1 | 1 | 5 | 3 | 1 | 3 | 2 | 3 | 1 | 3 | 2 | 2 | 2 | 2 |
| Replies intensity | 1 | 1 | 2 | 5 | 3 | 1 | 3 | 2 | 1 | 2 | 3 | 3 | 3 | 1 | 1 |
| Course activity | 1 | 1 | 1 | 2 | 3 | 3 | 2 | 1 | 2 | 1 | 2 | 2 | 2 | 1 | 2 |
| Invitations | 1 | 2 | 5 | 2 | 2 | 2 | 5 | 2 | 5 | 5 | 5 | 5 | 5 | 3 | 2 |
| Motivation | 2 | 1 | 2 | 3 | 4 | 2 | 3 | 2 | 2 | 2 | 3 | 3 | 3 | 2 | 2 |
| Video duration | 1 | 2 | 3 | 2 | 1 | 4 | 1 | 1 | 3 | 3 | 1 | 2 | 1 | 2 | 1 |
| Documents and scripts | 4 | 2 | 2 | 1 | 1 | 1 | 3 | 3 | 1 | 1 | 1 | 3 | 3 | 1 | 1 |
| Download feature | 2 | 1 | 5 | 2 | 2 | 1 | 2 | 2 | 2 | 2 | 2 | 1 | 1 | 1 | 2 |
| Self-assessment | 3 | 3 | 3 | 3 | 3 | 3 | 2 | 3 | 5 | 5 | 5 | 2 | 3 | 3 | 5 |
| Self-study plan | 4 | 1 | 1 | 1 | 1 | 1 | 1 | 1 | 1 | 3 | 1 | 1 | 1 | 1 | 3 |
| Quizzes level | 3 | 1 | 1 | 1 | 3 | 1 | 1 | 3 | 1 | 1 | 2 | 1 | 1 | 3 | 3 |
| Assignments level | 5 | 1 | 1 | 5 | 1 | 2 | 1 | 2 | 1 | 1 | 2 | 1 | 1 | 3 | 3 |
| Learning strategy | 3 | 3 | 2 | 2 | 2 | 1 | 1 | 1 | 1 | 1 | 2 | 2 | 1 | 2 | 2 |
| Transparency | 5 | 1 | 1 | 3 | 5 | 2 | 2 | 3 | 1 | 3 | 3 | 3 | 3 | 5 | 3 |
| Learning exp. integration | 1 | 2 | 1 | 2 | 3 | 1 | 2 | 1 | 2 | 2 | 2 | 2 | 2 | 1 | 1 |

## 3.4 Learning Analytics

Within the past few years, technology and the availability of the Internet have evolved so rapidly that the world of information has changed. Education has participated in this revolution and disciplines like e-learning, distance learning, and OER, etc. have made robust appearances. Learning can be considered as a product of interaction (Elias, 2011). Learners interact with teachers, educational resources, and their colleagues. Although the interaction between the teacher and the student can assess the quality of a course through evaluation or analysis of grades and completion rates, there are many questions that such interactions or interviews, surveys, and open discussions cannot answer.

Nearly all educational resources and higher education institutions, especially those with distance learning approaches, have moved online. Likewise, interactions have also been shifted online. Usually online systems are computer-mediated, and as a result, almost all interactions are recorded into what is called log files. For instance, through log files analysis in LMS or MOOC environments, we can track students who log in, who posted in a forum, who dropped out and even who watched a video.

Substantial technological advancements significantly redefined the domain of education. The majority of educational institutions adopted technologies to increase the quality of their teaching activities. With growth in the use of educational technology, researchers, practitioners, and administrators recognized huge potentials of the vast amounts of data produced and collected by these systems (Siemens, 2013b). Some of their potentials include increasing student retention and improving the admission process, improving student learning outcomes, personalizing learning and instruction, and improving the overall student learning experience. The broad potentials of the use of data led to the formation of the fields of educational data mining (EDM) (Romero & Ventura, 2010), academic analytics (Campbell, deBlois & Oblinger, 2007), and learning analytics.

Although they all share many of the same methods, techniques, and even goals, the distinctive feature of learning analytics is a holistic perspective to learning and environments in which learning happens (Baker & Inventado, 2014). As a consequence of the interest to improve courses, learning environments, teaching, and learning, learning analytics has witnessed unprecedented growth since 2011. The field mainly seeks to optimize learning and the environments in which it happens. Despite that learning analytics is a young field, it has strong connections to web analytics, EDM and academic analytics.

### 3.4.1 Deriving Learning Analytics

As previously stated, learning analytics is derived from related fields of EDM, academic analytics, and web analytics. Also, some other researchers in the field relate it to information visualization (Duval & Verbert, 2012). Duval and Verbert (2012) stated that in the domain of learning analytics, there are vast data repositories that contain traces of where people go and with whom they interact. Analytical applications then try to translate such interactions through data mining or information visualization.

**Academic Analytics**

Academic analytics was first used by Goldstein and Katz (2005). The authors stated that they used the "academic analytics" term to describe the intersection between information technology and business intelligence to manage the academic enterprise. Campbell and Oblinger (2007) described the benefits of academic analytics as it can help institutions to address student success and accountability. They added that academic analytics utilizes the generated data from academic systems with statistical techniques and predictive modeling to help stakeholders of academies succeed in their interventions (Campbell & Oblinger, 2007). Academic analytics are situated more at the level of higher education; learning analytics are more into translating the benefits of analysis to learning and teaching in every educational setting.

**Educational Data Mining (EDM)**

The EDM website (www.educationaldatamining.org) defines EDM thusly: "Educational Data Mining is an emerging discipline, concerned with developing methods for exploring the unique types of data that come from educational settings, and using those methods to better understand students, and the settings which they learn in." Learning analytics and EDM are alike in respect to analysis domain and objectives. Both fields are concerned with empowering learners and optimizing learning environments. However, EDM is more into using tools and techniques from data mining and machine learning disciplines (Baker & Yacef, 2009); specifically, clustering, decision trees, prediction models, artificial intelligence and classification (Baker & Yacef, 2009; Romero & Ventura, 2007; Romero & Ventura, 2010). Learning analytics draws on research and methods from a diverse set of disciplines including business intelligence, machine learning, data mining, statistics, online information science, educational psychology, education, information visualization and learning sciences (Gašević, Dawson & Siemens, 2015; Khalil & Ebner, 2016d).

There are some fruitful surveys and review studies that provide a holistic background of EDM such as the study by Romero and Ventura (2007; 2010) and Baker and Yacef (Baker & Yacef, 2009).

**Learning Analytics Definitions**

Learning analytics has a history of several definitions. Siemens (2010) defined it as “the use of intelligent data, learner-produced data, and analysis models to discover information and social connections, and to predict and advise on learning.” Elias (2011) has another definition, and described it as:

> Learning analytics is an emerging field in which sophisticated analytic tools are used to improve learning and education. It draws from, and is closely tied to, a series of other fields of study including business intelligence, web analytics, academic analytics, educational data mining, and action analytics. (p. 2)

Nevertheless, the Society for Learning Analytics Research (SoLAR) has adopted a final description of learning analytics. SoLAR followed the first Learning Analytics and Knowledge conference (LAK ’11) definition in which it was defined as “the measurement, collection, analysis and reporting of data about learners and their contexts, for purposes of understanding and optimizing learning and the environments in which it occurs” (LAK11, para. 44; SoLAR, 2013). This definition highlights two intertwined goals of learning analytics: a) from a theoretical perspective, to use the data to increase *understanding* of learning processes, and b) from the practical perspective, to use the data to *act on and intervene* in an individual’s learning, helping them to achieve defined learning outcomes and improve their overall learning experience. This makes learning analytics distinctive from, for example, academic analytics which primarily focuses on the use of data to support overall institutional business operations (e.g., increasing retention or improving admission) and business intelligence (e.g., saving costs, increasing revenues).

### 3.4.2 Learning Analytics in Higher Education

Siemens and Long (2011) stated that the aim of learning analytics is to evaluate user behavior in the context of teaching and learning, further, to analyze and interpret it to gain new insights, and to provide the stakeholders with new models for improving teaching, learning, effective organization, and decision making. Some key factors are the return of the resulting knowledge to the teachers and students to optimize their teaching and learning behavior, to promote the development of skills in the area, and to better understand education as well as the connected fields, e.g. university business and marketing. Available resources can be used more efficiently to provide better support and individual care to develop potential.

The combination of higher education and learning analytics has proven to be helpful to colleges and universities in strategic areas such as resource allocation, student success, and finance. These institutions are collecting more and more data than ever before in order to maximize strategic outcomes. Based on key questions, data is analyzed and predictions are made to gain insights and set actions. Many examples of successful analytics and frameworks use are available across a diverse range of institutions (Bichsel, 2012). Ethical and legal issues of collecting and processing student data are seen as barriers by the higher education institutions in learning analytics (Sclater, 2014).

In this part of the literature, we present a state-of-the-art review to evaluate the progress of learning analytics in higher education since its early beginnings in 2011. We conducted the search with three popular libraries: the LAK conference, the SpringerLink (from Springer Publishing), and the Web of Science (from Thomson Reuters) databases.

#### 3.4.2.1 Learning Analytics Background in Higher Education

In the area of higher education, institutions look to a future of uncertainty and change. In addition to national and global as well as political and social changes, competition at the university level is increasing. Higher education needs to increase financial and operational efficiency, expand local and global impact, establish new funding models during a changing economic climate and respond to the demands for greater accountability to ensure organizational success at all levels (van Barneveld, Arnold & Campbell, 2012). Higher education must overcome these external loads in an efficient and dynamic manner, but also understand the needs of the student body, who represents the contributor as well as the donor of this system (Shacklock, 2016).

In addition to the stiff competition, universities have to deal with the rapidly changing technologies that have arisen with the birth of the digital age. In the course of this, institutions collected enormous amounts of relevant data as a byproduct, for instance, when students take an online course, use an Intelligent Tutoring System (ITS), (Arnold & Pistilli, 2012; Bramucci & Gaston, 2012; Fritz, 2011; Santos et al., 2013) play educational games, (Gibson & de Freitas, 2016; Holman, Aguilar & Fishman, 2013; Holman et al., 2015; Westera, Nadolski, & Hummel, 2013) or simply use an online learning platform (Casquero et al., 2016; Wu & Chen, 2013; Santos et al., 2015; Softic et al., 2013).

In recent years, more universities use learning analytics methods in order to obtain findings on the academic progress of students, predict future behaviors and recognize potential problems in the early stages. Further, learning analytics in the context of higher education is an appropriate tool for reflecting the learning behavior of students and providing suitable assistance from teachers or tutors. This individual or group support offers new ways of teaching and provides a way to reflect on the learning behavior of the student. Another motivation behind the use of learning analytics in universities is to improve inter-institutional cooperation, and the development of an agenda for the large community of students and teachers (Atif et al., 2013).

On an international level, the recruitment, management, and retention of students have become high-level priorities for decision makers in institutions of higher education. Especially, improving student retention starts with understanding the reasons behind and/or prediction of the attrition which has become a focus of attention due to the financial losses, lower graduation rates, and inferior school reputation in the eyes of all stakeholders (Delen, 2010; Palmer, 2013).

#### 3.4.2.2 Research Questions and Methodology

This state-of-the-art research aims to elicit an overview of the advancement of the learning analytics field in higher education since its emergence in 2011. The proposed Research Questions (RQ) to answer are:

- **RQ1**: What are the research strands of the learning analytics field in higher education (between January 2011 and February 2016)?

- **RQ2**: What kinds of limitations do the research papers and articles mention?
- **RQ3**: Who are the stakeholders in the domain of learning analytics in higher education, and how should they be categorized?
- **RQ4**: What learning analytics methods have been used in their papers?

To perform this literature review study, we followed the procedure of Machi and McEvoy (2009). Figure 9 displays the six steps for a literature review that have been used in this process.

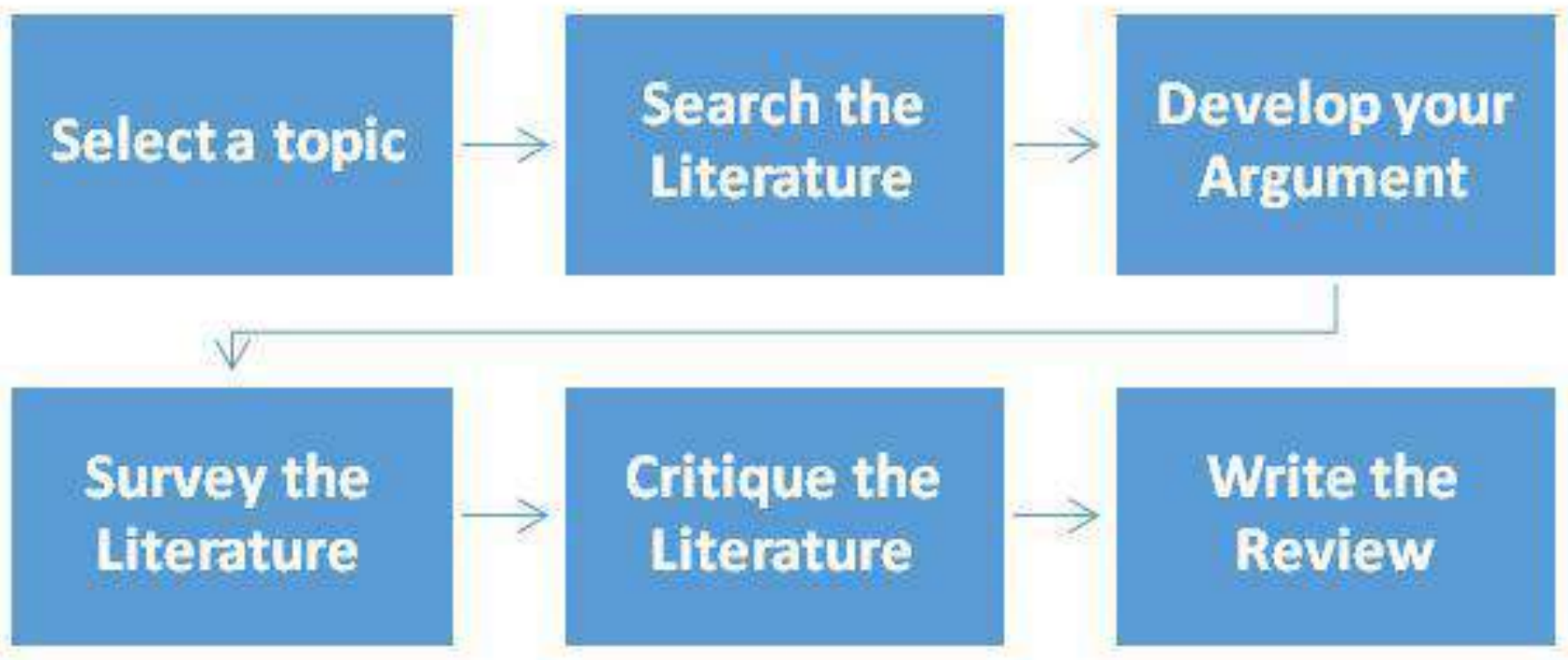

**Figure 9.** Learning analytics in higher education methodology- literature review: Six steps to success. (Machi and McEvoy, 2009)

After we selected our topic, we identified data sources based on their relevance in the computing domain:

- The papers of the LAK conference published in the Association for Computing Machinery (ACM) Digital Library,
- The SpringerLink, and
- The Thomson Reuters Web of Science database

The search parameters in the listed databases were per the following: in the LAK papers, we didn't need to search for the "learning analytics" term because the whole conference covers the learning analytics field. We searched the title, the abstract and the author keywords for "higher education" and/or "university." In the SpringerLink database, we searched for the "learning analytics" term in conjunction with either "higher education" or "university" ("learning analytics" AND "higher education" OR university"). In the Web of Science database, we searched for the topic “learning analytics” in conjunction with either “higher education” or “university” within the research domain of “science technology.”

The defined inclusion criteria of the retrieved papers from the libraries were set to be a) articles should be written in English, and b) published between 2011 till February 2016. We superficially assessed the quality of the reported studies, considering only articles that provided substantial information for learning analytics in higher education. Therefore, we excluded articles that did not meet the outlined inclusion criteria.

Our initial search results found a total of 135 publications (LAK: 65, SpringerLink: 37, Web of Science: 33). During the first stage, the search results were analyzed based on their titles, author keywords, and abstracts. After a quick look at the first stage publications, we excluded articles that did not meet our defined inclusion. As a result, the final considered publications were 101 papers. Our method then was to thoroughly read each publication and actively search for their research questions, techniques, stakeholders, and limitations. Additionally, we added to our spreadsheet the Google Scholar citation count as a measurement of article’s impact.

#### 3.4.2.3 Response to RQ1

To answer the RQ1, which corresponds to “What are the research strands of the LA field in HE (between January 2011 and February 2016)?” we tried to extract the main topics from the research questions of the publications systemically. We identified that many of the publications do not outline their research questions clearly. Many of the examined publications described their research as use cases. This concerns, in particular, the older publications of 2011 and 2012.

As a result, we did a brief text analysis on the fetched abstracts in order to examine the robust trends in the prominent field of learning analytics and higher education. We collected all the article abstracts, processed them through the R software (http://www.r-project.org), and then refined the resulted corpus. In the final stages, we demonstrated the keywords and chose the word cloud as a representation tool of the terms (see Figure 10). The figure was graphically generated using one of the R library packages called “wordcloud” (https://cran.r-project.org/web/packages/wordcloud/index.html, last accessed 01.August.2016).

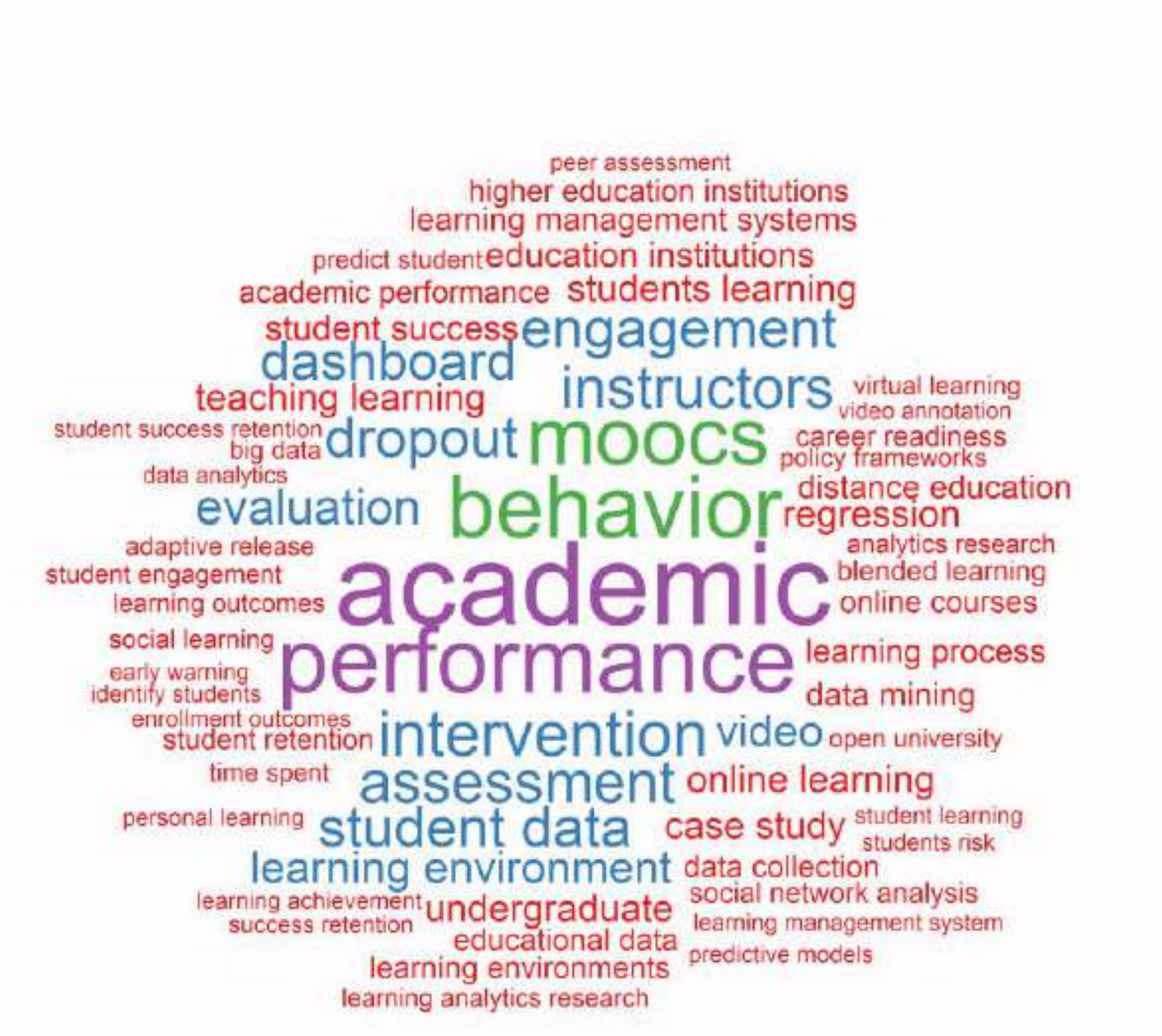

**Figure 10.** Word cloud of the prominent terms of learning analytics in higher education based on articles' abstracts

To ease reading the cloud, we adopted four levels of representation depicted in four colors. The obtained list of words was then classified into singular phrases, bi-grams, tri-grams and quad-grams. The most cited singular words were "academic," "performance," "behavior" and "MOOCs." "Learning environment," "case study" and "online learning" were the most repeated bi-grams. The highest tri-grams used in the abstracts were "learning management systems," "higher education institutions" and "social network analysis." Quad-grams were only limited to "massive open online courses" which were merged at the final filtering stage with the "MOOCs" term.

The word cloud gives a glance at the general topics when learning analytics is ascribed to higher education. Historically, learning analytics researchers focused on utilizing its techniques towards enhancing performance and students' behaviors. The popular adopted educational environment was MOOC platforms. Learning analytics was also used to perform practices of interventions, observing dropout, videos, dashboards and engagement.

In Figure 11, the collected articles from the library data sources are displayed as a bar graph. Results show an obvious increase in the number of publications since 2011. For instance, there were 32 papers in 2015, incremented from 26 articles in 2014 and 17 articles in 2013. However, there were only 5 articles in 2011 and 12 articles in 2012. Because February 2016 was the date of collection of the publications in this study, the 2016 year was not indexed with so many articles. Publications from the LAK conference proceedings were the highest in 2011, 2012 and 2013. This is explained because the LAK conference was the main promoter of learning analytics in e-Learning international forums. The figure further shows a remarkable involvement of journal articles from SpringerLink and Web of Science digital libraries from 2013.

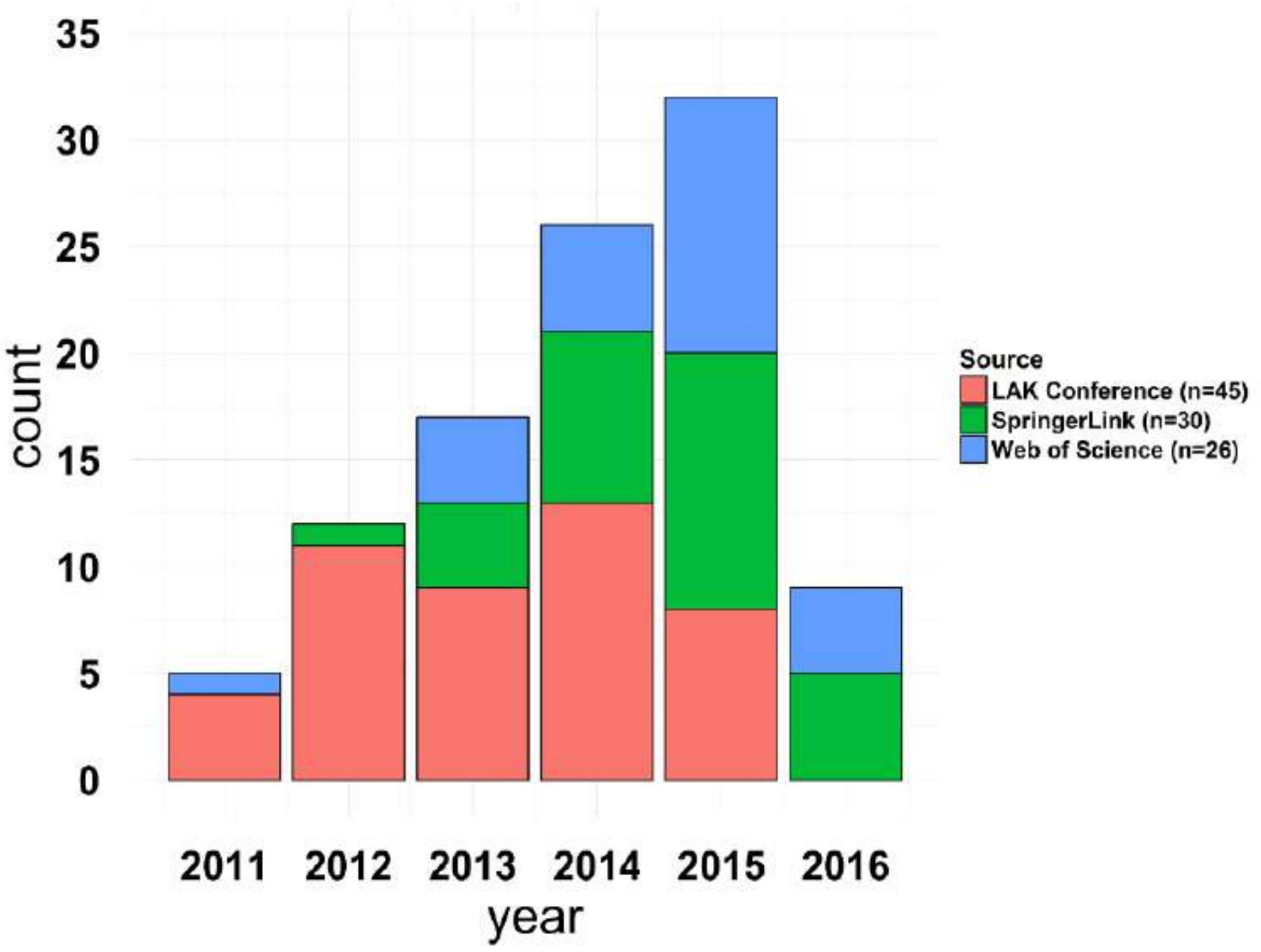

**Figure 11.** The collected articles distributed by source and year of the learning analytics in higher education

In Table 5, we cross-referenced the relevant publications with Google Scholar to derive their citation impact. The table shows a sample of the most cited publications. We noticed that the 2012 year has the highest impact publications followed by 2011 and 2013 respectively.

**Table 5.** The most cited publications of learning analytics in higher education

| Paper Title | Year of publication | Google Scholar Citation (Feb. 2016) |
|---|---|---|
| Course Signal at Purdue: Using Learning Analytics to Increase Student Success (Arnold & Pistilli, 2012) | 2012 | 164 |
| Social Learning Analytics: Five Approaches (Ferguson & Shum, 2012) | 2012 | 94 |
| Classroom walls that talk: Using online course activity data of successful students to raise self-awareness of underperforming peers (Fritz, 2011) | 2011 | 52 |
| Goal-oriented visualizations of activity tracking: a case study with engineering students (Santos et al., 2012) | 2012 | 46 |
| Where is Research on Massive Open Online Courses Headed? A Data Analysis of the MOOC Research Initiative (Gasevic et al., 2014) | 2014 | 46 |
| Course Correction: Using Analytics to Predict Course Success (Barber & Sharkey, 2012) | 2012 | 36 |
| Improving retention: predicting at-risk students by analyzing clicking behavior in a virtual learning environment (Wolff et al., 2013) | 2013 | 34 |
| Learning designs and Learning Analytics (Lockyer & Dawson, 2011) | 2011 | 33 |
| The Pulse of Learning Analytics Understandings and Expectations from the Stakeholders (Drachsler & Greller, 2012) | 2012 | 30 |
| Inferring Higher Level Learning Information from Low Level Data for the Khan Academy Platform (Muñoz-Merino, Valiente & Kloos, 2013) | 2013 | 28 |

#### 3.4.2.4 Response of RQ2

We identified for RQ2, which corresponds to “What kind of limitations do the research papers and articles mention?” three different limitations, either explicitly mentioned in articles or being implicitly within the context.

- Publications with limitations through size were reported aligned with the need for more detailed data like (Barber & Sharkey, 2012; Rogers, Colvin & Chiera, 2014). Other articles mentioned small group size like (Martin & Whitmer, 2016; Strang, 2016), while publications such as (Prinsloo et al., 2015; Yasmin, 2013) mentioned problems of generalization of the approach or methods, scalability, and wider context issues.
- Other limitations were identified in relation to time. Papers like (Ifenthaler & Widanapathirana, 2014; Lonn et al., 2012) stated that continuous work is needed because of a shortage of projects or not enough time to prove hypotheses.
- There were publications with limitations related to the culture. Many of such papers mentioned that their approach might only work in their educational culture and is not applicable somewhere else (Arnold et al., 2014; Drachsler & Greller, 2012). Additionally, ethics differ strongly around the world, so cooperative projects between different universities in different countries need different moderation; the use of data as well could be ethically questionable (Ferguson & Shum, 2012).

Furthermore, there were serious ethical discussions about data ownership and privacy which have recently arisen. Slade and Prinsloo (2013) pointed out that learning analytics touches various research areas and therefore overlaps with ethical perspectives in areas of data ownership and privacy. Questions about who should own the collected and analyzed data were highly debated. As a result, the authors classified the overlapping categories in three parts:

- The location and interpretation of data
- Informed consent, privacy and the de-identification of data
- The management, classification and storage of data

These three elements generate an imbalance of power between the stakeholders, which were addressed by proposing a list of six grounding principles and considerations, including learning analytics as moral practice, students as agents, student identity and performance are temporal dynamic constructs, student success is a complex and multidimensional phenomenon, transparency, and higher education cannot afford not to use data. (Slade & Prinsloo, 2013)

#### 3.4.2.5 Response to RQ3

To answer the RQ3, which corresponds to “Who are the stakeholders in the domain of learning analytics in higher education, and how could they be categorized?” we determined the stakeholders via the publications and categorized them into three types. We found that the examined papers in this study refer to four types of users when learning analytics is involved in higher education studies: 1) learners, 2) teachers, 3) administrators, and 4) researchers. We merged the Researchers and Administrators from the original classification into one distinct group. The aim of this step is to separate the academic analytics, which is represented by administrators/decision makers in educational institutions, from learners and teachers.

In Figure 12, we depict our defined learning analytics-higher education stakeholders as a Venn-Diagram.

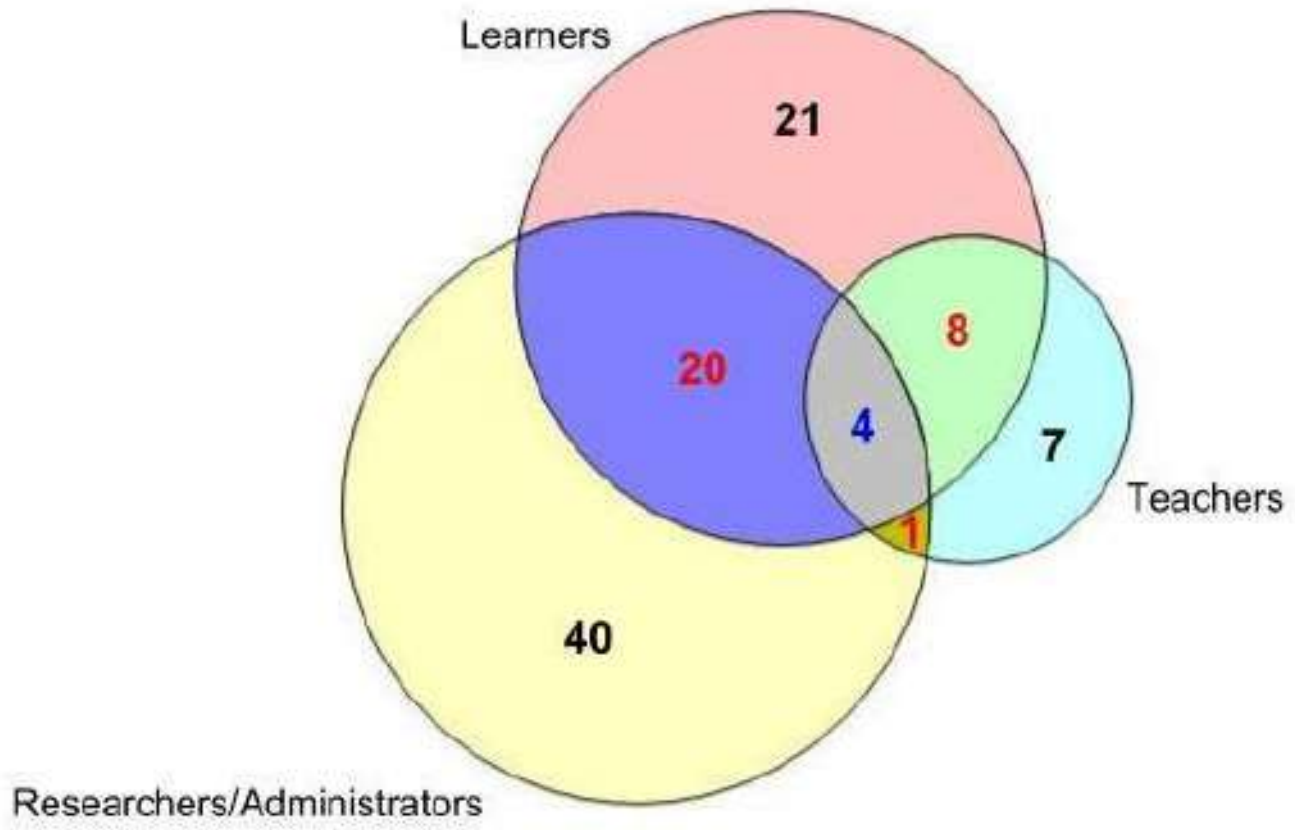

**Figure 12.** Venn-diagram of learning analytics-higher education stakeholders

The number inside each circle denotes the number of publications that considered each stakeholder. The figure shows that there was more research conducted concerning the researchers/administrators with overall 65 publications and 40 of them only concerning themselves than in the field of learners with a total of 53 publications and 21 single mentions. Also, it seems that teachers are only a “side-product” of this field with only 20 mentions and only 7 dedicated to them alone.

Most of the combined articles addressed Researchers/Administrators together with Learners (20 publications). Only eight articles can be found with an overlap between Learners and Teachers, which should be one of the most researched and discussed combinations within learning analytics in higher education. Nearly no work has been done by combining Researchers/Administrators with Teachers (in one publication) and only four papers connected all three stakeholders.

#### 3.4.2.6 Response to RQ4

By analyzing the selected studies to answer RQ4, which corresponds to “What techniques do they use in their papers?” we identified the techniques used in learning analytics and higher education publications. We took into account the methods presented by Romero and Ventura (2013), Khalil and Ebner (2016d), and Linan and Perez (2015). Our findings of the used techniques proposed the following:

1) **Prediction**: Predicting student performance and forecasting student behaviors

2) **Clustering**: Grouping similar materials or students based on their learning and interaction patterns
3) **Outlier Detection**: Detection of students with difficulties or irregular learning processes
4) **Relationship mining**: Identifying relationships in learner behavior patterns and diagnosing student difficulties
5) **Social network analysis**: Interpretation of the structure and relations in collaborative activities and interactions with communication tools
6) **Process mining**: Reflecting student behavior in terms of its examination traces, consisting of a sequence of course, grade and timestamp
7) **Text mining**: Analyzing the contents of forums, chats, web pages and documents
8) **Distillation of data for human judgment**: Helping instructors to visualize and analyze the ongoing activities of the students and the use of information
9) **Discovery with models**: Identification of relationships among student behaviors and characteristics or contextual variables; integration of psychometric modeling frameworks into machine learning models
10) **Gamification**: Includes possibilities for playful learning to maintain motivation; e.g. integration of achievements, experience points or badges as indicators of success
11) **Machine Learning**: Finds hidden insights in data automatically (based on models that are exposed to new data and adapt themselves independently)
12) **Statistics**: Analysis and interpretation of quantitative data for decision making

To analyze the most used learning analytics techniques in higher education through the examined papers, we counted each paper and its methodology in Figure 13. The results in the figure show that the research is focused mainly on prediction, followed by the distillation of data for human judgment. Other new techniques like gamification were found in few papers. Important techniques like clustering, machine learning, statistics, and SNA were identified in many papers.

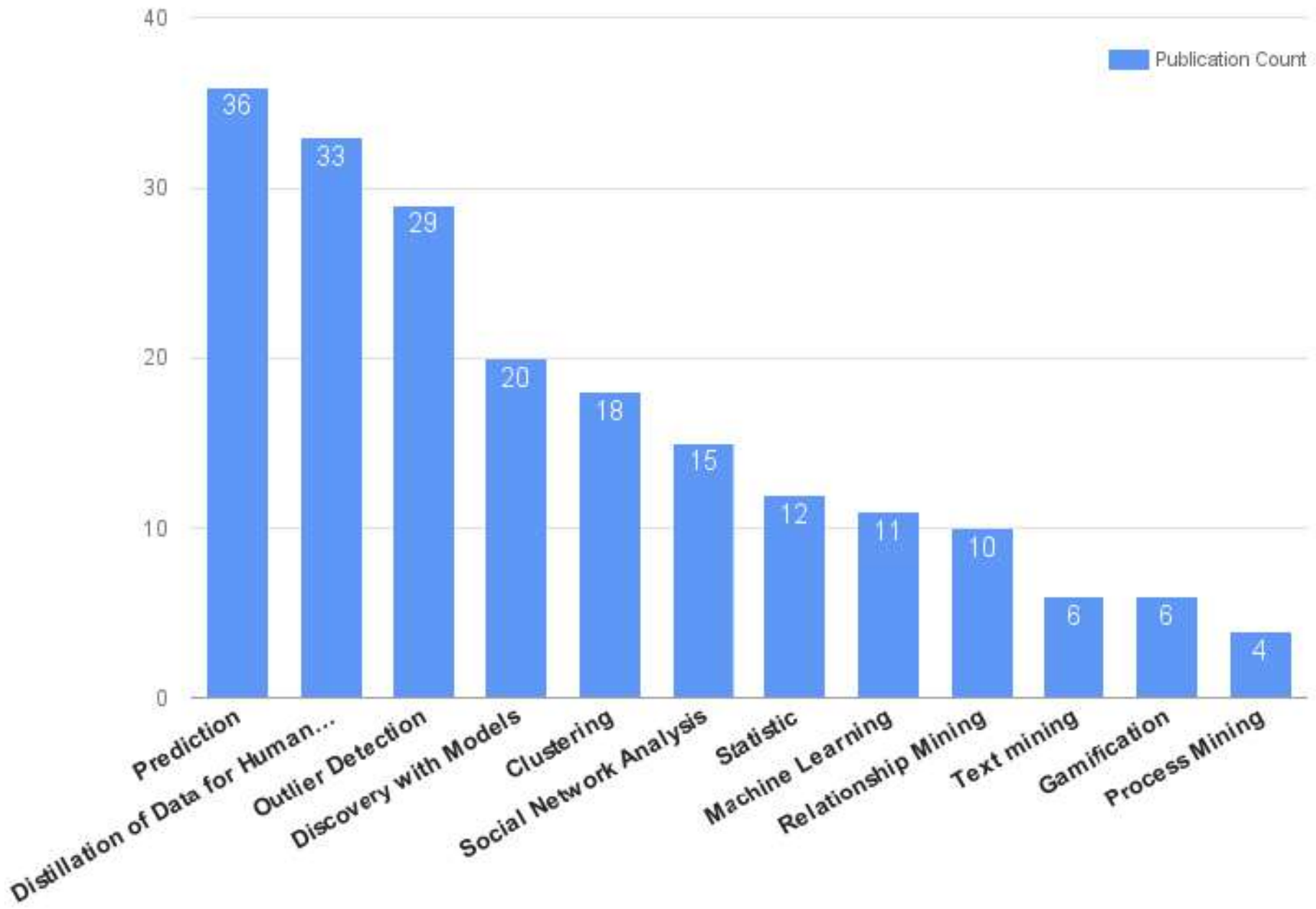

**Figure 13.** Learning analytics techniques in higher education grouped by the total number of publications

#### 3.4.2.7 Conclusions and Summary

In this part of the literature, we examined hundreds of pages to introduce a remarkable literature review of the learning analytics field in an important domain of education, the higher education sector. We presented a state-of-the-art study of both domains based on analyzing articles from three major library references: the LAK conference, SpringerLink, and Web of Science. The total number of relevant publications was equal to 101 articles. Using the big dataset of the included publications, we identified the main research strands. Most of the publications described use cases rather than comprehensive research, especially the prior publications, which is comprehensible because, at the time, the universities had to figure out how to handle and harness the abilities of learning analytics for their benefit. The literature answered four research questions that helped us perform further research in this doctorate thesis.

We identified that there was a clear increase in the number of publications since 2011 till 2015. Further, he involvement of journal articles from the SpringerLink and Web of Science libraries in 2013 and 2015 over the LAK conference publications was apparent.

Our analysis indicates that there was clamor regarding who are the main stakeholders of learning analytics and higher education. As the leading stakeholders of learning analytics should be learners and students (Khalil & Ebner, 2015b), we found that researchers play a major role in the loop between both domains. The direct overlap between learners and teachers was not evidently identified in our study.

On the other side of this study, results revealed that the usage of MOOCs, enhancing learning performance, student behavior, and benchmarking learning environments were strongly researched by learning analytics experts in the higher education domain. Papers of prediction like "Course Signals at Purdue: using learning analytics to increase student success" (Arnold & Pistilli, 2012) was one of the most cited articles of our inclusion. This concurs with our findings that the utmost adopted technique by learning analytics practitioners in higher education was prediction. Finally, we found that ethical constraints drive the limitations to the greatest extent of this literature review study.

### 3.4.3 Learning Analytics Methods Survey

The field of learning analytics provides tools and technologies that offer the potential to do proper interventions and improve education in general. For the time being, educational information systems, MOOCs for instance, hold "Big Data" of learners that create huge data repositories. According to learning analytics definition, the data need to be analyzed by typical methodologies in order to reflect benefits on learning and teaching. EDM and learning analytics are enriched by methods of data mining and analytics in general (Baker & Siemens, 2013). In its first stages, research studies of learning analytics frameworks and structure discussed methods such as visualizations, data mining techniques (Elias, 2011), SNA (Ferguson, 2012) and sentiment analysis (Siemens, 2012), in addition to statistics which was also mentioned as a required tool to build learning prediction models (Campbell, deBlois & Oblinger, 2007).

SoLAR brought to success the annual organization of LAK conferences since 2011. Accordingly, several categories of methods to analyze educational datasets were used. Most of these methods tend to process data quantitatively and qualitatively to discover interesting hidden patterns. Baker and Siemens (2013) mentioned that educational data is what drives new methods to be used in learning analytics. They stated, "The specific characteristics of educational data have resulted in different methods playing a prominent role in EDM/LA than in data mining in general, or have resulted in adaptations to existing psychometrics methods" (p. 4). In this section of the literature of this dissertation, we surveyed publications from the LAK conference from 2013 to 2015. The purpose is to list the most common methods used in the field of learning analytics in the recent years. As learning analytics was formally used as a field in educational technologies, we believed that the period of 2013-2015 would give us a light background on the learning analytics methodologies used. This survey is intended to:

- Identify the common methods used by learning analytics to reach its intended goals.
- Determine methods that are highly cited, e.g. by Google Scholar (http://scholar.google.com), and establish a future forecast towards new research work.
- Assist in comparing the beginning view of the field and the ongoing vision regarding methods.

#### 3.4.3.1 The Research Design of the Learning Analytics Methods Survey

As mentioned before, the conference of Learning Analytics and Knowledge is considered to be the first and the largest repository of learning analytics publications. We mainly focused on it and surveyed 91 papers from LAK 2013 (Suthers et al., 2013), LAK 2014 (Pistilli, Willis, & Koch, 2014) and LAK 2015 (Baron, Lynch, Maziarz, 2015). We excluded papers with topics philosophy, frameworks and conceptual studies of learning analytics because they address structures and do not accommodate a mechanism for revealing patterns. We also faced papers with unclear methods, and these were excluded as well. In the end, 78 publications were used for examination. This study was influenced by the work of Romero and Ventura (2007; 2010) and Dawson et al. (2014). The classification of methods was based on reading the abstract, keywords, general terms, methodology section and the conclusion of each paper. In some publications, we put more analysis into examining the literature and the reference list. Furthermore, we collected the total number of citations for each analyzed paper from Google Scholar and observed the trending topics.

#### 3.4.3.2 Learning Analytics Methods

Learning analytics is a combination of different disciplines like computer science, statistics, psychology, and education. As a result, we realized different analysis methods that tend to be too technical but rather pedagogical. Before classifying the analysis methods, we gravitated towards the beginning topic of the emergence of learning analytics, which briefly described methods and tools for collecting data and analyzing them (Ferguson, 2012; Siemens, 2012). However, the survey reveals more methods being used to examine learner data. Our main method categories, which will be explained in detail later on, are (a) data mining techniques, (b) statistics and mathematics, (c) text mining, semantics and linguistic analysis, (d) visualization, (e) social network analysis, (f) qualitative analysis, and (g) gamification.

Figure 14 shows the grouping of the methods used in learning analytics for LAK publications with the number of papers in each category. It should be noted that some publications may be referenced in a different category. Moreover, a paper could be referenced in multiple method categories. The bar plot in the figure shows that researchers of 31 publications used data mining techniques and 26 research studies used statistics and mathematics to analyze their data. This makes both methods the most employed techniques of analysis. We also see that “Text Mining, Semantics, and Linguistic” analyses, as well as visualizations, are being used in 13 LAK publications equally. However, social network analysis and qualitative analysis, as well as gamification, were the least used techniques.

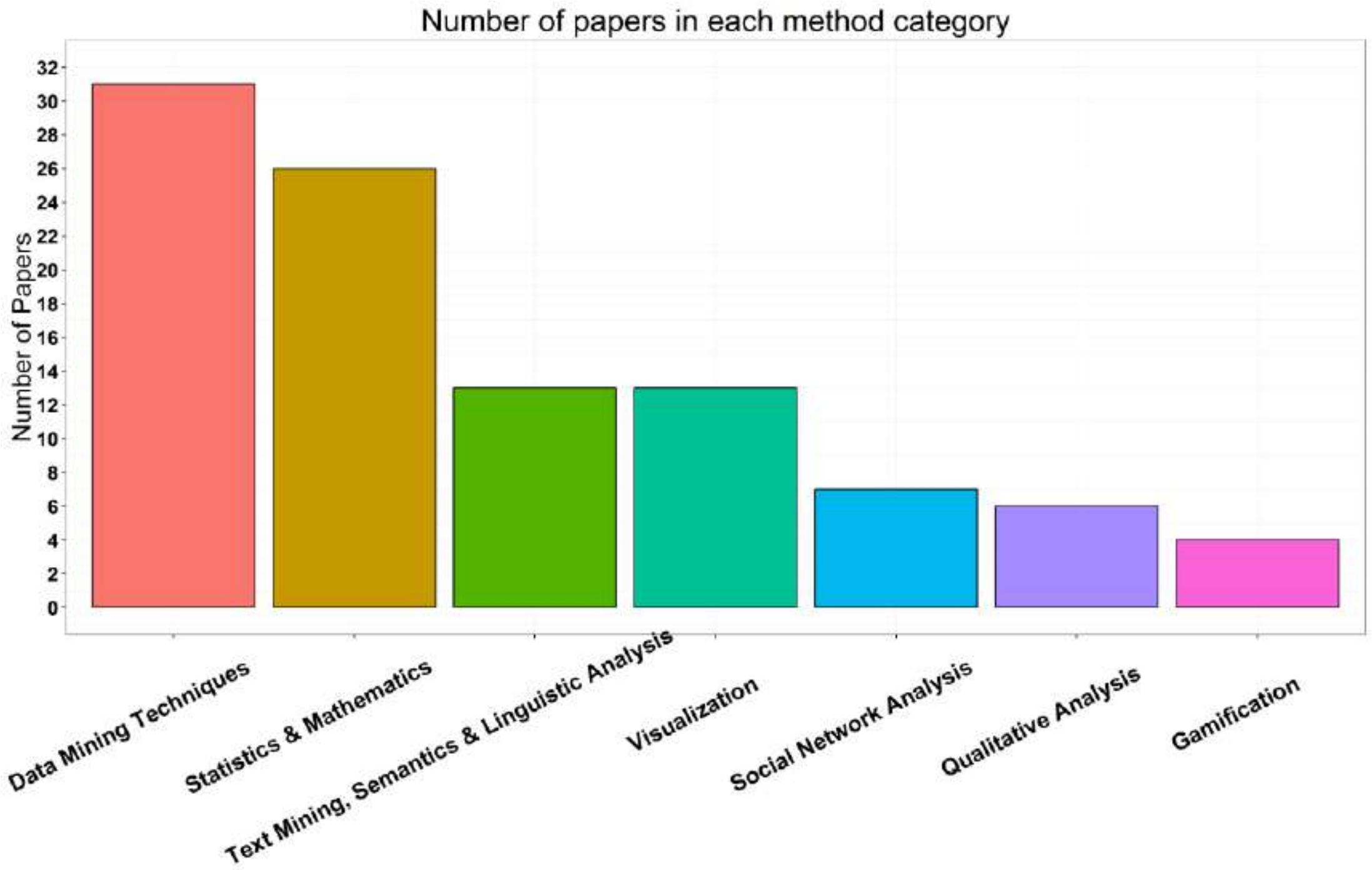

**Figure 14.** The number of the examined LAK papers grouped by methods. Some papers share more than single category

In the following, we categorize methods in more detail and state relevant publications under each method.

A. *Data Mining Techniques:* Data mining tends to make sense out of data. The definition of learning analytics cited a similar idea, namely understanding the data, but in this case, the data references learners. The survey shows that data mining techniques are the most used method for analyzing and interpreting the learners' log data. The decision tree algorithm was used to predict the performance drop and the final outcome of students in a Virtual Learning Environment (VLE) (Wolff et al., 2013). Other researchers used several classification techniques such as step regression, Naive Bayes and REP-Trees to study student behavior and detect learners who game the system (Pardos et al., 2013). Clustering was used to propose an approach for the purposes of enhancing educational process mining based on the collected data from logs and detecting students at risk (Bogarín, 2014). Discovering relations between two factors was accomplished by using multiple linear regression analysis to forecast the relation between studying time and learning performance (Jo, Kim, & Yoon, 2014). Moreover, data mining is used for assessment such as the work at the University of Missouri-Columbia, which proposed an automated tool to enable teachers to assess students in online environments (Xing, Wadholm, & Goggins, 2014). It was remarked that regression analyses were the common mechanism among data mining techniques.

B. *Statistics and Mathematics*: Statistics is the science of measuring, controlling, communicating and understanding the data (Davidian & Louis, 2012). The publications show that researchers use descriptive statistics and mathematics, such as the mean, median and standard deviation to signify their results. In addition, inferential statistics was used side by side with data mining in some of the publications. Markov chain was used to study student behavior in solving multiplication (Taraghi et al., 2014). Different statistical techniques were operated to build a grading system (Vogelsang & Ruppertz, 2015). Additionally, statistical discourse analysis with Markov chain was employed to study online discussions and summarize demographics (Chiu & Fujita, 2014), as well as examining student problem-solving behavior and adapting it into tutoring systems (Eagle et al., 2015).

C. *Text Mining, Semantics & Linguistic Analysis*: Publications which refer to ontologies, mining texts, discourse analysis, Natural Language Processing, or study of languages appear in this category. Some studies refer to text analysis for assessment purposes of short answer questions (Leeman-Munk, Wiebe, & Lester, 2014), to enhance collaborative writing between students (Southavilay et al., 2013) or to contextualize user interactions based on ontologies to illustrate a learning analytics approach (Renzel & Klamma, 2013). The linguistic analysis was clearly used in parsing posts from students for prediction purposes (Joksimović et al., 2015). Finally, online discussion forums were analyzed to pioneer an automatic dialogue detection system in order to develop a self-training approach (Ferguson et al., 2013).

D. *Visualization*: When the information is visually presented to the field experts, efficient human capabilities develop to perceive and process the data (Kapler & Wright, 2005). Visual representations create the advantage of expanding human decisions within a large amount of information at once (Romera & Ventura, 2010). There are several studies that cited visualization as a method to analyze the data and deliver information to end users, such as: building a student explorer screen to prepare meetings and identifying at-risk students by the teachers (Aguilar, Lonn, & Teasley, 2014), studying MOOC’s attrition rates and learners’ activities (Santos et al., 2014), building an awareness tool for teachers and learners (Martinez-Maldonado et al., 2015), and a dashboard for self-reflection goals (Santos et al., 2013). Information can be interpreted into heat maps, scatterplots, diagrams, and flowcharts which were observed in most of the statistical, mathematical and data mining based publications.

E. *Social Network Analysis (SNA)*: SNA focuses on relationships between entities. In learning analytics, SNA can be used to promote collaborative learning and investigate connections between learners, teachers, and resources (Ferguson, 2012). Moreover, it can be employed in learning environments to examine relationships of strong or weak ties (Khalil & Ebner, 2015b). This category includes network analysis in general and Social Learning Analytics (SLA). The survey observed researchers who built a collaborative learning environment by visualizing relationships between students about the same topic (Schreurs et al., 2013). A two-mode network was used to study students’ patterns and to classify them into particular groups (Hecking, Ziebarth, & Hoppe, 2014). It was also used with a grading system in a PLE to examine the centrality of students and grades (Koulocheri & Xenos, 2013). Again, not so far from this survey study, a network analysis was done to analyze citations of LAK conference papers (Dawson et al., 2014). The authors studied the degree centrality and pointed out the emergence and isolated disciplines in learning analytics. SNA was used to analyze data of connectivist MOOCs by examining interactions of learners from social media websites (Joksimović et al., 2015).

F. *Qualitative Analysis*: This category is related to the decisions based on explained descriptions of the analysts. For instance, 1) a qualitative evaluation of data mining techniques was made to understand the nature of discussion forums of MOOCs (Ezen-Can et al., 2015), 2) usage of qualitative interviews, which are answered with words to build a learning analytics module of understanding fractions for school children (Mendiburo, Sulcer, & Hasselbring, 2014), 3) qualitative meta-analysis to investigate teachers' needs in technology enhanced learning covered by the umbrella of learning analytics (Dyckhoff et al., 2013).

G. *Gamification*: Gamification is the use of game mechanics and tools to make learning and instruction attractive and fun (Kapp, 2012). This method is considered as a technique on its own because of its relevant appearance in educational workshops when requests to make learning entertaining are presented. Some examples are using rewards points and progress bars to enhance the retention rate and building a gamified grading system (Holman, Aguilar, & Fishman, 2013) or presenting a competency map with progress bars, pie charts, labels and hints to improve student performance (Grann & Bushway, 2014). A significant study on monitoring students in a 3D immersive environment (Camilleri et al., 2013) was also advised as another type of gamification technique.

#### 3.4.3.3 The Prominent Methods

In this last part of the survey, we consider learning analytics methods that have frequently been cited. We used Google Scholar as a foundation to check methods' popularity. All the data were collected recently and retrieved before the submission date. Figure 15 shows Google Scholar citations for the analyzed LAK conference papers based on the methods category. The publications with the method type Data Mining and Techniques were the most cited articles (452 citations). The maximum number of citations in this survey belongs to the paper of Kizilcec et al.

(Kizilcec, Piech, & Schneider, 2013) with 236 citations. Although we took into consideration the time span of publications, we saw that articles that belong to MOOCs are the most cited papers. Statistics and Mathematics publications were cited 363 times. Qualitative analysis and gamification publications were the least cited articles. In Figure 16, we show a density plot of publications' citations grouped by year. The x-axis records the number of citations converted into logarithmic scale for ease of reading. The y-axis records the density of publications per year. Since we did not survey a fair number of publications per year, we intended to use this plot instead of histogram plot, which is highly sensitive to bin size.

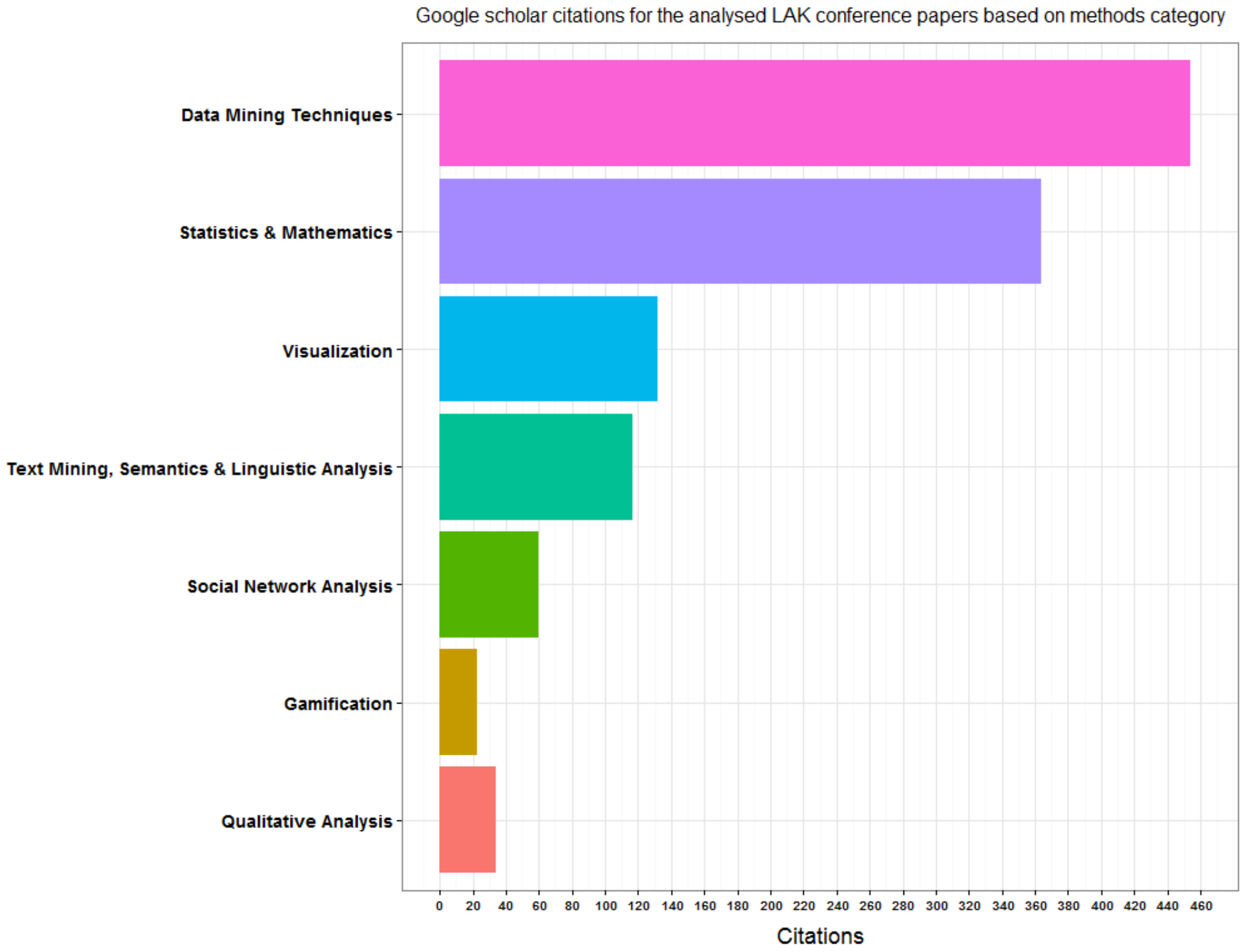

**Figure 15.** The number of Google Scholar citations of the examined LAK papers based on methods category. Retrieved on 26th October, 2015

Some of the 2013 articles attracted numerous citations which exceed the expectations such as (Kizilcec, Piech, & Schneider, 2013; Pardos et al., 2013). A descriptive analysis of the articles in 2013 leads to: (*median*= 8, *mean*= 24.37, *max*= 236); articles in 2014: (*median*= 5, *mean*= 5.56, *max*= 17); and articles of 2015: (*median*= 1, *mean*= 1.42, *max*= 4). The low number of citations for the 2015 publications is reasonable as the time span between this survey and 2015 LAK publication date is around six months.

In the end, we noticed that learning analytics researchers are adopting data mining and statistics more often than other techniques. Gamification as a method can be considered a newly emerging field that adds value to analyze student data. Social network analysis appears to be diminishing in the years to come, and information visualization as well as linguistic analysis still contain important methods to analyze educational datasets. We also noticed that the number of MOOC articles published in 2013 was significant by the distinct number of citations in Google Scholar. We also saw that some publications had a high impact on education with their peak Google Scholar score. In fact, the upcoming learning analytics events might show extinction of some methods and an upsurge in the appearance or emergence of new techniques, which can be allocated to our defined categories.

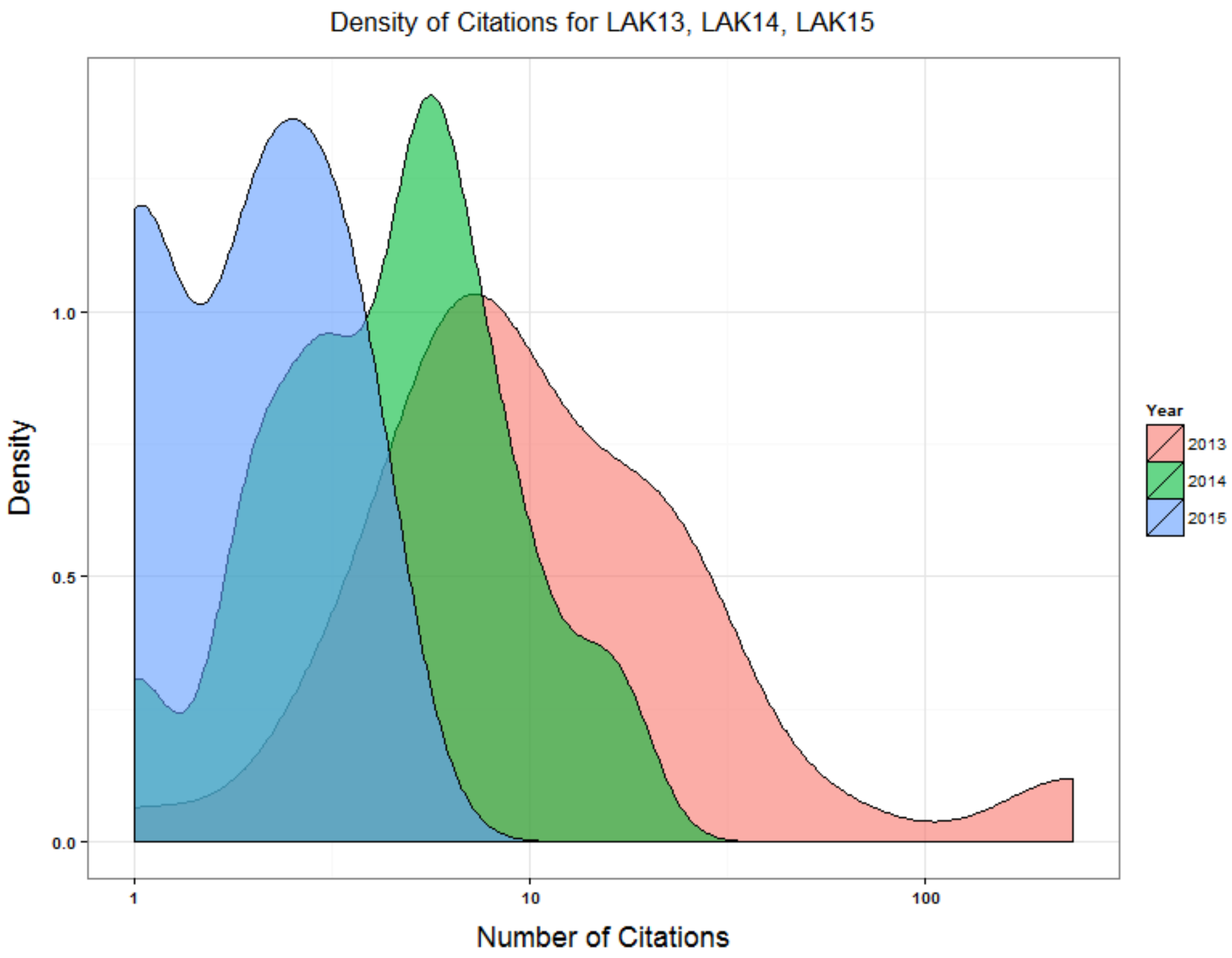

**Figure 16.** Density plot of the total number of Google Scholar citations for the examined articles grouped by year of publication

### 3.4.4 Key Trends of Learning Analytics[2]

Since the first edition of the Learning Analytics and Knowledge conference in 2011, learning analytics received much attention from educational researchers, practitioners, and administrators, resulting in numerous conferences, workshops, and journal publications dedicated to learning analytics. While there have been numerous empirical studies in the learning analytics domain, only a few studies looked at the state of the learning analytics field as a whole. Most of these studies focused on commonly used analytics methods and techniques and less so on the emerging trends in the published literature. In this section of the literature, we present results of a systematic review which looked at the current state of the learning analytics literature. Using automated topic modeling techniques, we systematically analyzed 1,315 learning analytics articles and identified key trends in the literature and how they changed over time. Our investigation produced 21 distinctive topics that together paint a comprehensive picture of the current state of the learning analytics field. Our investigation showed a steady growth in the number of learning analytics publications and revealed a growing interest in the topics such as predictive models, challenges of learning analytics adoption, and examination of students' study behavior. More details are presented in the following subsections.

#### 3.4.4.1 Overview of Learning Analytics Key trends

In emerging research fields, such as learning analytics, a systematic analysis of the published literature represents significant research activity in exploring the state of the field and identifying key challenges and directions for future research (Mulrow, 1994; Rowley & Slack, 2004). In the last few years, there have been several reviews of learning analytics and related fields such as educational data mining, technology-enhanced learning, and massive open online courses. One of the early reviews is conducted by Romero and Ventura (2007), who looked at the particular domains of EDM application (i.e., classroom setting and distance education, including Web-based courses, popular content management systems, and intelligent learning systems) and adopted methods and tools including statistics, visualization, web mining, clustering, classification, outlier detection, association rule mining, sequential pattern analysis, and text mining. The same authors produced an updated review five years later (Romero & Ventura, 2010) which provided slightly different and more detailed description of problems and methods in EDM research. These include, 1) analysis and visualization of data, 2) feedback provision, 3) recommendation of learning resources, 4) student performance prediction, 5) student modeling, 6) detecting desirable and undesirable student behaviors, 7) student grouping, 8) social network analysis, 9) development of concept maps, 10) assembling course materials, and 11) planning and scheduling of learning activities (Romero & Ventura, 2010).

[2] This systematic review study was carried out in collaboration with doctoral students from the University of Edinburgh:
Khalil, M., Kovanovic, V., Joksimović, S., Ebner, M., & Gasevic, D. Where are learning analytics today? Examining the key trends in the published literature from 2011 to 2016. (in preparation).

With the development of learning analytics as a new field focusing explicitly on data analytics of student learning processes, Baker and Siemens (2013) produced a review of EDM and learning analytics fields, their similarities and differences, as well as an overview of the key themes and methods in educational data mining/learning analytics research. Baker and Siemens (2013) identified prediction models, structure discovery, relationship mining, distillation of data for human judgment and discovery with models. A similar study by Baker and Inventado (2014) listed as key topics, 1) prediction methods (i.e., classification, regression, and latent knowledge estimation), 2) structure discovery (i.e., clustering, factor analysis, and domain structure discovery), 3) relationship mining (i.e., association rule mining, sequential pattern mining, correlation mining, causal data mining), and 4) discovery with models as key themes in EDM and learning analytics research.

A more focused analysis of studies presenting quantitative empirical evidence in EDM/learning analytics was done by Papamitsiou and Economides (2014). Through the analysis of forty journal and full-length conference papers between 2008 and 2013, Papamitsiou and Economides (2014) identified seven key themes in the published learning analytics and EDM research: learning management systems (LMS) analysis, MOOCs & social learning, Web-based education, cognitive tutors, computer-based education, multimodality, and mobile learning. Finally, a study by Slater et al. (2016) provided an extensive overview of the commonly adopted tools for feature engineering, algorithmic analysis, visualizations, and other specialized problems such as text mining, social network analysis, Bayesian knowledge-tracing, and process and sequence mining.

Particularly focusing on the learning analytics field, Nunn et al. (2016) conducted a systematic review of published learning analytics literature in order to investigate current learning analytics methods, key benefits of learning analytics, as well as the main challenges of learning analytics adoption. Nunn et al. (2016) identified the key benefits of learning analytics curriculum as personalization and improvement, prediction and improvement of student learning outcomes, improvement in instructor performance and personalization of learning experience, and the provision of employment opportunities and improvement in understanding of learning processes. Similarly, the main challenges of learning analytics implementation include the collection of heterogeneous data and their complex analysis, connection to pedagogy and learning sciences, and ethical and privacy issues. Further studies like the one done by us (2016d) (which was previously discussed in Chapter 3.4.3) surveyed the LAK conference proceedings between 2013 and 2015 to examine adopted learning analytics research methodologies. Khalil and Ebner (2016d) found a strong focus on data mining and statistical techniques, with other important topics in learning analytics literature being text analysis, learning analytics visualizations, social network analysis, qualitative learning analytics research, and gamification.

Lastly, another review that looked specifically at the learning analytics literature is a study performed by Dawson et al. (2014). Through citation network analysis of papers published in the first three years of the LAK conference (2011-2013) and in three special issues on learning analytics (i.e., *Journal of Asynchronous Learning Networks, 16(3),* 2012*; Journal of Educational Technology & Society 15(3),* 2012*; and American Behavioral Scientist 57(10),* 2013), Dawson et al. (2014) identified five key categories of learning analytics publications. Those include, 1) evaluation papers, 2) opinion papers, 3) papers reporting personal experience with learning analytics, 4) proposal of different learning analytics solutions, and 5) valuation research papers. Dawson et al. (2014) also identified the dominance of educational and computer science researchers and the dominance of quantitative research methods in the learning analytics literature.

In addition to studies that used traditional, manual systematic review methods, there have been several small-scale studies that use automated topic modeling techniques for the analysis of published learning analytics literature (Nistor, Derntl & Klamma, 2015; Derntl, Günnemann & Klamma, 2013; Hu et al., 2014).

The study by Derntl, Günnemann, & Klamma (2013) used dynamic topic modeling to examine LAK '13 data challenge dataset which consisted of papers published at the EDM conference (2008-2012, 239 papers) and the LAK conference (2011-2012, 66 papers) and a special issue on learning analytics of *The Educational Technology & Society Journal* (10 papers). Derntl,

Günnemann, & Klamma (2013) obtained a solution with 20 topics as the most important, such as learning modeling, predictive analytics, and social network analysis. Another study that used dynamic topic modeling was done by (Nistor, Derntl & Klamma, 2015) which examined papers published at the EC-TEL and LAK conferences between 2011 and 2014. Nistor, Derntl & Klamma (2015) identified as the main topics of learning analytics, visualization of learning progress, prediction of student dropout and success, assessment, and conducting interventions. Although it provides interesting findings, the study was limited given its focus on papers from 2014 only and detailed examination of only 19 papers which might not be indicative of the broader learning analytics field.

Finally, a study by (Hu et al., 2014) also looked at the papers in the LAK '13 data challenge. What is interesting is that the authors provided an interactive tool to browse learning analytics publications which can be used to identify important papers, authors and even geospatial representation of institutions involved in learning analytics research.

The difference between the study presented in this section and the previous related work is that our analysis dataset is much bigger, more than four times as big as the one used by Derntl, Günnemann, & Klamma (2013), the biggest of the currently published studies. This is not surprising as learning analytics has grown considerably since 2014 when the latest topic modeling analysis was conducted. As such, our analysis provides a more comprehensive overview of the learning analytics topics and trends, especially in the last three years. Likewise, the previous studies looked at the learning analytics literature together with related fields such as educational data mining or technology-enhanced learning. Given a more mature state of the learning analytics field today, we posit that a new analysis, focusing explicitly on learning analytics publications, is warranted.

#### 3.4.4.2 Research Questions and Methodology

Given the previous work in the field of learning analytics, the focus of this study is to examine the state of the learning analytics literature, including the key research topics and their change over time. As such, the main research questions in the presented section will answer:

- **RQ1**: What are the main topics in the learning analytics fields, as expressed by the published research literature (2011-2016)?
- **RQ2**: How did the identified learning analytics topics evolved over time?

- **RQ3**: What are the main venues hosting publications of learning analytics research (aside from the LAK conference and the *Journal of Learning Analytics*)?

By answering these questions, we aim to provide a systematic and comprehensive overview of the state of learning analytics literature as expressed by the published research articles. By looking at the key topics in learning analytics, the goal is to discover important trends in the field, and fruitful directions for future research.

**Dataset**

In order to address our research questions, we performed a computer-based search through the Scopus digital library (www.scopus.com), given its comprehensive coverage of a large number of venues related to education, educational psychology, and educational technology research. We searched for the "learning analytics" term in the title, abstract or keywords using the following search query[3]:

*TITLE_ABS_KEY({Learning Analytics})*

Overall, the search resulted in 1,226 results that satisfied the criteria for inclusion in the review:

- The study was published in a peer-reviewed journal or conference proceedings,
- The study was published between 2011 and 2016[4],
- The study was written in English,
- The study focus was on addressing challenges emerging from the learning analytics research field, and
- The study had abstract available.

We also included studies published in the *Journal of Learning Analytics* (JLA, http://www.learning-analytics.info), given that this particular venue was not indexed by the Scopus digital library and is one of the two major venues for learning analytics research (the other one being the International Learning Analytics & Knowledge Conference series which is indexed by the Scopus).

---

[3] Curly braces were used instead of double quotes as Scopus interprets double quotes more flexibly. For example, the phrase “Learning from analytics” and similar would have been returned if we had used the simple double quotes.
[4] The search was performed on Oct 2, 2016 to make results as up-to-date as possible

In total, we included 89 articles from the *Journal of Learning Analytics*, published in eight volumes within the three issues (each JLA issue has three volumes; at present, the latest is vol. 3, no 2.), which resulted in a total of 1,315 total articles included in our analysis. For each of the articles we included its: 1) publication type, 2) title, 3) author names, 4) publication year, 5) abstract, 6) publication venue name, 7) publisher name, 8) volume and issue number (for books, book chapters, and journal articles), 9) conference name and location (for conference papers), and 10) page numbering information.

**Analysis Procedure**

The overall outline of the analysis process is shown in (Figure 17).

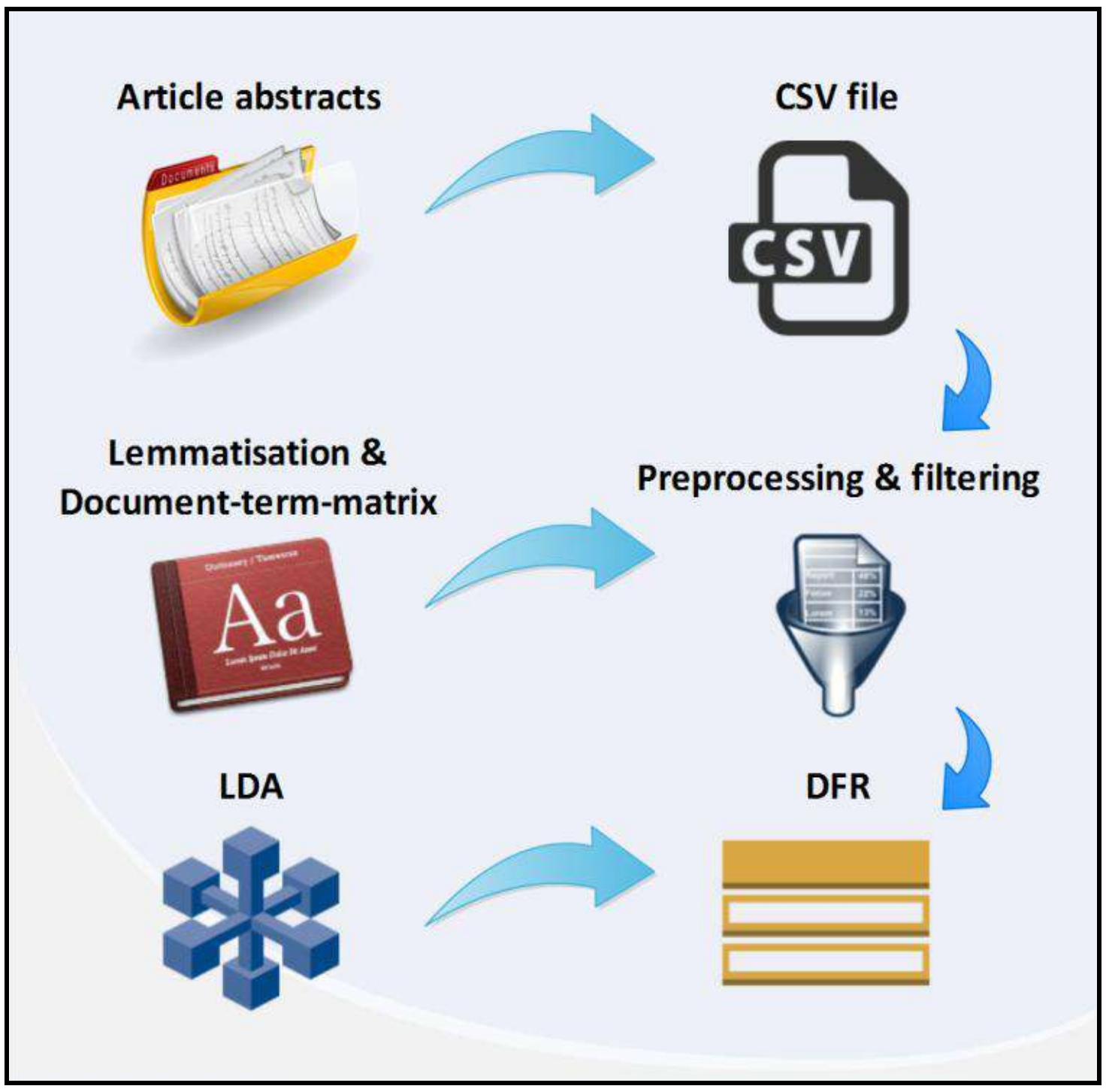

**Figure 17.** Outline of the document pre-processing and topic extraction process of the systematic review

To analyze our dataset and extract main themes of the learning analytics research field, we used Latent Dirichlet Allocation (LDA) (Blei, Ng & Jordan, 2003), a popular probabilistic topic modeling technique. LDA can be used to identify latent themes in the document corpora based on the word co-occurrences and are often used in social sciences and humanities. Due to its simplicity and usefulness of the discovered topics, LDA is becoming one of the key techniques for analysis of large document corpora (Crain et al., 2012). It has been used for automating the analysis of various types of textual data, including analysis of newspaper articles (Wei & Croft, 2006; Yang, Torget & Mihalcea, 2011), social media (Mehrotra et al., 2013), and online product reviews (Titov & McDonald, 2008).

In the context of educational and learning analytics research, topic modeling has been extensively used for a wide variety of problems, such as analyzing large corpora of MOOC news articles (Kovanović et al., 2015a; Kovanović, 2015b), student collaborative writing (Southavilay et al., 2013), student online discussions (Ming & Ming, 2012; Reich et al., 2014), course evaluations forms (Reich et al., 2014), and even LAK-related Twitter messages (Chen, Chen, & Xing, 2015). Finally, as previously mentioned, topic modeling has been used for analysis of learning analytics literature (Hu et al., 2014; Nistor, Derntl, Klamma, 2015; Derntl, Günnemann, & Klamma, 2013).

**Data Preprocessing**

Prior to use of LDA, we performed certain data pre-processing to improve the final results of the analysis, as commonly done in text mining research (Kovanović, 2015c). We first performed the lemmatization of the document abstracts in order to reduce the number of the distinct terms in the analysis, which increases the quality of the word co-occurrence estimates. Next, we removed all numbers, words shorter than three letters, and generic stopwords, using the list of English stopwords available in the Mallet toolkit (McCallum, 2002). We also removed some of the words that can be considered stopwords in our particular context as they do not provide much information about their use given the nature of our dataset. Those are: “result,” “show,” “datum,” “study,” “learning,” “analysis,” “academic,” “analytic,” “student,” “teacher,” “learner,” “question,” “course,” “use,” and “learn.” Next, we removed words that do not appear in at least 5% of the documents, which significantly reduced the number of words that are being analyzed. Likewise, we also removed words with low TF-IDF (term frequency–inverse document frequency) score, which indicates that the words appear too broadly across the corpus. Plus, we removed words that were slightly below median TF-IDF score (0.95 of the median TF-IDF score was the threshold) to remove the majority of the words that are occurring too frequently and thus substantially decrease the analysis space (i.e., the number of words for which co-occurrences are calculated).

Much like many of the clustering algorithms (e.g. *K*-means), LDA algorithm requires the number of topics that should be discovered to be specified a priori. In our study, we used a method proposed by (Cao et al., 2009) which looks at the density of the topic distance graph to identify an optimal number of topics. We used the implementation provided by *ldatuning R* library (Nikita, n.d.), and selected the optimal number of topics. The final topic modeling solution on the whole corpora was implemented using the MALLET topic modeling library (McCallum, 2002). The interactive visualization of the topic modeling was implemented through customization of output provided by the *dfrtopics R* library (Goldstone, n.d.).

#### 3.4.4.3 The Systematic Review Results

**Dataset Overview**

From the 1,315 included articles in our study, we can see that conference papers and journal articles represent a large majority of the learning analytics publications (see Table 6).

**Table 6.** Number of scholarly learning analytics publications over the years

| Year | Book | Book Chapter | Journal Article | Conference Paper | Total | Perc. |
|---|---|---|---|---|---|---|
| 2011 | 0 | 0 | 2 | 27 | 29 | 2.2% |
| 2012 | 0 | 0 | 20 | 75 | 95 | 7.2% |
| 2013 | 0 | 1 | 33 | 147 | 181 | 14% |
| 2014 | 1 | 11 | 67 | 199 | 278 | 21% |
| 2015 | 0 | 12 | 144 | 254 | 410 | 31.1% |
| 2016 | 1 | 5 | 135 | 181 | 322 | 24.4% |
| **Total** | 2 | 29 | 401 | 883 | 1315 | 100% |

Conference papers represent about two-thirds of the publications, while journal articles represent 30% of the articles. Books and book chapters are featured much less prominently, with only 29 book chapters on learning analytics and two full books being published (2.2% and 0.002% of the total number of publications, respectively).

If we look at the number of publications across the years, we see a very steady growth of the number of articles across the years, with the largest increases in the first two years. The number of articles more than tripled between 2011 and 2012, and almost doubled between 2012 and 2013. Although the number of articles for 2016 is smaller than for 2015, this is most likely due to the delay in article availability in indexing databases which mostly impacts 2016 articles. Finally, it should also be noted that at the moment of conducting the analysis (Oct 2016), year 2016 was ongoing, which rendered the number of 322 articles for 2016 incomplete.

Looking at the particular types of publications, we see that the majority of book chapters were published during 2014 and 2015. This is one of the signs that the field of learning analytics is slowly reaching a more mature state, with publications which reflect and synthesize already existing empirical evidence in the field. For journal articles, the most significant increase occurred during 2014 and 2015, with both years producing double the number of articles as in the previous year. Not surprisingly, this coincides with the establishment of the *Journal of Learning Analytics*, which had its first issue in 2014.

To answer the research question about outlets that host the learning analytics research, we calculated the number of papers from conference proceedings and journal articles. Table 7 shows the top venues with the corresponding number of publications.

**Table 7.** Most common venues publishing learning analytics research

| # | Source | No. of papers |
|---|---|---|
| **Journal Articles** | | |
| 1 | Journal of Learning Analytics | 89 |
| 2 | Computers in Human Behavior | 33 |
| 3 | Journal of Asynchronous Learning Network | 13 |
| 4 | Journal of Universal Computer Science | 12 |
| 5 | Educational Technology and Society | 11 |
| 6 | International Journal of Technology Enhanced Learning | 11 |
| 7 | British Journal of Educational Technology | 10 |
| 8 | Computers and Education | 10 |
| **Conference Proceedings** | | |
| 1 | Learning Analytics and Knowledge (LAK) | 249 |
| 2 | European Conference on Technology Enhanced Learning (EC-TEL) | 58 |
| 3 | Learning Analytics and Knowledge workshops | 49 |
| 4 | International Conference on Advanced Learning Technologies (ICALT) | 44 |
| 5 | International Conference on Computers in Education (ICCE) | 31 |
| 6 | ACM Conference on Learning at Scale (L@S) | 24 |
| 7 | International Conference on Multimodal Interaction (ICMI) | 23 |
| 8 | IEEE Global Engineering Education Conference (EDUCON) | 20 |
| 9 | Australasian Society for Computers in Learning in Tertiary Education (ASCILITE) | 20 |

As expected, the Journal of Learning Analytics and the LAK conference series as well as the LAK workshops are the main promoters of learning analytics research to the wider community. Besides those expected venues, the *European Conference on Technology and Enhanced Learning (EC-TEL)* is the second leading conference covering learning analytics research. This is aligned with the study by (Kawase, Siehndel & Gadiraju, 2014) who showed that learning analytics was one of the most discussed themes at the *EC-TEL '13* conference. This trend has continued, as our results showed that *EC-TEL '15* conference included more studies about learning analytics than any previous *EC-TEL* event.

Similar to *EC-TEL*, the *International Conference on Advanced Learning Technologies (ICALT)* featured 44 papers about learning analytics. The list of conferences with more than twenty learning analytics publications also includes the *International Conference on Computers in Education (ICCE)*, *Learning at Scale conference (L@S)*, *International Conference on Multimodal Interaction*, which makes sense for its acronym *(ICMI)*, the *conference of the Australasian Society for Computers in Learning in Tertiary Education (ASCLITE)* and the *Global Engineering Education Conference (EDUCON)*.

Peer-reviewed journals featured about a half of the number of papers that were presented at the conferences, which is expected given how young the field of learning analytics is. As shown by Bernard et al. (2004), research journals typically publish more mature research studies, which make them harder to develop and ultimately publish. After the *Journal of Learning Analytics*, the journal with the highest number of learning analytics publications is *Computers in Human Behavior* which published a significant number of 33 publications. The remainder of the journals published significantly fewer learning analytics articles, with the *Journal of Asynchronous Learning Networks* being in the third position with 13 published articles.

**Topic Modeling Results**

Our analysis of all possible numbers of topics from the range between 2 and 50 resulted in the optimal number of 21 topics by following the method proposed by Cao et al. (2009). This number of 21 topics was used in the rest of the analysis (see Figure 18).

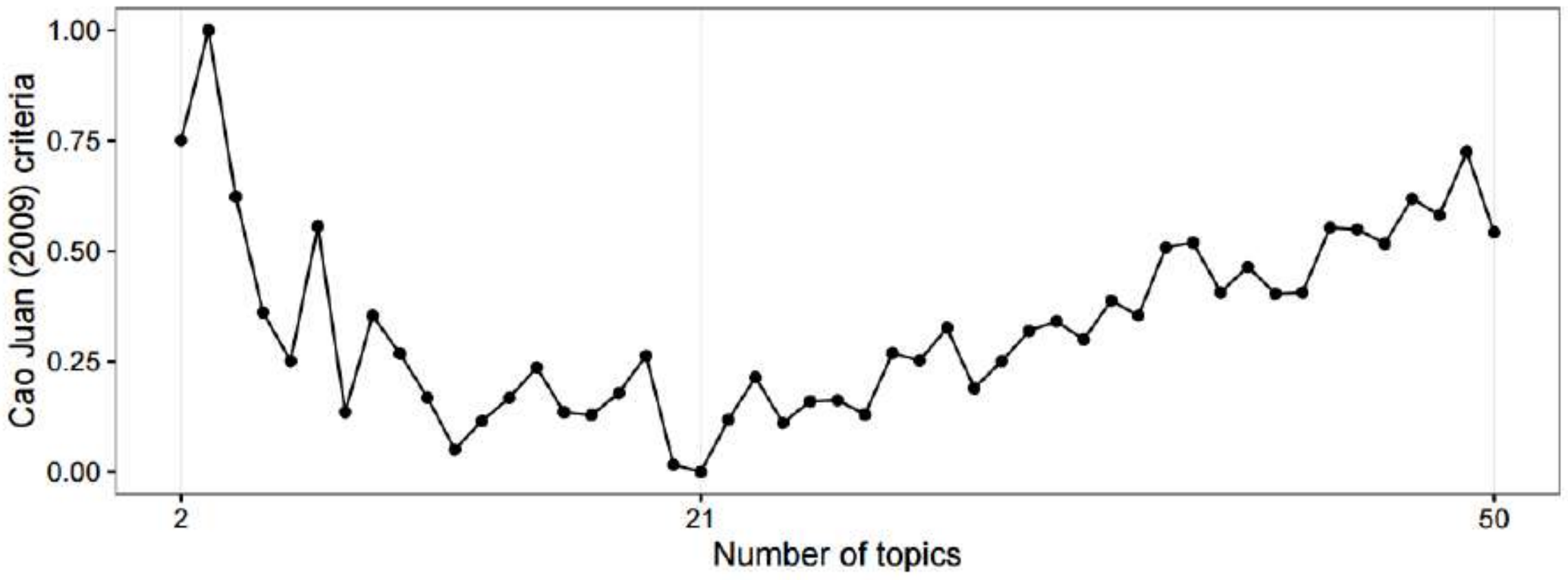

**Figure 18.** Selection of the optimal number of topics based on Cao et al. (2009) approach in topic modeling

The use of LDA algorithm to model 21-topic model solution resulted in the list of topics shown in Table 8 in the following page. The representative keywords of each topic revealed the main concepts of the learning analytics field in scholarly publications in the period of six years. Moreover, they reflect the recurrent themes that have been considered in the research state. The rank and frequency values in the table show the topic modeling correlation score between the number of publications and the keyword relevance in the abstracts' context. Although topic modeling is considered useful and meaningful in understanding hidden themes of a document text corpora, inferring topics is a hard task to perform (Chang et al., 2009). The unsupervised learning mode of the topic modeling approach required us to depend on exogenous data to understand what the main themes are, and to predict future trends. Thus, we further investigated the abstracts and the evolution of topics in the selected time period (2011-2016).

We extracted the top ten most distinctive words that represent each of the 21 topics identified, as well as the relative frequency of the topic in the corpora. Based on the analysis of the extracted words, their associations with each topic, and the list of highly relevant papers, we labeled each of the topics to summarize the primary theme within each topic. We can see that the most frequent topics relate to the field of learning analytics itself, modeling of student behavior and outcomes, log data analysis, data mining techniques, and provision of feedback.

**Table 8.** Topic modeling identified topics and their 10 most representative keywords

| ID | Frq. | Rank | Topic label | Topic keywords |
|---|---|---|---|---|
| 1 | 6.3 | 2 | Student behavior and outcomes | behavior test factor outcome grade type impact high relate log |
| 2 | 5.3 | 10 | Assessment | assessment task feedback generate outcome goal large indicator assess computer |
| 3 | 4.2 | 16 | MOOCs | mooc massive platform engagement pattern participant number rate discussion high |
| 4 | 5.7 | 8 | Social Network Analysis | social network content media interaction community relationship collaborative structure form |
| 5 | 6.0 | 4 | Data Mining Techniques | technique computer science mining human insight school algorithm gain area |
| 6 | 2.2 | 21 | Video Analysis | video classroom interaction behavior source capture record effectiveness user researcher |
| 7 | 5.3 | 9 | Learning analytics frameworks | theory evaluation framework goal implementation field example setting indicator key |
| 8 | 3.0 | 18 | Concept learning | concept domain structure factor content discussion relationship demonstrate visual real |
| 9 | 5.0 | 11 | Web analytics | user pattern service application web behavior create real usage visual |
| 10 | 4.6 | 13 | Learning interactions | interaction face virtual reserve rights classroom feedback |
| | | | | participant discussion instructor |
| 11 | 2.8 | 19 | Group-based problem-solving | group problem content domain digital resource high type theory action |
| 12 | 2.2 | 20 | Access pattern analytics | resource access pattern usage view large strategy number material outcome |
| 13 | 5.8 | 5 | Feedback | feedback skill quality educator progress class higher evidence evaluation enable |
| 14 | 4.3 | 14 | Predictive analytics | prediction feature predict algorithm behavior machine mining technique source early |
| 15 | 7.6 | 1 | Learning analytics field | field digital community researcher mining area science author issue literature |
| 16 | 6.0 | 3 | Log data analysis | log application publishing switzerland platform language record facilitate real call |
| 17 | 5.7 | 7 | Learning analytics adoption challenges | issue big higher institution application large framework address decision field |
| 18 | 4.3 | 15 | Collaborative learning | framework task collaborative interaction participant decision pedagogical computer evaluation app |
| 19 | 5.7 | 6 | Implementation reports | institution success program higher change intervention group impact create rate |
| 20 | 3.1 | 17 | Discussion interventions | intervention instructor discussion engagement strategy interaction assess participant pedagogical |
| 21 | 4.8 | 12 | Learning analytics visualizations | visualization project platform web decision implement pedagogical enhance article school |

Results show that the most relevant topics are related to the learning analytics field itself. These topics include keywords related directly to the community (e.g. researchers, authors) and the research state (e.g. literature, review, article). Findings reveal that there was a strong concentration over time on the general dimensions of the learning analytics field and (dis)similarities with other related branches (i.e. EDM, web analytics, and academic analytics). As expected of a young research field, there are several articles within this category. For instance, publications like Siemens' (2013b) establish a strong foundation towards the development of analytics models, while others (Chatti et al., 2012; Greller & Drachsler, 2012; Siemens & Baker, 2012) defined objectives, research directions, and identified stakeholders. Interestingly, the behavior modeling and outcomes analysis topic was the second most frequent topic. A typical example is a study by Lowes, Lin, & Kinghorn (2015) who examined the association between students' online behavior and their final outcome. Next in order are log data analysis and data mining techniques. The log data category papers consider log systems and applications. Articles in this topic such as the one by Khalil and Ebner (2016e) or Ogata and Mouri (2015) focused on adopting learning analytics to monitor and track learners as well as discovering hidden patterns.

Next, we examined the changes in topics distribution across the six years of our analysis (see Figure 19). We see a sharp drop in the web analytics and social network analysis topics, while the topics of MOOCs, predictive analytics, and learning analytics adoption challenges show an increasing trend over time. In the majority of the remaining topics we see a fluctuation in the topic distribution over time which might be an indication of a revival in some of the learning analytics interests (e.g., learning analytics field itself, provision of feedback, and visualizations) or the nuance of our particular dataset, which — despite our best efforts — is still only a subset of the all learning analytics research.

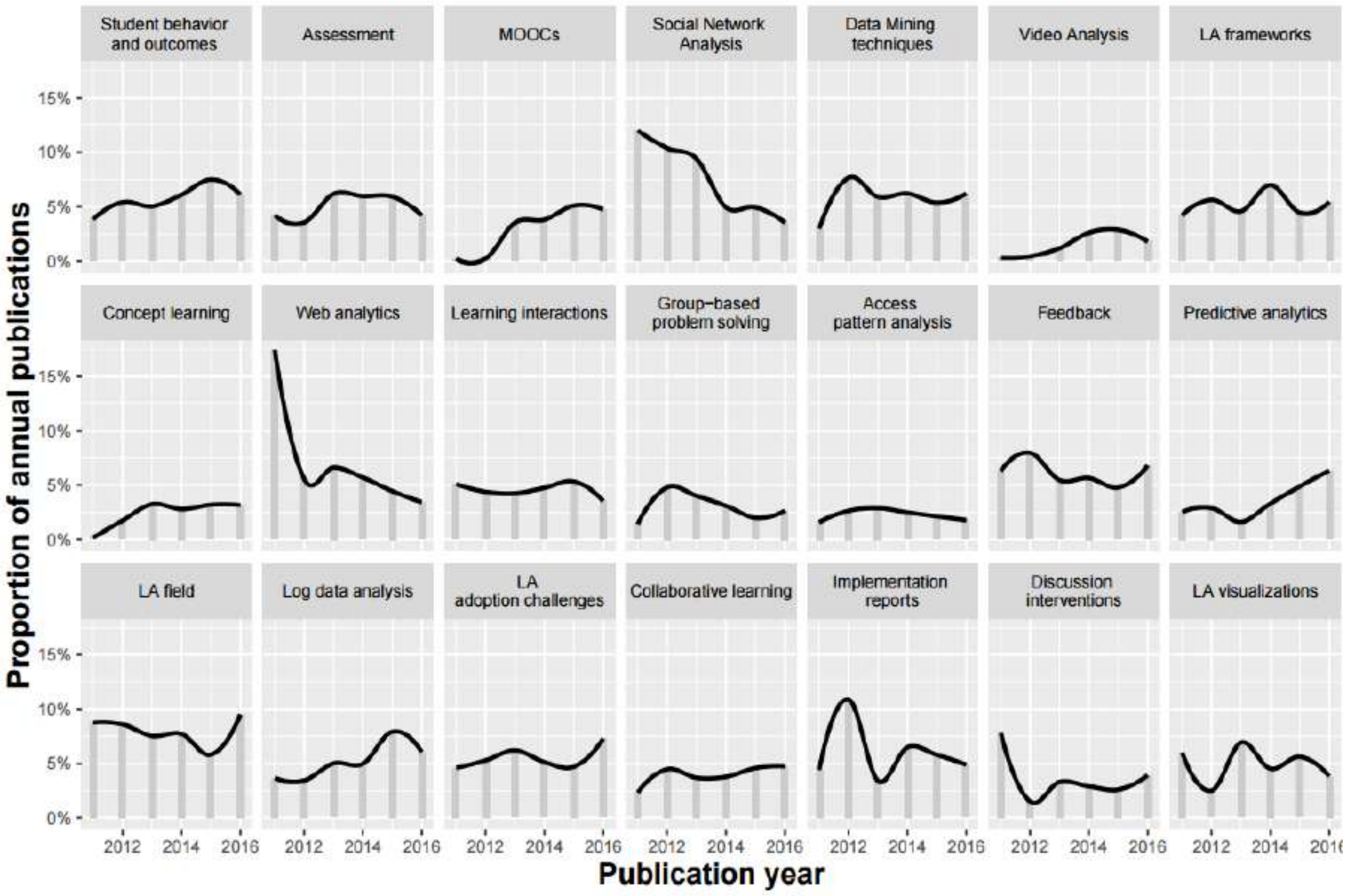

**Figure 19.** Changes in learning analytics topic distribution over the last six-year period

With the increased use of data, learning analytics research calls for the increased importance of learning theories in analysis and interpretation of large amounts of data (Wise & Shaffer, 2015; Dawson, Mirriahi, & Gašević, 2015). Thus, it is no surprise that discussion around building predictive analytics, providing feedback, challenges, and issues with learning analytics adoption or learning theories and theoretical frameworks, among others, is receiving significant attention in learning analytics research.

#### 3.4.4.4 Learning Analytics Topic Modeling Discussion

Perhaps one of the most striking findings of our study is the decreased focus on social network analysis (SNA) in studying learning and teaching practices in learning analytics research (Figure 19). As one of the hallmarks of learning analytics, SNA has been applied in a wide variety of settings, ranging from formal to informal educational contexts primarily studying interactions within discussion forums (Zhu et al., 2016; Koulocheri & Xenos, 2013; Schreurs et al., 2013). This finding, however, should be observed with respect to the total number of studies published in recent years. Specifically, the fact that SNA has been applied in 5% of the total number of studies published in 2015 (Table 6) still includes a considerably higher number of publications than was the case in 2011, for example. This could further mean that with the increased number of studies published in the learning analytics research field, other methods, such as machine learning or multimodal analytics as a broad term that captures advanced sensing and artificial intelligence technologies, are taking their place in the plethora of analytical approaches applied in understanding and improving learning and learning environments.

It is also interesting to note that the discussion of the main concepts of learning analytics attracted increased attention in recent years, after a period of noticeable decrease of focus. Specifically, in the first several years, researchers aimed at defining the field and establishing the main postulates within the learning analytics research (Siemens & Baker, 2012). Moreover, the link between learning analytics and other educational fields was quite normal at the early stages since it builds on and connects to other important domains of cyber-learning, user modeling, computer supported collaborative learning, and technology-enhanced learning (Gašević et al., 2015). In recent years, however, this focus has shifted towards addressing some of the identified challenges and a call to reexamine the position of theory in learning analytics research (Dawson, Mirriahi, & Gašević, 2015; Wise & Shaffer, 2015).

Likewise, the discussion about data mining techniques and computer algorithms/methods were reported widely in publication abstracts. Data mining has a strong interdependency with exploratory data analysis (Tukey, 1977) and is referred to as knowledge discovery that involves methods for relationships finding (Slater et al., 2016). It is also not surprising that many articles talked about data logs and the concurrent techniques from the data mining field. They are commonly practiced to identify patterns of log files to understand learner attitude (Baker & Yacef, 2009).

As noted by Dawson et al. (2014), much of the learning analytics research focuses on building predictive models for identifying students at risk or predicting final learning outcomes. This trend has also been observed in the topics extracted from the abstracts included in the analysis. It seems, however, that the increased interest in predictive analytics and examination of the association between the metrics of student behavior and learning outcome have been, to a certain extent, caused by advances in MOOC research. Moving from observational studies that aimed at investigating participation patterns in learning at scale, research on MOOCs shifted towards examining factors that might influence learning (Reich, 2015). Specifically, relying on various methods of learning analytics and educational data mining, this stream of research examined to what extent data collected by MOOC platforms can help predict learning outcomes and scaling learning interventions (DeBoer et al., 2014).

Besides previous observations, there are some remarkable annotations that were quite interesting to us. "Intervention" is mostly associated with institution and higher education. This reflects the value of employing data to guide decision-making in formulating higher education systems. Other keywords were completely absent because they were not prominent enough in abstracts. For instance, terms related to dropping out of courses, students at risk, and ethical and privacy issues were not present. Despite the extensive conversations around attrition and retention, there were few studies that highlight these problems. Yet, such topics could be strongly related to predictive analytics which is an increasing trend progress as shown in Figure 19.

It is also interesting to note that interest in visualizations in learning analytics research tends to vary over the years (Figure 19). Being discussed in the considerable number of studies in 2011 (Duval, 2011; Vatrapu et al., 2011), this number dropped the very next year. However, given the current patterns and increased tendency in utilization of data mining techniques, provision of formative and summative feedback, and predictive analytics, among others, it seems reasonable to expect an increased focus on the development of visualizations in learning analytics within the next several years. Specifically, visualization has been commonly applied to track and analyze data obtained from both students and instructors and to inform learning-related decisions and interventions. Focusing on novel methods of measuring and understanding learning could potentially result in solutions that would be integrated into the existing learning platforms. Moreover, developing data visualizations that promote learning and avoid negative effects identified in the literature is of critical importance.

In additional to the previous observations, we were enthused to look for security, ethical constraints and policy matters in the document corpora. Our topic modeling approach was incapable of identifying the “policy” keyword for instance. Still, the topic "learning analytics adoption challenges" located terms associated with issues and concerns of the learning analytics field. Identifying such crucial keywords might be absent at the current research time as there has been an increased coverage only in recent times. An example of such a forum is the series of Ethics and Privacy in Learning Analytics workshops by the LACE project and special issue in the *Journal of Learning Analytics* (Gašević, Dawson, & Jovanović, 2016).

#### 3.4.4.5 Learning Analytics Key Trends Conclusions

It is believed that what has been covered by learning analytics is just the “tip of the iceberg” in comparison to what occurs in other educational settings (Hershkovitz et al., 2016). In the topic modeling study, we introduced in the previous section, we employed an automated discovery model on a big dataset of learning analytics literature peer-reviewed publications from conference proceedings, journal articles, books and book chapters. Our topic modeling approach resulted in an optimal number of 21 main categories. The most dominated topics from these results concur with, to some extent, what Dawson and Siemens (Dawson & Siemens, 2014) previously defined: “The field draws on and integrates research and methodology related to data mining, social network analysis, data visualization, machine learning, learning science...” (p. 4). However, this research study shows a plethora of other far-reaching topics covered by the field.

We found that the general trend of MOOCs, challenges, and predictive analytics seem to be promising for future topics of learning analytics. Despite their growth in 2011, it was surprising that there was no coverage of learning analytics in MOOCs in the period between 2011 and 2012. Research on discussion forums might be on the rise in future publications. The topic of web analytics was conspicuous in 2011 but is dramatically decreasing. This is likely due to the development of specialized tools for data collection and analysis in learning analytics.

Our analysis also identified other frequent topics related to the research state of the field itself, student behavior modeling, log file analysis and data mining techniques. Although we identified some emerging themes such as predictive models, provision of feedback and an increased focus on learning analytics research in MOOCs, social network analysis seems to lose sight of future learning analytics research.

## 3.5 Learning Analytics of Massive Open Online Courses

The growth of MOOCs in the modernistic era of online learning has seen millions of enrollments from all over the world. As previously defined, MOOCs are online courses that are open to the public, with open registration options and open-ended outcomes that require no prerequisites or fees (McAulay, Tewart, & Siemens, 2010). These courses have brought drastic actions to higher education from one side and to elementary education from the other side (Khalil & Ebner, 2015a). Learners of MOOCs are not only considered as consumers, but they are also generators of data (Khalil & Ebner, 2015b). MOOCs have the potential of scaling education in different fields and subjects. The anticipated results of MOOCs varied between business purposes like saving costs and improving the pedagogical and educational concepts of online learning. Since MOOCs are an environment of online learning, the educational process is based on video lecturing. In fact, learning in MOOCs is not only exclusive to that, but social networking and active engagement are major factors, too (McAulay, Tewart, & Siemens, 2010). Contexts that include topics, articles or documents are also considered as supporting material in the learning process. Nevertheless, there is still disagreement about the pedagogical approach of information delivery to the students. The quality of the offered courses, completion rate, lack of interaction, motivation, engagement, and grouping students in MOOCs have been, in addition, debated recently (Clow, 2013; Khalil & Ebner, 2014).

While MOOC providers initialize and host online courses, the hidden part is embodied in recording learner activities. Nowadays, ubiquitous technologies have spread among online learning environments and tracking students online has become much easier. The pressing needs of ensuring that the audience of eLearning platforms is getting the most out of the online learning process and the need to study their behavior compelled the emergence of the field of learning analytics. Lately, the research portion in the study of the behavior of online students in MOOCs has become widely spread across journals and conferences. Our survey study on learning analytics (the full study available in Chapter 3.4.3) showed that the ultimate citation counts using Google Scholar (scholar.google.com) were relevant to MOOC articles (Khalil & Ebner, 2016d).

In fact, the use of learning analytics in MOOCs is still a very young research area. We saw in the earlier section of learning analytics trends in MOOCs that topics of both fields were nearly absent before 2013. There might be previous studies of data analysis in educational datasets of MOOCs with different names, but that does not ensure, to some extent, good coverage of learning analytics in massive open online courses. Also, some experts confirmed that there are few studies in both areas. Researchers like (Moissa, Gasparini & Kemczinski, 2015; Vogelsang & Ruppertz, 2015; Kloos et al., 2016; Ruipérez-Valiente et al., 2016) mentioned that learning analytics in MOOCs studies are still not deeply researched.

Learning analytics of MOOCs has also seen very few frameworks that connect both areas together. One of the few frameworks on learning analytics of MOOCs was presented by Drachsler and Kalz (2016). Drachsler and Kalz (2016) introduced a conceptual framework that provides the interplay between learning analytics and MOOCs, called the MOOC Learning Analytics Innovation Cycle (MOLAC). Their framework (see Figure 20) is divided into three levels: macro, micro and meso levels. The macro level represents interventions of learning and teaching. The work by Alario-Hoyos, Muñoz-Merino, Pérez-Sanagustín, Delgado Kloos and Parada G. (2016) gives an example of the micro level of the MOLAC framework where activities of learning analytics and data collection are focused on reflection and prediction. The research work by Rayyan et al. (2016) is an example of a meso level of the MOLAC cycle where insights about behavior are made for a group of students rather than individuals. Although the framework is considered an innovative step towards the research of learning analytics and MOOCs, it lacks scientific papers on the meso level. The authors did not provide enough explanation of the macro level as well.

Another simple framework called "the funnel of participation" by Clow (2013) connected MOOCs with learning analytics where the framework is adopted from the marketing funnel. The framework is summarized into four main stages, including awareness, interest, desire and action.

The author argued that the funnel of participation framework can define dropout in a different way than what is usually known in the traditional learning.

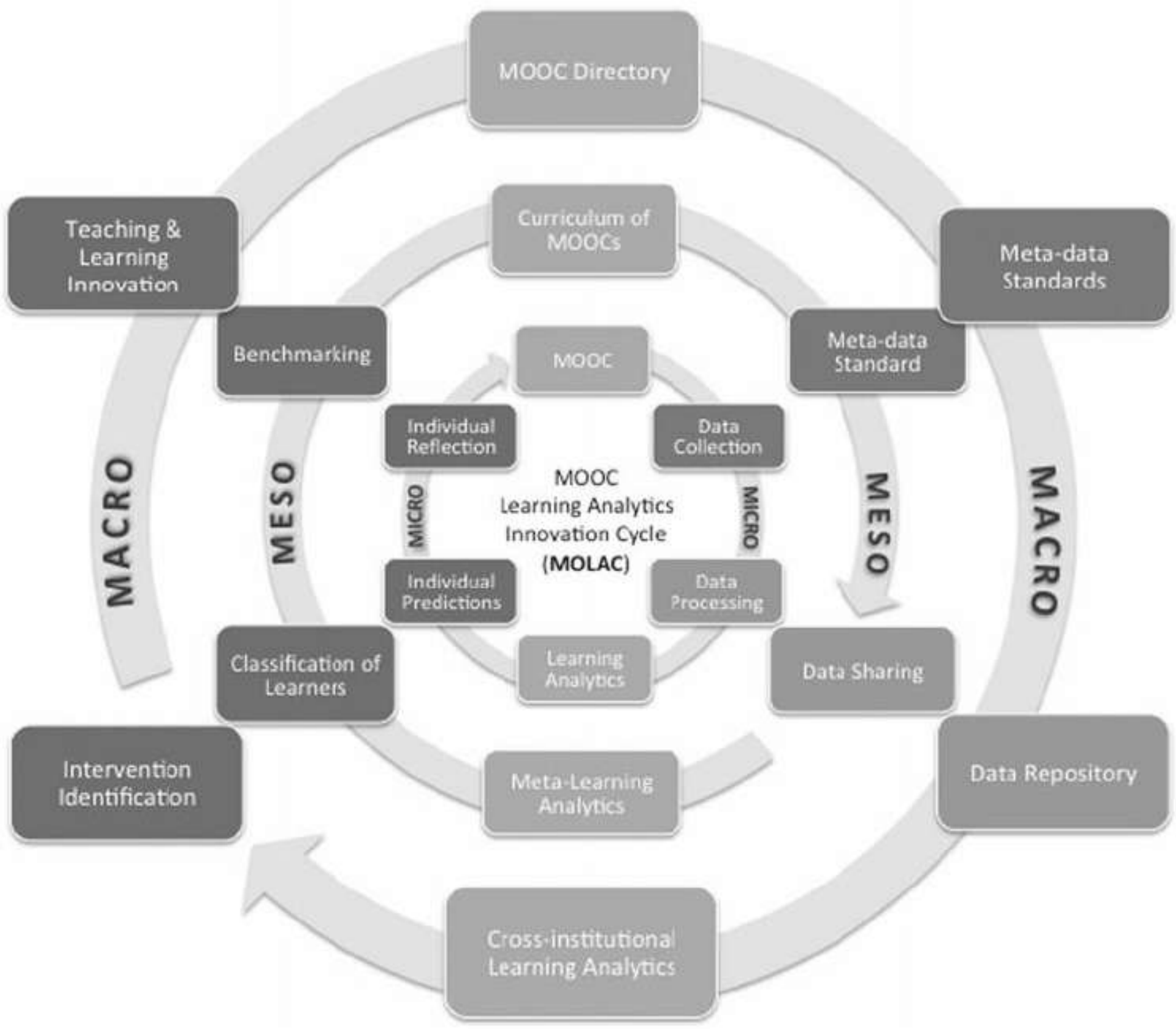

**Figure 20.** The MOLAC conceptual framework by Drachsler and Kalz (2016)

### 3.5.1 Systematic Key Trends of Learning Analytics of MOOCs

In this section, we did a brief text analysis and mapped the screening of the abstracts from the Scopus database in order to:

- Grasp what has been researched in learning analytics of MOOCs.
- Realize the main research trends of the current literature on learning analytics and MOOCs.

Scopus is a database powered by Elsevier Science. Our selection of this library is because of the valuable indexing information it provides and the usability of performing search queries. The conducted literature exploration was performed by searching for the following keywords: “Learning Analytics” and “MOOC,” “MOOCs” or “Massive Open Online Course.” The query used to retrieve the results was executed on 11- April- 2016 and is shown in Figure 21. The language was refined to the English articles only.

```
Your query : ((TITLE-ABS-KEY("Learning Analytics" AND "MOOCs") OR TITLE-ABS-
KEY("Learning Analytics" AND "MOOC") OR TITLE-ABS-KEY("Learning Analytics" AND
"Massive Open Online Course")) AND ( LIMIT-TO(LANGUAGE,"English" ) ) )
```

**Figure 21.** Search query to conduct the literature mapping of "learning analytics" and "MOOCs" from Scopus digital library

The returned results equaled 80 papers. Only one paper was retrieved in 2011, none from 2012, 11 papers from 2013, 23 papers from 2014, 37 from 2015 and 8 papers from 2016. Abstracts were then extracted and processed to a Comma-Separated Values (CSV) file. After that, we created a word cloud in furtherance of representing text data to identify the most prominent terms. Figure 22 depicts the word cloud of the extracted abstracts. We looked at the single, bi-grams, tri-grams and quad-grams common terms. The most repeated single words were "MOOCs," "education," and "engagement." On the other hand, "learning analytics," "online courses," and "higher education" were recorded as the prominent bi-grams. "Khan Academy platform" and "Massive Open Online Courses" were listed on the top of the tri-grams and quad-grams respectively. As long as massive open online courses were represented in different terms in the abstracts, we abbreviated all the terms to "MOOCs" in the corpus.

Figure 23 shows the most frequent phrases fetched from the text. Figure 22 and Figure 23 show interesting observations of the researched topics of learning analytics in MOOCs. By doing a simple grouping of the topics and disregarding the main phrases which are "Learning Analytics" and "MOOCs," we found that researchers were looking mostly at the engagement and interactions.

It was quite interesting that the dropout and the completion rate were not the major topics as we believed. Design and framework principles as well as assessment were ranked the second most referenced terms. Social factors and learning as well as discussions grabbed attention afterwards, while tools and methods were mentioned to show the mechanism done in offering solutions and case studies.

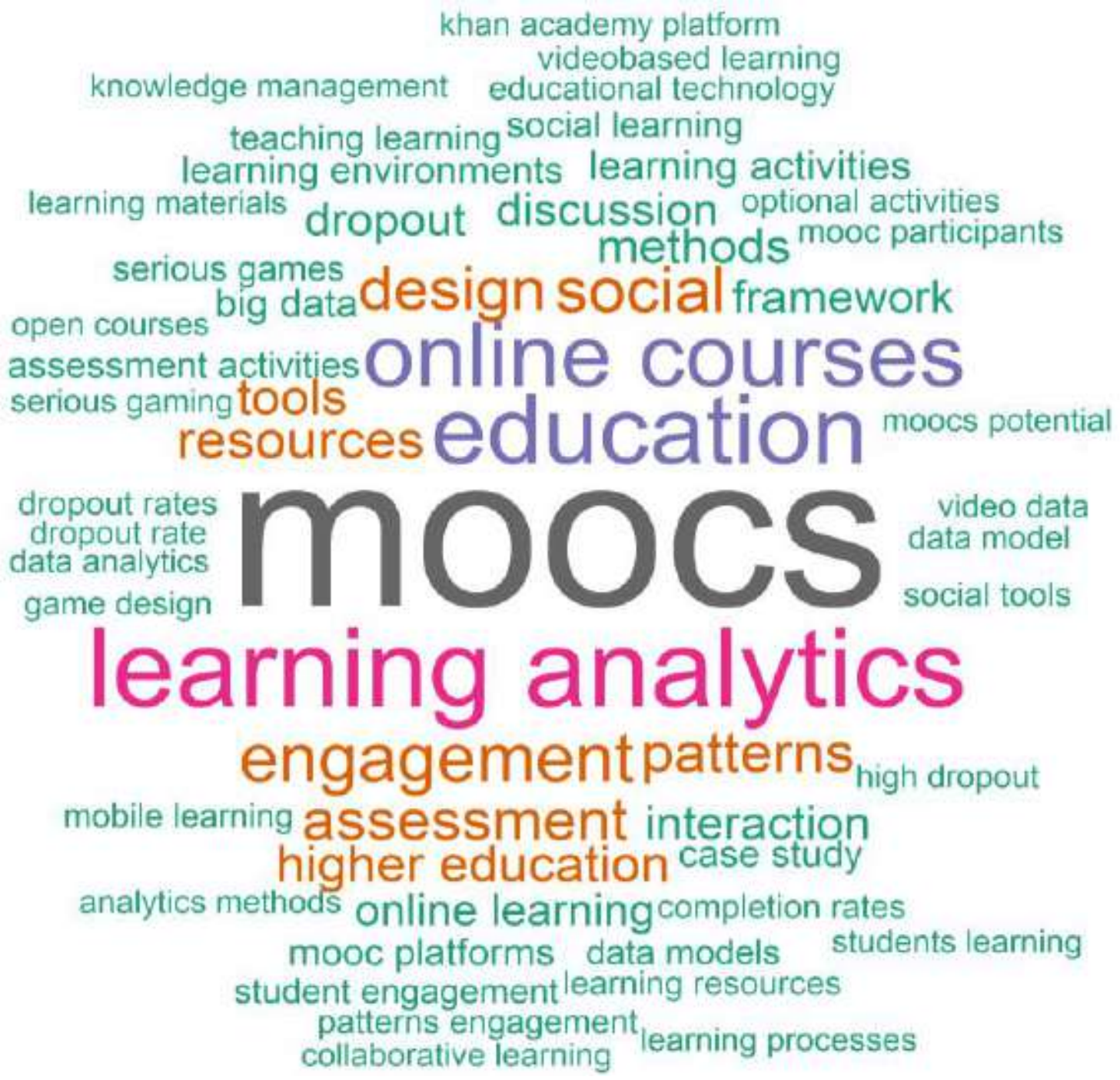

**Figure 22.** Word cloud of the most prominent terms from the abstracts of articles "learning analytics" and "MOOCs" from Scopus digital library

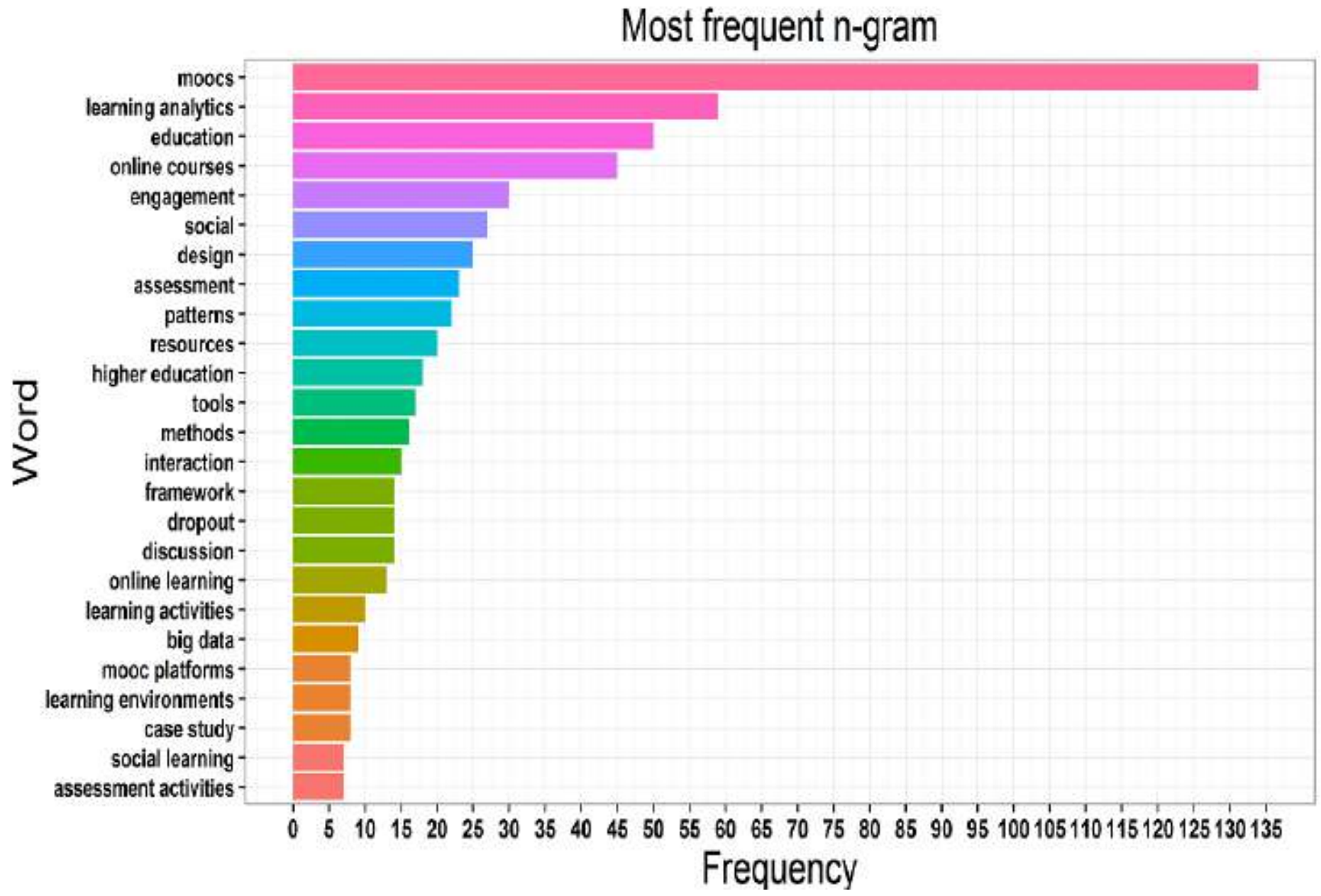

**Figure 23.** The most frequent terms extracted from the abstracts of articles "learning analytics" and "MOOCs" from Scopus digital library

It is believed that learning analytics can drive numerous benefits to MOOCs as an environment and to the learner as a stakeholder. Yet, advantages can go beyond that and cover other stakeholders like teachers and decision makers (Chatti et al., 2012). Knox (2014) mentioned that learning analytics brings great potential to MOOCs. He said that "Learning analytics promises a technological fix to the long-standing problems of education" (Knox, 2014).

Various related work and articles of learning analytics practices in MOOCs will be briefly discussed in Chapter 5 prior to each case study.

## 3.6 Chapter Summary

This chapter provided a broad literature review study and a background of the main content of this thesis. We first talked about how e-learning emerged, followed by distance learning development and open educational resources emergence. The literature, thereafter, introduced what are MOOCs and added our experience in evaluating multiple MOOCs from various MOOC-providers. Our evaluation contributed to an *xMOOCs evaluation grid* which can be used to investigate other extended massive open online courses in the future. Next, we talked about learning analytics, showed how learning analytics was derived from other fields of web analytics, academic analytics, and educational data mining. Furthermore, we elaborated the connection between learning analytics and higher education, who the stakeholders are and what methods have been used to do learning analytics. We then introduced our learning analytics method survey in the period between 2013 and 2015 from LAK papers; the results show that learning analytics use methods from statistics, data mining, and NLP the most. Likewise, our extended systematic review of 1,315 papers from different sources showed that predictive modeling as well as research in MOOCs and feedback were the main key trends of the research of learning analytics.

Finally, we talked briefly about learning analytics of MOOCs and revealed that learning analytics has the potential to tackle MOOC dilemmas. Although learning analytics is thought to provide solutions to the dropout problem of MOOCs, the systematic abstract scanning of 80 articles demonstrated that learning analytics has the potential to deal with motivation, engagement and interaction issues of MOOCs.

# 4 LEARNING ANALYTICS FRAMEWORK AND PROTOTYPE DESIGN[5]

It is now obvious that the Internet has altered the learning models of educational institutions in schools, academies, and universities. Learning through technology, and specifically online learning, offers the flexibility of access anytime and anywhere (Cole, 2000). At the moment, students can access learning materials, take quizzes, ask questions, engage with their colleagues and watch learning videos through the Internet. On the other hand, teachers can examine their students' performance through different applications which ease their supervision duties.

Concepts of traditional learning have changed, and the upcoming technologies have created new learning environments that did not exist previously, like MOOCs. Despite the massive quantity of learning contexts, each learning environment is a unique system by itself.

MOOCs have reserved a relevant and valuable position in educational practice from various perspectives. The further collected data from MOOCs open the doors to analyze their data to improve learning, teaching and the environment itself. This chapter introduces the Austrian MOOC platform (iMooX) and describes our implementation of employing learning analytics through developing a tool that will construct the foundation of collecting, processing, and revealing hidden patterns of student data for research purposes in this dissertation. Later, the tool will be evaluated based on short use cases.

## 4.1 The iMooX MOOC Platform

Since education in the countries of Central Europe is primarily offered as face-to-face classes, the need for distance learning was a must, especially for a country like Austria (Neuböck, Kopp &

[5] Parts of this chapter have been published in:

Khalil, M., & Ebner, M. (2015b). Learning Analytics: Principles and Constraints. In *Proceedings of World Conference on Educational Multimedia, Hypermedia and Telecommunications* (pp. 1789-1799). AACE

Khalil, M. & Ebner, M. (2016e). What Massive Open Online Course (MOOC) Stakeholders Can Learn from Learning Analytics? In Spector, M., Lockee, B., Childress, M. (Eds.), *Learning, Design, and Technology: An International Compendium of Theory, Research, Practice, and Policy*, Springer International Publishing. (pp. 1-30). Doi: 10.1007/978-3-319-17727-4_3-1

Ebner, 2015). As long as MOOCs provide an excellent opportunity for online learners and a strong basis of distance education, this paves the way to encourage decision makers to adopt a project to establish a MOOC platform in Austria. iMooX is an online learning stage and the first Austrian xMOOC platform was founded in 2013 as a result of collaboration between the University of Graz and Graz University of Technology. The platform went online in February 2014 (see Figure 24).

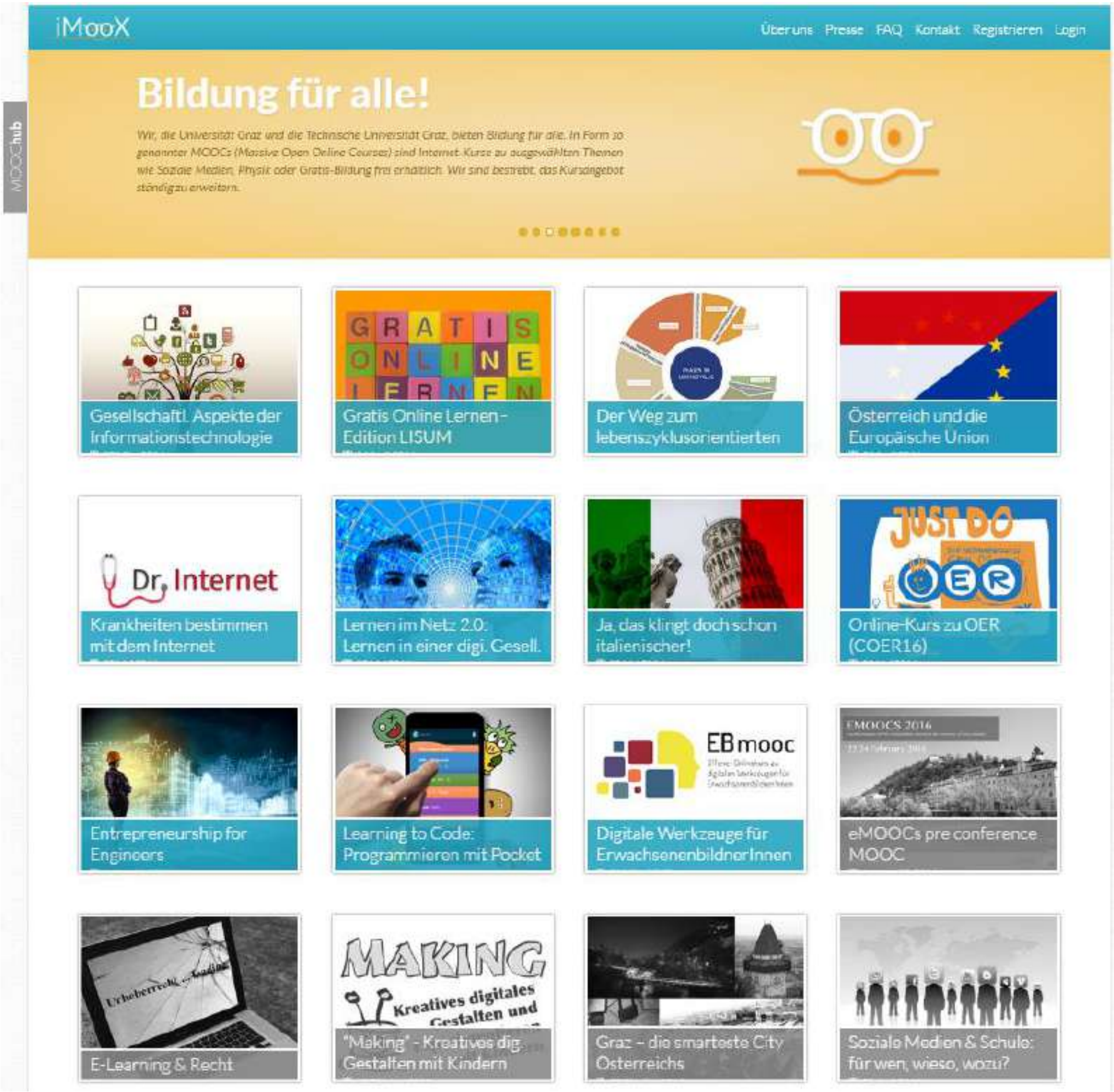

**Figure 24.** iMooX MOOC platform homepage

iMooX has enthralled over 5000 users from different participant target groups. The main idea behind the platform was to introduce explicit Open Educational Resource (OER) courses, keep pace with Open Education and lifelong learning tracks, and to attract a public audience extending from schoolchildren to elderly people, or to academic degree holders (Fischer et al., 2014).

A recent study done in 2015 based on three offered courses revealed some demographic information about iMooX participants (Neuböck, Kopp & Ebner, 2015). The research study showed that 65% of learners were male, 44% were aged between 20-34 years, and 25% were over 50 years old. On the other hand, the educational level status showed that most participants already had an academic degree, whereas less than 10% of students had no graduation or completed a primary school education.

The pedagogical approach of iMooX consists of offering courses to students on a weekly basis. One or more videos are presented each week in diverse styles (see Figure 25). In addition, documents, interactive learning objects, reference to topics in forums and articles on the Web are also offered. Usually, the duration of each course does not exceed more than an eight-week period with a convenient workload.

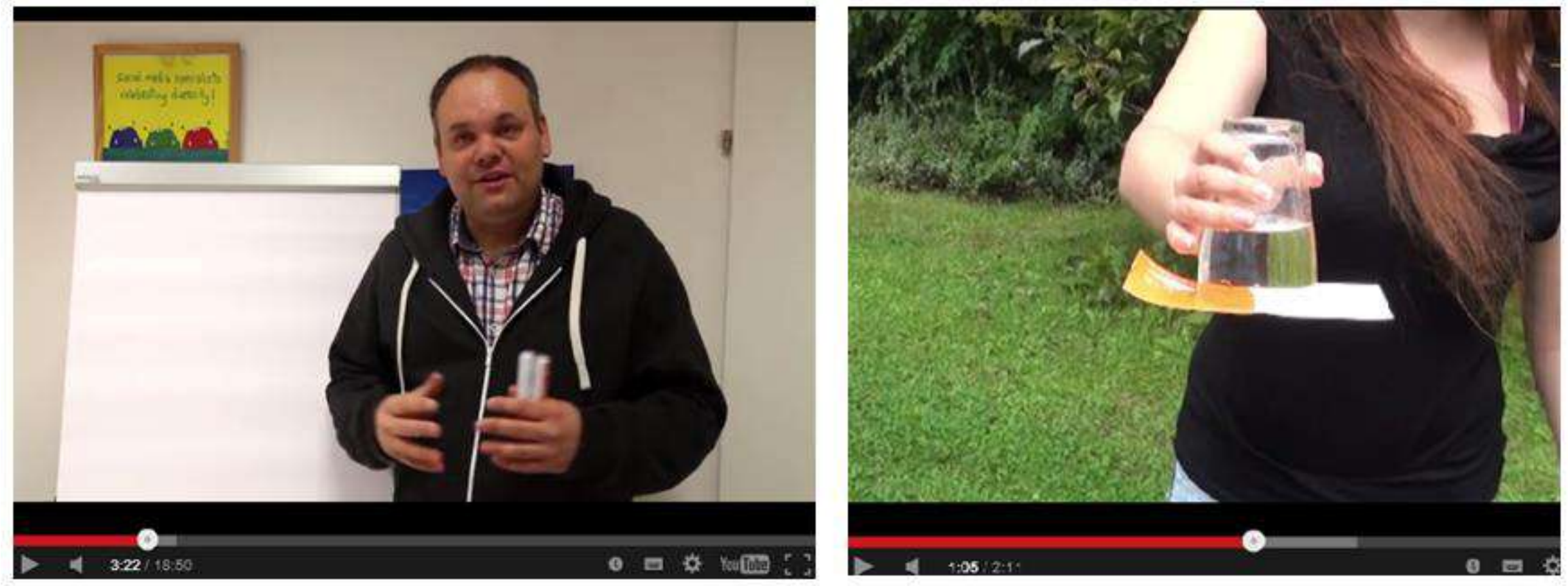

**Figure 25.** Video lectures in iMooX are presented in diverse styles. Left: personal presentation. Right: experiment presentation

The design of the iMooX platform endeavors to follow the cognitive-behaviorist pedagogy theme concepts of Gagne (1965):

- Acquiring the learners' attention by providing them the correct steps to gain the learning theory through the online education system
- Listing the objectives and learning goals of each online course
- Demonstrate the stimulus by presenting active online learning videos
- Giving feedback through discussion forums and regular emails

- Assessing performance through computerized assessment of the exams
- Providing guidance, which usually depends on the learners themselves where self-learning is imperative due to the online learning environment conditions

Furthermore, the platform also supports social-constructivist pedagogy. It proposes social discussion forums where learners get in touch with instructors as well as an information exchange that takes place between the students themselves (Khalil & Ebner, 2013).

German is the primary communication language of all courses provided. The online courses are presented on a weekly basis and vary in topics between Science, Technology, Engineering, Mathematics (STEM) as well as history and human sciences. Every week of each course consists of short videos and multiple choice quizzes. The quiz system is fairly different in the iMooX platform in which each student has the option to do five attempts per quiz and the system automatically picks the highest grade. There were two main reasons behind this; from the psychological point of view, the student is less stressed and behaves in a more comfortable manner, whilst researchers can study the participant's learning behavior based on the number of attempts made by the student (Khalil & Ebner, 2015a). The iMooX platform offers certificates to participants completely for free; it is only required that students successfully finish the quizzes and fill out an evaluation form at the end of each course in which they assess their own experience with the enrolled MOOC.

## 4.2 Learning Analytics Framework

In this part of the study, we present our conceptual framework that helped us summarize the learning analytics lifecycle and develop the iMooX Learning Analytics Prototype (iLAP). The learning analytics framework comprises processional steps, starting from the learning environment and ending with interventions and optimizations. We based this foundational framework on our literature studies in 2015 of what learning analytics frameworks and lifecycles are available. It is noteworthy to mention that the generated framework is a general learning analytics scheme that absolutely inspired us to develop the design ontology of the iLAP.

We gathered information about learning analytics from conference proceedings, workshop results as well as publications in different journals over the last four years. Furthermore, we looked at the current available frameworks and reference models. This motivated us to investigate further to propose an approach that presents a framework as well as a life cycle. We took into account the idea of closing the cycle loop (Clow, 2012), the current models by (Greller & Drachsler, 2012) and (Chatti et al., 2012) as well as the latest updates of learning analytics in the area. Afterward, we modeled our approach (see Figure 26).

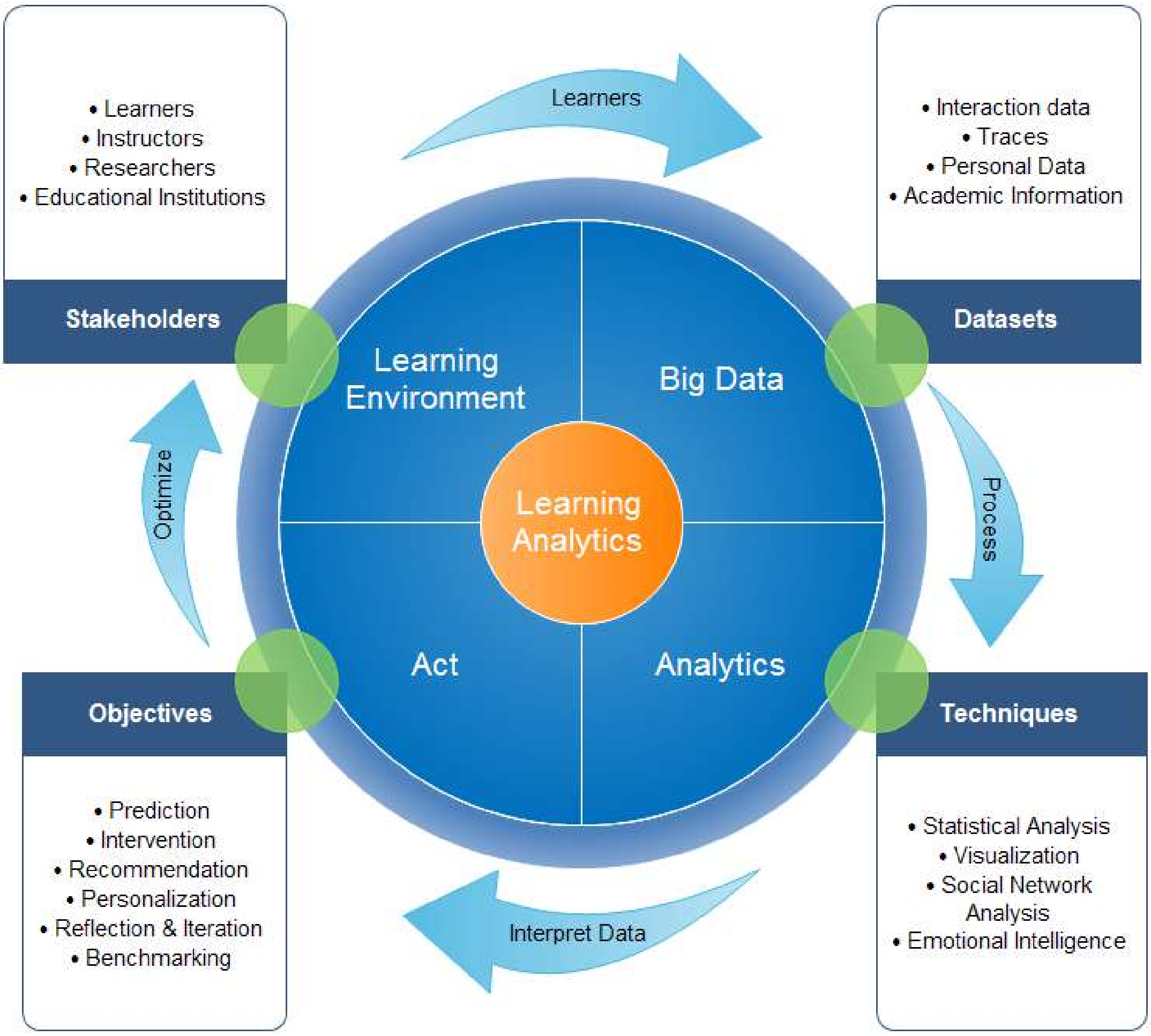

**Figure 26.** Learning analytics framework and lifecycle

The learning analytics framework in Figure 26 considers of four main parts: 1) Learning environment where stakeholders produce data; 2) Big Data, which consist of massive amounts of datasets and large repositories of information; 3) Analytics, which comprises different analytical techniques; 4) Act, where objectives are achieved to optimize the learning environment. Next, we talked about these main parts in detail.

### Learning Environment

With the ubiquitous technologies spread among education, there is a large collection of educational and learning environments involved, such as: Personal Learning Environments (PLE), Adaptive Hypermedia educational systems, Interactive Learning Environments (ILE), LMS, and MOOCs. All these learning environments are a goldmine of data that learners leave behind (Romero & Ventura, 2010). For example, logging a mouse click by its x and y coordinates, or the menu items times clicks or the time a student spends on a question can produce a huge amount of data that can be analyzed to provide information about the students' motor skills (Mostow & Beck, 2006). The learning environment has many aspects, but in this proposed learning analytics cycle, the focus will be on the actors/stakeholders.

There are different groups who are engaged in learning analytics. Each group can benefit according to their vision and mission. For instance, learning analytics is advantageous to support people in clarifying and relating information, peer learners and digital artifacts and to support people in pursuing their learning (Fournier, Kop & Sitlia, 2011). Taking into account the stakeholders' roles from our literature review studies, the focus of learning analytics and learning environments like MOOCs is mainly around learners. Nevertheless, other actors like teachers, decision makers, and researchers are also involved. Learners look to enhance their performance, personalize online learning, and get recommendations. Instructors look to enhance their teaching methods, provide real-time feedback to students and monitor the learning progress of their students. Researchers evaluate courses, improve course models and discover new ways of delivering educational information through analysis of data. Finally, educational institutions/decision makers look at supporting decision processes to achieve higher educational goals like increasing completion rate and monitoring courses.

### Big Data

Learning analytics consider mining learner activities. Most learning analytics definitions reference learners as the main actor in the learning analytics process (Siemens, 2010; Duval, 2011; Ebner & Schön, 2013; Taraghi et al., 2014). Learners leave a lot of data behind them when using any learning environment. In the old educational methodologies, the learner is considered as a consumer. She/he has no possibility to be an active participant in the education process. On the other hand, with learning analytics, learners are not only consumers but have also become producers of data.

In educational environments, there are different types of data to be processed. These data are restricted to the area of education and therefore have authentic semantic information (Romero & Ventura, 2010). Manyika et al. (2011, p. 1) defined "Big Data" as "the reference to datasets whose size is beyond the ability of typical database software tools to capture, store, manage, and analyze." While learners are using the educational platforms, they generate data. This yields to repositories of datasets. These datasets include, but are not limited to:

- Interaction data; such as the data that is related to visualizations and forums discussions
- Traces; which can be number of logins, mouse clicks, number of accessed resources, number of finished assignments, videos accessed, documents accessed, files downloaded, questions asked, discussions involved, and social network activities; such as tweets, blogs, and comments
- Personal data: name, date of birth, local address, email address, personal image, ID or any other personal related information
- Academic Information; these are courses attended, grades, graduation date, exams taken, certificates, etc.

While Big Data includes this large amount of educational information, it should be searched, processed, mined and visualized in order to retrieve a meaningful knowledge.

**Analytics**

There are different methods to analyze data in the atmosphere of education. These analytics methods seek to discover interesting patterns hidden in the educational datasets. Learning analytics techniques use various types of analytical methodologies. These techniques and methods have been deeply discussed and explained in our literature review studies in the related work chapter (*Chapter 3.4.2 and Chapter 3.4.3*).

**The Act**

In this stage, the analysis outcome is interpreted to achieve desired objectives of learning analytics. The greatest value of learning analytics comes from optimizing the objectives, as interventions that affect the learning environment and its stakeholders (Clow, 2012). Goals vary between performing prediction of performance and dropout, making an intervention such as preventing drop-out, determining students at risk and advising students who might need additional assistance to improve student success,recommending and personalizing,enhancing reflection and iteration through a self-evaluation process by the students themselves,benchmarking and evaluation of courses and identification of weak points in learning environments and course instructional design. Further different actions can be found in our survey studies in the literature chapter.

## 4.3 iMooX Learning Analytics Prototype (iLAP)

A MOOC platform cannot be considered as a real modern technology enhanced learning environment without a tracking approach for analysis purposes. Tracking student-left traces on MOOC platforms with a learning analytics application is essential to improve the educational environment and understand students' needs. Therefore, our task was to make iMooX anticipate the steps by having an analytical approach named as the "iMooX Learning Analytics Prototype." iLAP intends to track students for research purposes. It embodies the functionality to interpret low-level data and present them to administrators and researchers. The iMooX learning analytics tool was developed through a master thesis research project (Moser, 2015).

The iMooX Learning Analytics Prototype is built based on the learning analytics framework introduced in the previous section. The lifecycle was adopted in order to enhance the framework and to apply it successfully to the MOOC platform to glean the educational context of the courses directed toward the benefit of various types of learners. The overall goal of this prototype is to integrate a real analytics tool into a MOOC platform and to render useful decisions based on educational and pedagogical approaches. The iLAP followed the "Prototyping" methodology discussed in chapter 2 by which different versions was tested before the present version. Currently, the prototype is available for use by administrators, researchers, and decision makers. Instructors can ask for student results regularly upon request. The iMooX managing institution dedicates diligence to the ethical and security dilemmas and constraints due to the extreme restrictions on student privacy regulations in Austria. According to the European Law Directive 95/46/EC[6], there are restrictions on the information disclosure of students until a clear consent or a truly anonymous technique is applied.

The overall design of the iMooX Learning Analytics Prototype was to propose a tool that provides the MOOC administrators with a proper interpretation of the bulk data that is generated by the learners. The complexity of log files that the web server produces has been taken into account, and which is responsible for passing the student-left traces to the learning analytics server. A proper processing method with the particularity of being reliable, fast and safe was therefore required for reviewing the log files in order to present them as readable information. The prototype was developed in virtue of four main stages with a reflective concept of optimizing the learning environment, which is the MOOC platform, and improving the MOOC stakeholders gained benefits, specifically learners and teachers. Figure 27 shows the main stages of the iLAP design architecture.

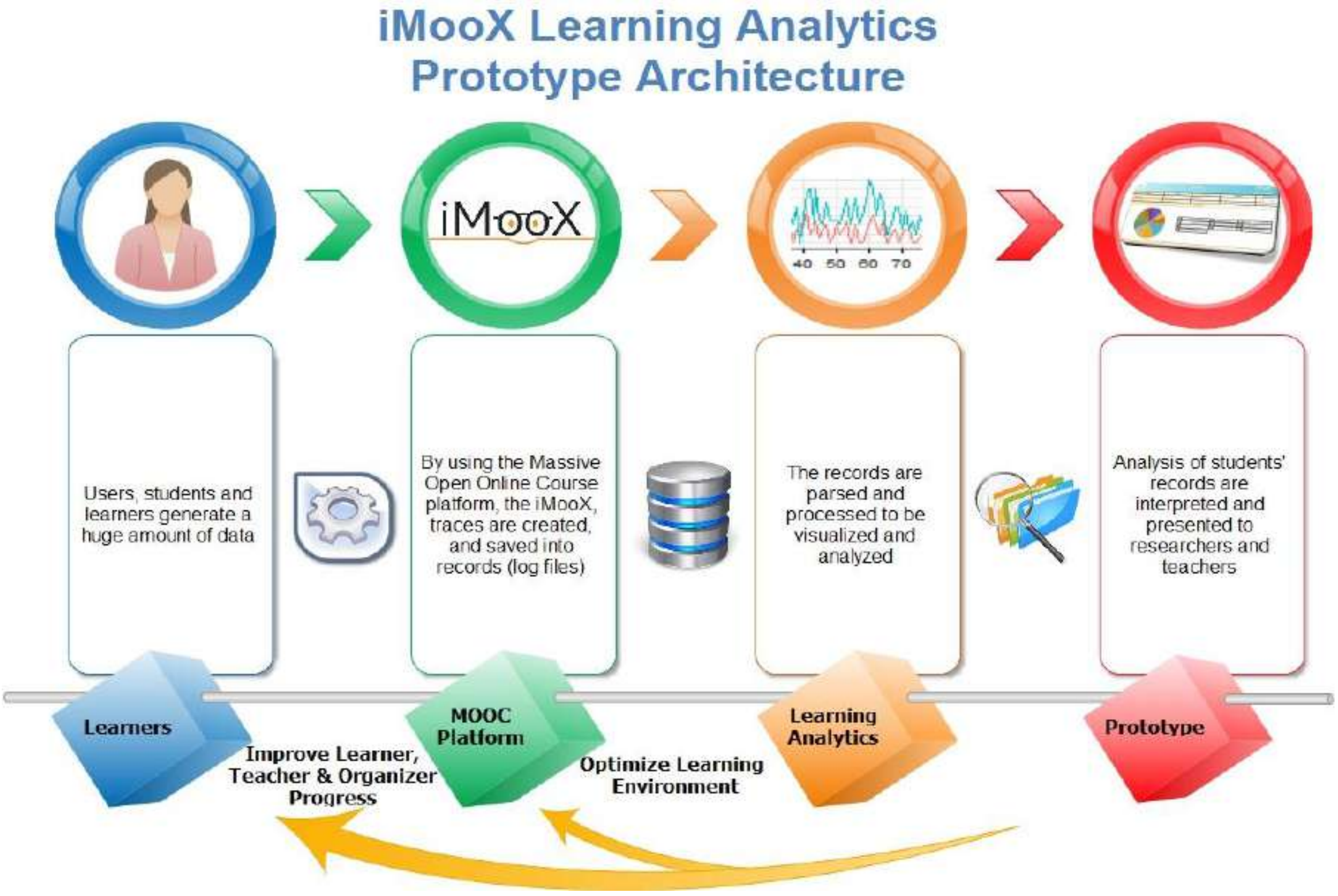

**Figure 27.** The iMooX Learning Analytics Prototype (iLAP) design ontology

---

[6]http://eur-lex.europa.eu/LexUriServ/LexUriServ.do?uri=CELEX:31995L0046:en:HTML (last access August 2015)

The first stage of the design architecture of the learning analytics tool begins by generating data in the learning environment of the MOOC platform. Whenever a user registers an account, enrolls in a course, watches a video or withdraws from a course, this is recorded and results in the generation of log files. A mass amount of log files creates what is called “Big Data.” Wherever large log files require better management, suitable data management and administration has been taken into account with the prototype framework. The next step is resolved by the web server which is responsible for collecting student information. Gathering user information is accomplished through tracking users on the MOOC platform. Traces of the students create time-referenced descriptions and accurate content that are gathered for designating features of learners and their interaction activities (Perry & Winne, 2006). In this stage, the system records several interactions such as the logging frequency, the total number of the course document downloads, number of readings in forums, the summation of posts per user, video interactions, total number of quiz attempts and quiz scores with the time frame manner of all activities. With all these activities, the stream of information flows to the main database to be parsed and processed in furtherance of becoming visualized to the end user.

In the third stage, learning analytics operations are performed by parsing the logs and processing them to filter the noisy data, since the data in the log files is unstructured, duplicated and not regularly formatted. The learning analytics server is thus programmed to synchronically organize log files and operate semantically to pick up keywords that help in detecting students activity inside the bulk text file, the log file. These keywords are relevant to what has been coded in the backend to pick the appropriate phrases to distinguish between student interactions. The collected data and the process of transforming it should cut the edge into meaningful MOOC indicators that reflect the activities of the users. Figure 28 shows a sample of a raw log file before being processed by the learning analytics server.

Finally, the collected and organized data are brought forward to be interpreted and visualized to the end user. In this stage, the learning analytics prototype is presented as a user interface for monitoring purposes and observation. The prototype user interface is only accessible by researchers and administrators at the moment. All the educational data sets collected by the prototype are secured by a Virtual Private Network (VPN) in order to enhance data protection against unauthorized access. The perception of the visualized results should guide the MOOC stakeholders to (1) benchmark the learning environment and its courses and (2) improve learner, teacher, and administrator progress for meeting the pedagogical practices of iMooX.

Learning analytics should provide powerful tools to support awareness and reflection (Verbert et al., 2013; Chatti et al., 2014). From the software side, iLAP is intended to show visualizations and to provide noiseless data for researchers, and from the awareness side, reflection of the conclusions of observations on the course developers, learners and teachers is contemplated.

```
Mon Mar 16 2015 06:47:44 GMT+0100 (CET)%:%ulistahl%:%https://online.tugraz.at/tug_online/visitenkarte.show_vcard?pPersonenId=99E141532528D1D7&pPersonenGruppe=3%;%[|||]Mon Mar 16 2015
08:20:47 GMT+0100 (Mitteleurop&#228;ische Zeit)%:%kathrinrefrei%:%http://elearningblog.tugraz.at%;%[|||]Mon Mar 16 2015 08:22:56 GMT+0100%:%martinlukas%:%http://www.abendblatt.de/kultur-
live/article2308283/Digitalitaet-ist-Chance-und-Bedrohung.html%;%[|||]Mon Mar 16 2015 08:30:26 GMT+0100%:%wald%:%http://www.abendblatt.de/kultur-live/article2308283/Digitalitaet-ist-Chance-und-
Bedrohung.html%;%[|||]Mon Mar 16 2015 08:30:43 GMT+0100%:%wald%:%http://de.wikipedia.org/wiki/Digital_Native%;%[|||]Mon Mar 16 2015 08:30:59 GMT+0100%:%wald%:%https://digitalegesellschaft.de/
%;%[|||]Mon Mar 16 2015 08:31:19 GMT+0100%:%wald%:%https://books.google.at/books?id=vZMCBQAAQBAJ&printsec=frontcover&dq=Die+digitale+Gesellschaft:+Netzpolitik,+B%C3%BCrgerrechte+und
+die+Machtfrage&hl=de&sa=X&ei=1ODLVOKkNlPqas7ngrgG&ved=0CDEQ6AEwAw#v=onepage&q&f=false%;%[|||]Mon Mar 16 2015 08:33:36 GMT+0100 (CET)%:%durany%:
%http://www.abendblatt.de/kultur-live/article2308283/Digitalitaet-ist-Chance-und-Bedrohung.html%;%[|||]Mon Mar 16 2015 08:37:48 GMT+0100 (CET)%:%durany%:%http://www.abendblatt.de/kultur-
live/article2308283/Digitalitaet-ist-Chance-und-Bedrohung.html%;%[|||]Mon Mar 16 2015 08:38:49 GMT+0100 (CET)%:%durany%:%http://www.faz.net/aktuell/feuilleton/debatten/frank-schirrmacher-ueber-
den-digitalen-wandel-und-die-offene-gesellschaft-12836746.html%;%[|||]Mon Mar 16 2015 08:39:13 GMT+0100%:%walterwoi%:%https://online.tugraz.at/tug_online/visitenkarte.show_vcard?
pPersonenId=99E141532528D1D7&pPersonenGruppe=3%;%[|||]Mon Mar 16 2015 08:41:08 GMT+0100 (CET)%:%durany%:%https://books.google.at/books?
id=vZMCBQAAQBAJ&printsec=frontcover&dq=Die+digitale+Gesellschaft:+Netzpolitik,+B%C3%BCrgerrechte+und+die
+Machtfrage&hl=de&sa=X&ei=1ODLVOKkNlPqas7ngrgG&ved=0CDEQ6AEwAw#v=onepage&q&f=false%;%[|||]Mon Mar 16 2015 08:54:02 GMT+0100 (Mitteleurop&#228;ische Zeit)%:%kathrinrefrei%:
%http://www.e-health-com.eu/who-is-who/einzelne-eintraege/schreier-dr-ing-guenter/%;%[|||]Mon Mar 16 2015 08:58:28 GMT+0100%:%walterwoi%:%http://www.abendblatt.de/kultur-
live/article2308283/Digitalitaet-ist-Chance-und-Bedrohung.html%;%[|||]Mon Mar 16 2015 08:59:43 GMT+0100%:%walterwoi%:%http://glossar.sozialebewegungen.org/digital-natives/%;%[|||]Mon Mar 16 2015
09:00:46 GMT+0100%:%walterwoi%:%http://www.marcprensky.com/writing/Prensky%20-%20Digital%20Natives,%20Digital%20Immigrants%20-%20Part1.pdf%;%[|||]Mon Mar 16 2015 09:04:17 GMT+0100
(CET)%:%evaseiler%:%http://www.e-health-com.eu/who-is-who/einzelne-eintraege/schreier-dr-ing-guenter/%;%[|||]Mon Mar 16 2015 09:04:36 GMT+0100 (CET)%:%perca%:
%https://online.tugraz.at/tug_online/visitenkarte.show_vcard?pPersonenId=99E141532528D1D7&pPersonenGruppe=3%;%[|||]Mon Mar 16 2015 09:04:37 GMT+0100 (CET)%:%perca%:
%http://elearningblog.tugraz.at%;%[|||]Mon Mar 16 2015 09:04:42 GMT+0100 (CET)%:%perca%:%http://www.abendblatt.de/kultur-live/article2308283/Digitalitaet-ist-Chance-und-Bedrohung.html%;%[|||]Mon Mar
16 2015 09:04:44 GMT+0100 (CET)%:%perca%:%http://glossar.sozialebewegungen.org/digital-natives/%;%[|||]Mon Mar 16 2015 09:04:46 GMT+0100 (CET)%:%perca%:
%http://www.marcprensky.com/writing/Prensky%20-%20Digital%20Natives,%20Digital%20Immigrants%20-%20Part1.pdf%;%[|||]Mon Mar 16 2015 09:04:47 GMT+0100 (CET)%:%perca%:
%http://www.faz.net/aktuell/feuilleton/debatten/frank-schirrmacher-ueber-den-digitalen-wandel-und-die-offene-gesellschaft-12836746.html%;%[|||]Mon Mar 16 2015 09:04:49 GMT+0100 (CET)
```

**Figure 28.** A sample log file that includes students' interactions before being processed through the iLAP

### 4.3.1 Implementation Framework

In this section of this chapter, the implementation framework of the learning analytics prototype is presented in Figure 29. Simply said, the implementation encompasses five steps. The first step starts with the MOOC platform where the learners initiate activities. The students' discussions and their interactions with learning videos as well as their progress in quizzes are noted in the log files. These log files are generated by the web server shown in the figure as the second step. The main structure of the web server belongs to the Apache HTTP Web Server family (http://httpd.apache.org/). With its convenient Graphical User Interface (GUI), error management tool and powerful security features, the working environment is pertinent to the desired needs. In the third step, the process proceeds to transfer the log files to the Log Files Management tool. The noisy data is filtered according to the description noted in the previous section, and the flood of logs is organized.

In the fourth step, where the core of the implementation framework resides, the learning analytics server parses the incoming log files from the management stage and differentiates between the learners' activities and extracts their timing frames. The server-side code is written in Python programming language. Whenever an activity is detected, the information is stored in an intelligent programmed database storage in which researchers have the option to browse it and operate different analysis or educational data mining techniques with high authentication and authorization criteria. This enhances the resilience for additional data processes to be added in the future either to the front end user or for research purposes.

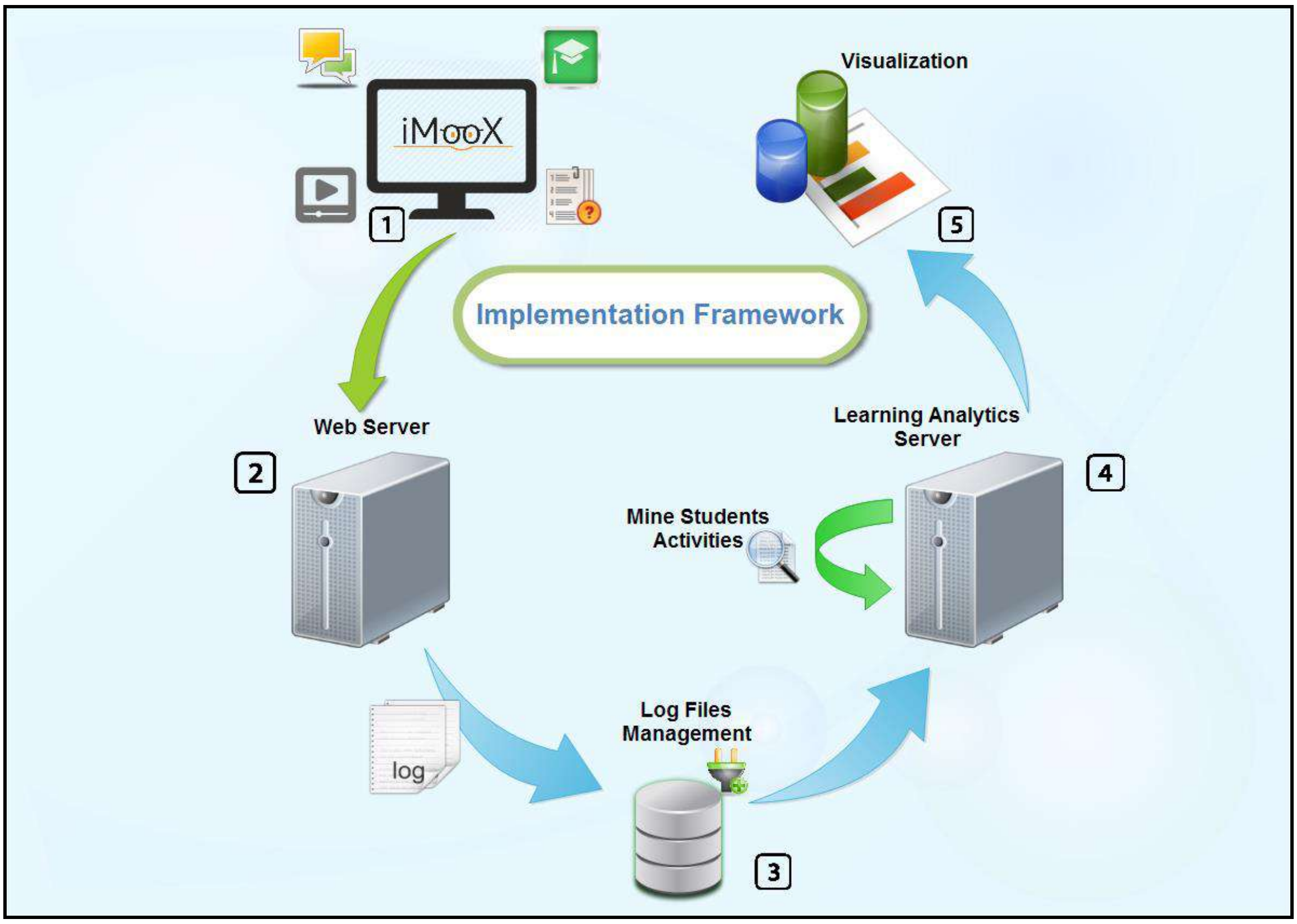

**Figure 29.** iMooX Learning Analytics Prototype (iLAP) implementation framework

Finally, the fifth step is the visualization and the user interface presentation part of the learning analytics prototype. At this stage, the processed data that come from the learning analytics server indicating the model learners' MOOC activities are now appropriate to be visualized to the end user. The data are presented in a textual format and chart forms, e.g. pie charts, scatter plots, line plots, bar charts, etc. The user can display a full statistics screen of each student and each course.

#### 4.3.1.1 Privacy consideration

The collection and processing of student information in learning analytics applications could comprise ethical issues in the context of their private data. Basically, issues fall into subjected categories as the following: A) Data accessibility and accuracy, B) Privacy and identification of individuals, C) Disclosure of processed and analyzed information, D) Achieving the Confidentiality, Integrity and Availability (CIA) of data in each learning analytics phase, E) Possession and ownership of data. (Khalil & Ebner, 2015b)

In the learning analytics prototype project, the main concerns were to preserve learners' sensitive information. It is a familiar demand that institutions or teachers ask for further information about the analyzed results from the educational datasets. The requests for a broader information range of educational datasets analysis may lead to ethical breaches of students' personal information (Greller & Drachsler, 2012). Some other studies draw attention to guaranteeing student anonymity in order to avoid embarrassments and exposure of data misuse (Baker, 2013; Slade & Prinsloo, 2013)

Thus, we tried to build an elastic tool that aims to sustain student privacy as well as keep learning analytics operations functional. We proposed a de-identification and anonymization system to preserve the ongoing process of the analysis model while minimizing the risk of harmful privacy information disclosure incidents. The anonymization system is built based on the European Data Protection Directive 95/46/EC law of privacy. More details about the built anonymization framework can be found in Chapter 6.

### 4.3.2 User Interface and User Dashboard

One of the aimed purposes of the iLAP is to develop an easy-to-read dashboard. The intended plan was to make visualizations that ease actions taken by researchers and decision makers. Visualizations should not only be connected with meaning and facts (Duval, 2011), but also with decision making. The user interface version of iLAP is only accessible by researchers and administrators. A teacher version, however, is attainable in a static format which shows general statistics about his/her teaching course. The Dashboard shows various MOOC objects which we usually refer to in this thesis as "MOOC indicators" or "MOOC variables." The Dashboard offers to search for any specific user in a particular period. The returned results cover the following:

- Quiz attempts, scores, and self-assessment

- Downloaded documents from the course (if applicable)
- Login frequency
- Forums reading frequency
- Forums posting frequency
- Watched videos

Figure 30 shows the *user dashboard* where administrators can view student progress in every course they are enrolled in. The examiner can observe quiz attempts, student performance as well as the logging frequency in a specified time frame as required. The user interface provides the opportunity to track student activities with all MOOC variables. On the top part of the dashboard, the admin can choose a specific username, identify the exact date and choose the MOOC. On the left side, the dashboard shows user quiz information. For instance, in the figure, the user made three quiz attempts in one of the offered courses. The line chart shows his/her score in every attempt. At the center of the dashboard, the downloaded files by the user are shown. Usually, these documents are uploaded by the teacher and are mostly *PDF* supplementary files. On the right hand of the dashboard, the iLAP dashboard shows activities of the user. It shows login frequency, forum readings, and forum posting. Finally, at the bottom of the dashboard, video interactions are shown for every video a user watched. The video section is an interactive *JavaScript* module. The administrator can click on each video ID and then get additional comprehensive details about how frequent the student used pause and/or play clicks.

In Figure 31, iLAP *course dashboard* is displayed. Similar to the *user dashboard*, the admin can pick up the course and the time period. General statistics are shown thereafter. For instance, the figure shows a course called *Lernen im Netz* 2014 with the top downloaded files, number of written posts, number of reads in forums as well as the number of quizzes and watched videos. Figure 32 shows other visualizations related to quizzes and discussion forums monitoring.

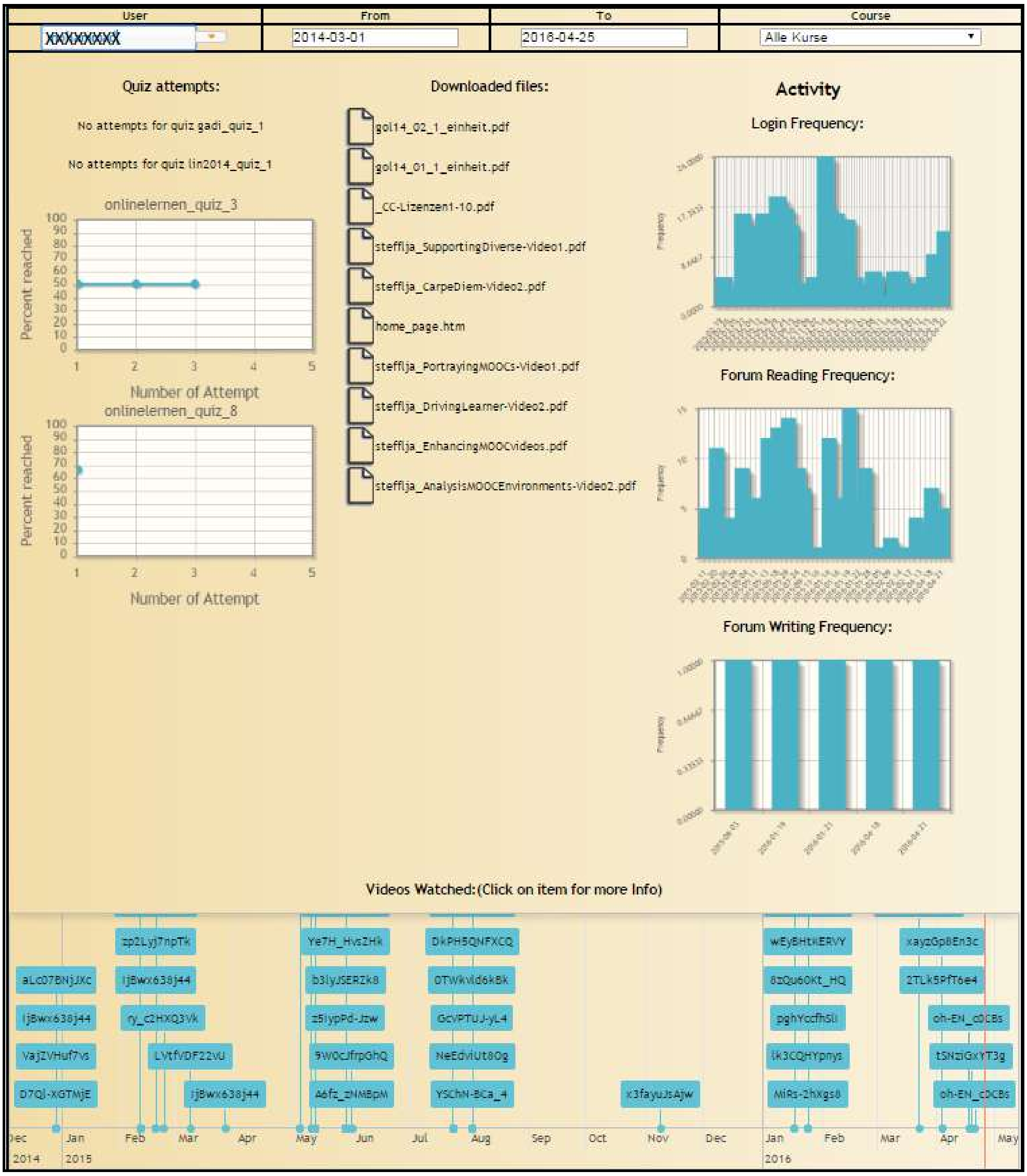

**Figure 30.** iMooX Learning Analytics Prototype (iLAP) user dashboard – admin view

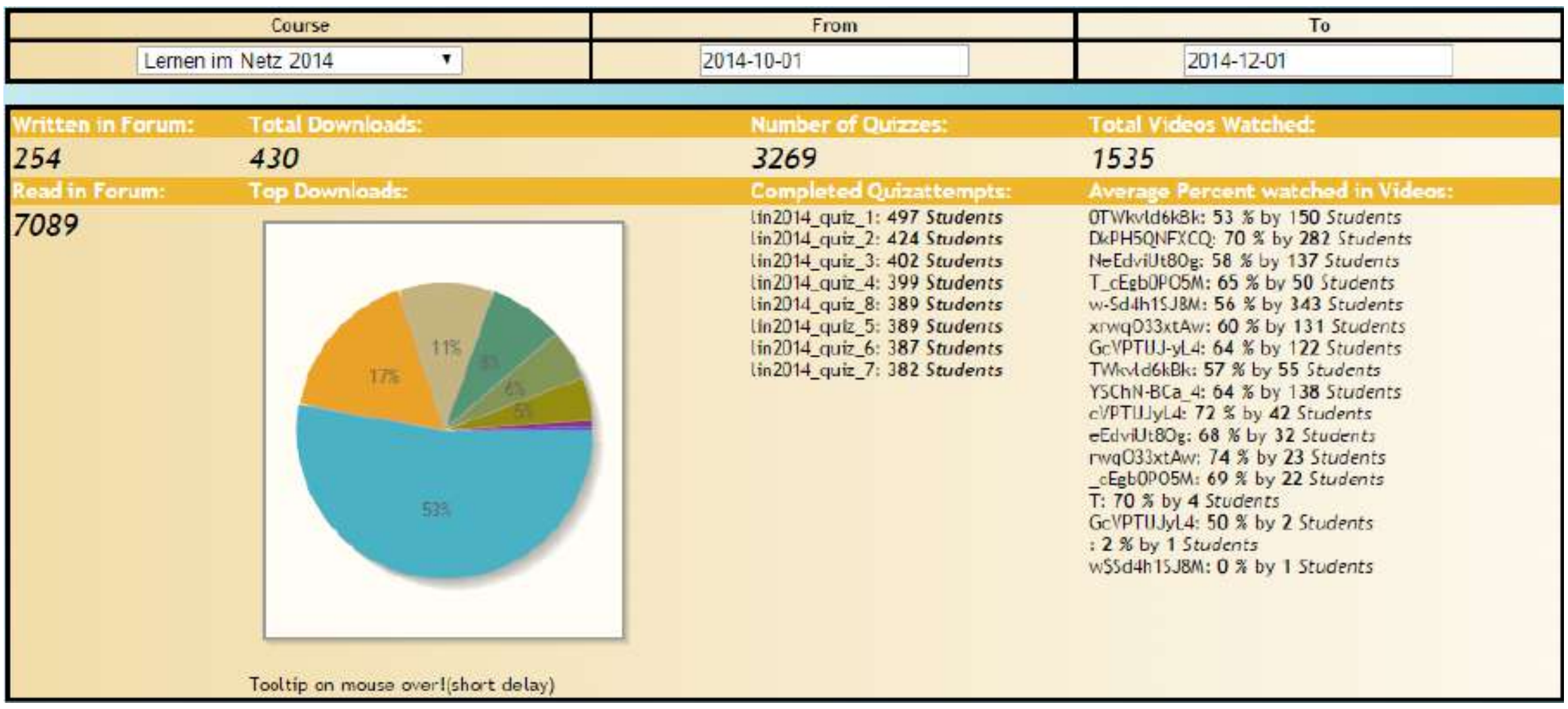

**Figure 31.** iMooX Learning Analytics Prototype (iLAP) course dashboard

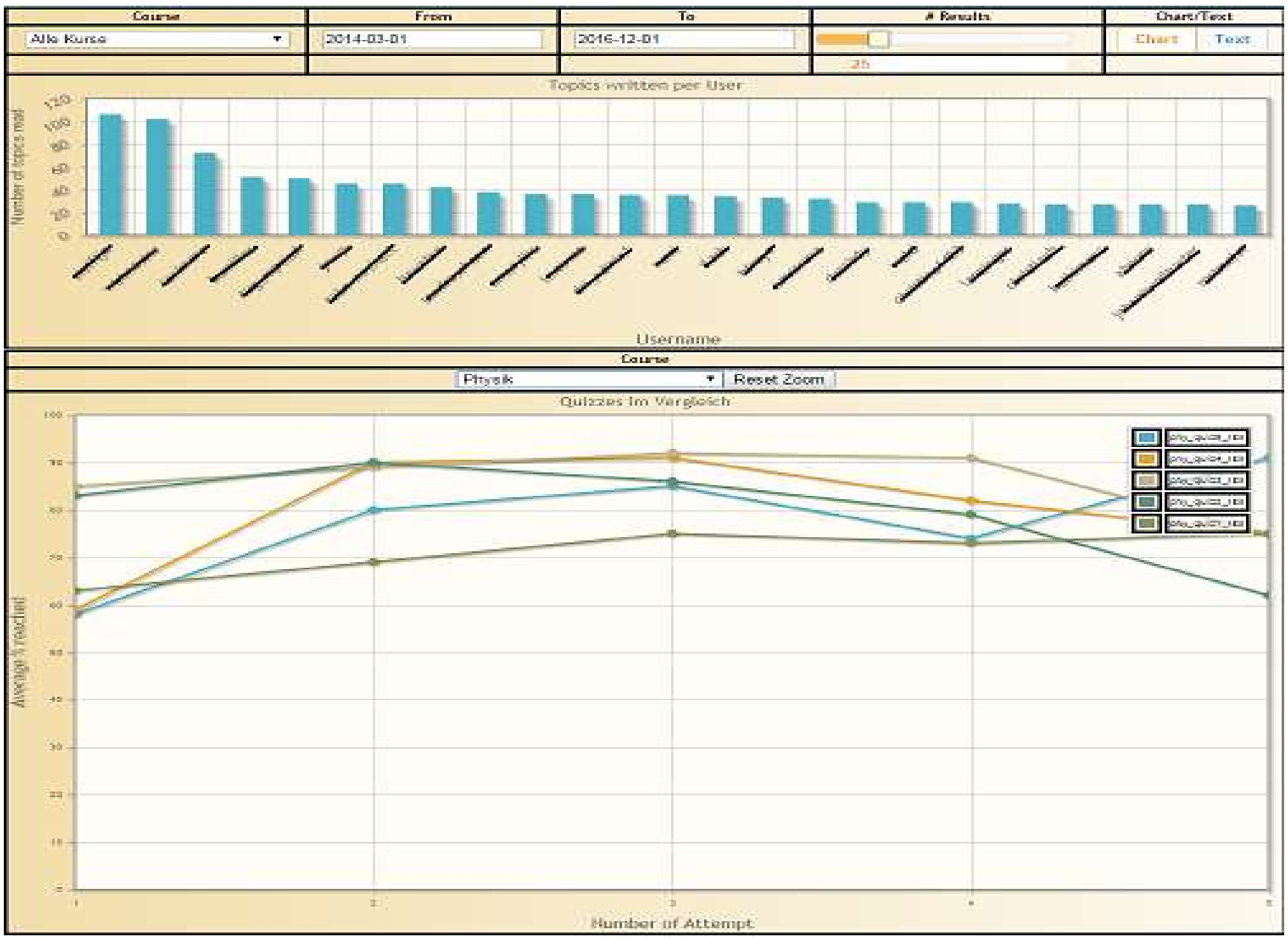

**Figure 32.** iMooX Learning Analytics Prototype (iLAP) monitoring of some MOOC variables

The structure of the user interface in iLAP is distinct to supporting an interactive working area by providing a *parameter dashboard* as shown in Figure 33. The layout of the parameter dashboard tab allows the user to compare two parameters. For instance, relations can be elicited between total posts in the discussion forums and the score of the exams as a meta-statistical case. In addition, the user interface provides the feature of exporting the results as a document, making it applicable to be printed or emailed upon request.

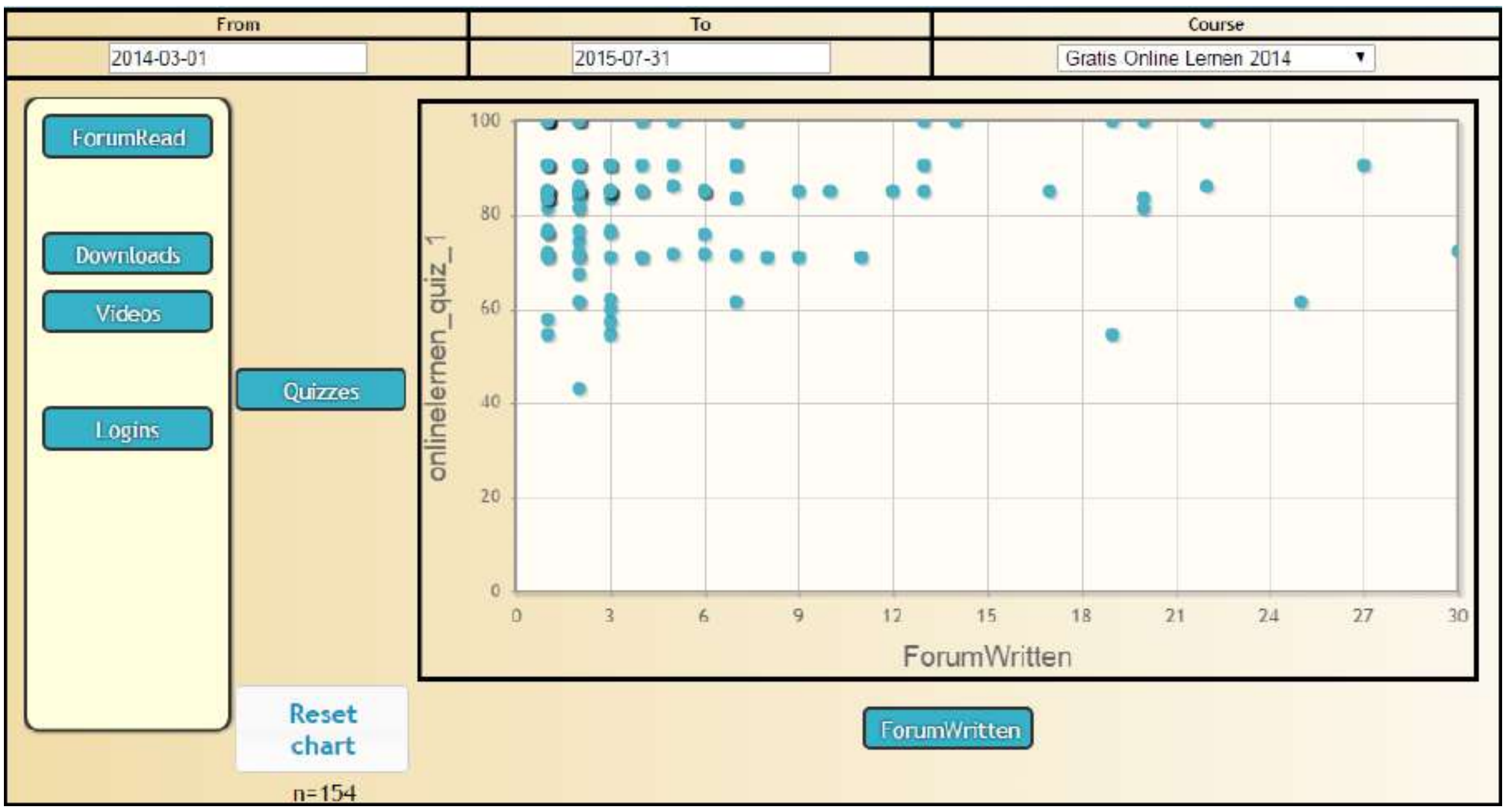

**Figure 33.** iMooX Learning Analytics Prototype (iLAP) parameters dashboard

### 4.3.3 Log File System

The need for extra data analyses necessitated the adoption of a separate database that aims to save all student information. The iLAP implementation framework at step-4 is processed by the learning analytics server in order to finally visualize the data to decision makers and admins. From then on, the information in the database which has the mined students' information has records of MOOC variables that we will use later to do more studies on engagement and behavior of iMooX students. The detailed records in this database helped us to do more research since we can manipulate data according to our needs. For instance, we can export data records to comma-separated values (CSV) files and then do professional data analysis using the *R* software or *SPSS*. The table structure of the iLAP log file system has 12 main records as shown in Figure 34. Main records include: courses, files, downloaded files, forums, reads in forums, posts in forums, quiz attempts information, login information, general quiz details, student information, individual video details and finally all video information.

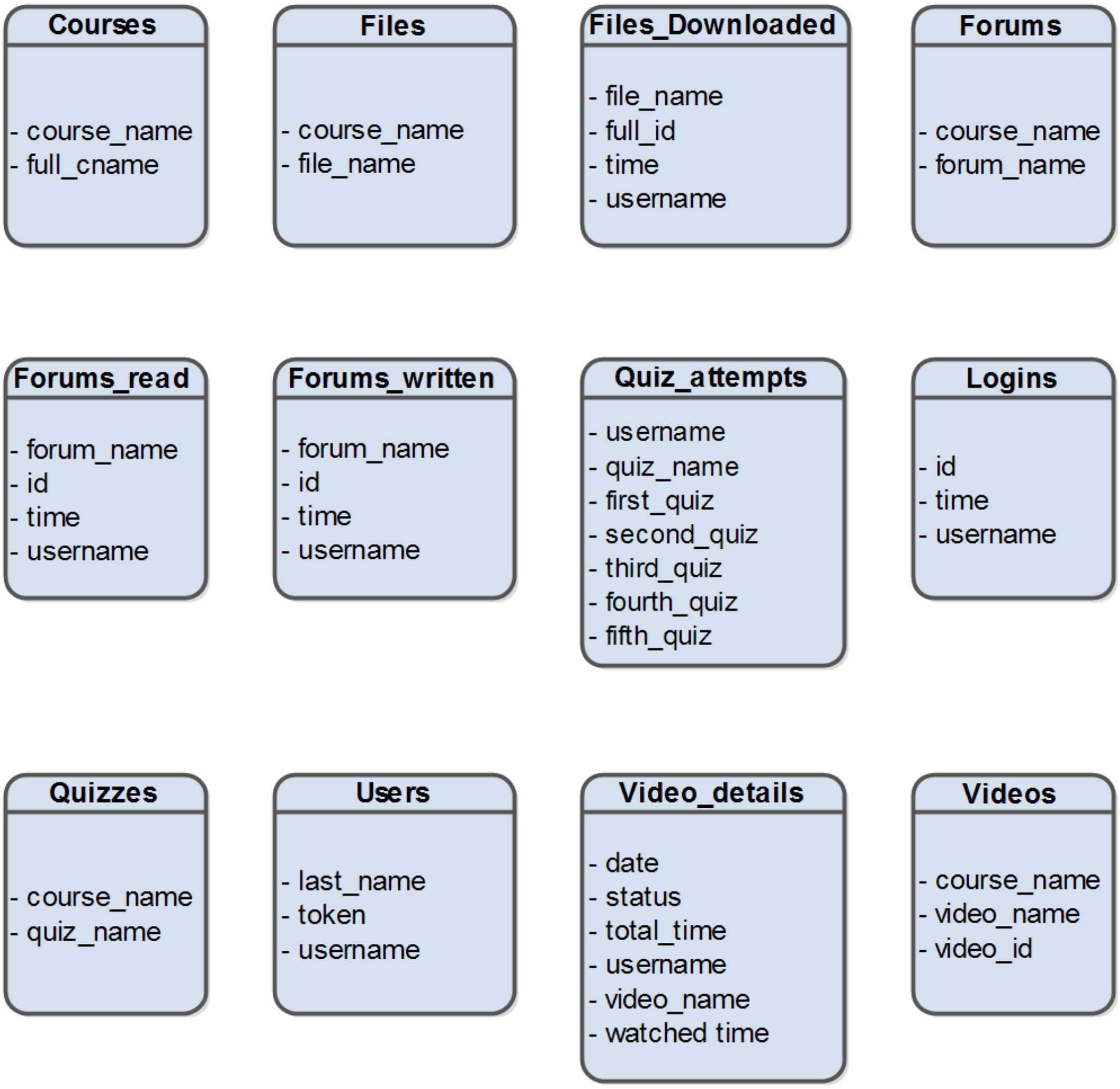

**Figure 34.** Table structure of the iLAP log files system

The exported datasets from these tables are usually followed by filtering where we identify duplicate records and thereafter content analysis by which units of analysis (MOOC variables) are measured and benchmarked based on either quantitative or qualitative decisions. An advanced usage of these variables is explained in Chapter 5.

## 4.4 iLAP Evaluation

As a means of evaluating iLAP, the followed method consists of the application potential of collecting data, the processing of them, and finally storage in the database. After that, we looked at examining datasets based on quantitative analysis followed by qualitative decisions in order to extract useful information. Extrapolating beneficial information from learners' traces is a challenge and requires exploratory analysis (d'Aquin and Jay, 2013). Visualizations and descriptive statistical models were mainly used to outline different characteristics of the Learning Analytics Prototype.

The evaluation process started with implementing the tool in two offered courses on the iMooX platform. The investigated courses were: *Gratis Online Lernen*, in which means "Free Online Learning" in English, abbreviated as (GOL-2014), and *Lernen im Netz*, or in English "Learning Online," abbreviated as (LIN-2014). Both of these courses were lectured to students in German and offered in 2014. Courses were presented within a rich content that included all the MOOCs interactive components: forums, documents to download, learning videos, and multiple-choice quizzes. The GOL-2014 course workload was set to be 2 hours/week, starting on 20$^{th}$ October 2014, and ending on 31$^{st}$ December 2014. The lead instructor was a faculty member of the Graz University of Technology. The LIN-2014 workload was set to be 5 hours/week, starting on 13$^{th}$ October 2014, and ending on 31$^{st}$ December 2014, and the course's instructor was a faculty member of the University of Graz.

The GOL-2014 was a free course open to anyone and without previous knowledge. The course content was about free education through the Internet with tips and tricks on how it can be done. On the other hand, the LIN-2014 was not only a free MOOC, but also a university course, counting the students coming from the University of Graz. Its main subject was giving an overview of trends in learning through mobile, social media and the principles of Open Educational Resources. Every week, a batch of short videos was released for both courses and suggested articles to read were posted on the course's homepage wall. A student had to score at least 50% in each GOL-2014 quiz and 75% in LIN-2014 quizzes in order to successfully pass the course, with the ability to repeat a quiz up to five times. The iMooX platform was planned out to consider the highest grade of the five attempts.

iLAP provides us with a tremendous amount of information from the MOOC platform. The data were directly collected from both of the examined courses through the process described in Figure 27 and Figure 29. The examined MOOCs educational data sets include over 100,000 records of events with 1,530 students registered. These records contain activities related to discussion forums, documents, videos statistics and quiz scores of each student in each course. In order to start evaluating the collected data from the learning analytics prototype, organizing the records and carrying out data transformation and manipulation was required to fulfill the principles of “tidying the data,” such as cleaning the messy data sets and mutating them into an easily visualized and structured form (Wickham, 2014). It is worth mentioning that the data manipulation in the evaluation process is different from that in the implementation stage. The data that is processed in this evaluation phase is taken directly from the Learning Analytics server, while the data manipulation in the implementation framework is required for the end user visualization phase where the user interface layout is presented.

Different use cases will now be presented to point out the potential of the learning analytics prototype for MOOC stakeholders. These use cases will also provide us with a background of what learning analytics can reveal from the iMooX data.

### 4.4.1 Use Case: Defining Participants and Dropout

Previous research studies on the iMooX platform were carried out using surveys and questionnaires (Neuböck, Kopp & Ebner, 2015). However, after the implementation of the iLAP, gathering information about participants in every course offered becomes much more intense. One of the first steps in this evaluation was to generate a general description about the MOOC platform participants. In the first analysis of counting the number of students who were certified and who were registered for both courses, we were able to generate a bar graph that shows the differences between these two groups (Figure 35).

The summary showed that there were 1012 registrants in the GOL-2014 course and 177 students who were handed a certificate, which means a ratio of 17.49% of the total registrants. The LIN-2014 course included 519 registrants and 99 certified students, which make them 19% of the total course registrants.

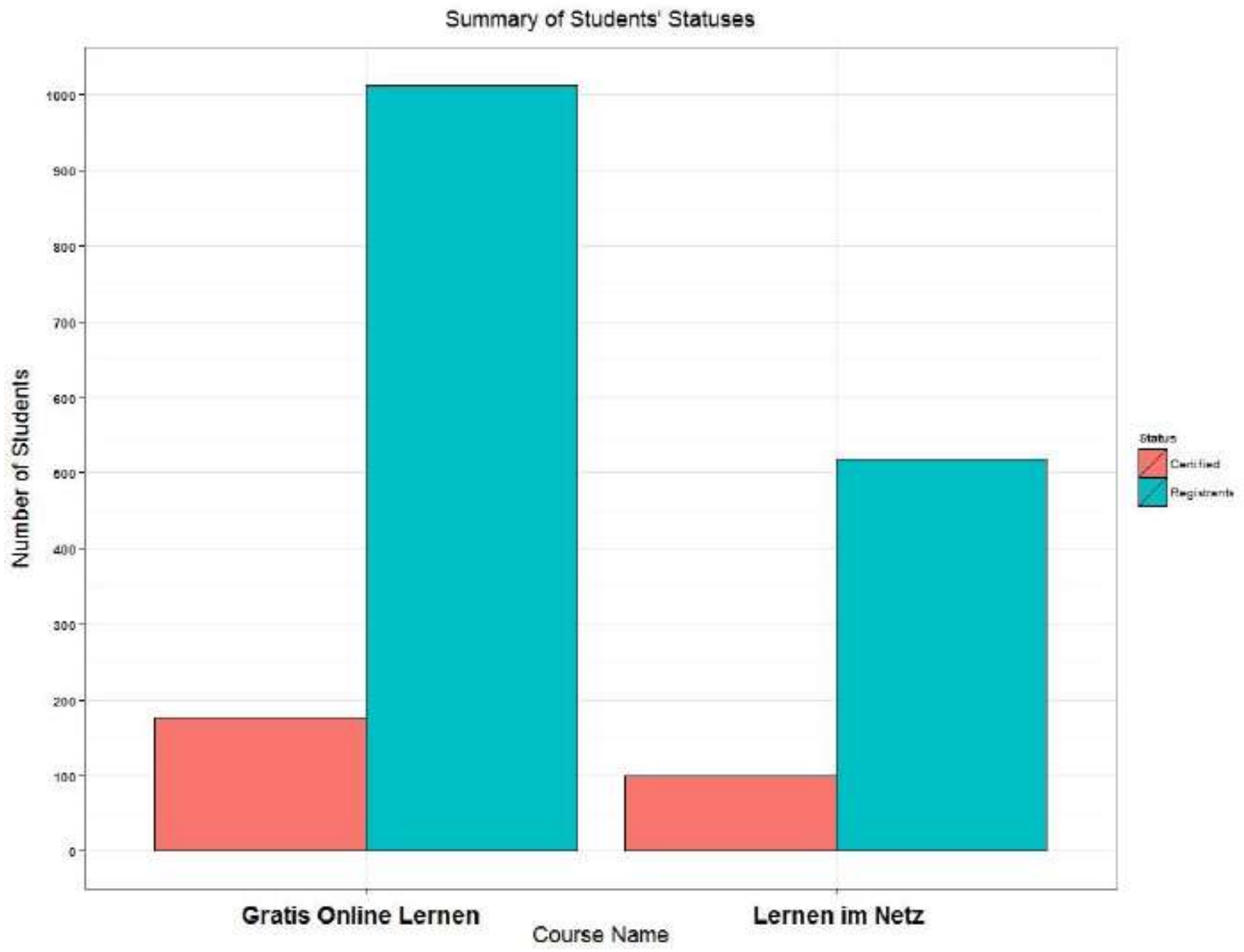

**Figure 35.** General description that summarizes students' status of the examined MOOCs

Our next intention was to typecast students to more than just two groups of students. Categorizing online participants in MOOCs has been a hot topic since 2008. Various studies mentioned categorizing the students based on their engagement and motivation (Kizilcec, Piech, & Schneider, 2013; Hill, 2013, Assan et al., 2013; Tabaa and Medouri, 2013). By advancing within the same route, and based on the data sets collected from both of the examined courses, the division of participants based on their *general activity* was as follows:

- Registrants: the students who enroll in one of the available courses
- Active learners: the students who at least watch a video, post a thread in the discussion forums or attend a quiz
- Completers: those who successfully finish all the quizzes, but do not answer an evaluation form at the end of the course
- Certified learners: completers, but in addition, they review their learning experience through an evaluation form at the end of the course.

By gathering the data from the iLAP, clustering them as above and visualizing the results in Figure 36, the analysis showed that both courses have 1531 registrants, 1012 registrants in the GOL-2014 and 519 registrants in LIN-2014. Data also showed that there were 812 active learners in both courses, 479 active students in GOL-2014 and 333 active students in LIN-2014; 348 completers in which GOL-2104 had 217 students who completed the course and 131 completers in LIN-2014. There were 276 certified learners in both courses, 177 in GOL-2014 and 99 students in LIN-2014.

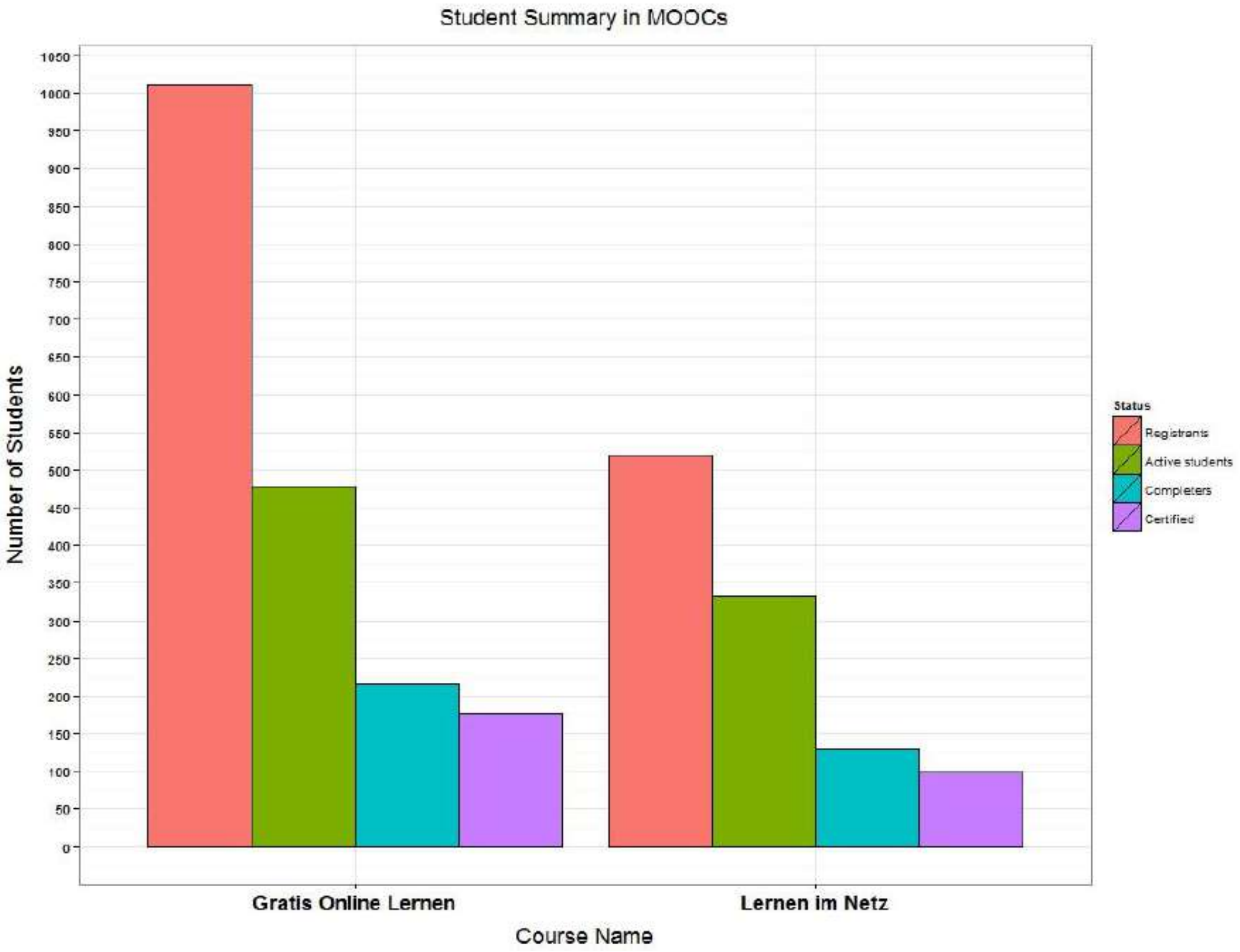

**Figure 36.** Summary of students' status after categorizing them as registrants, active students, completers and certified

The evaluation shows a remarkable controversy between registrants and active students. GOL-14 had 47.3% active students, while LIN-2104 had 64.16% active students. The higher completion rate in the LIN-2014 can be explained by those students who belong to the University of Graz who can obtain a total of 4 ECTS points if they achieve a pass, which will be added to their university records.

The new categorization of MOOC students in the iMooX platform opened the window for a new definition of dropout in MOOCs. Jordan (2013) found that 7.6% is the average completion rate in MOOCs. MOOCs are familiar with high attrition rates and a low motivation environment for learners (Khalil and Ebner, 2014). (Rivard, 2013) stated that a Coursera MOOC called “Bioelectricity” lost 80% of its students before the course actually began, and the course finished up with 350 certified students out of 12,700 registrants. Whether the students who gain certificates are to be considered as perfect students still remains an ambiguous question.

In our case, it is still unclear whether the dropout rate should be referenced to the ratio between completers and registrants or to completers and active users. On the other hand, what about the ratio of certified students to either registrants or active students? According to Rodriguez (2012), participants in MOOCs can go two different ways, as either lurker or active. Table 9 is thus introduced to show different definitions of dropout rate and their percentages based on the new categories of MOOC participants in the iMooX platform.

**Table 9.** Different dropout rate definitions based on participant categories in the examined MOOCs

| MOOC | Dropout rate definition | | | | |
|---|---|---|---|---|---|
| | certified to registrants | certified to active stud. | completers to registrants | completers to active stud. | active stud. To registrants |
| GOL-2014 | 82.50% | 63.04% | 78.55% | 54.69% | 52.67% |
| LIN-2014 | 80.92% | 70.27% | 74.75% | 60.66% | 35.84% |

The definitions above can help us enhance the completion/certification ratio in iMooX. Nevertheless, the dropout rate of students who registered and then fell back shows that the students who enrolled (registrants) and became active in the LIN-2014 course were 64.16% with a dropout rate of 35.84% while registrants in the GOL-2014 course dropped by 52.67% to reach 479 active students out of 1012 registrants. As a result, motivating registrants to be active students is critical to improve the retention ratio in MOOCs. Similar categorization is used in Chapter 5.6 for purposes of increasing the ratio of this class.

### 4.4.2 Use Case: Video Patterns

Like any other MOOC platform, iMooX depends on video lectures as an elementary approach to delivering learning content to the students because of the significant role of the video content in the MOOC platforms. The video lectures are hosted on YouTube. iLAP mines when a student clicks play, stop or when the video is fully watched. Figure 14 shows a graph line of learner interaction during four weeks learning videos in the GOL-2014 MOOC. The turquoise line shows the number of students who pause or skip a segment of the video on a specific second. The red line shows the number of students who replay the video at a specific second. A similar case study is Brooks, Thompson and Greer (2013) by which the authors classified video watchers as engaged rewatcher, regular rewatcher, and pauser rewatcher, depending on the number of pauses and replays.

Figure 37A and Figure 37B represent videos during week1 and week2. Figure 37C and Figure 37D correspond to videos of the 7th and 8th weeks respectively. It is evident that the activity is much higher in the first two weeks than the last two weeks.

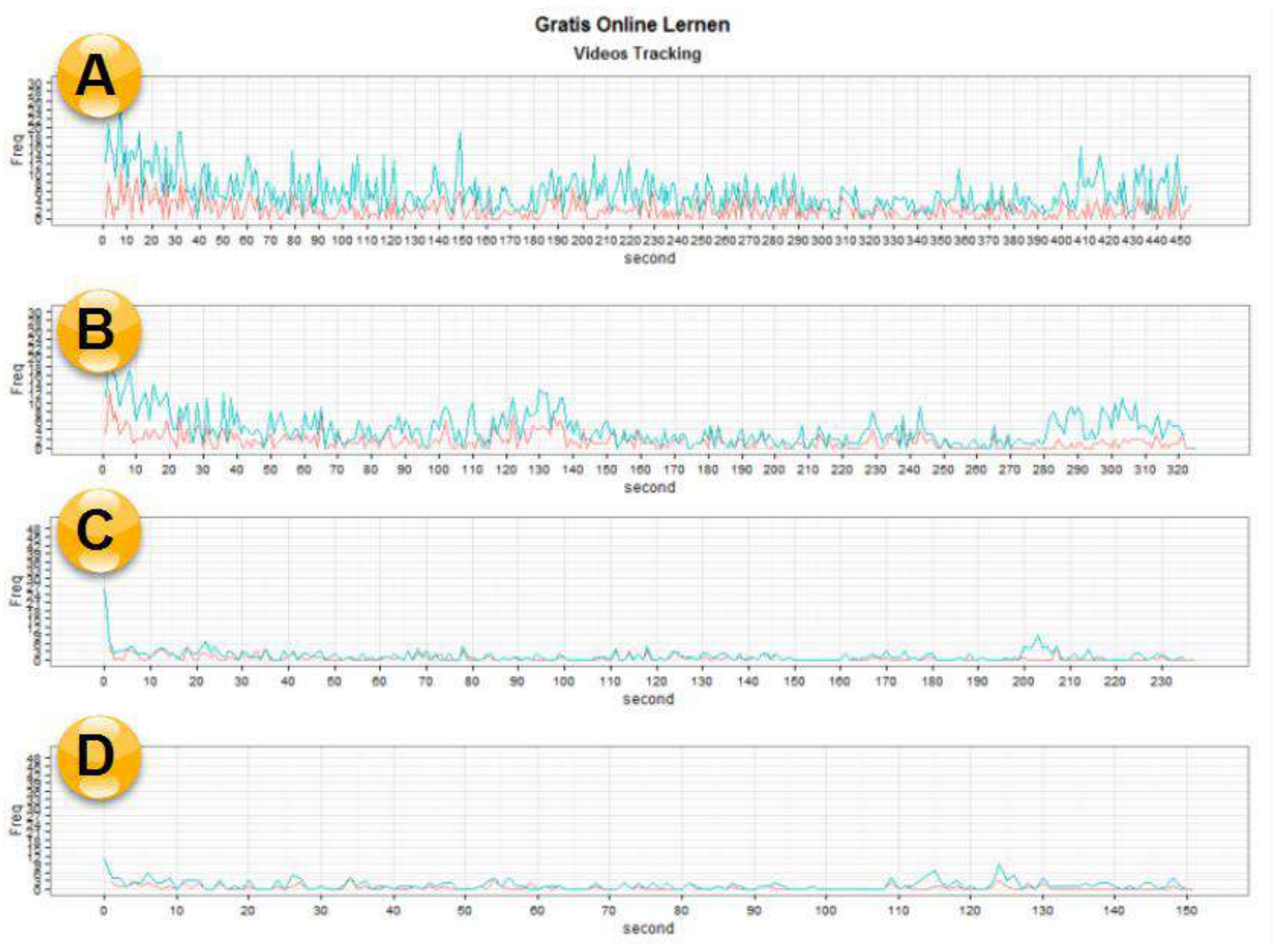

**Figure 37.** GOL-2014 MOOC video tracking. From top to bottom (A) week1 videos; (B) week2 videos; (C) week7 videos; (D) week8 videos

It was also remarked that most video activity happens during the first and the last minutes as well as throughout intensive learning content segments. By contrast, video activity decreases through time; it has been noticed that there is a drop in video viewing after the first three weeks in both MOOCs, GOL-2014 and LIN-2014.

Detailed video analysis is carried out in Chapter 5.2 "*Case Study: Tracking Videos Activity in MOOCs.*"

### 4.4.3 Use Case: Discussion Forums Patterns

This use case concerns analyzing the discussion forums MOOCs indicator, which refers to user readings and writings. The Learning Analytics Prototype mined the discussion forums activities and split them into forum posts and forum reads. The analysis pushed the pedagogical hypothesis, which shows that the more interactive the modes of student engagement are, the better student learning performance is (Waldrop, 2013). Several research studies have drawn attention to the significant effect of MOOC discussion forums on providing an enhanced adaptive support to students and groups in MOOCs like the one by (Ezen-Can et al., 2015).

During the course sessions, there were 21,468 readings in the GOL-2014 forums and 9136 readings in the LIN-2014 forums. In respect to posts, there were 834 posts in the GOL-2014 and 280 posts in the LIN-2014 course. Figure 38 demonstrates reading in both of the course forums. On the left, Figure 38A, the visualization employs a line graph to show reading activity in the LIN-2014 course. It is obvious that students become less interested in reading in the discussion forums after the first weeks. In Figure 38B, the total number of reads reached the highest in the first two days of the GOL-2014 course. The topmost count of reads was on 21$^{st}$ October, which is the first day when videos and content were released. The first week collected 6708 reads, the fourth week gathered around 1700 views and the last week garnered only 1414 reads.

In summary, it was interesting to find that nearly 50% of both courses' forums readings happened by the end of the first two weeks. However, only 10% was the share of readings in forums in the last two weeks.

By the same token, writing in forums did not present a different picture. Figure 39 is a dot plot showing that students wrote more often in the first two weeks and that this period therefore takes the lion's share of the whole number of posts. Each point in the plot represents a student. The maximum number of posts in GOL-2014 was on the first day of the course, with 64 posts. The total number of posts during the course period was 834, with an average of 27.57 posts and a median of 26 posts and there were only 6 posts when the course ended. The LIN-2014 course collected 280 posts, with an average of 21.12 posts, a median of 5 posts and only 2 posts after the course ended.

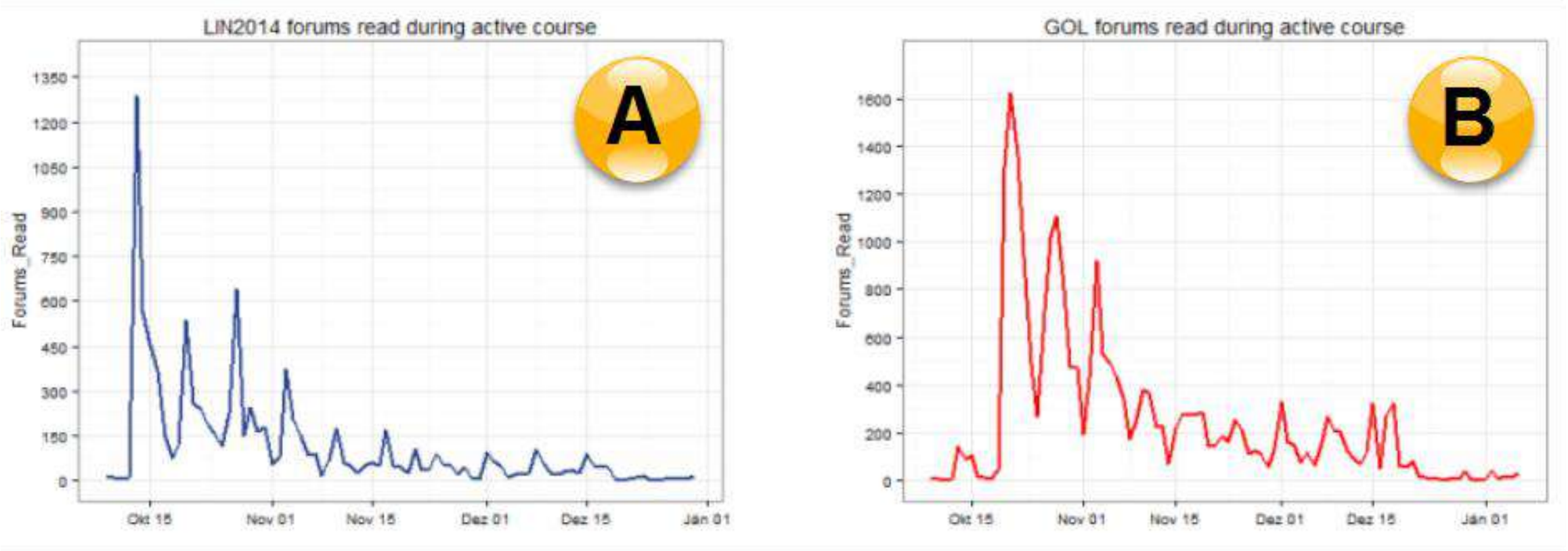

**Figure 38.** Students' readings in MOOCs discussion forums. From left to right (A) LIN-2014 course forum; (B) GOL-2014 course forum

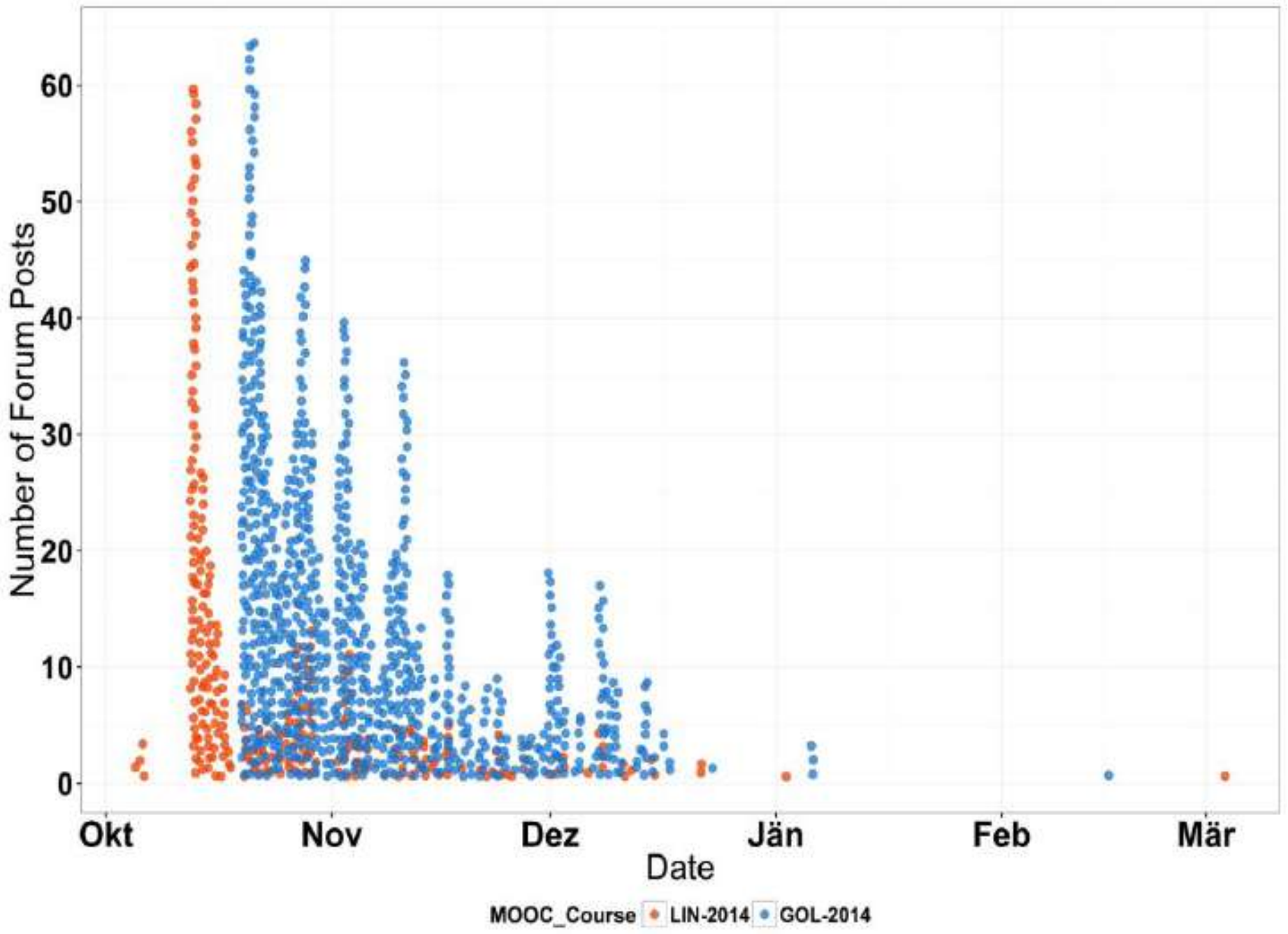

**Figure 39.** Students posts in MOOCs forums of LIN-2104 and GOL-2104

The lead management of iMooX thereafter prepared plans to improve the social communication between instructors and students through the platform discussion forums because of the expected relation between motivation, engagement and discussion forums.

### 4.4.4 Use Case: Quizzes and Grades

Almost all MOOC platforms offer quizzes and exams for students to check their learning understanding. The use of learning analytics illustrates the analysis of student behavior and performance. As stated above, iMooX proposes quizzes but in a different form than the traditional method. The students have the opportunity to improve their skills by allowing them to do five self-assessment quiz attempts. In Figure 40, iLAP data shows the total number of quiz attendance in the GOL-2014 MOOC. The total number of trials apparently decreases strongly in the first four weeks. However, the last four weeks indicates more stability in the dropout rate.

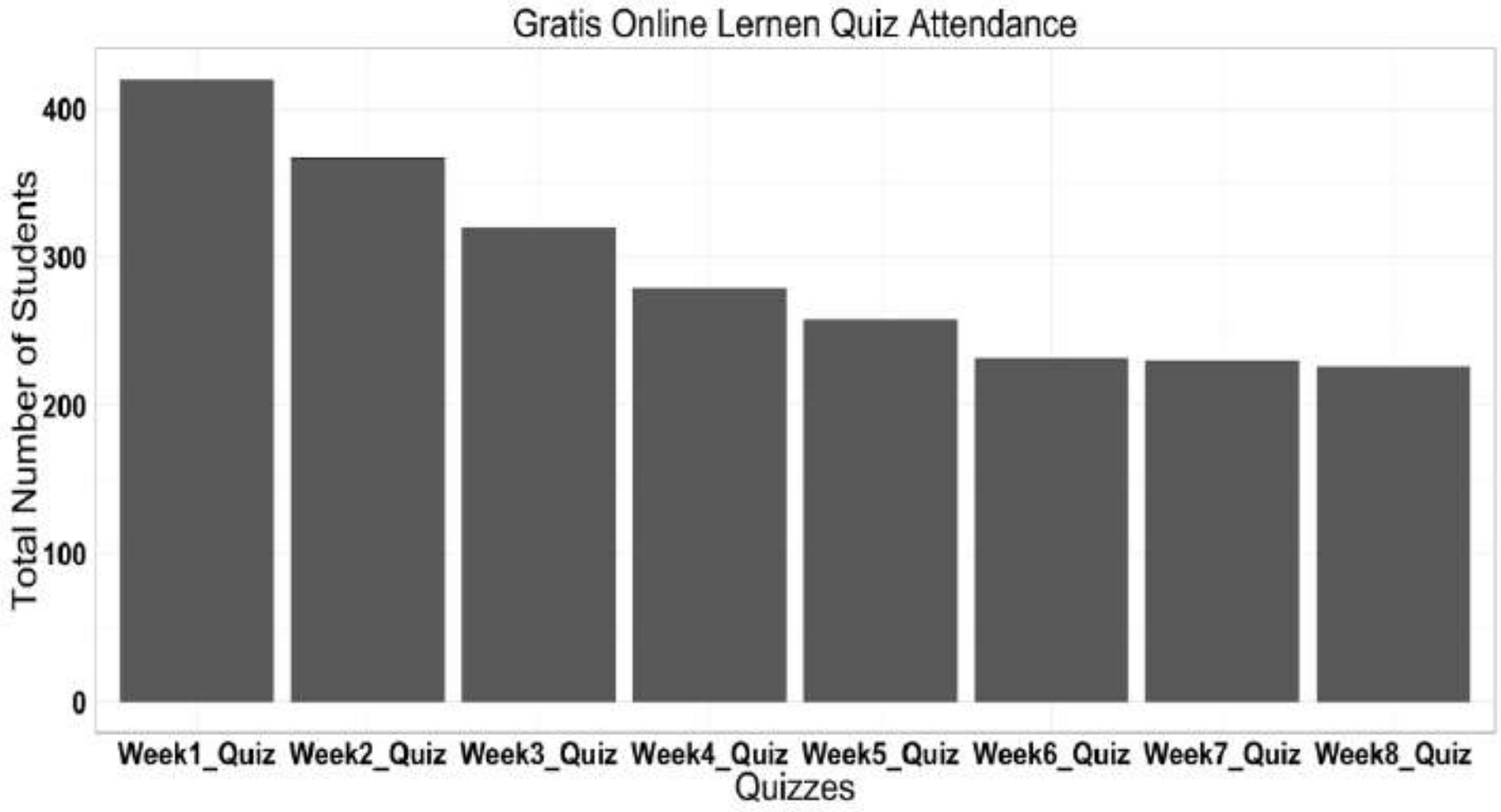

**Figure 40.** Quiz attendance in GOL-2014 MOOC (eight weeks)

According to a research study by (Ye & Biswas, 2014), lecture watchers and quiz attendees play a major role in defining student performance in MOOCs. Quiz performance accompanied with downloaded documents and readings in the discussion forums were analyzed for the GOL-2014 MOOC.

In Figure 41, which shows a portion of the GOL-2014 quiz analysis, a perceptible correlation between students who downloaded documents for the week and their quiz grades for the first attempt was observed. The y-axis depicts the grade; the x-axis displays file names of each week. Each colored point represents one student. In the top section, the students who downloaded both files scored higher than those who did not download any. The first week quiz average score for the group that downloaded files (337 users), was 80.7% with a median of 85%, while the mean for the other group who did not (100 users), was 74.12% with a median of 71%.

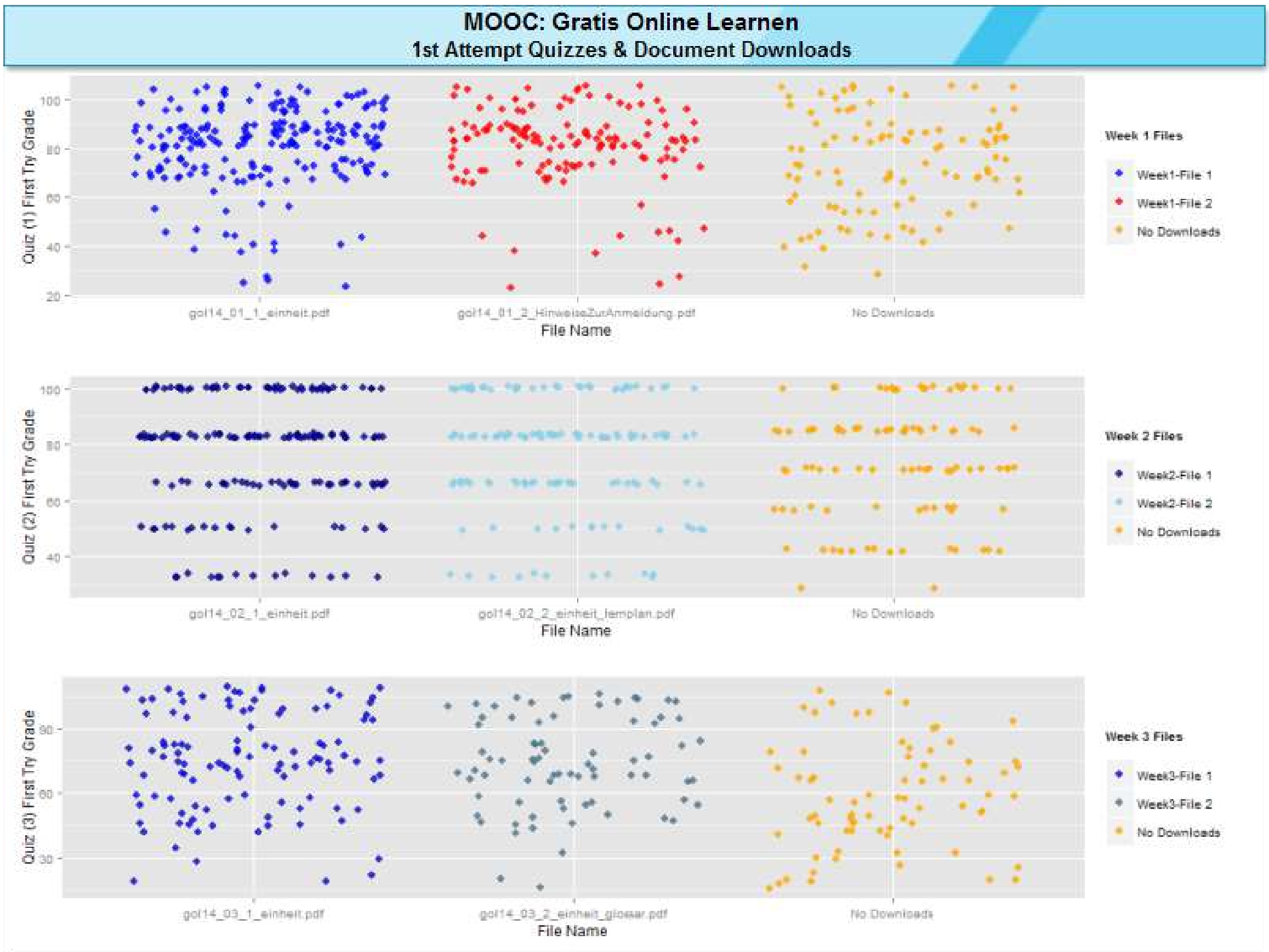

**Figure 41.** Analysis view of GOL-2014 MOOC first attempt quizzes compared to file downloads in week1, week2 and week3

In order to maintain student performance, their grades in parallel with their social activity were analyzed. Students of MOOCs, who are engaged in forums, have been found to score better in the exams than who were less active (Cheng et al., 2011; Coetzee et al., 2014). Consequently, a correlation test to compare the students who read in forums and who did at least one quiz (active students) in GOL-2014 MOOC was done as an evaluation example of the iLAP quizzes data. Figure 42 is a scatter plot which reveals a relatively weak relationship between both factors.

The Y-axis shows a number of readings but with the use of square root in order to attain an ease of pattern recognition. The x-axis records the average score of all quizzes taken by students. The blue line represents a smooth linear regression line while the gray area around it is the standard error. Students with high performance (grade > 90) have a "reading in forum" median score of 21 reads. On the other hand, there were still students who read more than 20 times, but failed to pass some of the quizzes. Respectively, the standard error area is wider when the grades are less than 60. It is hard to fully prove that students who read in forums score better on quizzes, because there are still other factors that influence overall performance.

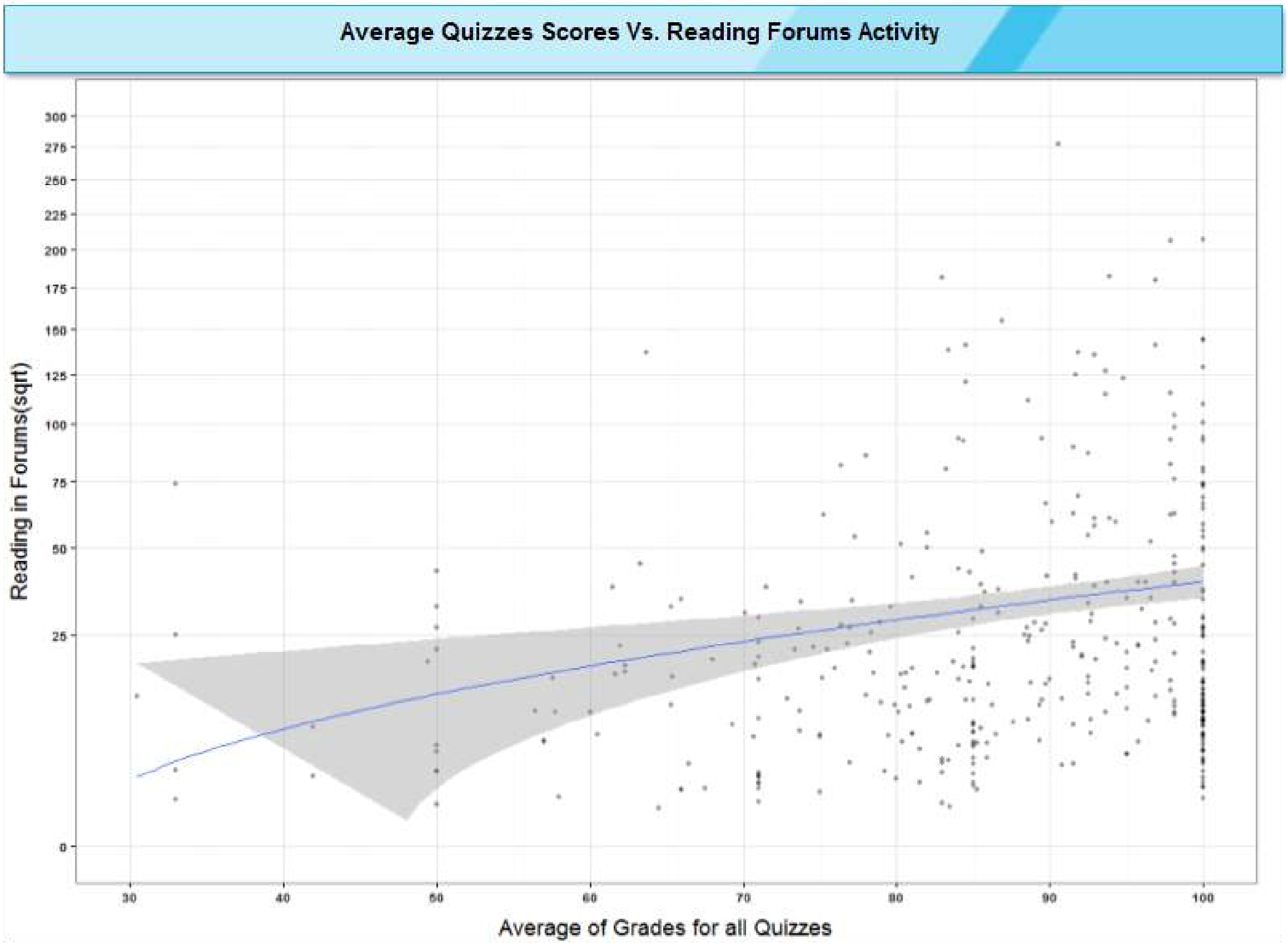

**Figure 42.** The relation between readings in forums and student performance in GOL-2014 MOOC

## 4.5 iLAP Related Work

Various applications were developed to solve the pressing needs of understanding learners and enhancing online learning environments similar to the iLAP. It is realized that most learning analytics applications focus on Learning Management Systems (LMS). However, there have not been many research studies on learning analytics practices in MOOCs as already discussed. For example, Tabaa and Medouri (2013) presented a Learning Analytics System for MOOCs (LASyM), which analyzes the huge amount of data generated by MOOCs in order to reveal useful information that can help in building new platforms and assist in reducing the dropout rate.

LASyM lacks consideration of privacy and the extensive analysis can exceed the limits to be reached in the personal student-level data.

Dyckhoff, Zielke, Bültmann, Chatti, and Schroeder (2012) introduced the Learning Analytics Toolkit for Teachers (eLAT) with a simple GUI that requires no knowledge in data mining or analysis techniques. The tool can be used by the teachers to examine their teaching activities and to enhance the general assessments. It can be implemented on MOOCs, Moodle and other learning systems. Yousef, Chatti, Ahmad, Schroeder, and Wosnitza (2015) built a learning analytics application based on learners' perspective surveys to enhance personalization in learning analytics practices. Yet, the application does not mention if the developers took the personal information of students into consideration. LOCO-Analyst (Learning Object Context Ontology Analyst framework) is another tool that provides teachers with feedback about students and their performance based on a semantic web (Jovanovic et al., 2008). Additionally, analysis of student patterns in MOOCs was recently mentioned in different studies such as (Ferguson & Clow, 2015) and (Joksimović et al., 2015). iMooX Learning Analytics Prototype shows promising features to discover and examine the behavior of MOOCs students.

It should be noted that several studies analyzed MOOC components and learner engagement. But ultimately, the iLAP differs from the previous tools and research studies because it was introduced into the area of student performance, based on relations with indicators from online learning environments, focusing in particular on the MOOC platform. Furthermore, the logging system of iLAP is also very important to evaluate and measure student behavior and engagement.

## 4.6 Chapter Summary

MOOCs and learning analytics appear to be well suited to each other. MOOCs offer large education datasets, and learning analytics offer discovery of hidden patterns from these datasets. In this chapter, we presented the following:

1. An introduction of the Austrian xMOOC platform. iMooX platform is an online learning stage which offers free online courses to students from schools, students from universities (higher education institutes) and students from the general public. Our research in this thesis is mainly built on case studies from the iMooX MOOC-platform.

2. Our proposed learning analytics framework. The approach portrays a learning analytics lifecycle. This provides a complete overview consisting of learning environment, Big Data, analytics, actions, and optimization.
3. The further design, development, and implementation of iLAP: a learning analytics tool in the iMooX MOOC-platform. The main objective of this tool is to provide detailed information about courses and students of iMooX. Further, we looked to having a log files database that will help us understand the behavior of students in terms of their quizzes, social activities, video interactions and performance.
4. The evaluation process of iLAP based on two offered MOOCs.

The impact usage of iLAP will be further interrogated based on many case studies in the following chapters.

# 5 CASE STUDIES

After implementing the learning analytics tool in iMooX, we started collecting raw data from the offered MOOCs. Thus, in this chapter, we present several case studies that mainly aim at describing the potential of learning analytics in every different learning scenario from iMooX. In general, the listed case studies follow the timeline of the offered MOOCs.

The chapter includes six case studies that summarize our experiment. Chapter 5.1 is titled "Activity Profile and Self-Assessment Quizzes." Chapter 5.2 is titled "Tracking Videos Activity in MOOCs." Chapter 5.3 has a title of "Dropout Investigation." Chapter 5.4 is the case study of "Clustering Patterns of Engagement to Reveal Student Categories." Chapter 5.5 is presented as "Case Study: Fostering Forum Discussions in MOOCs," and Chapter 5.6 has a title of "Fostering Student Motivation in MOOCs."

## 5.1 Case Study (1): Activity Profile and Self-Assessment Quizzes[7]

The availability of MOOCs for school pupils has been obtained on different platforms in the USA and Europe. The well-known MOOCs provider, edX, continues to offer numerous high school courses for students to help them prepare for college (http://www.edx.org/high-school, last accessed December 2016). In this case study, we will examine school pupils' attitudes in one of the provided courses in Science, Technology, Engineering and Mathematics (STEM) at iMooX. The exploratory educational dataset analysis will help us to build an activity profile for the students enrolled and to examine their performance in the self-assessment quizzes.

### 5.1.1 Methodology

This case study research employed compound analysis methods on the collected data. With the large educational datasets collected by the iLAP, we applied different methods of tidying up the

---

[7]Parts of this section have been published in:

Khalil, M. & Ebner, M. (2015a). A STEM MOOC for School Children – What Does Learning Analytics Tell Us? In *Proceedings of the International Conference on Interactive Collaborative Learning* (pp. 1217-1221), Florence, Italy. IEEE.

data (Hadley, 2014) as well as classifying it into categories, in order to ease the processing and visualization phases. Finally, we generalized the outcomes and summarized the results.

The analysis invoked file permissions and authorized a safe operating environment. Furthermore, all the research phases and the results were restricted to the research members. We considered privacy constraints when learning analytics was applied. We ensured that student data were kept confidential. The educational dataset was secured by a local Virtual Private Network (VPN) to protect the information from any unauthorized access.

### 5.1.2 MOOC Structure and Overview

While the majority of MOOCs in the world are available to Higher Education (HE), school pupils do not really have such a global focus (Boxser & Agarwal, 2014). In its turn, iMooX has targeted some of its courses especially to this group of learners. Two of those courses are "Mechanics in Everyday Life" and "The Circle." Our research study examined the first course, "Mechanics in Everyday Life." It discussed physics, mechanics and aerodynamics sciences on a high school level. The course requires no previous knowledge, but the participant should know at least the basics of school science. This research study targeted the high secondary school pupils who finished their primary schools and had the essential knowledge of physics and science fundamentals.

Similar to other courses provided, German is the communication language of the course, with ten week periods. Each week consists of a couple of videos, forum discussions, as well as quizzes. The course included in summary 46 learning videos of less than 5 minutes each and 10 quizzes. Participants had to pass each quiz with at least 75% to successfully complete the course.

The planned workload was set to be 3 hours per week. At the final stage, completers, and those who successfully pass all the quizzes are heartened to inquire for a certificate after answering an evaluation questionnaire where they review their experience within the course.

Unlike other MOOCs, a student has the option to do five trials of the weekly quiz. The quiz approach in iMooX is fairly different than the traditionally known quiz systems, where multiple chances are available to try, due to the fact that it is intended to be a kind of self-assessment learning guide. The system is set up to record the highest grade the student scores. From the pedagogical point of view, we aim to study the participants' learning behavior over a number of trials each week, and from the psychological point of view, it is expected that such a system reduces stress and, therefore, makes the students behave more comfortably.

### 5.1.3 Case Study Results and Discussion

To start analyzing the collected students' data and build the activity profile, we planned to track the activities students processed in the course's online environment. Such activities create footprints that can assist in detecting student progression in the whole learning mechanism (Verbert et al., 2013). Within the educational datasets, the aim is to discover patterns that if combined with pedagogical approaches would lead to improving students' learning behavior from one side, and embellish the teaching methods on the other side. Accordingly, the analysis consisted of tracking most of the learners' operations in the course environment. We took into account ethical factors by respecting the privacy of our students' personal information.

We observed student activities in discussion forums, their performance, quiz trials and their interaction while watching the videos. The first analysis result embodied tracking students in the form of an activity profile. Any possible information regarding their commitment to learning through watching videos and exchanging arguments with the instructor in the discussion forums would allow researchers to evaluate the course or improve the course's model. Besides, it can support stakeholders with their decisions to achieve different educational goals. To take a better look at the data, we classified enrollees in this MOOC into two main categories: their class and their status. Figure 43 illustrates the activity profile of all the students enrolled in the course.

Unlike the classical way of assigning the timeline on the x-axis, we assigned the activities on the horizontal coordination line and the course's time frame on the y-axis. The tracked activities include, from left to right, a) Students who post in the discussion forum, b) Students who clicked on any thread in the forum, c) Number of quiz trials each student did, and d) Students who clicked on any of the learning videos. All activities were separated by the course's time length from week-1 till week-10. The school pupils have been specifically indicated by providing them special usernames ($N$= 27), while we could easily point out the other participants ($N$= 242) as long as the course was open to the public.

We categorized certified participants ($N$= 18) and those who passed all the quizzes and claimed a certificate and denoted them in red, while others and those who were not qualified for a certificate were symbolized in turquoise.

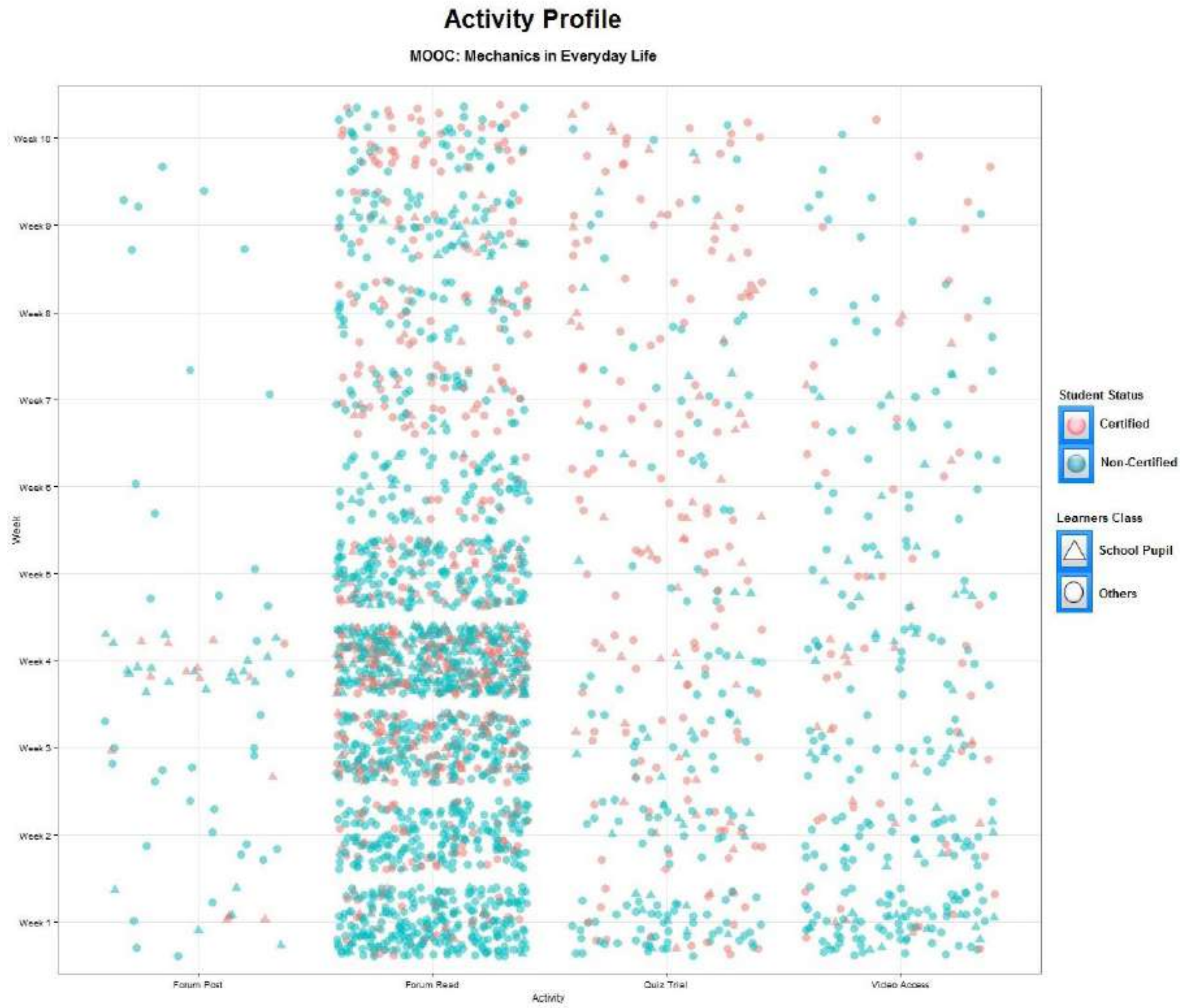

**Figure 43.** Activity profile of the examined MOOC's students. The activity profile is divided into learner classes and learner status.

The figure shows school pupils with a triangle symbol spread among all activities with a noticeable action in the quiz trials compared to the other participants with a square symbol. The school students were posting their reviews and questions more often in the first four weeks supported by the class teacher who encouraged them to use the online discussions. This proves the fact that the interaction between the students themselves and the instructor guarantees a friendly social atmosphere and fosters the overall exchange of communication (Ebner, Lackner & Kopp, 2014). On the contrary, due to the reason that this MOOC is not obligatory, the pupils were mostly motivated to complete the course either by their inner intention of interest or because the class teacher induced them to do so. According to in-the-class observations, the teacher was motivating the students by offering them some extra participation points. Nevertheless, the MOOC was not completely integrated with this science class, but as a collateral plan of study. In addition, the quiz scores were not related to their final class outcome, and that would explain the low student participation in quizzes.

The dropout rate is obvious in the activity profile. The observation shows passive interactions on week-5 in the video and forum posts activities. However, reading in the forums was flowing and was at peak on week-4. This is explained by the teacher answering students' questions through the online forum. 18.5% is the percentage of certified school pupils (*N*= 5). While (*N*= 18), 6.67%, was the certification rate among all who participated in this online course.

An interesting observation was made with the certified school pupils' attitudes amid the MOOC activities. Their participation in posting in the forums completely halted after week-4, and their reading in forums was barely seen in week-6, week-8, and week-10. However, the analysis shows an unexpected correlation between the learning material, which is the videos, and the completion rate. The pupils successfully passed the quizzes without even watching a single segment of any of the learning videos. It looks like the students were practicing the quizzes questions in trials and had been doing self-assessment, regardless of the fact that these questions were randomly changing every time they applied for the same quiz. For instance, it can be seen in Figure 43 that the certified pupils were doing more quiz trials when they didn't access the learning videos.

Different scenarios could be explained by these analysis results. The first explanation is the easy exam questions, which was nullified by looking at the number of quiz trials the pupils did and comparing them with the other learners. The second explanation clarifies that class pupils had enough knowledge to answer the questions without watching videos by asking the class teacher or looking for answers on the Internet and therefore they were not taking the quizzes seriously to test their consideration of the course's subject.

We believe that this MOOC wasn't considered more than typical homework the pupils liked to do. With the motivated teacher and having a sense of contest among the students to see who scores higher in the quizzes, we cannot proclaim more than the consideration that this MOOC as presented in the educational system and learning was happening by the teacher's instructions and endorsements.

Having looked broadly at the activity profile analysis, the evaluation of the students' behavior was expanded to cover the critical part of this study which is the quiz trials accompanied by the learning progression. Despite the fact of the traditional learning process, when a student solves a problem alone without a direct assessment or teacher's assistant, this would lead to a negative productive manner on the student (Merril et al., 1995). However, this was sidestepped by providing multiple choice questions with the possibility of repeating them up to five times and getting their results directly afterward.

In Figure 44, we show the performance progression of each participant during the course. Each plot in this figure represents a trial of all quizzes. The x-axis shows the trial score, while the y-axis records the final score of the quiz. The blue dots denote school pupils, and the orange ones denote other participants. Larger dots depict those who successfully passed a quiz with a score above 75%, while the smaller ones symbolize who couldn't.

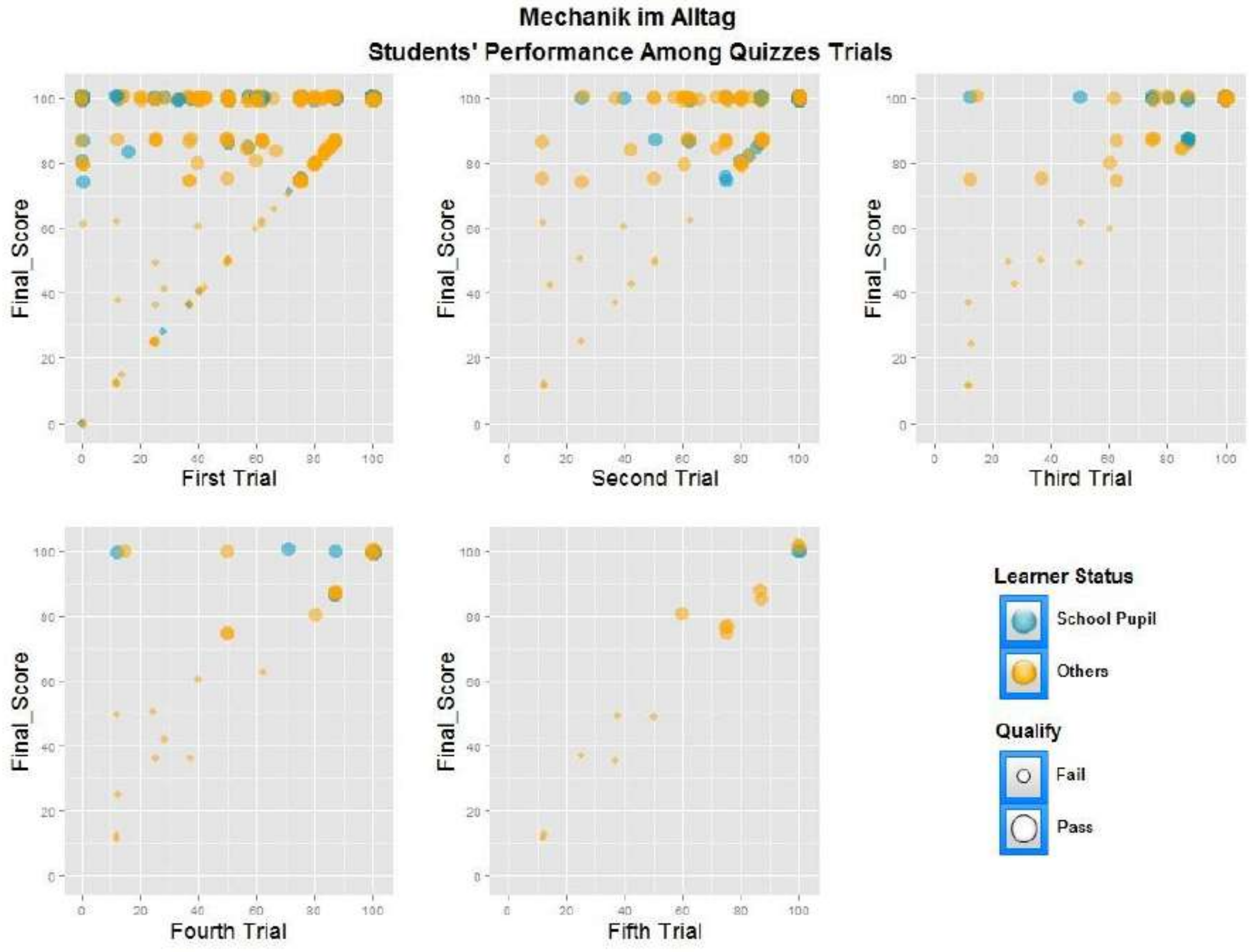

**Figure 44.** The MOOC quiz trials in relation to the students' performance

There were (*N*= 168) of quiz trials that school pupils did; first trial (*N*= 81), second trial (*N*= 55), third trial (*N*= 22), fourth trial (*N*= 8) and fifth trial (*N*= 2). However, there were (*N*= 532) of trials done by the other participants. That concludes that there were 6.22 trials per school pupil and 2.19 trials for each other participant. As a consequence of the plots in the figure being shaped into squares, any participant whose score is shown on any of the plot diagonals means that his/her score on that plot is his final recorded degree, and s/he stopped on that trial.

Getting more involved in the performance, we noted that the pupils whose highest score on the first trial were (*N*= 28). Yet, the majority of participants preferred to try the quiz another time, anticipating a higher score. For instance, the pupils whose second trial was their highest score were (*N*= 36) and whose third trial was their highest score were (*N*= 15), and that can be easily explained by the guided learning they gained through each trial. However, it was obvious that the secondary school pupils tended to do more quiz trials than other participants as a part of a competition portrayal.

Lastly, our observations show that the majority of participants performed better after each quiz trial. Usually, they stopped when their score met the required passing grade. Nevertheless, others liked to do the challenge and took the chance to grasp the full mark.

### 5.1.4 Case Study Summary

This case study discussed the learning analytics application applied to a STEM MOOC. The section showed different patterns extracted from the data generated by the participants. We explained the built activity profile that displayed the dropout rate and the pupils' total engagement during the course. Moreover, we discussed student performance during the self-assessment quiz trials and elucidated the benefits of guided learning on pupils' outcomes during the course.

## 5.2 Case Study (2): Tracking Video Activity in MOOCs[8]

This case study is established based on the data fetched from iLAP and the LIVE (Live Interaction in Virtual Learning Environments) system. LIVE is a web-based information system that provides interactive learning videos in iMooX. It offers the possibility to embed different forms of interaction in videos (e.g. multiple-choice questions). Since interactions, as well as communications, are very important in influencing factors of students' attention, LIVE has been developed to carry out some of these features in the iMooX platform.

In general, attention is considered as the most crucial resource for human learning (Frohlich, 1994). Due to this, it is of high importance to understand and analyze this factor. The results of such an analysis should be used to further improve the different methods of attention enhancement (Helmerich & Scherer, 2007). Moreover, learning analytics plays a major factor in enhancing learning environment components such as the video indicator of MOOCs and finally acts as a reflection of and benchmarking tool for the whole learning process (Khalil & Ebner, 2015b). In this case study, a MOOC named *Making – Creative, Digital With Children* is analyzed based on the collected data from iMooX Learning Analytics Prototype (iLAP) and LIVE. This case study aims to show how learning analytics can be used to monitor the activity of the students within videos of a MOOC.

### 5.2.1 MOOC Structure and the LIVE System Overview

The *Making* MOOC was offered in October 2015. It consisted of 11 videos. The MOOC was attended by both schoolteachers as well as people who educate children in non-school settings. It was scheduled in seven weeks with at least one video per week. The workload was assigned to be 2 hours/week and was proposed in the German language.

In general, the functionalities of LIVE could be categorized by the tasks of three different types of users. The first ones are normal users who could be seen as students. They are only allowed to watch the videos and to participate in the interactions. Figure 45 shows a screenshot of

---

[8] Parts of this section have been published in:

Wachtler, J., Khalil, M., Taraghi, B., & Ebner, M. (2016). On using learning analytics to track the activity of interactive MOOC videos. In M. Giannakos, D.G. Sampson, L. Kidzinski, A. Pardo (Eds.), *Proceedings of the LAK 2016 Workshop on Smart Environments and Analytics in Video-Based Learning* (pp.8–17) Edinburgh, Scotland: CEURS-WS.

a playing video which is currently paused and overlaid by an interaction. To resume playing, it is required to respond to the interaction which means that the displayed multiple-choice question must be answered in this example.

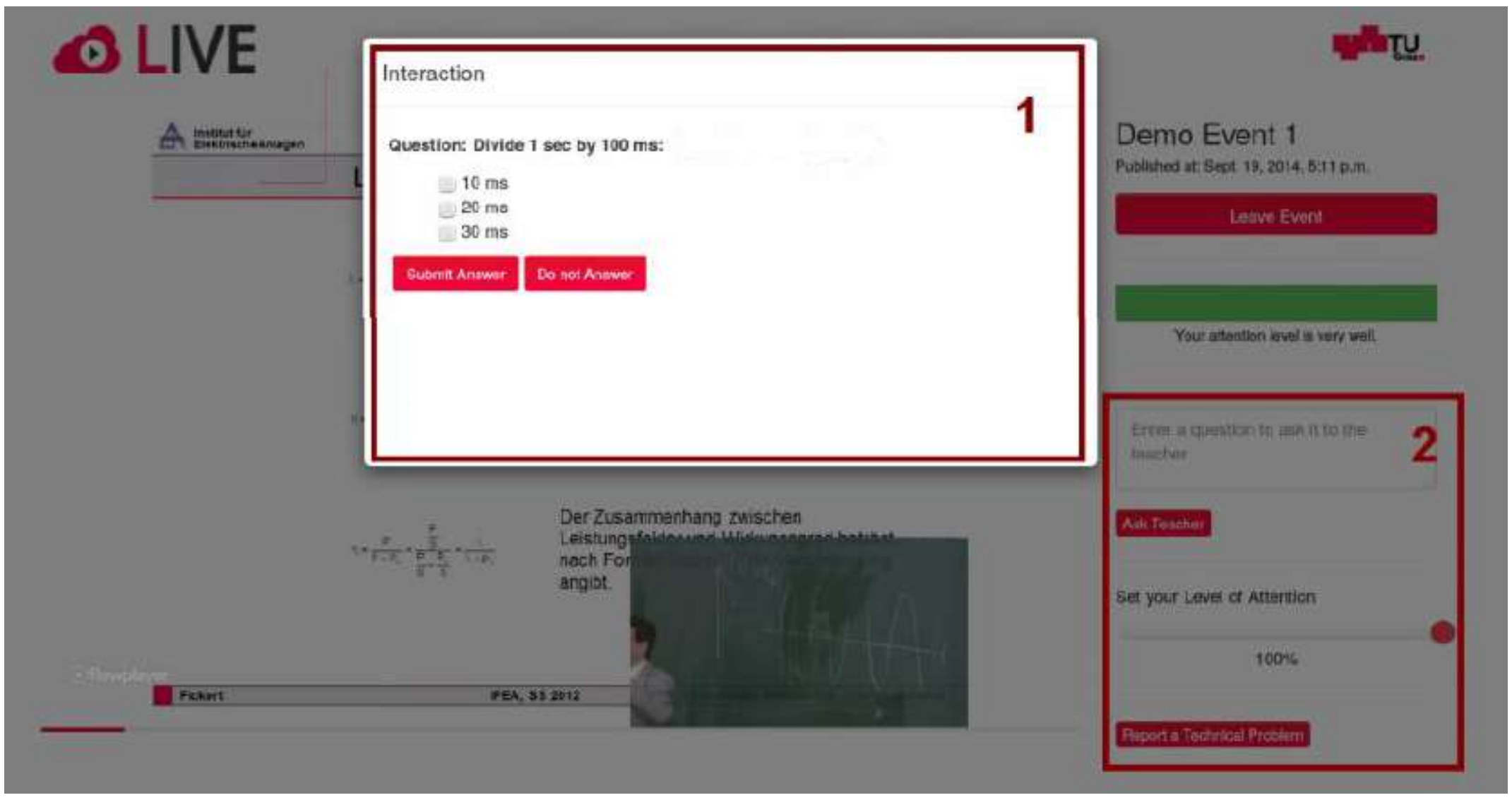

**Figure 45.** A screenshot of LIVE shows a video interrupted by a multiple-choice question

In comparison, the users of the second group are equipped with teacher privileges. They are additionally able to embed interactions in the videos as well as to view different forms of analysis. The third group of users is researchers. They can download different forms of analysis as raw data. This means that they can select a video and obtain the data as a spreadsheet (as CSV files). Our role here was to track students in the *Making* MOOC. Therefore, we acted as researchers and collected LIVE data for advanced analysis.

Finally, the following list aims to give a summary overview of the features of LIVE (Wachtler & Ebner, 2014):

- Offers different methods of interaction, like questions, Completely Automated Public Turing Test to Tell Computers and Humans Apart (captchas), multiple-choice questions
- Provides a logging system for research purposes
- The ability to download raw data for researchers; opportunities like iLAP or other tools

The followed methodology in this case study was carried out using two data repositories from iLAP and LIVE. The data provided by LIVE was evaluated using visualizations and exploratory analysis. Most of the data taken from LIVE belongs to the interactive multiple-choice questions with detailed time stamps. On the other hand, the data provided by iLAP is evaluated to measure the total activity of the weekly videos.

### 5.2.2 Case Study Results and Discussion

This section presents a very detailed analysis of the videos of the MOOC as well as the multiple-choice questions from LIVE. First, the delay of response to the questions provided by LIVE in the MOOC videos during the seven weeks is demonstrated by two figures. The first one is Figure 46 which visualizes a box plot. The x-axis records MOOC videos during the duration of the course, while the y-axis shows students' delay of response in seconds. This period was limited to 60 seconds. Students are categorized as certified students, who finished the course successfully and applied for a certificate, and non-certified students. In this figure, we tried to study the difference in behavior between both categories.

In some of the weeks, certified students took more time to answer the questions such as in week 4 and week 7. For instance, the certified student median in week 4 was 15 seconds, while the median for the non-certified students was 13 seconds. Furthermore, there was a 3 second difference in the median between certified and non-certified students in week 7. Additionally, the median in week 1 and week 5 were typically the same with an insignificant variation between the first and the third quartiles.

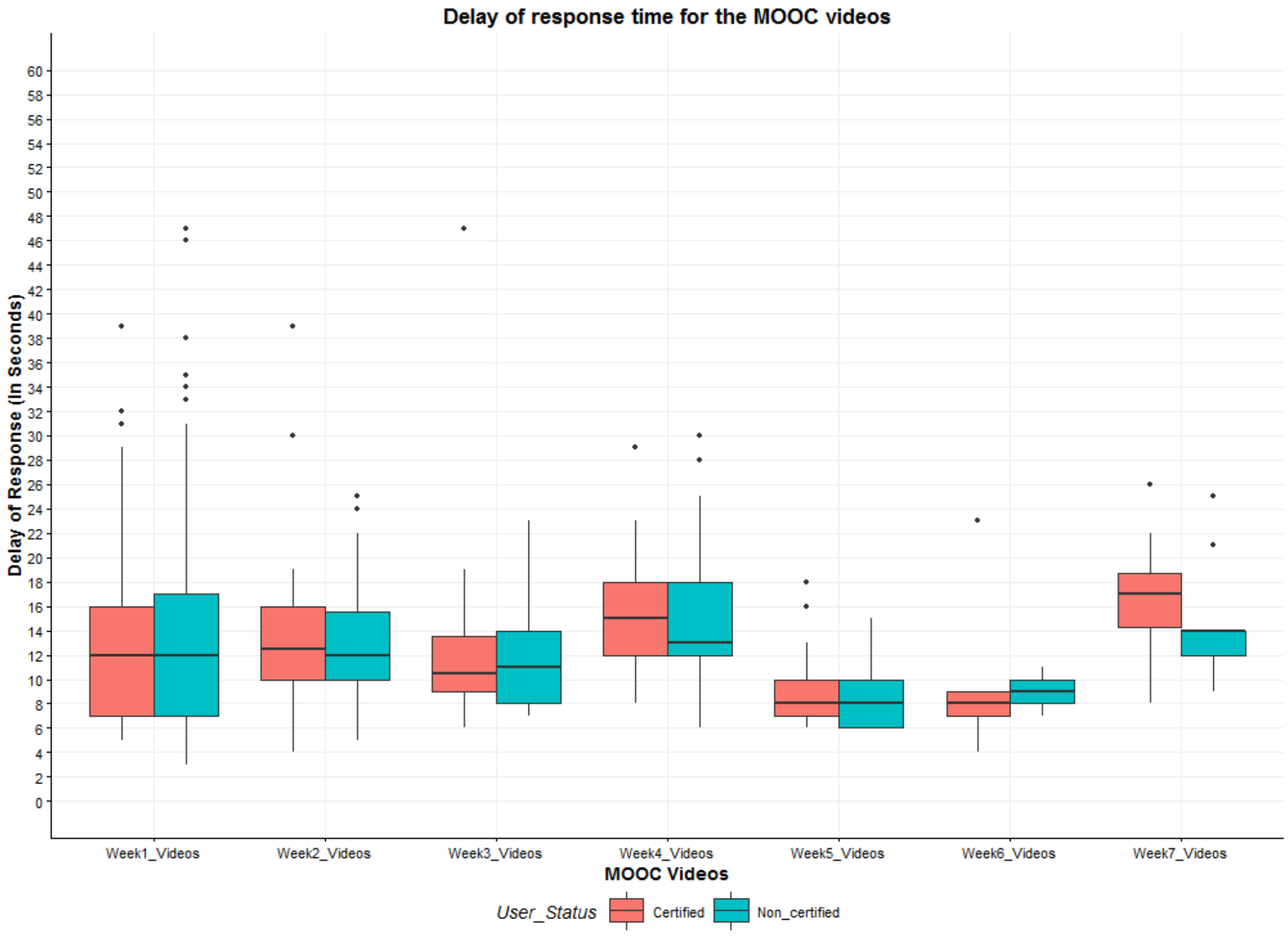

**Figure 46.** A box plot showing the reaction delays to multiple-choice questions. Red indicates certified students, blue indicates non-certified students.

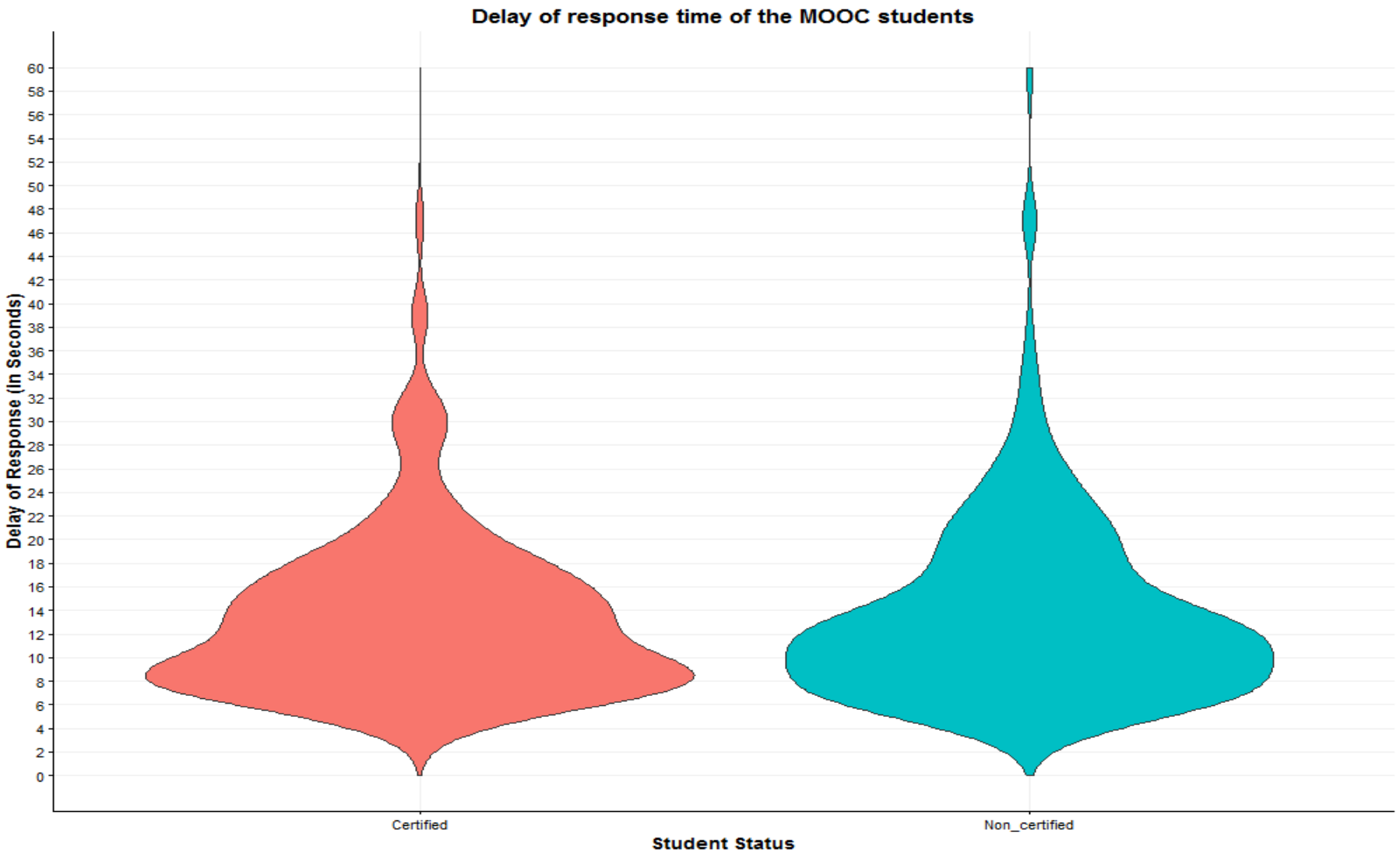

**Figure 47.** A violin plot summarizes the reaction delays to the multiple-choice questions of the examined MOOC. Red indicates certified students, blue indicates non-certified students

In comparison, Figure 47 visualizes a violin plot. The x-axis indicates student status. This visualization summarizes student status and the delay of response time to the multiple-choice questions in all the MOOC videos. The thickness of the blue violin shape is slightly wider than the red one in the (8-13) seconds range, which indicates there was more time needed to answer the questions. In addition to that, the non-certified violin shape holds more outlier attributes than the certified division.

We believe from the previous two observations that certified students took less time in answering the questions in general. This case can be explained in that the questions were easy to answer if the students were paying enough attention to the video lectures.

Figure 48 displays the timespan division in percentage and the timing of the first multiple-choice question represented as a vertical dashed line.

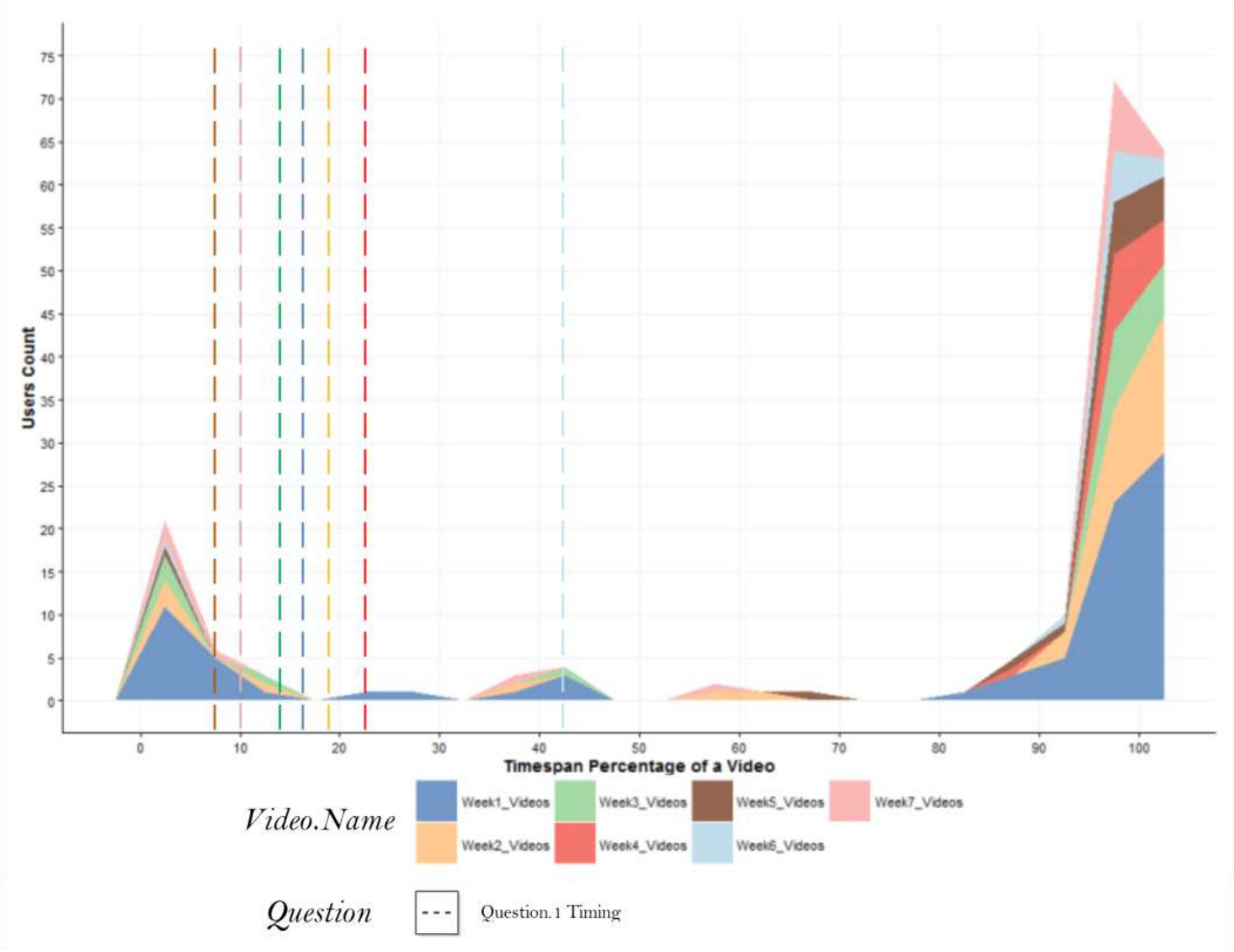

**Figure 48.** Timespan division in percentage and the timing of the first multiple-choice question

Using this visualization, we can infer the relevance timing of the first question to describe the drop rate during videos. The questions were programmed to pop up after 5% of any MOOC video.

Students may watch the first few seconds and make skips or drop out after that (Wachtler & Ebner, 2015), and this can be seen in the plot where students are dropping in the early 15% of the videos. To grab the attention of the students and maintain a wise attrition rate, the multiple-choice questions were intended to be shown randomly in the high drop rate scope. Further, week 6 was tested to check the postponed question effect on the retention rate. The data in the figure also shows that students do not drop out of a learning video in the range between 20%-80% unless they replay it on that period and spend time on a particular segment to understand complex content. The promising outcomes are seen with a stable attrition rate in the last four weeks when students are offered interactive content during the video indoctrinate process.

In Figure 49, the data is displayed in order to trace the video drop ratio of each second in every video.

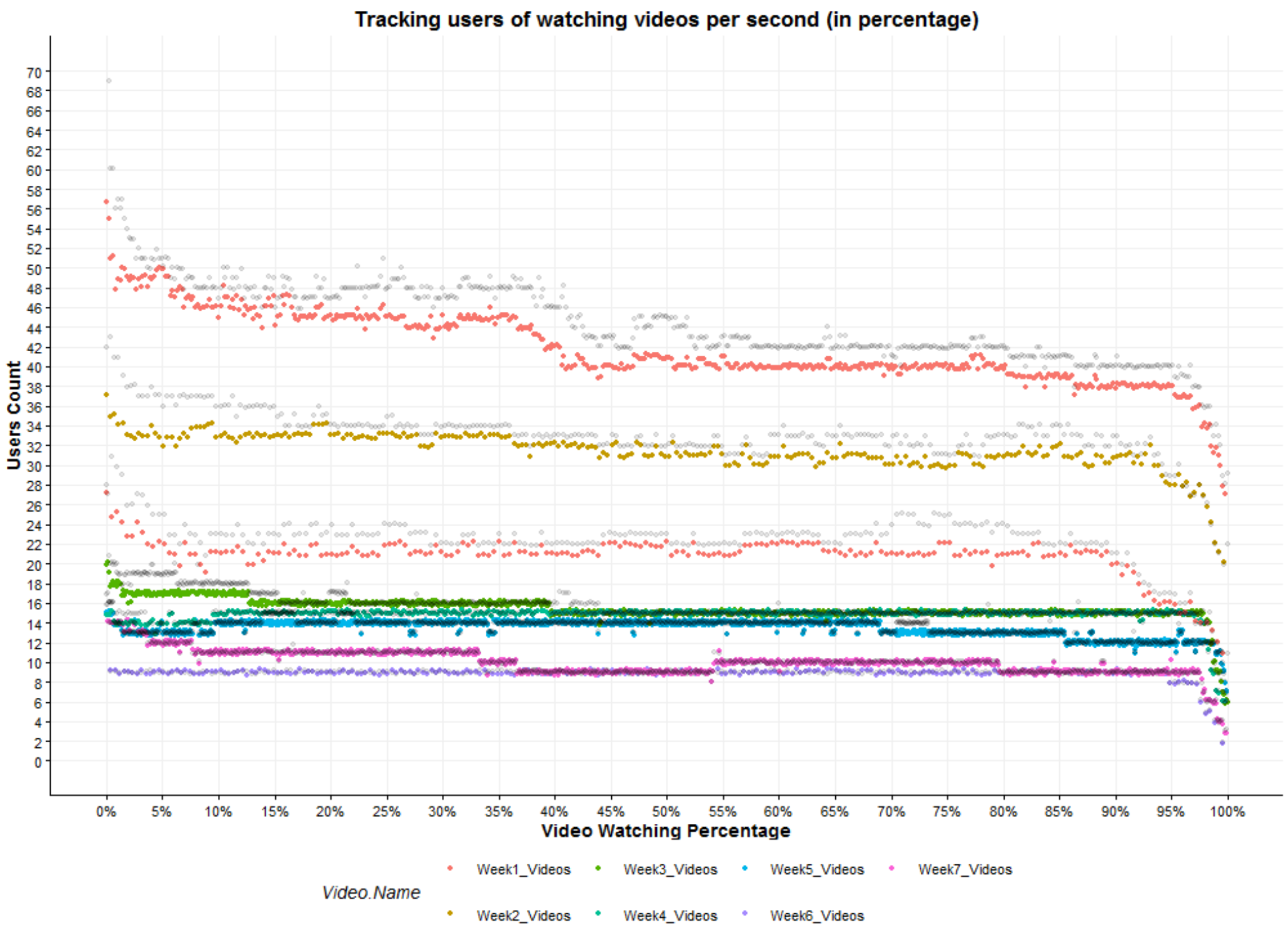

**Figure 49.** Trace of the video drop ratio of each second in every MOOC video

The x-axis displays the percentage of videos. The colored points specify the video name and the watcher count, while the black shadowed points indicate the number of views. For instance, it is obvious that the data in the first three weeks shows more views per user which can be explained as an initial interest of the first online course weeks. On the other hand, the views nearly equaled the number of users from week 4 to the last week.

Another interesting observation is the slow drop rate during the videos in all the weeks despite the high drop in the last 2-3% of every video. A clarification of such attitude is due to the closing trailer of every video which most students jump over.

Due to the independence of the examined MOOC, each video of this course does not rely on the previous one. The activity of every video varies in every week. For this reason, Figure 50 shows the activity of the total number of stop and play actions in the MOOC videos.

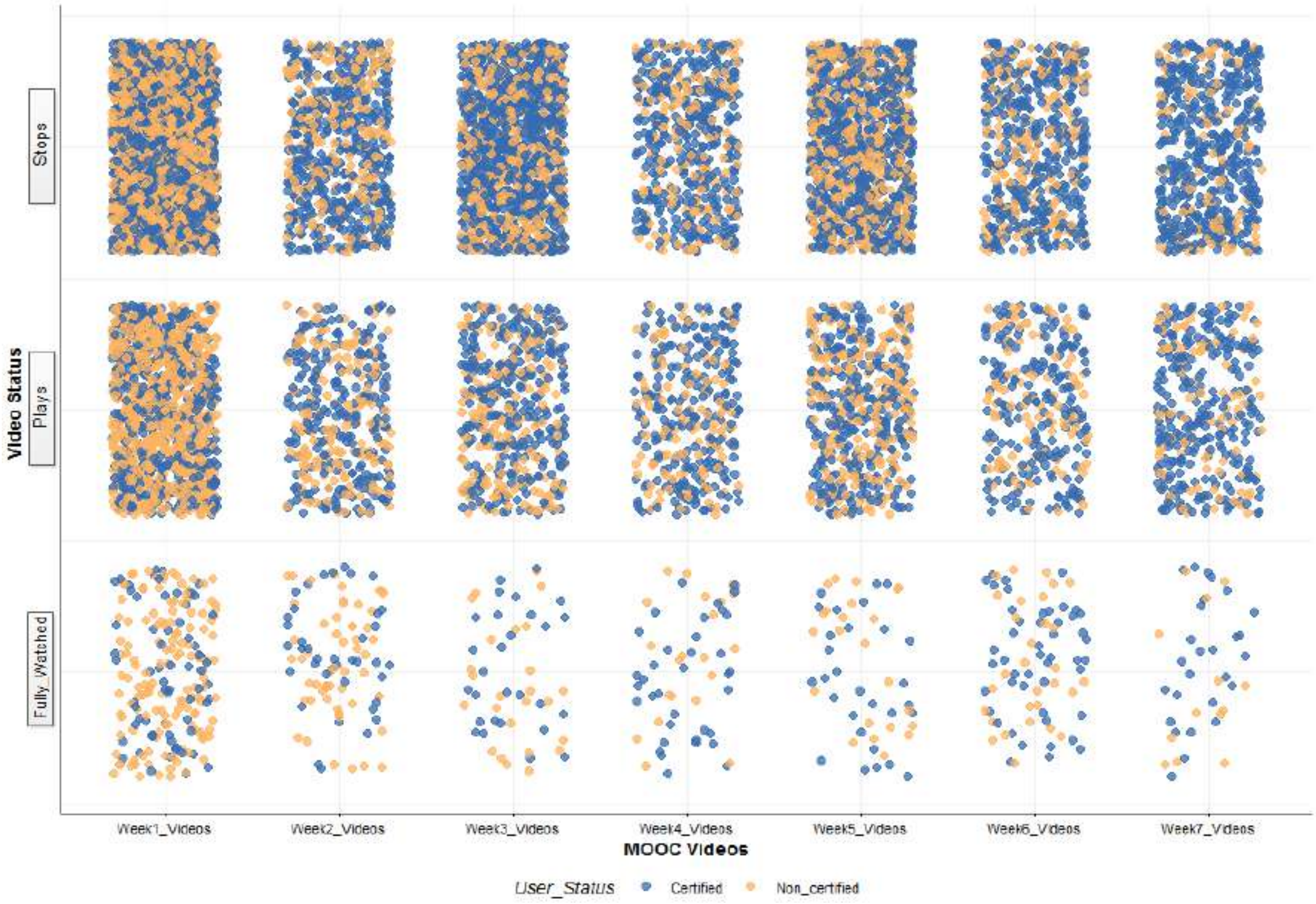

**Figure 50.** The activity of the total number of stops and plays in the MOOC video. Blue indicates certified students, orange indicates non-certified students.

The blue points denote the certified students while the orange ones denote the non-certified students. In fact, the first three weeks reflect proper enthusiastic action count. We realized that there was distinct activity by the non-certified students in week 5. A reasonable clarification is because of the interesting topic of that week which was about 3D-Printing. However, their engagement becomes much less in the last two weeks, as this was proven in other MOOCs' video analysis (Khalil & Ebner, 2016e).

### 5.2.3 Case Study Summary

Capturing the attention of learners in online MOOC videos is an intriguing obstacle across learning analytics discussions. In this case study, the use of an interactive video platform presenting videos of a MOOC was shown. Video interaction data collection was done using the LIVE system and iLAP. Data analysis points out the main functionalities of the LIVE platform as well as the participation and the activity of the students in one of the iMooX offered MOOCs. Additionally, we demonstrated an evaluation of this system in order to examine the behavior of students. Finally, the results showed that the main solution behind latching onto the students' attention is through evaluation of the questions' content and the timing of interactions. In addition, we conclude that by observing MOOC video activity statistics, we can show student behavior in every week of the MOOC and what attitude differences are noticed between certified and non-certified students.

## 5.3 Case Study (3): Dropout Investigation[9]

MOOCs are open platforms for everyone regardless of their location, age, sex, and education. Thus, the design of a MOOC must respect this unpredictable heterogeneity which results in a balancing act between multiplicity and unity in respect to resources and prior knowledge of students. Jasnani (2013, p. 7) mentions a "lack of professional instructional design for MOOCs" can be cited as one of the reasons for the low completion rates MOOCs suffer from. Others relate it to motivation and interaction (Khalil & Ebner, 2014). Many researchers like Jasnani (2013), Jordan (2013) and Hollands and Tirthali (2014) stated that MOOC completion rates usually range from around 3% to 15% of all enrollees.

The literature studies in this dissertation and the previous case studies clearly agree when stating that MOOCs suffer from dropout. Thus, we look during this Ph.D. dissertation into investigating the dropout issue of MOOCs using learning analytics. This case study examined some MOOCs offered by iMooX to look into what learning analytics can tell us regarding this issue.

### 5.3.1 Background

Several investigations have already been conducted to identify reasons for high drop-out rates that lead to low completion rates, such as Khalil and Ebner (2013a; 2013b) who worked out the importance of interaction to guarantee participants' satisfaction in a MOOC which increases the probability of course completion. Colman (2013) conducted a web-survey and asked for reasons why participants dropped a MOOC; amongst others, these six are named (Table 10):

**Table 10.** Reasons for dropping a MOOC according to Colman (2013)

| Reason | Classification |
|---|---|
| "Takes Too Much Time" | Intrinsic factors |

[9] Parts of this section have been published in:

Lackner, E., Ebner, M., & Khalil, M. (2015). MOOCs as granular systems: design patterns to foster participant activity. *eLearning Papers*, *42*, 28-37.

Khalil, M. & Ebner, M. (2016c). "Learning Analytics in MOOCs: Can Data Improve Students Retention and Learning?". *In Proceedings of the World Conference on Educational Media and Technology, EdMedia 2016*, pp. 569-576, Vancouver, Canada. AACE.

| | |
|---|---|
| “You’re Just Shopping Around” | |
| “You’re There to Learn, Not for the Credential at the End” | |
| “Assumes Too Much Knowledge” | Extrinsic factors |
| “Lecture Fatigue” | |
| “Poor Course Design” | |

The reasons for dropping a MOOC can, thus, be classified into two categories: internal or personal and external or imposed. Dealing with the second category, it can be observed that some MOOCs “are headlined by prominent professors in their respective fields” (Hay-Jew, 2014, P. 614), so it is then the university’s or the professor’s prestige that leads to high registration rates. An inappropriate course design or the lack of a clear course structure can be identified as main reasons for drop-out. Regarding the first category, it is more difficult to identify and validate the personal purposes as they are personal motivations. Sometimes MOOC participants just “shop around,” pick up different elements of a course but do not want to finish the course itself. Sometimes it is not the whole course that seems to be interesting but only parts of it that are new, innovative or simply appealing.

In this case study we implemented a comprehensive automatic tracking system (learning analytics) using iLAP tool for user activities within each single examined MOOC. Afterward, the data was thoroughly processed and interpreted for further recommendations and conclusions.

### 5.3.2 The Examined MOOCs Structure

Within the Austrian MOOC-platform iMooX, we scrutinized three different MOOCs focusing on student activity and completion rates: *Gratis Online Lernen* (‘Free Online Learning’), *Lernen im Netz* (‘Learning online’) and *Soziale Medien & Schule* (‘Social Media & School’). All three courses were held in the German language, were delivered at iMooX and were each eight-weeks long. The workload as defined before the course was known from the beginning: *Gratis Online Lernen* (2 hrs/week), *Lernen im Netz* (5 hrs/week) and *Soziale Medien & Schule* (3 hrs/week).

Participants who passed every quiz with at least 75% of the maximum points could download a certificate at the end of the course. However, *Lernen im Netz* was a special course as it was not only a MOOC but also a university lecture at the University of Graz. Students of the University of Graz could attend the MOOC as a free course worth 4 ECTS points but had to pass a supplementary electronic exam at the end of the semester.

### 5.3.3 The Case Study Analysis

First, we analyzed students based on their type (*Similar to our categorization in Chapter 4.1.1*). Regarding the course participants, a differentiation had to be made between participants who are registered, those who work actively, as well as those who just completed the course and those who also downloaded the certificate, as Figure 51 shows.

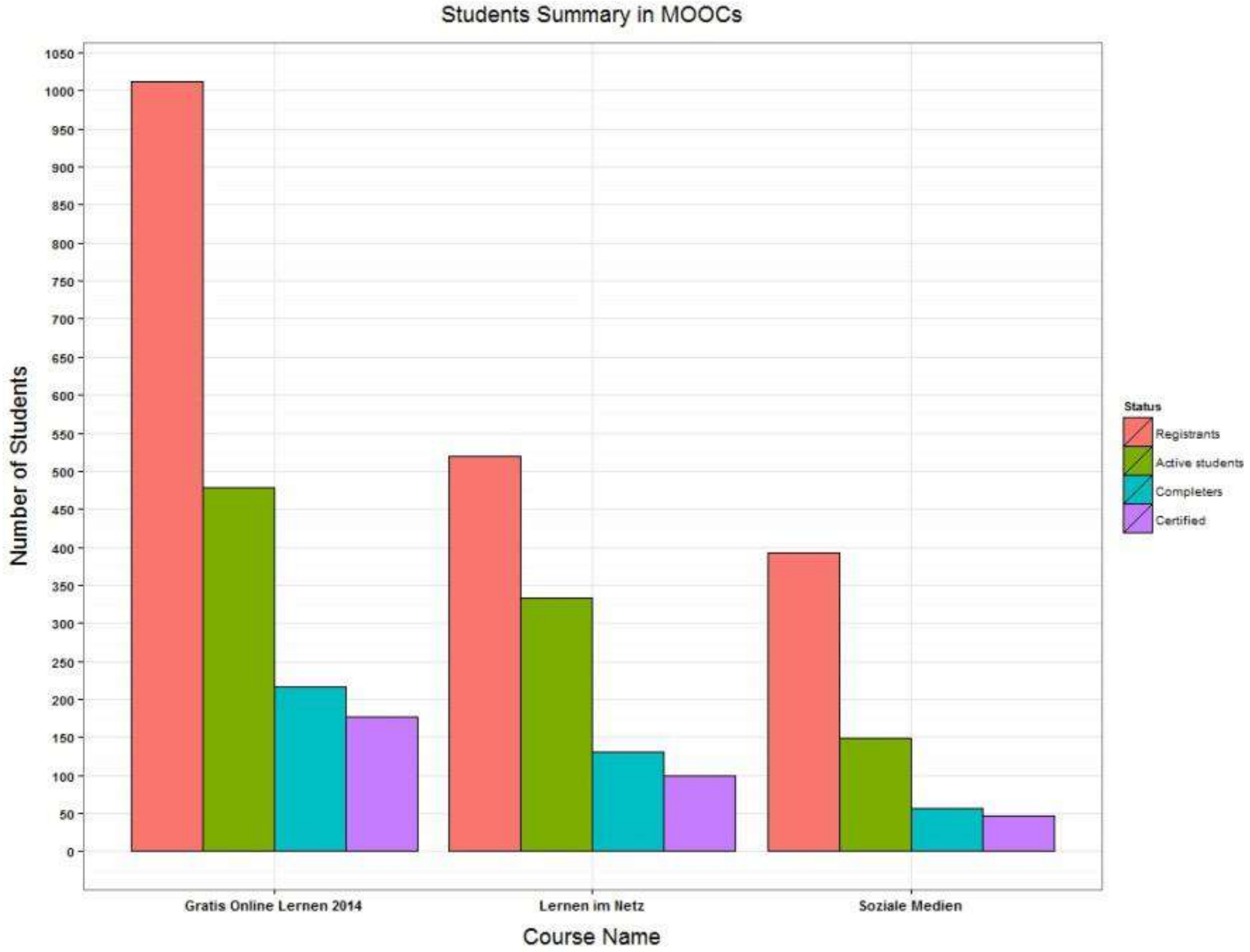

**Figure 51.** Statistics that show the number of students who enrolled in the three examined courses based on their type of participation

As the figure illustrates, an obvious gap between registered and active students can be observed in all three courses. Active students are those who wrote at least one forum post or did at least one quiz. In *Gratis Online Lernen*, 1012 students registered, but only half of them, 479 students, were active (47.33%). *Lernen im Netz* shows a higher percentage concerning the correlation between registered and active students (64.16%), *Soziale Medien & Schule* a lower percentage (37.9%). As in "traditional" lectures at brick and mortar universities the number of interested people who "pass by" and do not start a course is high. They register without planning to do the course, want to get to know the professor or the course, or are interested in just one unit or aspect of a course. They are more or less like tourists that "shop around" as Colman (2013) showed in his web-survey. Calculating the completion rates on the basis of these registration rates, *Gratis Online Lernen* had a completion rate of 21.44% and a certification rate of 17.49%, 25.24% completed *Lernen im Netz* and 19% downloaded the certificate, whereas *Soziale Medien & Schule* had a completion rate of 14.2% and 11.95% got certificated.

When calculating the completion rates on the basis of the active participants the numbers increase:

**Table 11.** Active participants who completed or downloaded the certificate

| MOOC | Active & completed | Active & certificate |
|---|---:|---:|
| Gratis Online Lernen | 45.3% | 36.95% |
| Lernen im Netz | 39.33% | 29.72% |
| Soziale Medien & Schule | 37.58% | 31.54% |

Table 11 illustrates that the completion rates are on a level that is comparable to our experienced traditional university lectures. It shows nevertheless that more than half or two-thirds of the active participants at a certain point of the MOOC lose interest, start lurking or become passive consumers. To identify this point the participants' activity should be taken into account. For the three Austrian MOOCs, the quiz trials, the reading of forum posts and the writing of forum posts can be investigated.

Figure 52 shows the number of quiz trials within the three courses since the quiz completion is crucial for obtaining a certificate.

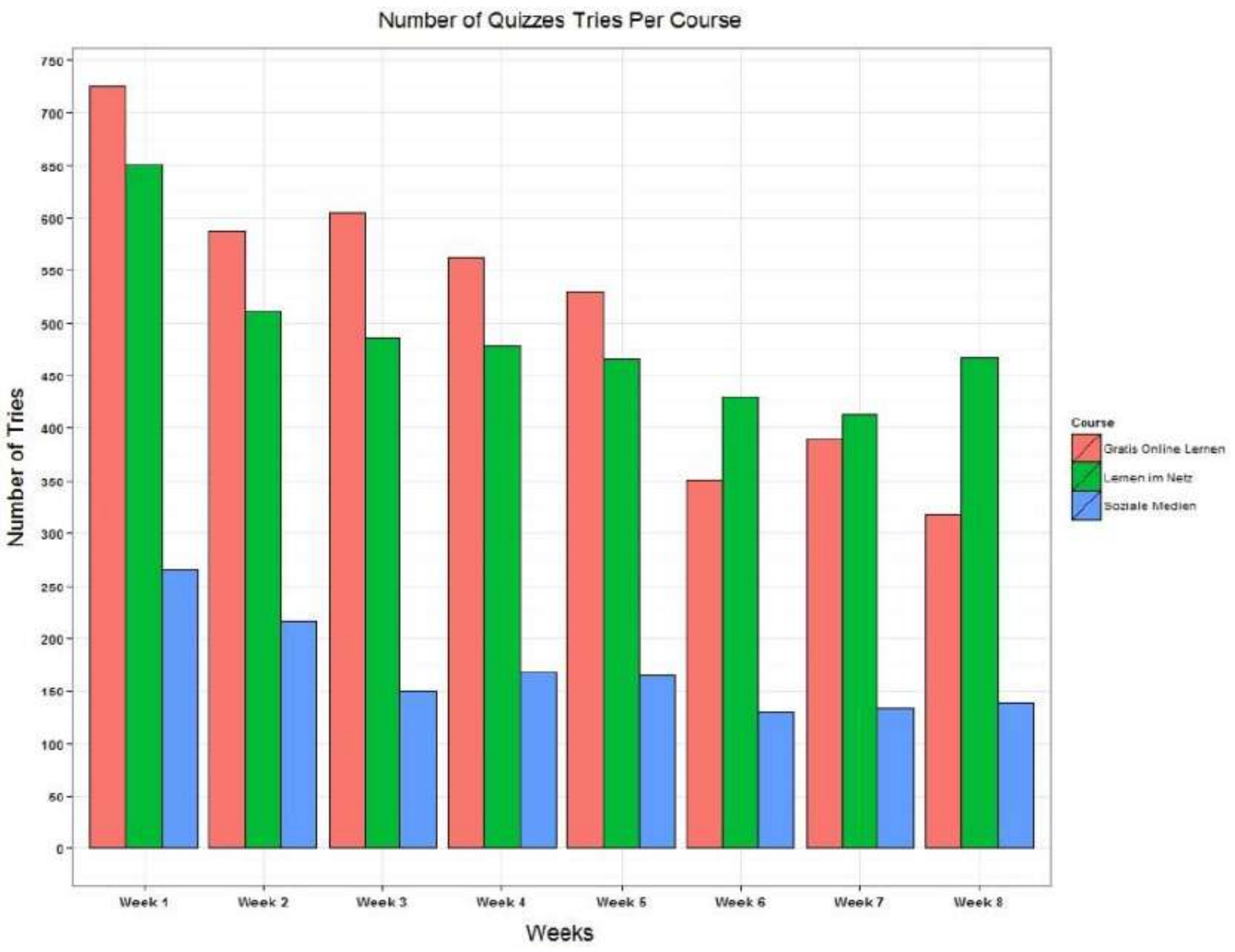

**Figure 52.** Quiz trials per week for the three examined MOOCs

Taking a closer look at participant quiz activity shows a tendency that the “drop-out point” can be identified between Week 4 and 5; week 8 in *Lernen im Netz* has to be seen as an outlier. In this course the last course week treated MOOCs as the subject and it can be presumed that the participants were highly interested in this topic or the quiz was very difficult, so more tries were needed.

This split in the middle of the course can also be found within the forum reads and the forum posts as Figure 53 and Figure 54 show, using the example of *Gratis Online Lernen*.

The participants read the forum postings, but the frequency diminishes within the course after the third week. Whereas in the first week 6706 reads can be counted, in week 4 there are only 1760 reads. *Lernen im Netz* has 1714 reads in week 1 and 465 in week 4. As the course had a first week that was intended to socialize the participants (Salmon, 2007) to make them familiar with the platform and each other, the readings in this pre-week are extremely high (2970 reads). *Soziale Medien & Schule* has 186 reads in the first and 153 in the fourth week; week 2 and week 8 can be seen as outliers as the forum reads are significantly higher than in the others weeks (i.e. 299 in the last week).

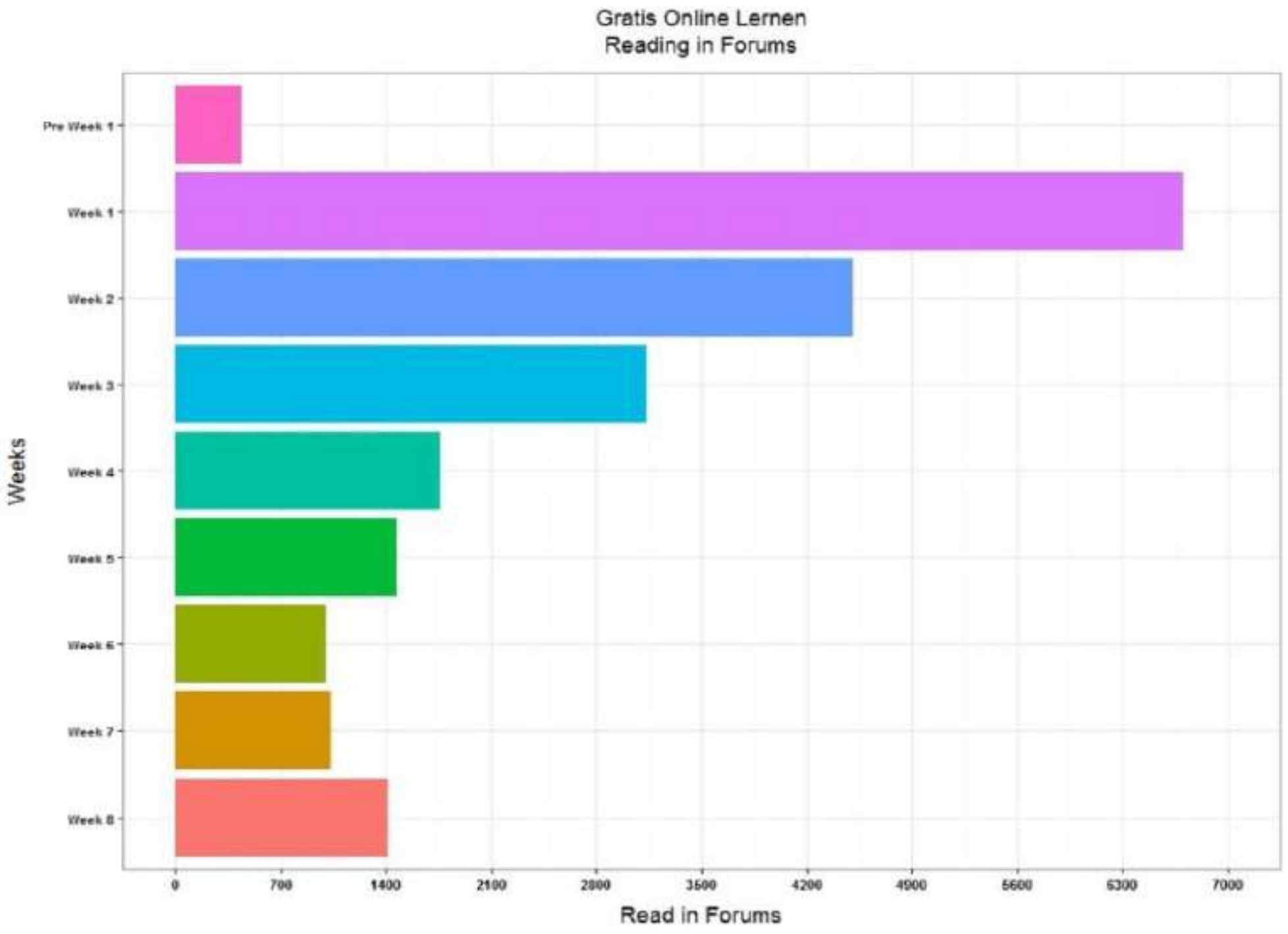

**Figure 53.** Forum Readings in *Gratis Online Lernen*

Understanding the active forum participation in terms of writing forum posts, the “drop-out point” can be seen in *Gratis Online Lernen* after the fourth week with 95 posts, whereas in week 1,251 postings were written. In the last week, still 50 postings were added to the forum. In *Lernen im Netz* (see Figure 54), the gap is even clearer: in the first week, the participants posted 169 entries, in the fourth week 20 and in the last week 9. The course *Soziale Medien & Schule* cannot be scrutinized in this context, as the number of posts is too low: one post in the first week, five in week 2, two posts in week 4 and no more posts after the fifth week.

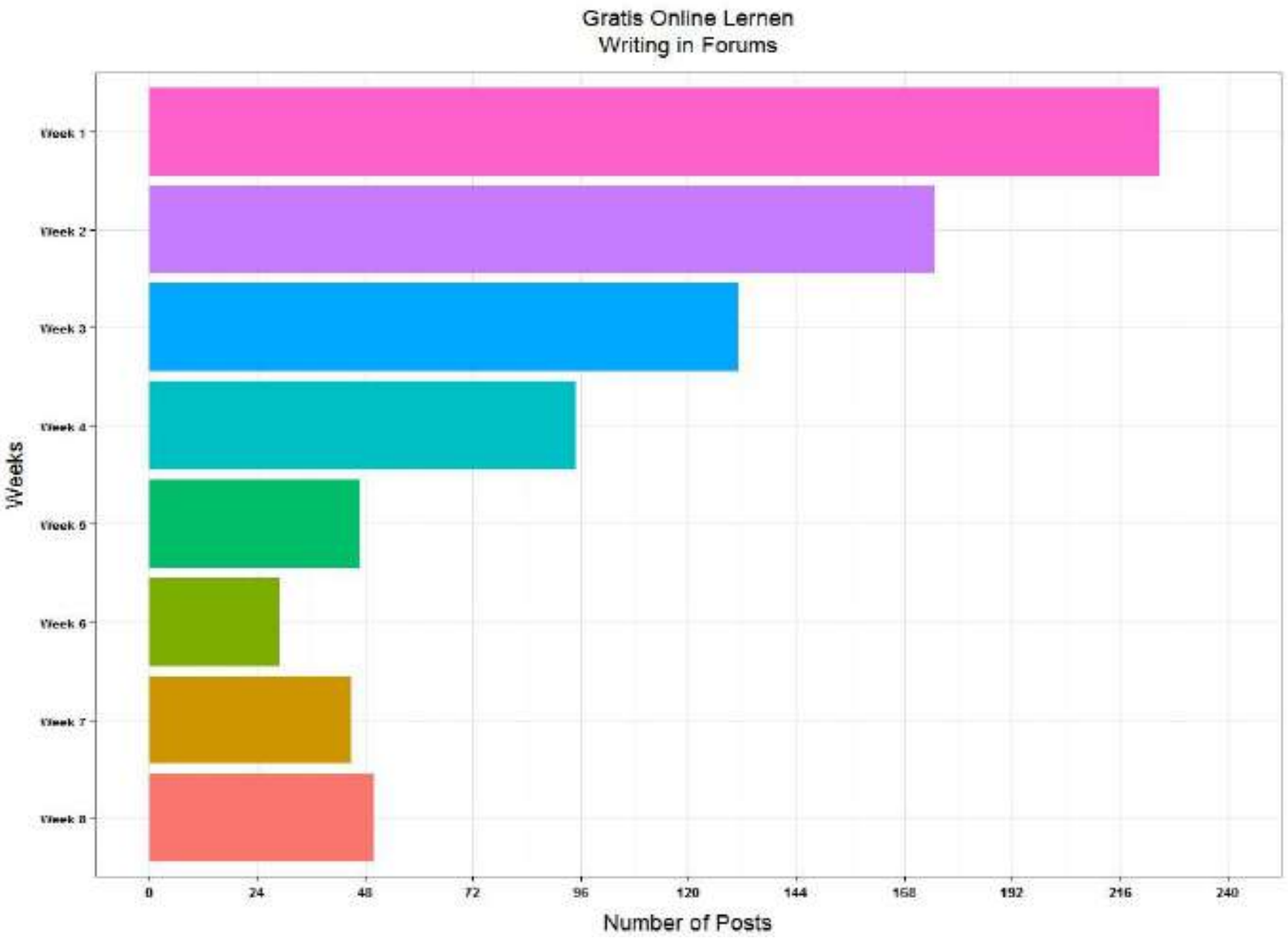

**Figure 54.** Forums posts in *Gratis Online Lernen* MOOC

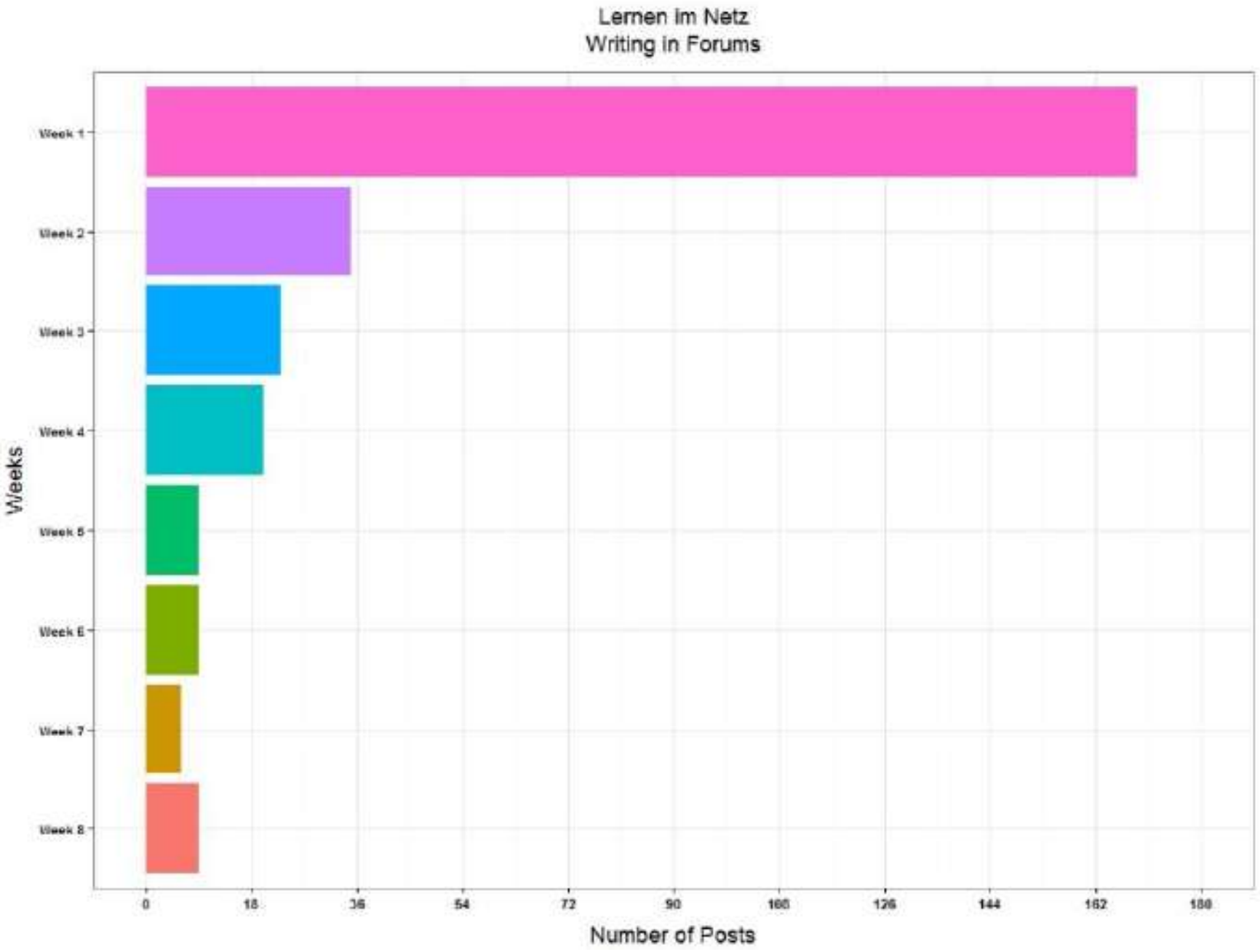

**Figure 55.** Forums posts in *Lernen im Netz* MOOC

#### 5.3.3.1 MOOC Variables Impact and Dropout

Next, we tried to measure through learning analytics the weight of effect of MOOC variables that leave more impact on dropout. To do so, we examined he *Gratis Online Lernen* and *Lernen im Netz* MOOCs since they are very active. We made two main groups of students: completed students, and those who successfully completed the course and did all the tasks. The second category is the students who dropped out during the course. The activities of each group are then assorted and the prevailing behaviors are analyzed. In the end, we observed the differences of each variable to observe which has more impact than the other.

The available MOOC variables that we were able to analyze were: a) quiz attempts, b) discussion forum readings, c) discussion forum writings, and d) login frequency. The MOOC video variable was excluded since the available data on videos was not useful to measure with its data structure. Each student profile was then dedicated with these indicators separately, similar to what we did in case study (1). In addition, all the collected data were distributed based on a weekly scale. Respectively, we calculated the total interactions for the completed students, who successfully finish MOOCs, and for the students who dropped and calculated the average.

Figure 56a shows the behavior of students who completed the course and the ones who dropped out in *Gratis Online Lernen* MOOC. The left figure displays the average of all interactions in the whole course period, while the right figure shows the average of interactions from week3 to week6. The reason for analyzing that period is due to the stability of the dropout rate during that duration as we saw in the previous analysis results of this case study. In the same fashion, Figure 56b demonstrates the behavior of both student types in the *Lernen im Netz* course. By observation of the line graph in both figures, students behaved nearly identical.

The figures show that the difference between quiz attempts is obvious between both student types. Under that circumstance, the quiz attempts factor was pulled to be the second critical MOOC variable.

However, the greatest difference gap among MOOC indicators was the forums reading activity. This explains several studies such as the one by Ezen-Can et al. (2015), which have drawn attention to the significant effect of MOOCs forum interaction to improve learning and attrition. Furthermore, the third remarkable difference is the login frequency which is noted as a decisive player in determining at-risk students, and this was highlighted in several studies like the one by Balakrishnan and Coetzee (2013). In contrast, the writings were not as efficient as readings; hence, the allocated weight for forums posts is the lowest.

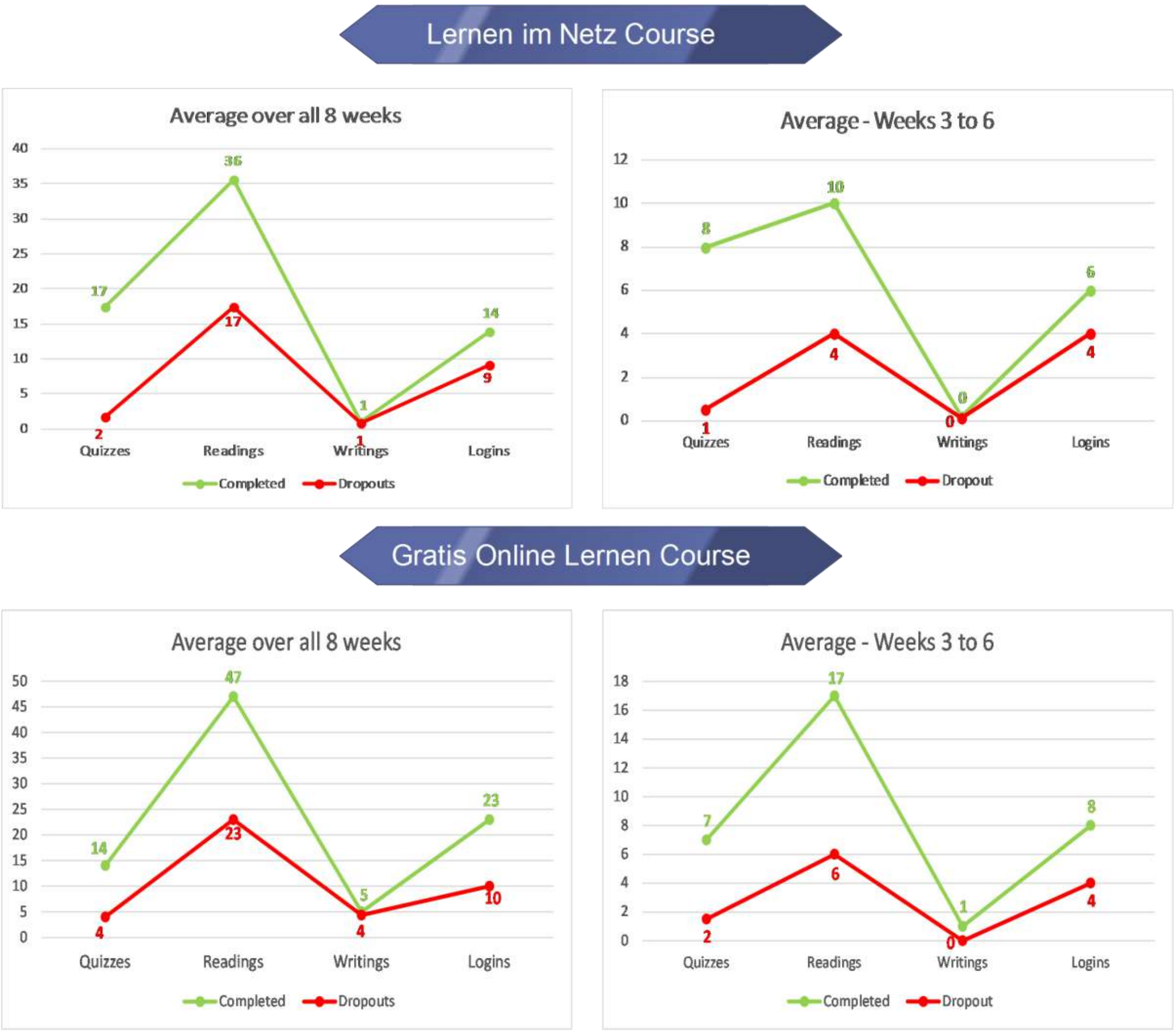

**Figure 56.** The average of MOOC interactions for completed and dropout students. Top (a) *Gratis Online Lernen* MOOC; bottom (b) *Lernen im Netz* MOOC

Generally after all, each MOOC indicator was weighted and calculated based on the difference between the activity performance of students who dropped and the ones who completed in both of the two cases of MOOCs. Subsequently, we proposed a simple equation that defines weights according to their adequate significance to (W1, W2, W3, and W4), by which W1>W2>W3>W4. The equation is articulated as the following:

$$Success\ Rate\ (SR/Week) = W1.Readings + W2.Quiz_Attempts + W3.Login_Frequency + W4.Writings$$

### 5.3.4 Case Study Discussion

The role of forum activities for MOOCs has already been subject to research providing a deeper understanding of communication and collaboration processes within the courses and the participant's community (Gillani et al., 2014; Khalil & Ebner, 2013a). The quality of the forum activity must be further analyzed for the above-mentioned courses. These results on a quantitative basis help to identify a tendency within xMOOCs regarding the probable "drop-out point." All shown figures recall somewhat the so-called Long-Tail effect often discussed in terms of Web 2.0 (Bahls & Tochtermann, 2012). Many learners begin the MOOC and only a few of them complete. On the basis of the presented data, the "drop-out point" for these three courses can be identified between the fourth and the fifth course week. At this point, the participants' activity decrease stops and stays more or less constant. This implicates those participants who are still active in week 5 are more likely to complete the course.

#### 5.3.4.1 Shorter MOOCs

As we could see in Table 10, there are two main categories of reasons why participants do not finish or even drop a MOOC: internal and external forces influence their decisions. The crucial point of decision whether to become passive or leave the course can be seen in course week 4. As a matter of fact, course developers should react to this phenomenon and adjust the instructional design of MOOCs. MOOCs tend to last four to eight or even twelve weeks. If the participants activity decreases dramatically till week four, course developers should consider to plan and design shorter MOOCs. For instance, a longer course can be broken down into several courses focusing on different aspects of a topic as the BBC does for the general topic World War I on the FutureLearn platform (https://www.futurelearn.com/organisations/bbc, last accessed 16.12.2016).

#### 5.3.4.2 Granular Certificates

Responding to the different orientations and motivations towards any learning system, a different certification process or attitude should be developed e.g. by awarding badges (Schön et al, 2013) or gamification elements. These badges can be seen as a possibility to make the informal learning process visible as they can be displayed in professional social networks, e.g. LinkedIn

(www.linkedin.com). Within a longer MOOC, different badges – according to topics, projects or special achievements – can be acquired, the collecting process and the prospect of the next badge could increase, or renew the extrinsic motivation.

### 5.3.5 Case Study Summary

MOOC critics often refer to the low completion rates and the high drop-out rates when disputing xMOOCs. As research has proven, at the beginning of a course the number and motivation of the participants is higher than in the last week. Using learning analytics to get a deeper understanding of the logic of xMOOCs, this case study has shown that the fourth and the fifth week of an eight-week-course are crucial in terms of participant motivation and vision toward course completion. This so-called “drop-out point” adjudicates on whether participants continue and most likely complete or whether they drop the course. We also tested the impact of MOOC variables and dropout and therefore realized that forums activity is a vital measurement of engagement. Therefore, we suggest that shorter MOOCs improve social communication. Motivating students through gamification elements like progress or badges elements can help, to some extent, the extrinsic motivation and engagement of MOOC students.

## 5.4 Case Study (4): Fostering Forum Discussions in MOOCs[10]

Forum discussions and their role in and for MOOCs have been widely scrutinized so far. The results are ambivalent. Researchers underline that only a few participants seem to actively contribute to forums. They are, however, highly important for a course's success and positive impression. In the previous case study, we saw that discussion forums make a huge difference in defining students at risk. Therefore, we thought that improving the discussion forums at iMooX might increase the retention rate and increase student motivation. As interaction in forums among the participants is crucial to foster their motivation to engage in a course (and not to drop out), a closer investigation into the forum was done in this case study. In this case study, we deduce design recommendations that help establish individualized support for participants, foster interaction and collaboration among learners, and thus, support them in self-regulated learning. We follow the Gilly Salmon's *Five Stage Model* that was designed for traditional online learning settings. We will also implement a new redesign of the discussion forums as a step to motivate students extrinsically using badges and rewards and evaluate the results.

### 5.4.1 Background

Discussion forums are, besides the videos and different assessment methods, a fixed part of both xMOOCs and cMOOCs (Jasnani, 2013). Their role within MOOCs has already been scrutinized from different perspectives and with ambivalent results. Huang et al. (2014, p. 125) summarized that participants use the forum for different purposes according to their personal needs and interests, "which appears to be more an inherent than an extrinsic trait." There are only a few participants that become active in a forum but play an important role in the construction of knowledge. Breslow et al. (2013, p. 22) showed in their study, which analyzed the first MOOC on edX, that "only 3% of all students participated in the discussion forum" and "52% of the certificate learners were active on the forum." Huang et al. (2014, p. 125) identified and

---

[10] Parts of this section have been published in:

Lackner, E., Khalil, M. & Ebner, M. (2016). How to foster forum discussions within MOOCs. A case study. *International Journal of Academic Research in Education*, 2(2). doi: 10.17985/ijare.31432

Reischer, M., Khalil, M. & Ebner, M. Does gamification in MOOC discussion forums work? In *Proceedings of the European Stakeholder Summit on experiences and best practices in and around MOOCs (EMOOCS 2017)*, Madrid, Austria. (in press) scrutinized the so-called *superposters*, i.e. learners who actively participate in forums, and highlight the important role that forums play for MOOCs.

On the other side, Onah et al. (2014) again showed that the interaction level in forums is generally low. They resumed (2014, p. 4): “In general, more active engagement strategies and the introduction of tasks related to forum posts are needed to encourage users both to initiate threads and to post replies to others.” Thus, an appropriate setting is needed to help establish individualized support for participants, foster interaction and collaboration among learners, and, thus, support self-regulated learning.

### 5.4.2 Gilly Salmon’s Five-Stage Model

More than ten years ago, Gilly Salmon introduced a book called *e-moderating: The Key to Teaching and Learning Online.* It has become a benchmark in the field of online learning and teaching. According to the author, e-moderation is crucial to establish a setting that offers individualized support for learners that fosters interaction and collaboration; hence the construction of knowledge, and that, finally, supports self-regulated learning.

In order to achieve these objectives, i.e. to guide and support the group and initiate learning processes, Salmon (2007, p. 28-50) suggests thinking in five different stages as Figure 57 illustrates. It is obvious that the individual stages are built on each other. At each stage, learners and teachers stay individually in terms of time and dedication. The first stage is summarized by making the learners familiar with the learning environment and the course design. In this stage, it is emphasized to let students get used to how to use the discussion forum, how to make a post and encourage them to read. Motivation flattens from time to time. Therefore, the design suggests using an e-moderator who would help students and motivate them to keep the knowledge construction going without being too instructional. At these first two stages, the forum might play an important role as a place to collect and publish all information that seems to be necessary to get used to the course (design), and to get to know each other (e.g. in a first introductory post) as well as to assist the participants in case of technical issues. Previous experience has shown that participants in forums have specific problems and concerns. Nevertheless, a communication channel between learners and instructor(s) may help them overcome earlier obstacles.

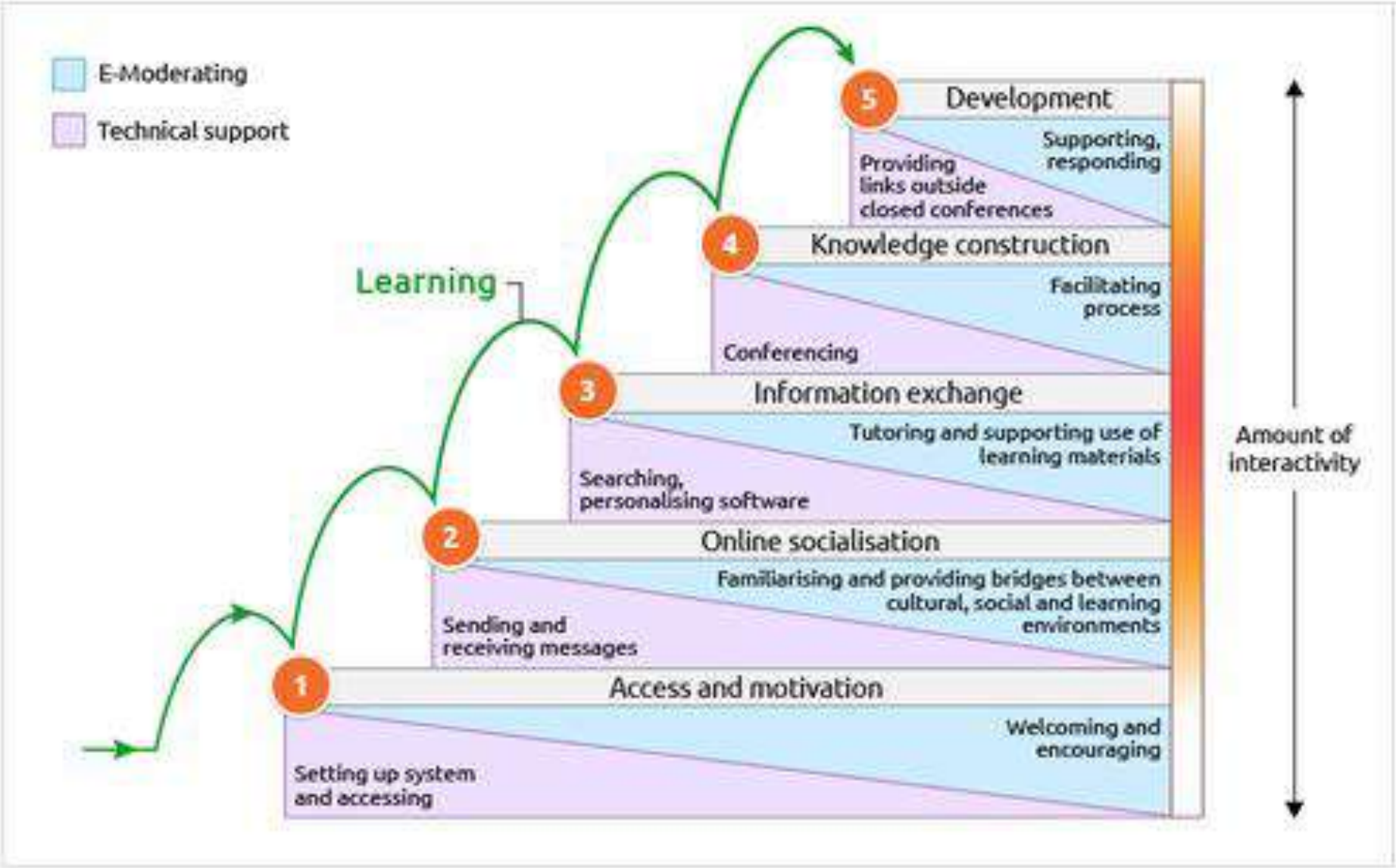

**Figure 57.** The five stage model by Gilly Salmon (Source: http://www.gillysalmon.com/five-stage-model.html)

At stage 3 the e-moderator has to focus on the participants' activity. At this stage, the e-moderators have to pay attention to the group dynamics and identify so-called lurkers (Beaudoin, 2002; Ebner & Holzinger, 2005) or browsers, i.e. participants who do not actively take part in a course but stay passive content consumers (Salmon, 2007). In the last two steps, which are knowledge construction and development, the learner's role becomes more vital and therefore, the learner increasingly becomes able to take control of his/her own learning. The e-moderator can support with some guidance hints. However, it is expected that online learners become confident and able to build on ideas by themselves.

### 5.4.3 Case Study Object (MOOC Structure)

In this case study, learning analytics was used in order to track user activities within the forum of an xMOOC called *Gratis Online Lernen* ('Free Online Learning') (GOL) which ran in autumn and winter 2014 at the Austrian MOOC-platform iMooX (Ebner, Schön, & Käfmüller, 2015). The MOOC was used in previous case studies and is again used here because of the large number of students enrolled. Once again, the MOOC structure followed a duration of eight weeks with an average workload of two hours per week, and the course language, like the other most offered MOOCs at the iMooX platform, was German. Every week consisted of videos, further reading resources (e.g. documents, web links), and a quiz. Moreover, a forum accompanied the course in which instructors and students could actively participate and discuss the topics. The course was intended for people not familiar with virtual space and can be seen as an introduction to finding one’s way on the World Wide Web.

The first descriptive analysis results show that there were 1012 registered participants, 479 of them labeled as *active* (47.33%). ‘Active participants’ are those who write at least one forum post, read a number of forum threads, or do at least one weekly quiz. If the completion rates are calculated on the basis of the registration rate, 21.44% of the participants completed the course, and 17.49% downloaded a certificate. Calculating the completion rate by the active participants, the percentage doubles: 45.30% were active and completed the course, whereas 36.95% were active *and* downloaded a certificate, too.

### 5.4.4 Case Study Analysis

We started the analysis with the overall forum activity in this MOOC. It can be stated that the number of forum reads was very high within the first four weeks as Figure 58 shows.

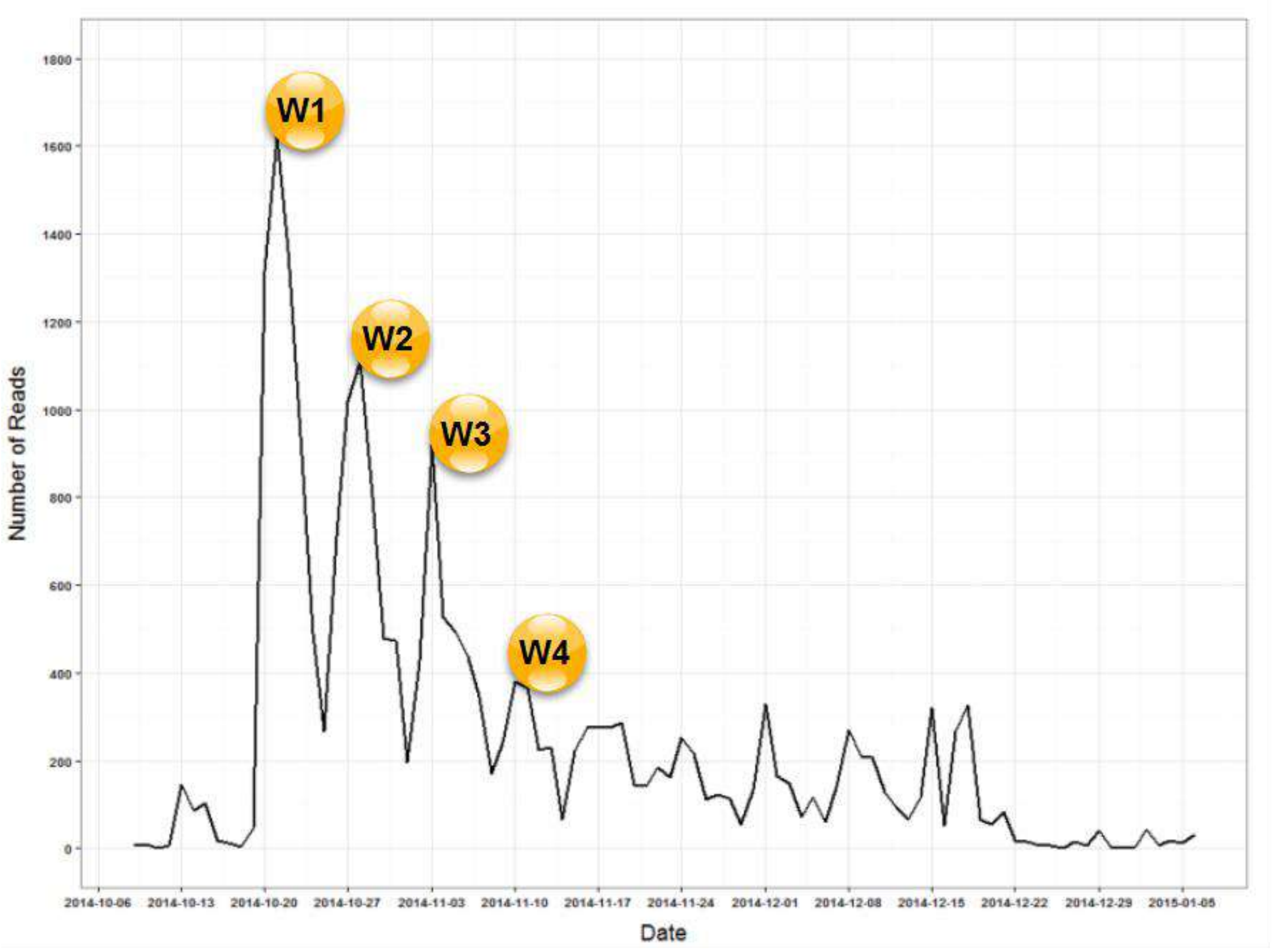

**Figure 58.** The total of GOL MOOC forum readings

Figure 58 illustrates that the participants had already started a forum activity before the course had started in detail. The first three weeks in the course were highly active with 6,706 reads. The frequency diminishes continuously to 1,760 reads for the whole period of week 4. It then stays more or less constant. The referent participants' number, in this case, is 1012 registered participants, as the reading of forum posts is not included in the definition of *active participation*. The first week was characterized by a round of introductions, where the participants were asked to introduce themselves and to react to their colleagues' introductions. The following weeks, the instructors posted discussion prompts on a regular (weekly) basis in order to foster interaction within the course and to start discussions in the forum.

As MOOCs are mostly an informal learning setting (Kop & Fournier, 2010), the learning rhythm is expected to be different than traditional working hours. MOOCs enable participants to learn according to their personal learning attitudes and preferences, independently in time and space. There are no fixed lessons, where participants have to come to class in terms of a synchronous meeting. There are deadlines that have to be met and the participants learn in a highly self-regulated way, at their own pace. To get more information about the working and learning rhythm within MOOCs, the reading time and rhythm spent in the forums are visualized in Figure 59 and Figure 60 for the *Gratis Online Lernen* MOOC.

Figure 59 shows the number of reads against the time of day. It can be clearly seen that the participants mostly read between 8 a.m. and 10 p.m. and that there is a higher number of reads between 6 p.m. and 10 p.m. Only a few participants are highly active in the morning, between midnight and 8 a.m.

The same holds for the time spent reading in a forum, as Figure 60 makes clear. It can be seen that the morning time seems to be used to quickly check the messages, whereas the slot between 8 a.m. and 11 p.m. is used to spend more time in the forum. It must be pointed out that these visualizations are based on quantitative data and it cannot be said that the participants read the postings (attentively) but only spent more time in the forum. The qualitative dimension cannot be deduced from the data available from the learning analytics database. On the other side, it can be assumed that just clicking on all the threads without reading does not make any sense. It can nevertheless be stated that there are peaks regarding the time spent in the forum around noon (10 a.m. – 2 p.m.), even though a lull around lunch time is apparent, and in the evening (6 p.m. – 10

p.m.) followed by suddenly falling till midnight. Between 3 a.m. and 5 a.m., almost no activity can be identified.

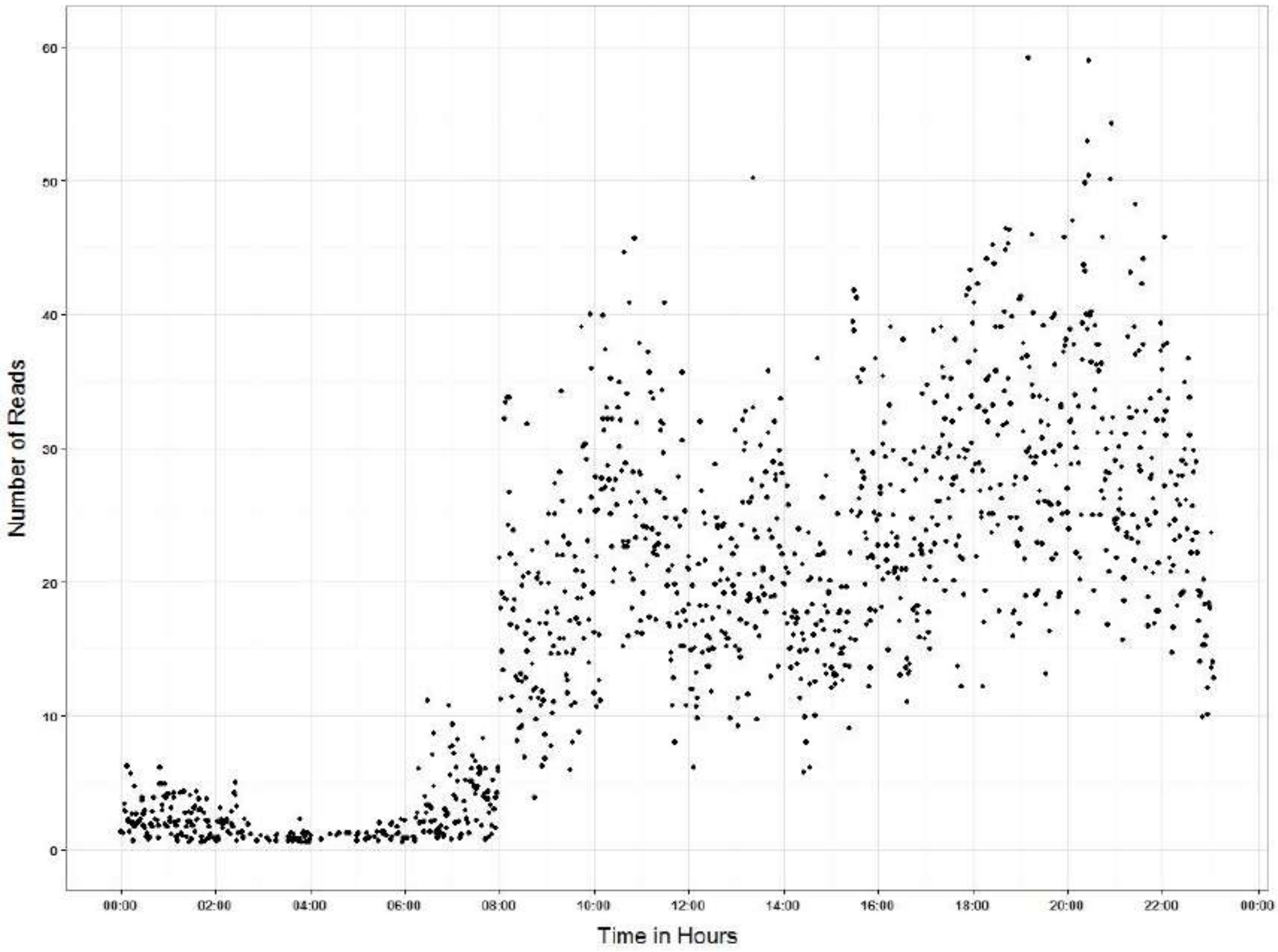

**Figure 59.** When do participants read in forums?

As reading forum posts is meant for consumption rather than to contribute actively, i.e. produce content, a closer look at the active participation within the GOL MOOC was added. Therefore, the number of written posts has been analyzed and visualized as shown in Figure 61.

In this case, the 476 active participants are the reference, but the tendency is the same. Participants and instructors are highly active in weeks 1 to 4, thus become more reticent from week 5. The reasons for this decline might be seen in the fact that the participants' questions were answered by the instructors (Onah et al., 2014), the forum was too confusing due to the high number of posts which led to reluctance (McGuire, 2013; Salmon, 2007) or other intrinsic motives (Huang et al., 2014).

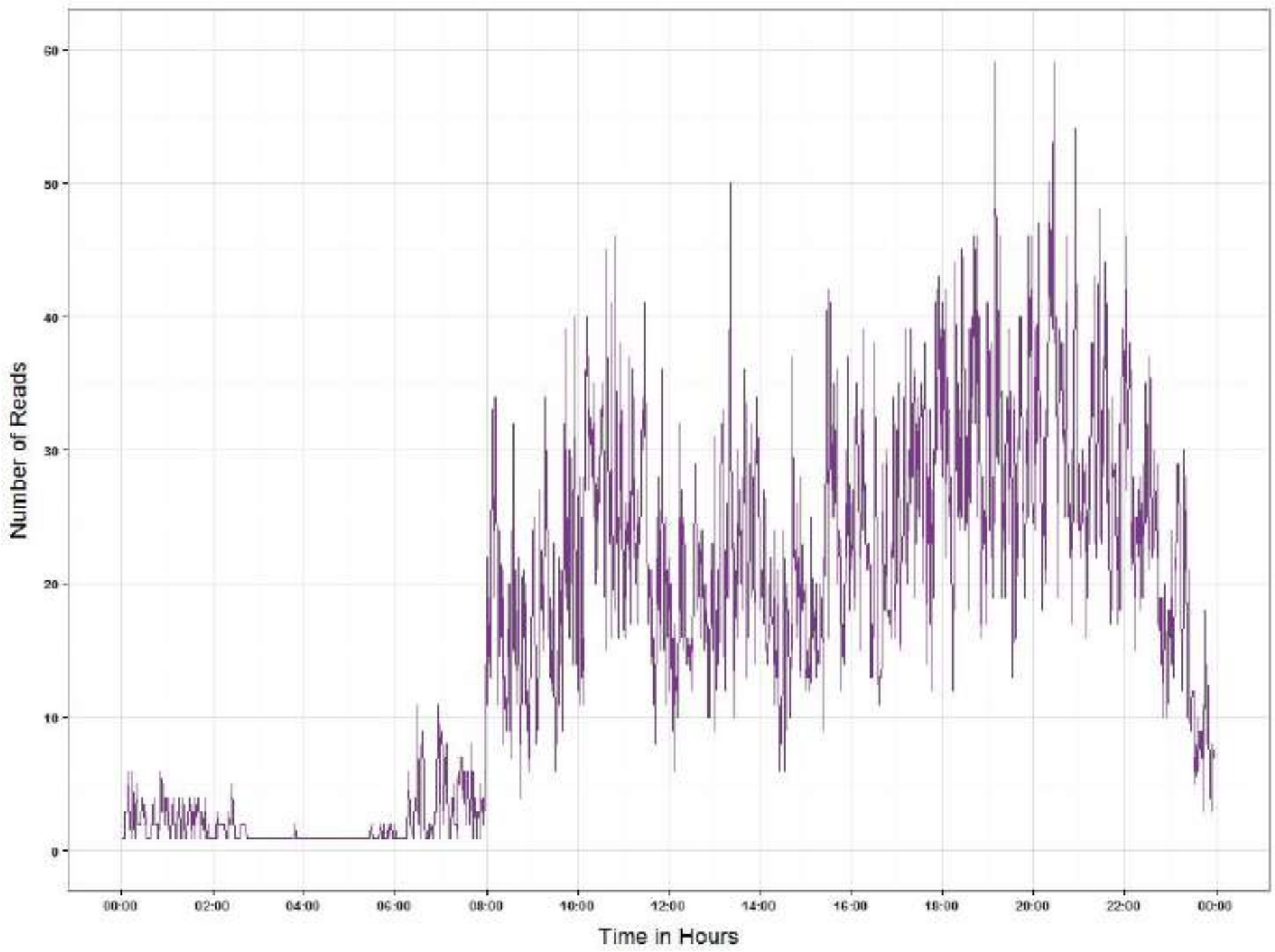

**Figure 60.** Time spent reading in the discussion forums

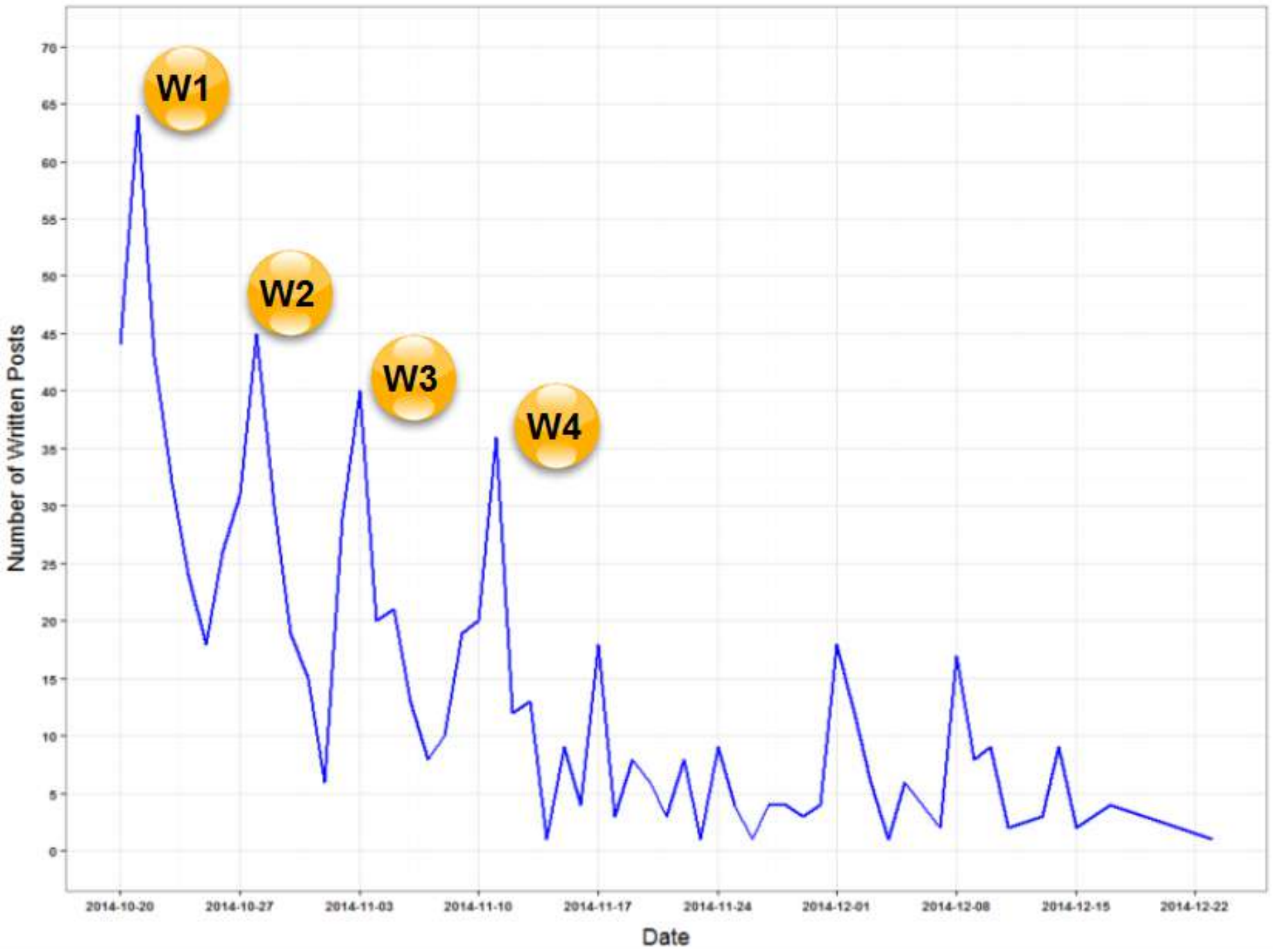

**Figure 61.** Number of written forum posts in GOL MOOC

Nevertheless, a logical and non-surprising correlation between reading and writing can be identified. To check this correlation, all participants who read and those who posted in the forum from the start of the MOOC till its end were retrieved from the database. By merging both datasets, a noticeable relation between reading and writing could be seen, as Figure 62 depicts.

Figure 62 depicts a linear correlation between reading and posting in the discussion forums. A sample of the dataset was tested randomly; the result is a Pearson product-moment correlation coefficient of 0.52, which indicates a moderate positive relation, as a 95% confidence interval between 0.46 and 0.57 can be identified, which leads to the correlation of 0.52. The figure visualizes the points in an upward shape. Some outliers or *superposters*, according to Huang et al. (2014), exist in our case study. For example, the highly active instructor is shown on the top right of the pane. To check the validity of the correlation, a second step linear test was done between reading and writing using proportions and not the original numerical values. The correlation result equaled to a value of 0.58 with $p$-value $< 0.10$.

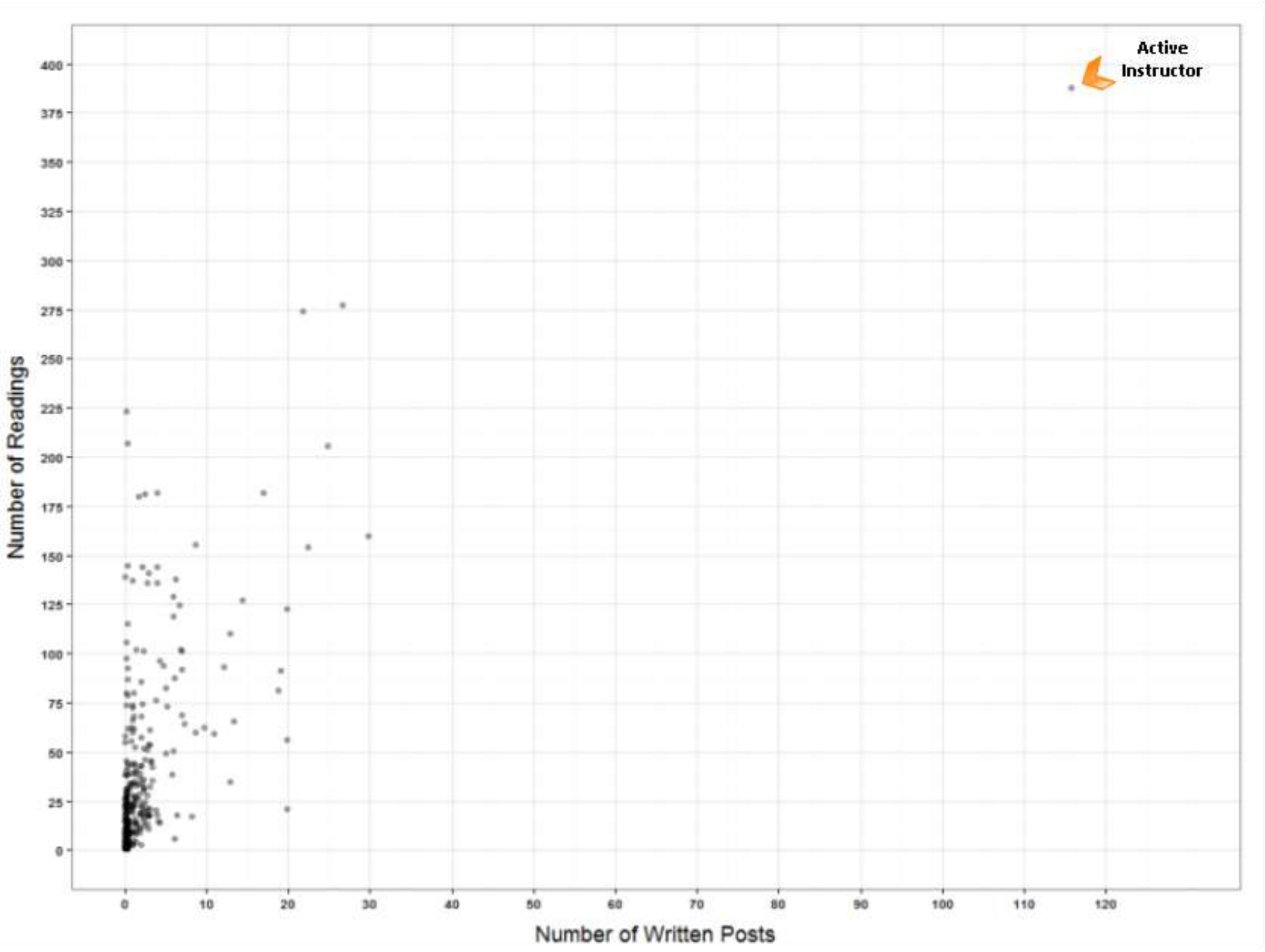

**Figure 62.** Relationship between the number of readings and writings in GOL MOOC forums

It may, therefore, be supposed that the main time spent passively consuming from one side and actively producing from the other side correlate to each other. Also, we noted that there are two main time frames around noon and in the evening as Figure 59 and Figure 60 have illustrated where participants find the time to read posts and write down their comments and posts.

### 5.4.5 Case Study Discussion and Evaluation

Bringing together the insights of the three mentioned parts of this paper, it can be said that forum discussions play an important role within MOOCs in order to enable the participants to communicate and collaborate with each other. This personal interaction among participants and between participants and instructors is crucial. Thus, there are several reasons to consider when designing a MOOC in order to actively and consciously foster forum discussions.

Gilly Salmon’s (2007) e-moderation model may be the first indication of these design processes but needs to be revisited as it was conceptualized for smaller groups. According to the author, a forum without moderation may not work or flatten. Thus, her focus lies on the interaction between teachers and students. The e-moderator’s role embraces activities to motivate learners on an individual level, as the author acts on the assumption that communication has to be fostered and is not a self-selling item. Therefore, e-moderators have to identify lurkers (Beaudoin, 2002; Ebner & Holzinger, 2005) or browsers (Salmon, 2007, pp. 36f.), and should name them personally according to Salmon. This includes that e-moderators know their target group by name. To do so the group must be rather small. Otherwise an individual support cannot be handled. Given the fact that the group of participants in MOOCs is massive and heterogeneous (Gaebel, 2014; Hollands & Tirthali, 2014), the instructors are not able to know every single participant and to witness their learning processes. Nevertheless, some crucial points mentioned by Salmon (2007) hold for traditional settings as well as for MOOCs. Taking into account the five stages, the following design recommendations may be deduced. Two units have to be distinguished due to the different learning processes and paces of the heterogeneous learning group.

Stage 1 (Access and Motivation) and stage 2 (Socialization) should be seen as the first unit. Participants should know how to enter the course (platform) and should be motivated from the beginning. For that reason, we redesigned the iMooX discussion forums using gamification elements (see the subsection below). At this point, the instructors make clear what the purpose is for taking part. Due to the group's heterogeneity, they will have to focus on different purposes and personal issues that motivate learners to take part. The first week of a MOOC may be used to get to know each other, as well as to get familiar with the platform and the course design (MoocGuide, 2015). This community building can be obtained via a round of introductions within the forum. In order to get a quick overview of the heterogeneous target group, a supplementary short demographic survey can be introduced at the beginning of the course, and in general, students can be motivated by extrinsic factors like badges or progress bars.

The first unit (the first two stages) should cover the week before the start of the MOOC and the first weeks of the MOOC. As Lackner, Ebner and Khalil (2015) show, the participant's activity diminishes from week 4 on. It is in the first four weeks that participants mostly perceive the feeling of being overwhelmed by an information overload. This overload might also be seen as one of the reasons for participants to abandon a course (McGuire, 2013; Salmon, 2007, p. 39). The moderators or active instructors should be prepared to cope with the high number of posts, thus input, in the first four weeks as our analysis in Figure 58 shows. During this period, it might be necessary to work in a team to help participants on a technical-administrative and content-related level, e.g. regarding registration, first attempts in the forum or with quizzes and content-related questions. This holds especially for the peak consumption and production times as illustrated in Figure 62. Instructors should decide beforehand if they react to the messages in a specific time frame or according to their personal time resources. In both cases, they have to announce their strategies in advance, i.e. at the beginning of the MOOC, in order to abate the participants' frustration level that normally grows when they are waiting for an answer. The installment of a FAQ forum thread might help as well.

Stage 3 (Information Exchange), stage 4 (Knowledge Construction) and stage 5 (Development) are the second bundle. The instructors should then keep in mind that not all registered participants become active and just a small percentage completes the course. It is not everybody's intention to complete a MOOC as Colman (2013) discusses. Several inducements and lurking or browsing participants are a normal phenomenon in MOOCs; the certificate that requires active participation is not the most important reason to register for a MOOC. Nevertheless, the instructor should schedule a weekly post that announces the program of the week and the upcoming deadlines in order to help students following the different discussion threads. Our research study by Khalil, Kastl, and Ebner (2016) is considered as supporting evidence. We clustered MOOC participants and found a group of students who were more involved in discussions than the others. These students were named *sociable* students.

In order to organize forum discussions, rules of conduct for the forum should be set and clearly communicated before the course start. These rules might cover the prohibition of violent, politically incorrect, homophobic, racist, illegal, or pornographic contributions as well as an explanation of the forum's structure. Moreover, rules should be stated inside forums about how to open a thread and when, so that the instructor can get in contact with the learners without delays.

Finally, the forum's borders should be open as the forum is often limited to the course and the integration of supplementary resources is difficult. A hashtag created for the course, such as #GOL2014 for the scrutinized MOOC, helps participants connecting outside the course form smaller networks according to, amongst others, personal interests, level of expertise, and geographical background (Guàrdia, Maina, & Sangrà, 2013). These smaller groups help them to organize themselves and foster interaction, collaboration and communication.

#### 5.4.5.1 Implementation of Gamification Elements in MOOC-Discussions Forums

As an interpretation of the first two stages, we thought of enhancing the motivational factor of discussion forums by redesigning the iMooX discussion forums graphical user interface. This implementation is a part of a master thesis carried out at the Graz University of Technology (Reicher, 2016). Our new design of the discussion forums followed the "gamethinking" process that implores users to solve problems using gamification elements (Zichermann & Cunningham, 2011). We implemented the following four main mechanisms of gamification a) rewards, b) badges, c) points, and d) leaderboard (Gunawardhana & Palaniappan, 2015). Gamification has recently emerged as a new method to motivate students in online learning platforms like MOOCs. In fact, in the learning analytics method survey in the literature chapter of this dissertation, we saw that the use of gamification is increasingly becoming popular and has been used in learning analytics in the last 3 years.

Our new design of the discussion forums used JavaScript (http://www.w3schools.com/js/, last visited August 2016), HTML (http://www.w3schools.com/html/, last visited August 2016), and CSS (http://www.w3schools.com/css/, last visited August 2016). JavaScript is very suitable to execute user interactions as well as manipulate HTML and CSS. HTML was used to structure the forum (hyperlinks, pictures). In order to design the page in an appealing way, we greatly benefitted from the use of CSS. Furthermore, the new discussion forum design was prepared to be suitable for different formats, i.e. desktop, smartphone or tablet to increase the overall user experience.

The interface was restructured as shown in Figure 63 and Figure 64. The old design is based on a tree view of clickable threads. The design is basic and any usage of images or other interactions (like private messaging, profile pictures, emoticons…etc.) could not be possible. Nevertheless, the new design in Figure 64 offers several interactive features. The users have the ability to make likes, set a favorite thread, upload a profile picture, set up their own profiles, or set their personal preferences.

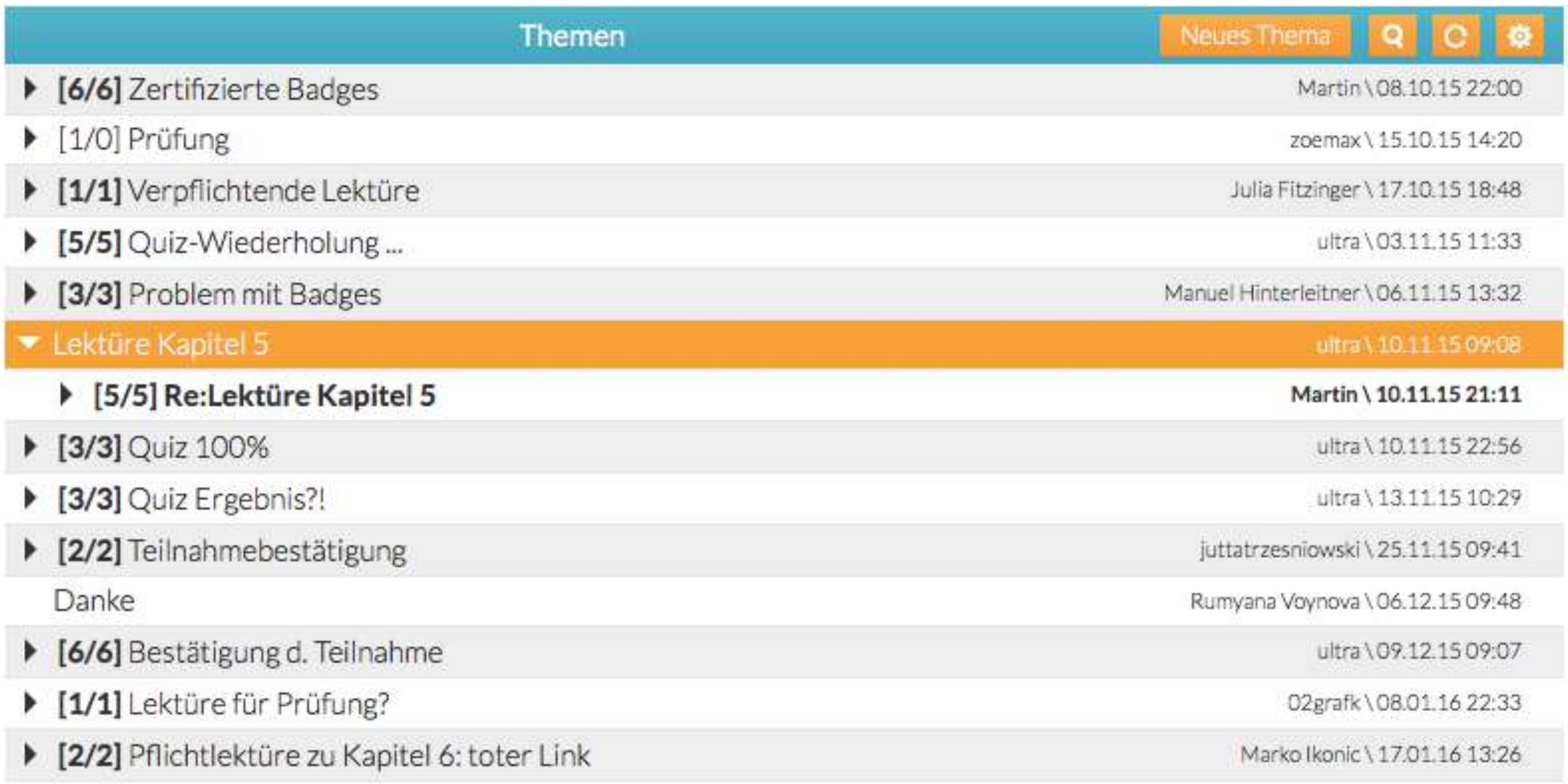

**Figure 63.** The old iMooX discussion forum design (Reicher, 2016)

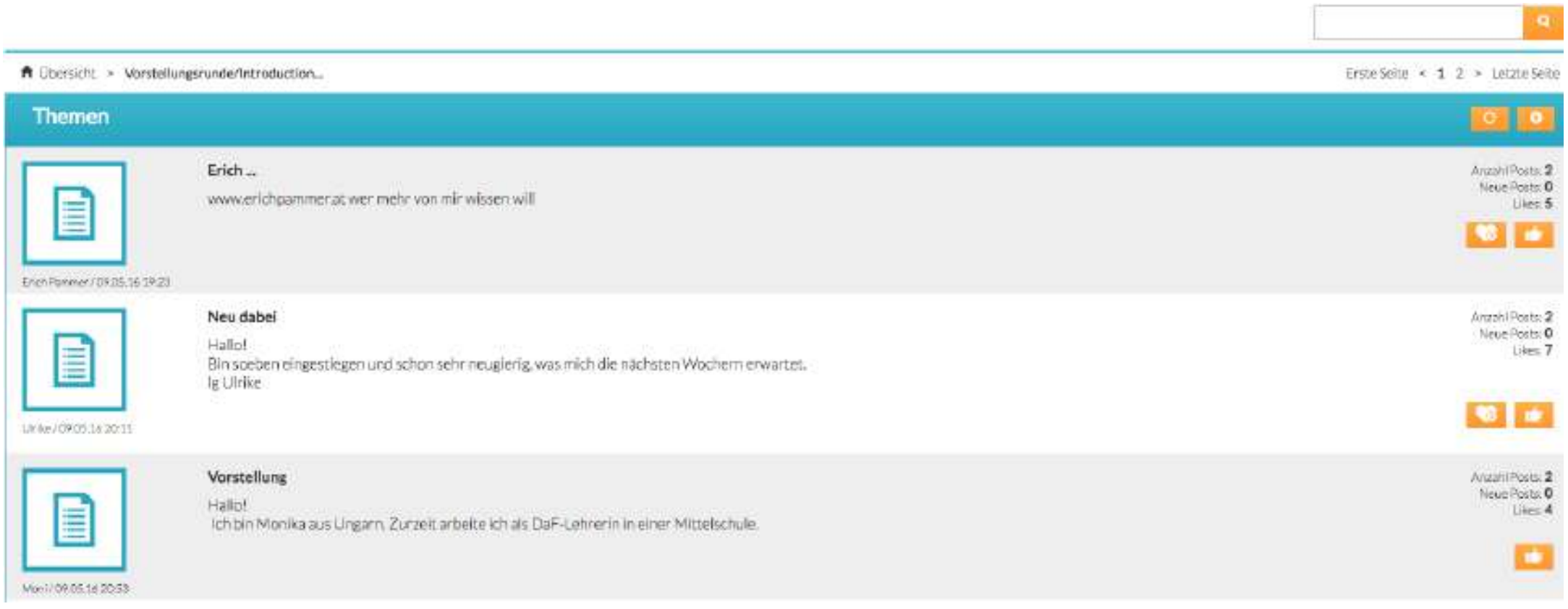

**Figure 64.** The new iMooX discussion forum design (Reicher, 2016)

In addition, the new design provides various gamification elements (see Figure 65) that are intended to increase user engagement, enhance the user-friendly environment, and add fun to the communication channels of the MOOC platform. Forum users can obtain badges as well as see their progress bar and level. For example, if a user likes a thread for the first time, they obtain the badge "Like." Upon reaching level 2 they can upload a profile picture and have a different username color. To encourage students to post, we created a badge called "5 helpful answers created." This badge labels a student as a problem solver to students who seek help or instruction.

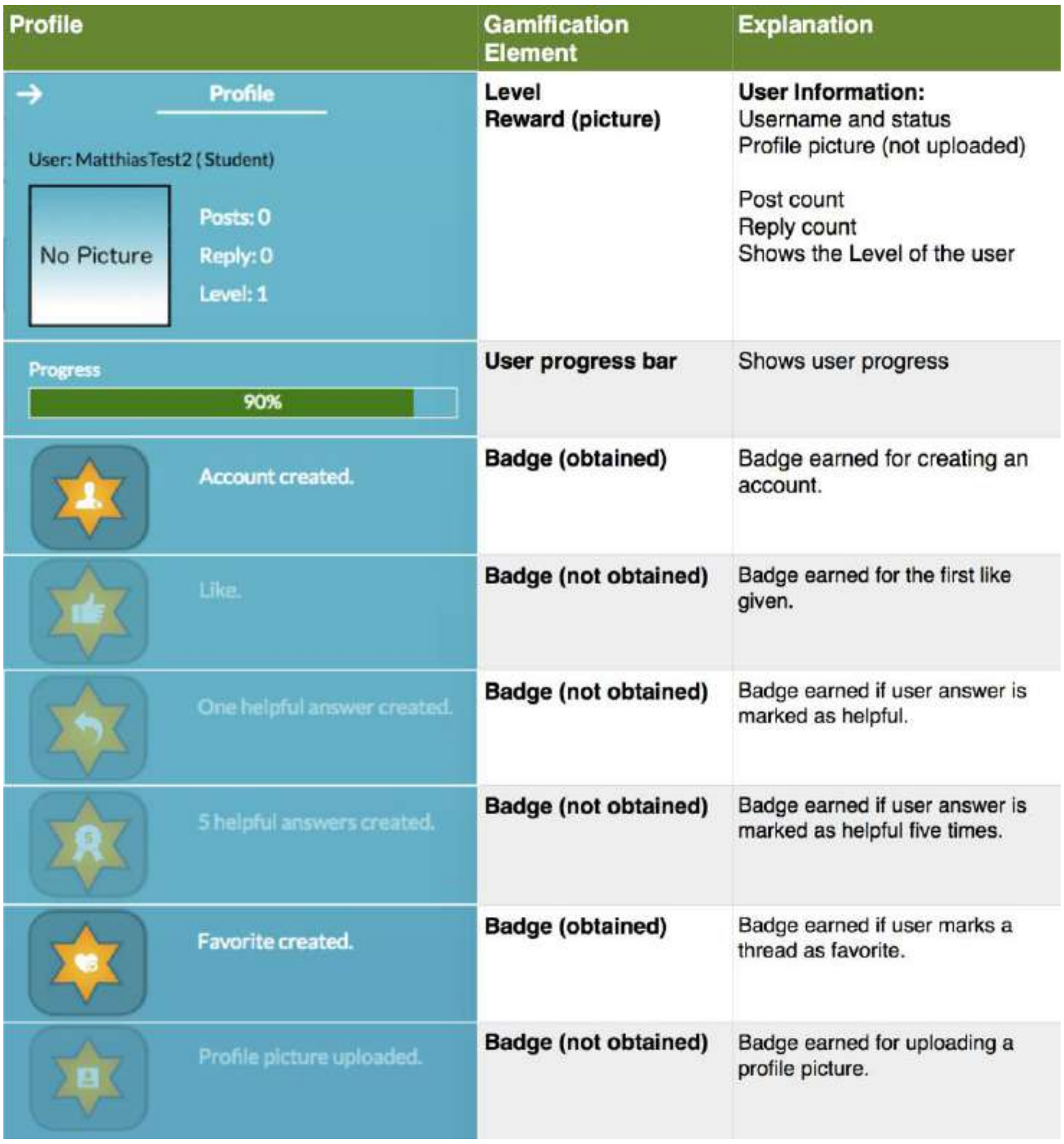

| Profile | Gamification Element | Explanation |
|---|---|---|
| Profile<br>User: MatthiasTest2 (Student)<br>No Picture<br>Posts: 0<br>Reply: 0<br>Level: 1 | **Level**<br>**Reward (picture)** | **User Information:**<br>Username and status<br>Profile picture (not uploaded)<br>Post count<br>Reply count<br>Shows the Level of the user |
| Progress<br>90% | **User progress bar** | Shows user progress |
| Account created. | **Badge (obtained)** | Badge earned for creating an account. |
| Like. | **Badge (not obtained)** | Badge earned for the first like given. |
| One helpful answer created. | **Badge (not obtained)** | Badge earned if user answer is marked as helpful. |
| 5 helpful answers created. | **Badge (not obtained)** | Badge earned if user answer is marked as helpful five times. |
| Favorite created. | **Badge (obtained)** | Badge earned if user marks a thread as favorite. |
| Profile picture uploaded. | **Badge (not obtained)** | Badge earned for uploading a profile picture. |

**Figure 65.** A general overview of the available gamification elements that are offered in the new forum design in iMooX (Reicher, 2016)

On the other hand, the design of the progress bar expresses a dynamic animation that shows the student's development to level up. The concept of the progress bar is to inform users on how convenient they are to completing tasks with a stamp of a percentage token, in this case, tasks like creating an account, updating personal information, making a couple of threads, likes, and having helpful answers. That is to say, each level gets harder and harder to reach and requires extra tasks to complete.

To evaluate the new discussion forum, we used the data from iLAP to support our results. We investigated two respectively offered courses, *Lernen im Netz* 2014 and *Lernen im Netz* 2016. Both courses lasted eight weeks, and students who completed courses were credited with university ECTS points.

As described in Table 12, there were 605 registered users in the 2016 MOOC and 519 participants in the 2014 MOOC. In *Lernen im Netz* 2016," 39.83% of users didn't use the forum at all. In comparison, only 33.53% participants did not use the forum in 2014. Furthermore, 76 certified participants in 2016 (12.56%) stand against 99 (19.07%) in 2014.

**Table 12.** Discussion forums usage for *Lernen im Netz* 2014 and *Lernen im Netz* 2016 MOOCs

| Type of Users | Lernen im Netz 2016 | Lernen im Netz 2014 |
|---|---|---|
| Registered users | 605 | 519 |
| Certified Users (%) | 76 (12.56%) | 99 (19.07%) |
| Never used the forums (%) | 39.83% | 33.53% |

We recorded the user actions from the new forum in the *Lernen im Netz* 2016 MOOC. Users were given the ability to create threads, post answers to created threads as well as like them and mark answers as helpful. Further, users were able to create favorites and save them for later use. As seen in Figure 66, the new forum has 50 threads created, 67 answers written, 88 likes given and 25 helpful answers posted.

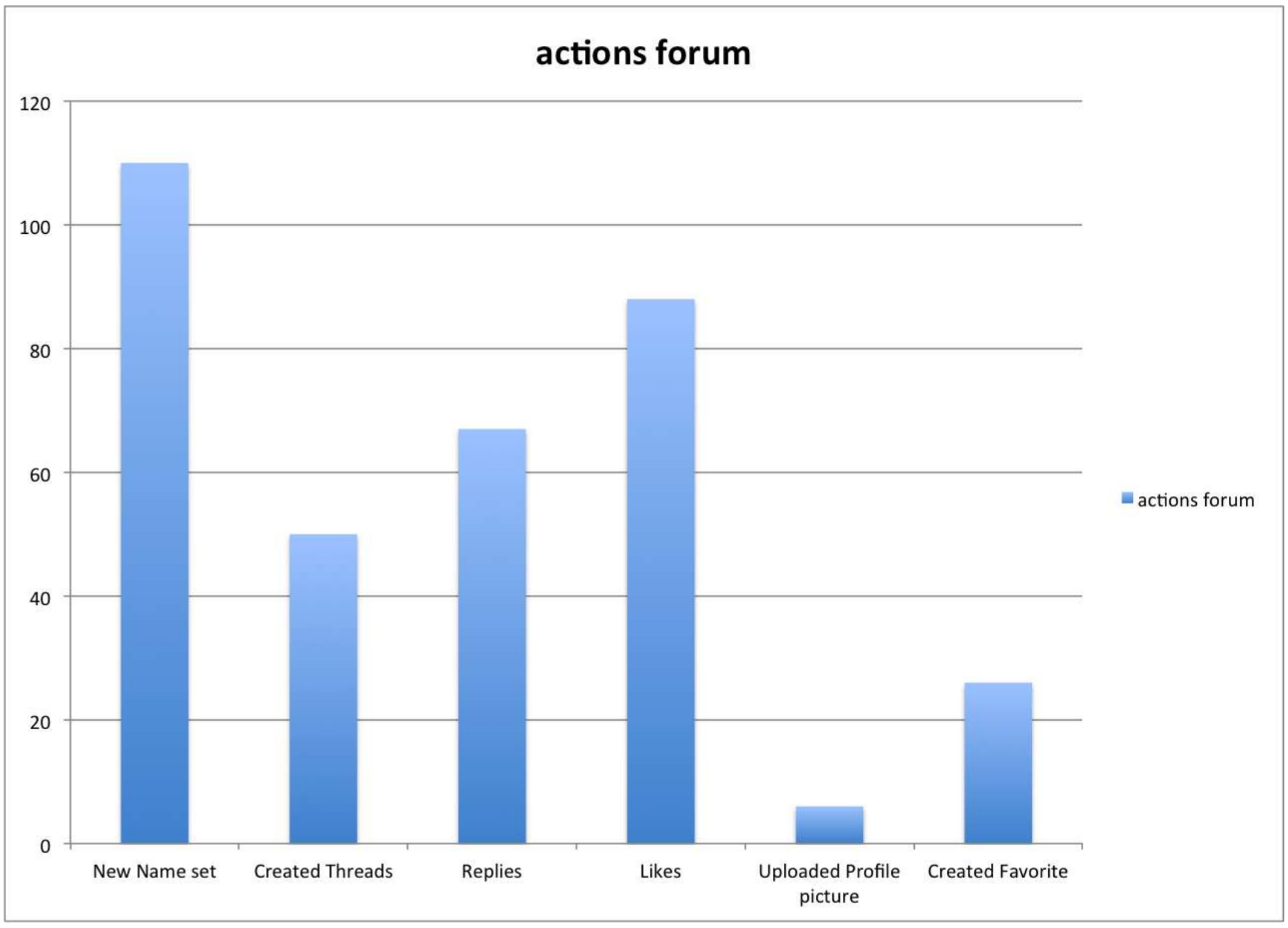

**Figure 66.** The new discussion forum actions stats

Next, we compared both MOOC forum activities during the weeks as shown in the line plots of Figure 67 and Figure 68. The red line depicts the activities of *Lernen im Netz* 2014 MOOC, while the blue line depicts the activities of *Lernen im Netz* 2016 MOOC.

Reading and writing activities are nearly identical for both MOOCs. However, there is a huge peak in the pre-MOOC of week1 credited to *Lernen im Netz* 2014. Moreover, Figure 67 shows that readings in the last week of the 2016 MOOC nearly tripled the readings of the seventh week. Another notable item is that Figure 68 shows the writing in the second and third week of the 2014 MOOC was high in comparison to the following weeks. In fact, our forum engagement results are in line with a recent MOOC study by Coetzee et al. (2014), where the authors found that 68% of students interact with videos or quizzes but never visit the MOOC forum.

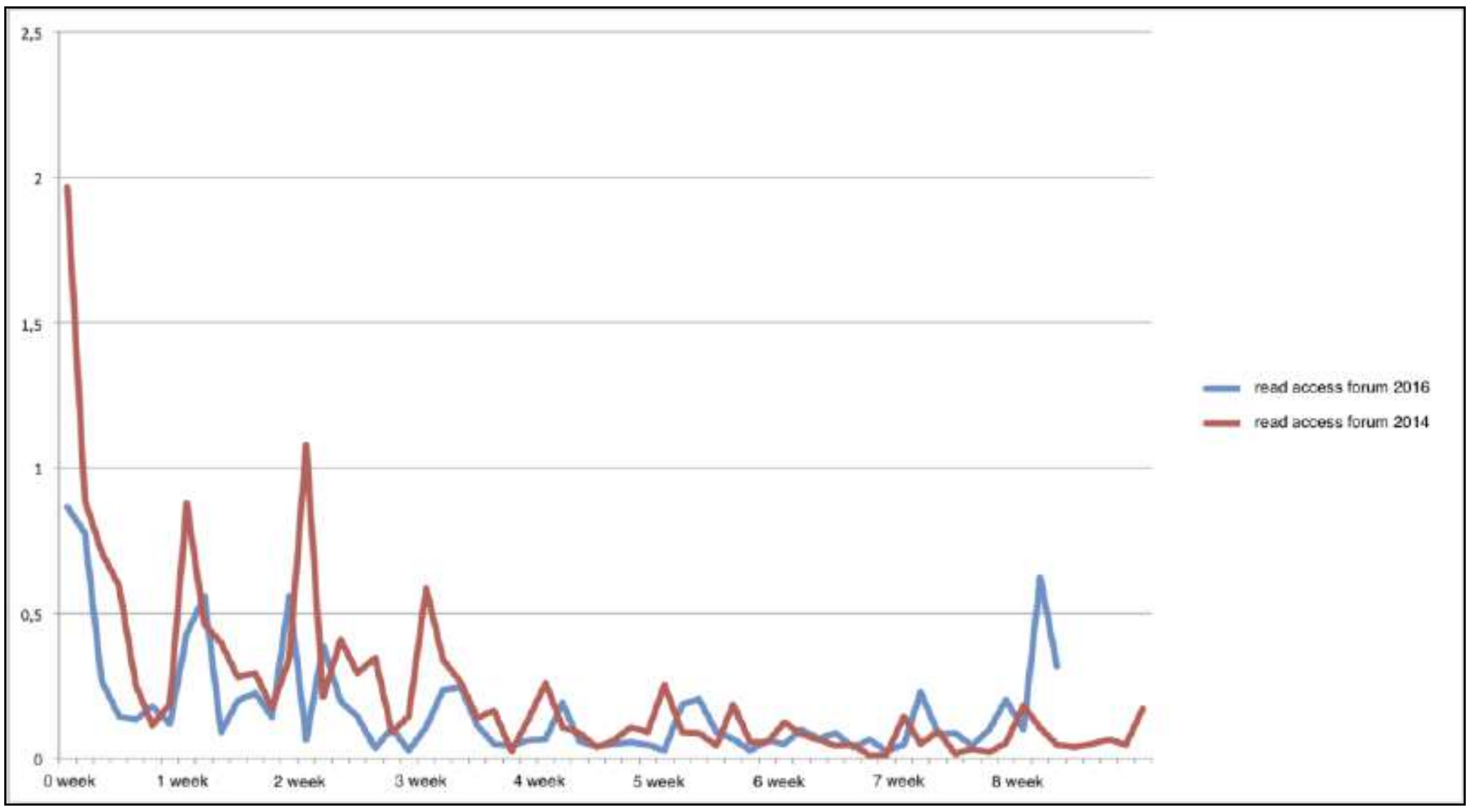

**Figure 67.** Comparison of reading in forums of 2014 and 2016 *Lernen im Netz* MOOCs

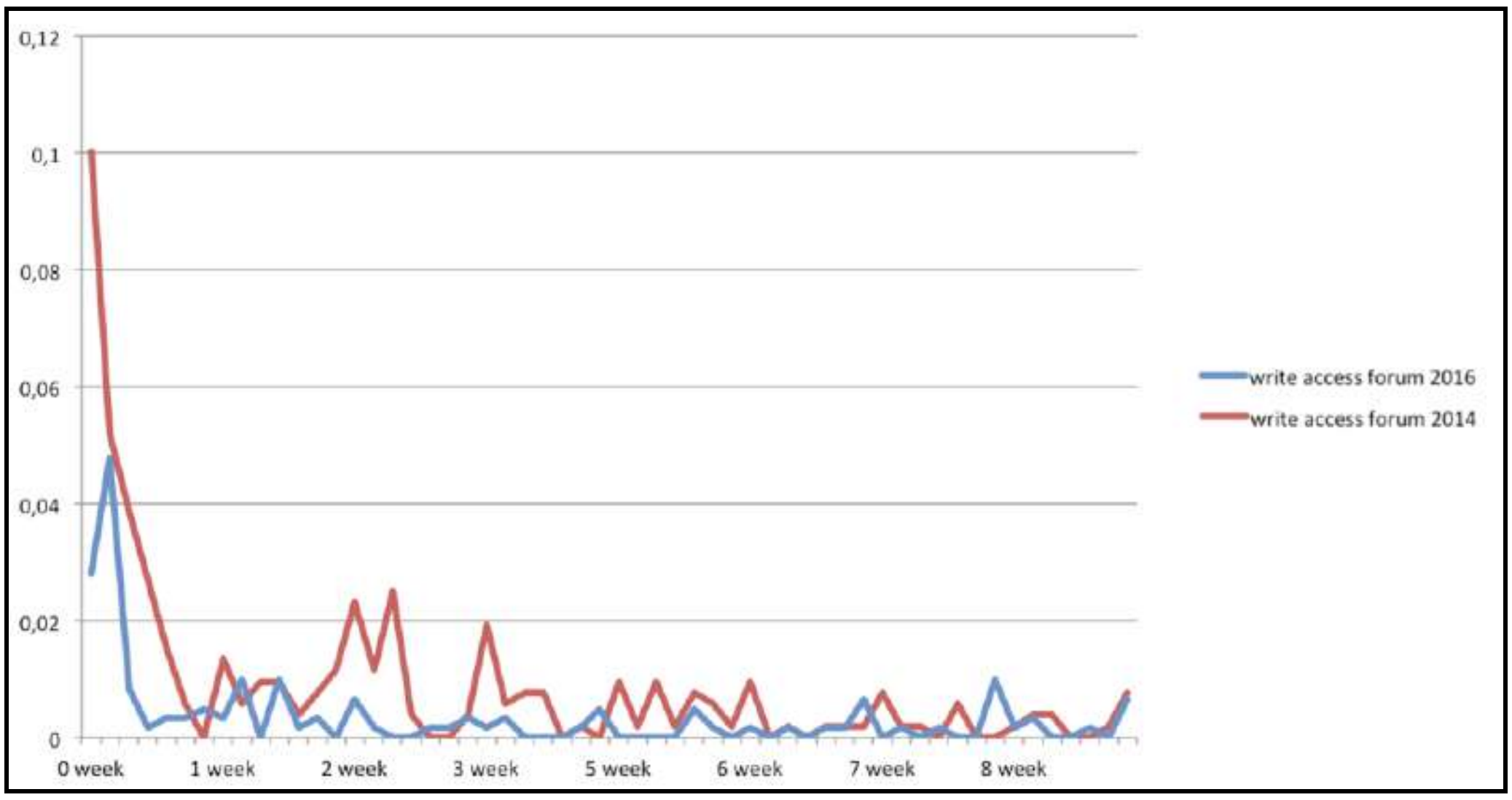

**Figure 68.** Comparison of writing in forums of 2014 and 2016 *Lernen im Netz* MOOCs

The previous analysis showed that forum activity was nearly the same for both MOOCs. In fact, the offered MOOC in 2014 was accredited with 4 ECTS points, while students who successfully finished *Lernen im Netz* 2016 MOOC got 3 ECTS points. This might have left a psychological impact on the students' activity in the MOOC in general and in the forums in particular. Another explanation for the slightly lower involvement might belong to the iMooX huge advertisement back in 2014 when iMooX was first launched. That is, *Lernen im Netz* 2014 MOOC was one of the few offered courses at that time.

In addition, we saw that writing in forums in 2016 was, to some extent, lower than in the 2014 MOOC. After a quick comparison of the content between the two forums, we saw that students of the 2014 MOOC were more active, by asking technical questions. As a matter of fact, this is attributed to the dissatisfaction of use of the old forum.

Although the results were not very promising, we believe that the new design is better than the older design since the students and the teacher of the examined MOOC agreed that the new design offers usability, flexibility, and is fun. They further said that the gamification elements of the new discussion forum increased their motivation and proposed exciting features. However, further evidence of the new forum design can be collected in *Lernen im Netz 2017* MOOC.

### 5.4.6 Case Study Summary

Forum discussions are an important part of each online learning setting as they help to transfer the feeling of being part of a learning community from analog to digital learning settings. Despite this positive aspect, one of the disadvantages of online learning is the lack of face-to-face communication.

Gilly Salmon's traditional *Five Stage Model* regarding e-moderation is primarily tailored to small learning groups in a rather traditional, private course setting. As MOOCs differ from traditional courses in terms of their instructional design (Kopp & Lackner, 2014), the moderation concept needs to be adapted to the MOOC-specific requirements highlighted in this case study. With this adaptation, forum discussions might help establish individualized support for participants, foster interaction and collaboration among learners, and, thus, support self-regulated learning. The lead management of iMooX implemented a new design for the discussion forums in order to motivate students, using gamification elements. The new design offered usability and user personalization; however, the results were not very promising as far as students becoming more involved in the forums. Nevertheless, future examination of new, forthcoming MOOCs might show potential in motivating students and fostering forum discussions. Also, appointment of a candidate to be an e-moderator might take place in the near future.

## 5.5 Case Study (5): Clustering Patterns of Engagement to Reveal Student Categories[11]

In this case study, our intention is to classify MOOC students into appropriate categories based on their level of engagement. Student activities in MOOCs reflect their motivation and engagement (Xu & Yang, 2016). The results of this case study will be in a model of clusters by which clustering was done on two subpopulations: I) Undergraduate from the university and II) External students from the public. Each cluster has students with certain behaviors based on their engagement level with the MOOC variables. The clusters afterward are compared with another classical education scheme called the (Cryer's scheme of Elton) for further examination and comparison. The intrinsic and extrinsic factors were considered for MOOC learning improvement.

### 5.5.1 Background

In this section of this case study, we list the related work of clustering students based on their engagement and motivation.

MOOCs have the potential to scale education in disparate areas. Their benefits are found to be an educational program in terms of outcomes, extending accessibility, and reducing costs. Motivation scientists have discussed that learning is strongly connected to intrinsic motivation as it is an important construct that reflects the natural human tendency towards being educated (Elliot & Harackiewicz, 1994; Ryan & Deci, 2000). On the other hand, extrinsic motivation plays a major role for learners in which they gain a separable outcome. Ryan and Deci (2000) said that ‘‘extrinsic motivation refers to doing an activity simply for the enjoyment of the activity itself, rather than its instrumental value (intrinsic motivation).’’ In MOOC learning environments, students enjoy either one type of these motivations or acquire them both. A simple example from

---

[11] Parts of this section have been published in:

Khalil, M., Kastl, C., & Ebner, M. (2016). Portraying MOOCs Learners: a Clustering Experience Using Learning Analytics. *In Proceedings of the European Stakeholder Summit on experiences and best practices in and around MOOCs (EMOOCS 2016)*, 265-278.

Khalil, M. & Ebner, M. (2016a). Clustering Patterns of Engagement in Massive Open Online Courses (MOOCs): The Use of Learning Analytics to Reveal Student Categories. *Journal of Computing in Higher Education*. DOI: 10.1007/s12528-016-9126-9

MOOCs on intrinsic motivation is when a student enrolls in a course purely out of curiosity (Wang & Baker, 2015). On the other hand, students can be extrinsically motivated by tangible rewards like certificates, which are offered by MOOC providers.

One of the prominent research studies in the LAK conference is the work of Kizilcec et al. (2013). Our survey study in the literature review of this dissertation found that the ultimate number of citations in the conference proceedings between 2013 and 2015 belongs to this article. The authors mainly focused on analyzing different subpopulations in MOOCs based on the level of engagement in MOOCs. Their results showed four types of students. (1) Completers: are the students who completed the courses, (2) Auditors: refer to those who completed the assessments infrequently, but were more interested in watching videos, (3) Disengaging students: students who dropped out after being engaged in the first third of the class, (4) Sampling: learners watched videos for only the first two weeks.

Their research results influenced other researchers like Ferguson et al. (2015) who replicated the same clustering methodology. Their research analyzed five MOOCs from the FutureLearn (http://www.futurelearn.com) MOOC platform. The authors concluded that clustering subpopulations of one MOOC is not always applicable to other MOOCs. For example, they clustered long MOOCs into seven groups, whilst shorter MOOCs were clustered into four groups. This conclusion indicates that different approaches may work with different MOOCs when it comes to cluster analysis.

Other types of clustering were used to examine assignments and lecture views in MOOCs. Anderson et al. (2014) clustered student engagement between these two factors into five subpopulations based on quantitative investigations. (1) Viewers: watch lectures, hand in few assignments, (2) Solvers: hand in assignments, but view few lecture videos, (3) All-rounders: balanced between the two groups, (4) Collectors: download lectures, hand few assignments, and finally, (5) Bystanders: registrants who never show up again.

Kovanovic et al. (2016) employed *k*-means clustering on 28 MOOCs from the Coursera platform and concluded five clusters. (1) Enrollees: students who are not active, (2) Low Engagement: students with very low activity, (3) Videos: students who primarily watch videos, (4) Videos & Quizzes: students engaged in videos and do quizzes, (5) Social: students who participate actively in discussion forums.

Finally, our early clustering which was done when evaluating the iLAP classified four types of students which are (1) Registrants: students who just enroll in a course and never show up, (2) Active learners: students who do some type of activity like watching a single video or attending one quiz, (3) Completers: students who successfully finish all quizzes, but do not ask for certificates, and the fourth group, (4) Certified students: concerns the Completers who ask for the certificate letter. However, in this case study, we go further, using detailed analysis and machine learning.

### 5.5.2 Methodology

Data collection and parsing are performed using the iMooX Learning Analytics Prototype (iLAP). The generated large amount of records enables us to analyze and classify learners. By tracing their left behind footsteps, the tool stores learner actions. It fetches low-level data from the different available MOOC indicators. Videos, files download, reading in forums, posting in forums, quiz results, and logins are such obtained information. The analyzed dataset in this study derives from a MOOC offered in the summer semester of 2015 by Graz University of Technology. The collected dataset was then parsed to refine the duplicated and unstructured data format. Data analysis was carried out using the R software. An additional combined package called NbClust (Charrad et al. 2013) was used for implementing the *k*-means clustering algorithm.

We followed the content analysis methodology in which units of analysis (MOOC indicators) get measured and benchmarked based on qualitative decisions (choosing *k* partitions in clustering, comprehensive decisions, survey…etc.) (Neuendorf 2002). These decisions were founded on sustained observations on a weekly basis and an examination of surveys at the end of the course.

### 5.5.3 The Examined MOOC Overview

The analyzed course is titled ‘‘Social Aspects of Information Technology’’ and is abbreviated in this article as GADI (Ebner & Maurer, 2008). We have selected this course because it is typical of being mandatory to university students and was at the same time open to the local and international general public. The university students came from different majors such as Information and Computer Engineering (Bachelor-6th semester), Computer Science (Bachelor-2nd Semester), Software Development and Business Management (Bachelor-6th semester) as well as for the Teacher Education of Computer Science (Bachelor -2nd Semester).

The course lasted ten weeks long. Every week included video lectures, discussion forums, readings and a multiple choice quiz. GADI contents were mainly formed on interviews with experts, with 21 video lectures in summary of about 17 min duration on average. The evaluation system followed the self-assessment principle in which each quiz could be repeated up to five times. The system is programmed to record the highest grade; however, the student had to score at least 75% of every single trial in order to pass the course. The teaching staff predefined the workload of 3 h per week. Students of Graz University of Technology gained 2.5 ECTS (European Credit Transfer and Accumulation System) points towards their degree if they successfully completed the MOOC; however, they still had to do an essential practical work, additionally.

### 5.5.4 Case Study Analysis

#### 5.5.4.1 Demographic Analysis

The GADI MOOC certification state is depicted in Figure 69. The number of students from the university was quite unlike any student number from previously offered courses in the university halls. The reason is clearly reasonable because the 2015 course was offered on the iMooX platform for the first time. There were 459 matriculated undergraduates and 379 external students. Because this MOOC is compulsory to pass the university class, the completion ratio was quite high. The general certification rate (who gained a certificate) of this MOOC was 49%. Specifically, 79.96% was the certification ratio of the undergraduates, and 11.35% of the external students were certified.

Candidates who successfully completed all the quizzes were asked to submit answers on an evaluation form at the end of the course. Questions varied between satisfaction factors and demographic information. Figure 70 reports general information about the certified students ($N$= 410). The x-axis depicts the student type and the y-axis records their age. Demographic analysis reveals that the majority of university students ($N$= 367, *female*= 40, *male*= 327) were men. The average age was 23.1 years-old, and the standard deviation was ($\sigma$= 2.94). On the other hand, we see a big difference in age and gender of the certified external students. The sample showed that there were ($N$= 43, *female*= 20, *male*= 23) students. The average age was 46.95 years-old, and the standard deviation was ($\sigma$= 10.88).

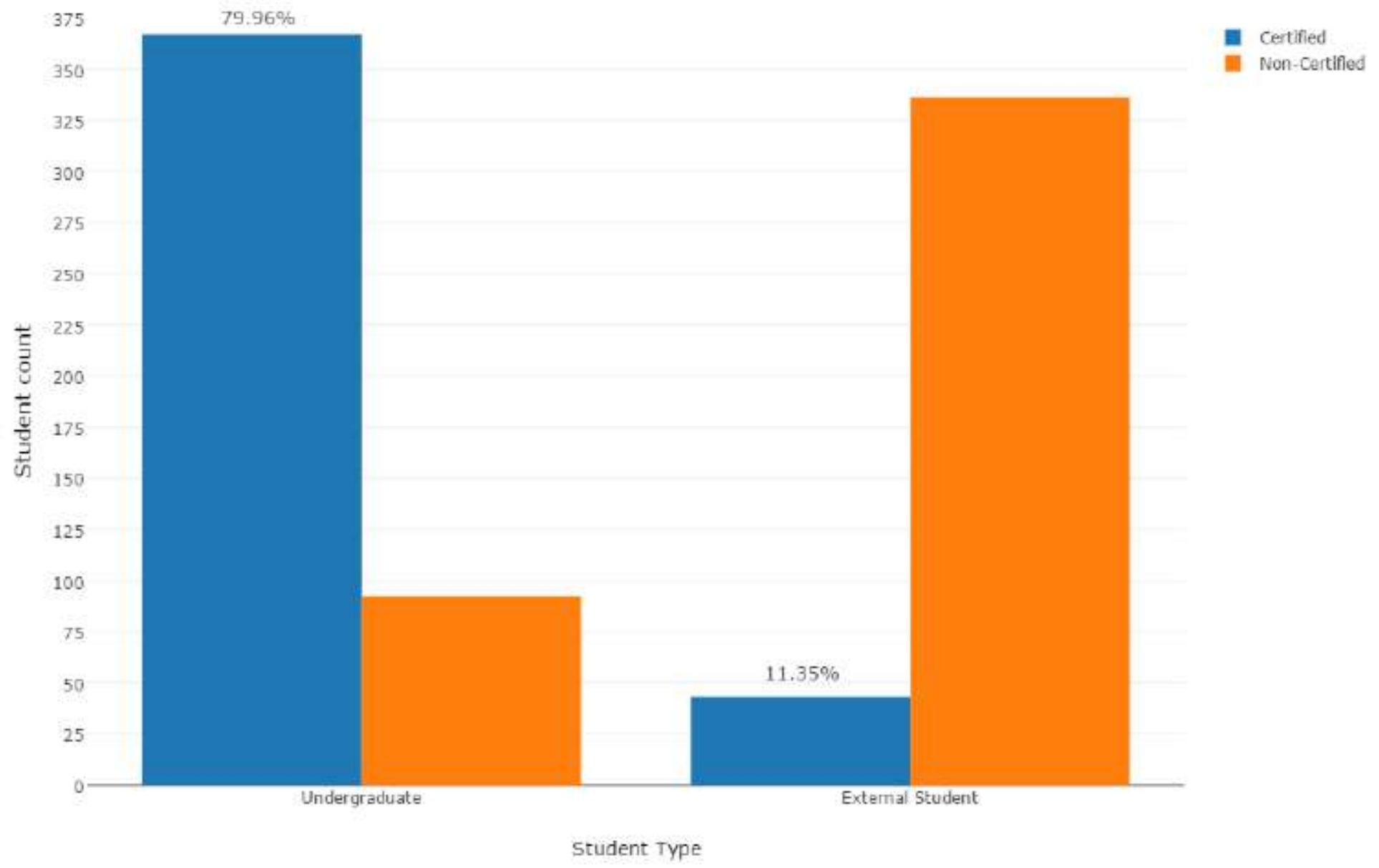

**Figure 69.** The total number of students (*N*= 838) in the GADI MOOC group by the status of certified or non-certified.

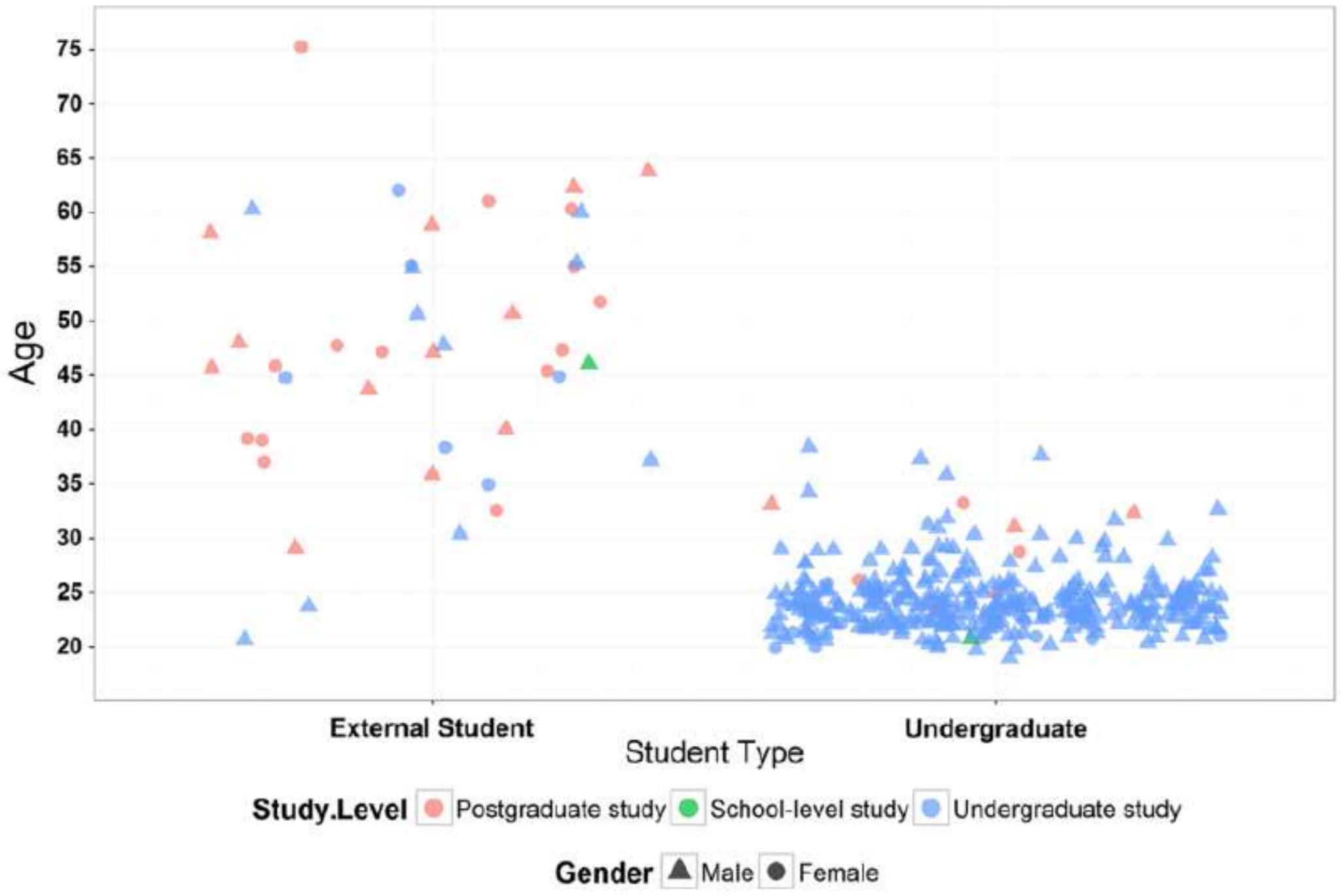

**Figure 70.** Certified students (*N*= 410) of the GADI MOOC, grouped by gender and study level

The overview of the population has further shown that 60% of the certified external students (*N*= 26) held bachelor, master or Ph.D. degrees. These results meet the conclusion of the Guo and Reinecke (2014) demographic research study on a bigger sample from the edX platform. They found that most students who earned a certificate held a postgraduate degree.

In Figure 71, we distributed certified students on the map. (*N*= 375) participants filled out a valid city/country information. More precisely (*N*= 337) students were from Austria while there were (*N*= 38) from other German-speaking countries like Germany and Switzerland. It was quite interesting to the instructional teacher to notice that students enrolled in the course came from other cities than Graz (University's hometown).

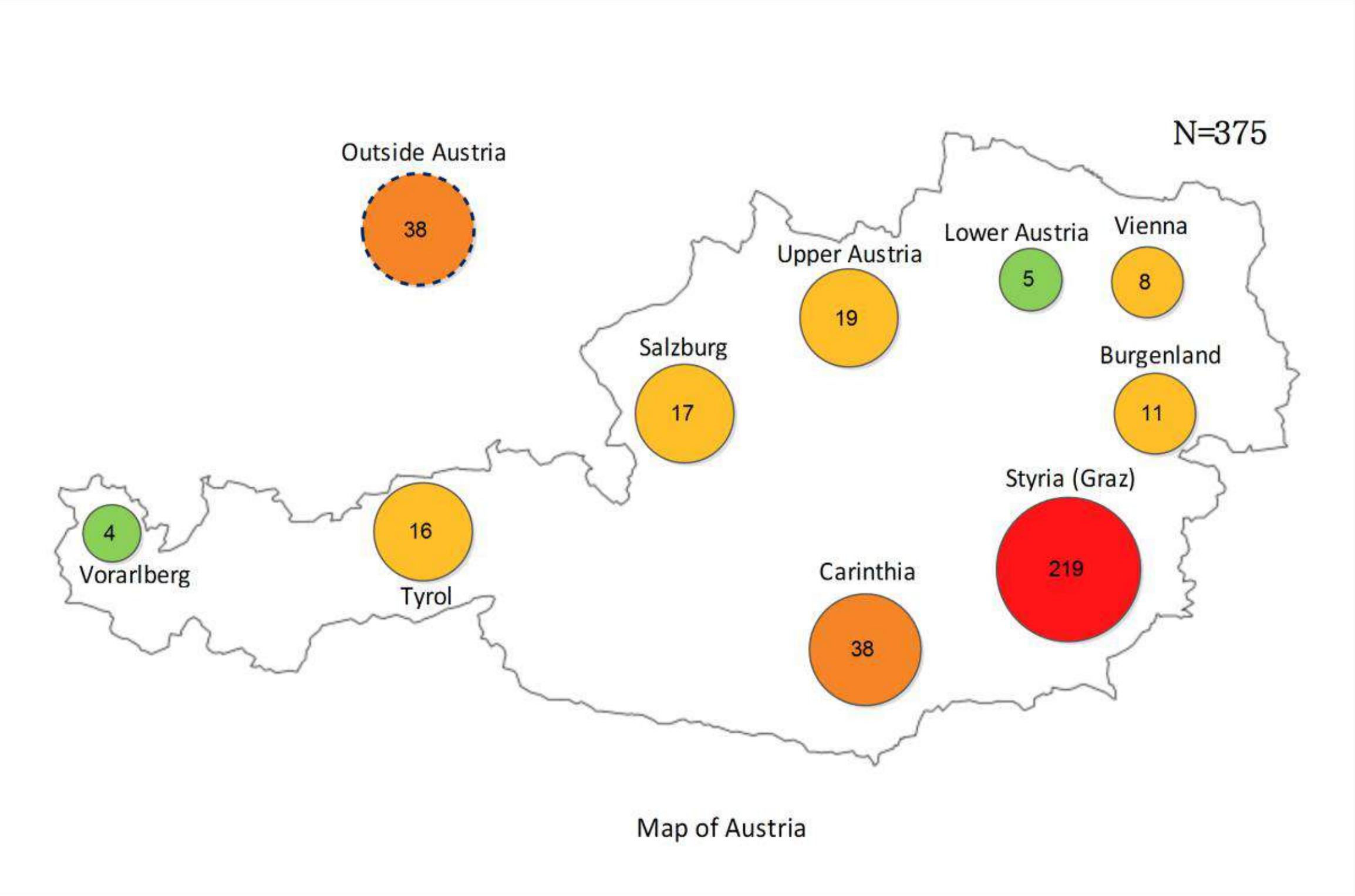

**Figure 71.** Certified students who filled out city information (*N*= 375), distributed on the map of Austria. From university's hometown (*N*= 219), other cities within Austria (*N*= 118), Outside Austria (*N*= 38)

### 5.5.4.2 Survey Analysis

Students who completely finished the MOOC had to answer an evaluation form before obtaining the certificate. Before going further to the cluster analysis, we collected valuable questions that gave us an overview about the certified population. The survey distinguished between (*N*= 43) external students and (*N*= 364) valid university students' input. A summary of the survey results is presented in Table 13.

The psychometrics of this survey were based on a Likert scale going from 1 (strongly agree) to 5 (strongly disagree). One interesting observation is identified in question 4 in which participants from both cases disagree about the discussion forums value in engaging them positively in the course. Furthermore, University students were more pessimistic towards the desire to learn than the external students, whereas both agreed about being satisfied with the weekly quizzes.

**Table 13.** Survey description of the certified students from the GADI MOOC

| Question | University students (*N*= 364) Mean ± SD | External students (*N*= 43) Mean ± SD |
|---|---|---|
| 1 Desire to learn throughout the course* | 2.92 ± 1.01 | 2.14 ± 0.96 |
| 2 Being disciplined to the course* | 2.72 ± 1.14 | 2.23 ± 1.06 |
| 3 Weekly quizzes satisfaction* | 2.28 ± 0.97 | 2.20 ± 0.94 |
| 4 Discussion forum actively engaged you in the course topic* | 3.97 ± 1.16 | 3.41 ± 1.19 |
| 5 Browsed materials outside the MOOC-platform (Wikipedia, external links…etc.)** | 2.45 ± 0.97 | 2.07 ± 0.96 |

SD standard deviation
* 1. Strongly agree … 5. Strongly disagree
** 1. Does not need … 5. More than 3 h a week

#### 5.5.4.3 Clustering

Clustering is about classifying a dataset represented by several parameters into a number of groups, categories or clusters. Estimating the number of clusters has never been an easy task. It was considered a complicated procedure for experts (Jain & Dubes, 1988). This section focuses on the experiment we did to cluster the GADI MOOC students. As conceded before, there was a fair gap between university and external students in this course, each having a particular purpose. We believe there is a fine portion of university students who were attending the course only to pass and transfer the allocated ECTS to their degree profile. Besides, other curious behaviors exist in the dataset we wish to portray. In order to answer the first research question, our prospect of using the k-means clustering was significantly promising.

The certification ratio of the university students, as shown in Figure 69, was much higher than Jordan's findings (Jordan, 2013). The enrollment intention of university students was still debatable as to whether they were attending the MOOC for learning or if they were only looking for the grade. The survey in Table 13 shows an average score of 2.92 regarding learning desire. However, we were more interested in investigating the learning analytics data. Accordingly, the clustering was done independently on both groups. We believe it is not compelling to combine both groups and then classify one single batch of students.

One of the main purposes of this study was to assign each participant in the MOOC to a relevant group that shares a common learning style. Each group should be distinct in full measure to prevent overlaps and the cluster elements should fit as tight as possible to the defined group parameters. To set up our experiment, we used the *k*-means clustering algorithm. The scheme of measuring distance to compute the dissimilarity was set to ‘‘Euclidean.’’ Selecting this method reduces the variability inside one cluster and maximizes the variability between clusters (Peeples, 2011). In order to begin clustering, we labeled the variables that were referenced in the algorithm. The expected results should be clusters with activities and learning objects that distinguish the MOOC participants.

iLAP mines various MOOC indicators such as login frequency, discussion forum activity, watching videos, quiz attempts, etc. Due to the relationship between these indicators, we excluded the highly correlated ones. Their impact, accordingly, will not affect the grouping sequence in the cluster classification, for instance, login frequency indicator and reading in forums indicator correlation valued to ($r$= 0.807, $p$-value < 0.01). As a result, our absolute selection of the MOOC variables in this clustering algorithm was:

- Reading Frequency: This indicates the number of times a user clicked on posts in the forum
- Writing Frequency: This variable determines the number of written posts in the discussion forum
- Video Watched: This variable contains the total number of videos a user clicked
- Quiz Attempts: Calculates the sum of assessment attempts in all weeks

**Use Case: University Students (Undergraduates)**

In this case, the *k* was assigned with a priori assumption of a value between 3 and 6, since we do not really want more than six groups. In order to pick an optimal *k*, we used the NbClust package to validate the *k* value. This package strongly depends on over 30 indices to propose the best clustering scheme (Charrad et al., 2013). To do so, we used a scree plot to visualize the sequential cluster levels on the x-axis and the groups sum of squares on the y-axis, as shown in Figure 72a. The optimal cluster solution can be identified at the point where the reduction in the sum of squares slows dramatically (Peeples, 2011). The vertical dashed red line depicts a critical point (*k*= 4) where the difference in the sum of squares becomes less apparent and respectively creates an ‘‘elbow’’ or a ‘‘bend’’ at cluster 4.

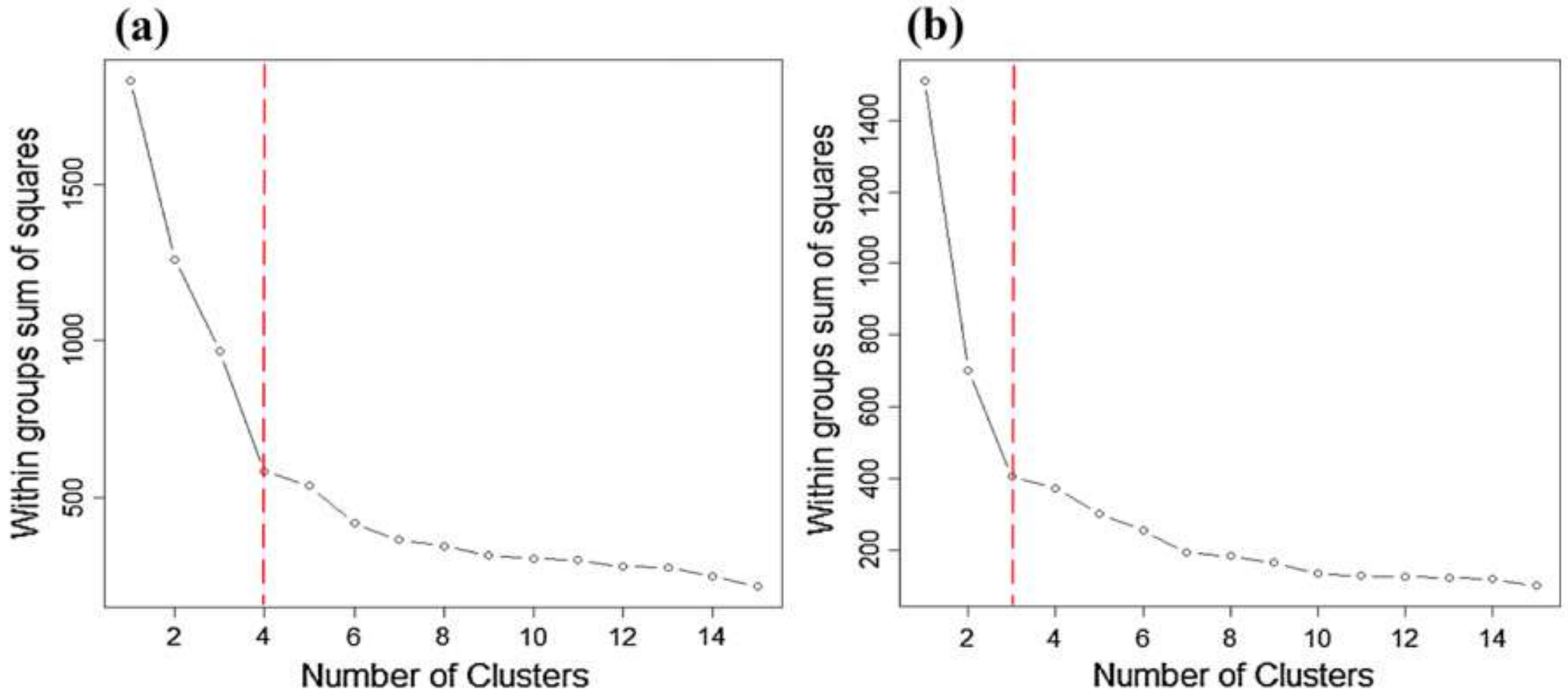

**Figure 72. (**a) Left: the optimal number of clusters of the GADI MOOC undergraduates (b) Right: the optimal number of clusters of the GADI MOOC external students

Afterward, we applied the *k*-means clustering algorithm by setting *k*= 4 for the university students. The outcome in Figure 73 depicts the generated clusters of the first case of the GADI MOOC participants. The clustering visual interpretation usually follows either hierarchical or partitioning methods. We used the partitioning method since it makes it easier for us to display each cluster in a two-dimensional plot rather than dendrogram plot. The x-axis and the y-axis show the first two principal components. These components are algorithmically calculated based on the largest possible variance of the used variables so that it shows as much flaw in the data as possible (Pison, Struyf, & Rousseeuw, 1999). Figure 73 shows a scatter plot distribution of four clusters. Two of the groups, which are the blue and the green, are overlapping. The relation between both principal components in x-axis and y-axis is valued to 67.76%. This percentage means that we have nearly 70% of unhidden information based on this clustering value[12]. This value would be higher in other circumstances when substantial overlap does not exist. However, this hardly meets our main goal of categorizing learners in the undergraduate GADI MOOC.

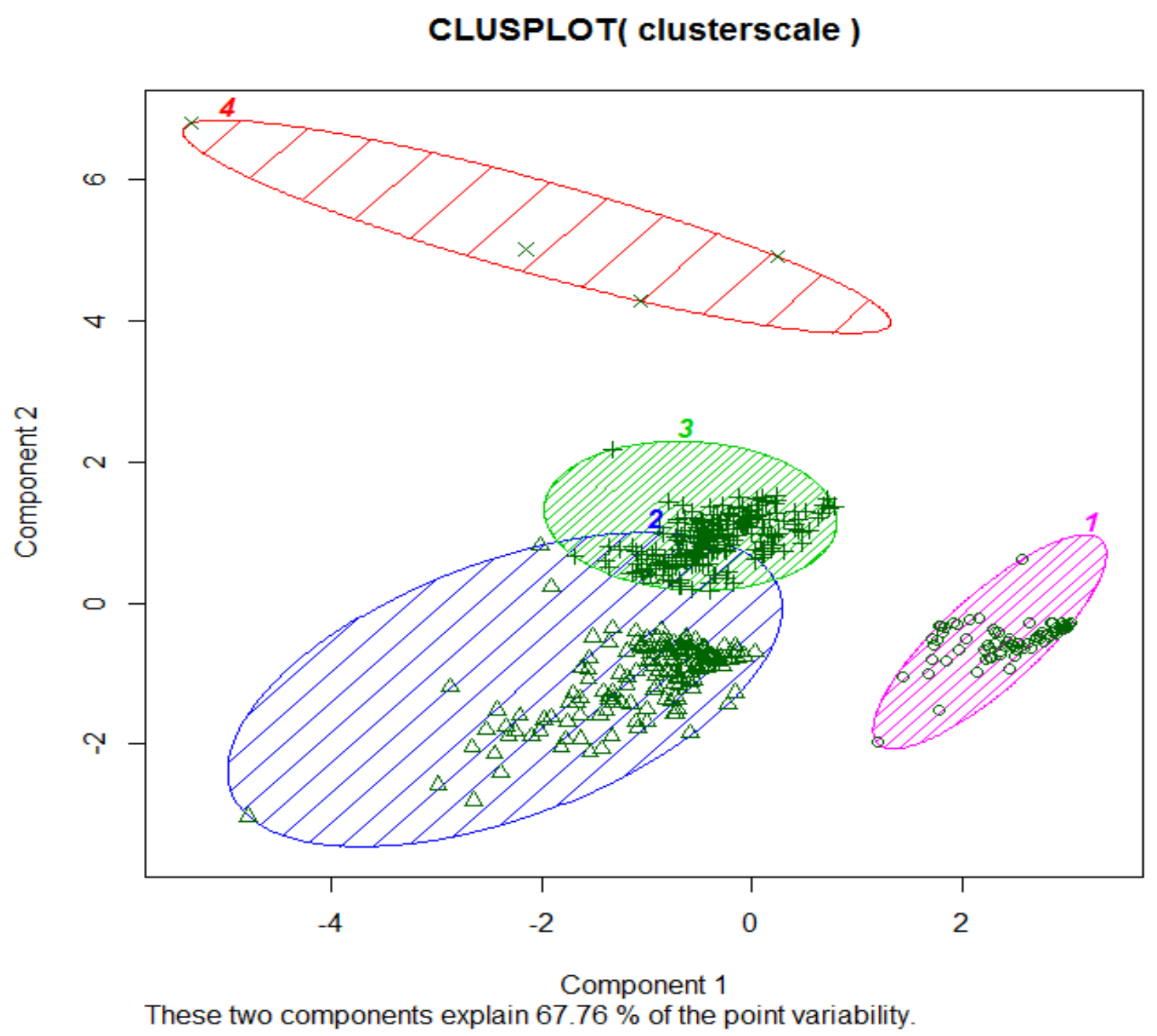

**Figure 73.** University students main clusters (*N*= 459, *k*= 4)

In the following, we show the formed clusters. A rough description for each describes their mode of engagement:

Cluster (1) or "Dropout," with the pink oval contains 95 students (20.69%). This group has low activity among the four MOOC variables. Only 10 students were certified. The attrition rate was high.

Cluster (2) or "Perfect Students," with the blue oval contains 154 students (33.55%). Most of the participants in this group completed the course successfully. Certification ratio was 96.10%. This group was highly engaged in reading in the discussion forum and accessing video lectures.

---

[12] Explanation: http://stats.stackexchange.com/questions/141280/understanding-cluster-plot-and-componentvariability (Last accessed, 11th May 2016).

Cluster (3) or “Gaming the system,” with the green oval has 206 participants (44.88%). Certification ratio was 94.36%. Students in this group shared the same learning style as in Cluster (2), however, it was notable that the rate of watching videos was quite low. The level of engagement in the number of quiz attempts was exceptionally higher than the other clusters.

Cluster (4) or “Social,” is the smallest group and has four participants only (<1%) depicted as the red oval. We noticed that the students in this cluster were the only ones that were writing in the discussion forum. The total amount of certified students in this cluster is 50%.

**Use Case: External Students**

The same methodology used in the previous section was applied to this case. The range of the *k* was assigned in a range between 3 and 6, and the NbClust package was used again to validate the *k* value. Figure 72b shows the suggested number of clusters. The vertical dashed red line depicts a critical point ($k$= 3) where the difference in the sum of squares becomes less apparent and respectively creates an ‘‘elbow’’ at cluster 3.

By applying the *k*-means clustering algorithm, we set *k* to 3 for the external students sub-dataset. Figure 74 depicts the clustering results. The first two principal components variability shows a competitive rate of 88.89%, which indicates a fair clustering validation.

In the following, we give a rough description of the clusters and list their characteristics:

Cluster (1) or “Gaming the System,” with the blue oval holds 42 students (11.08%). Certification ratio of this group was 76.20%. The social activity and specifically reading in forums was moderate compared to the other clusters. The level of engagement of quiz attempts was exceptionally higher than the other clusters.

Cluster (2) or “Perfect Students,” is represented by the red oval and has only 8 students (2.11 %). The certification rate in this group was 100%. Participants showed the highest number of written contributions and reading frequency in the forum, as well as an active engagement in watching video lectures.

Cluster (3) or “Dropout” with the pink oval includes all the other participants (86.80%). The group’s completion rate was very low (close to 1%).

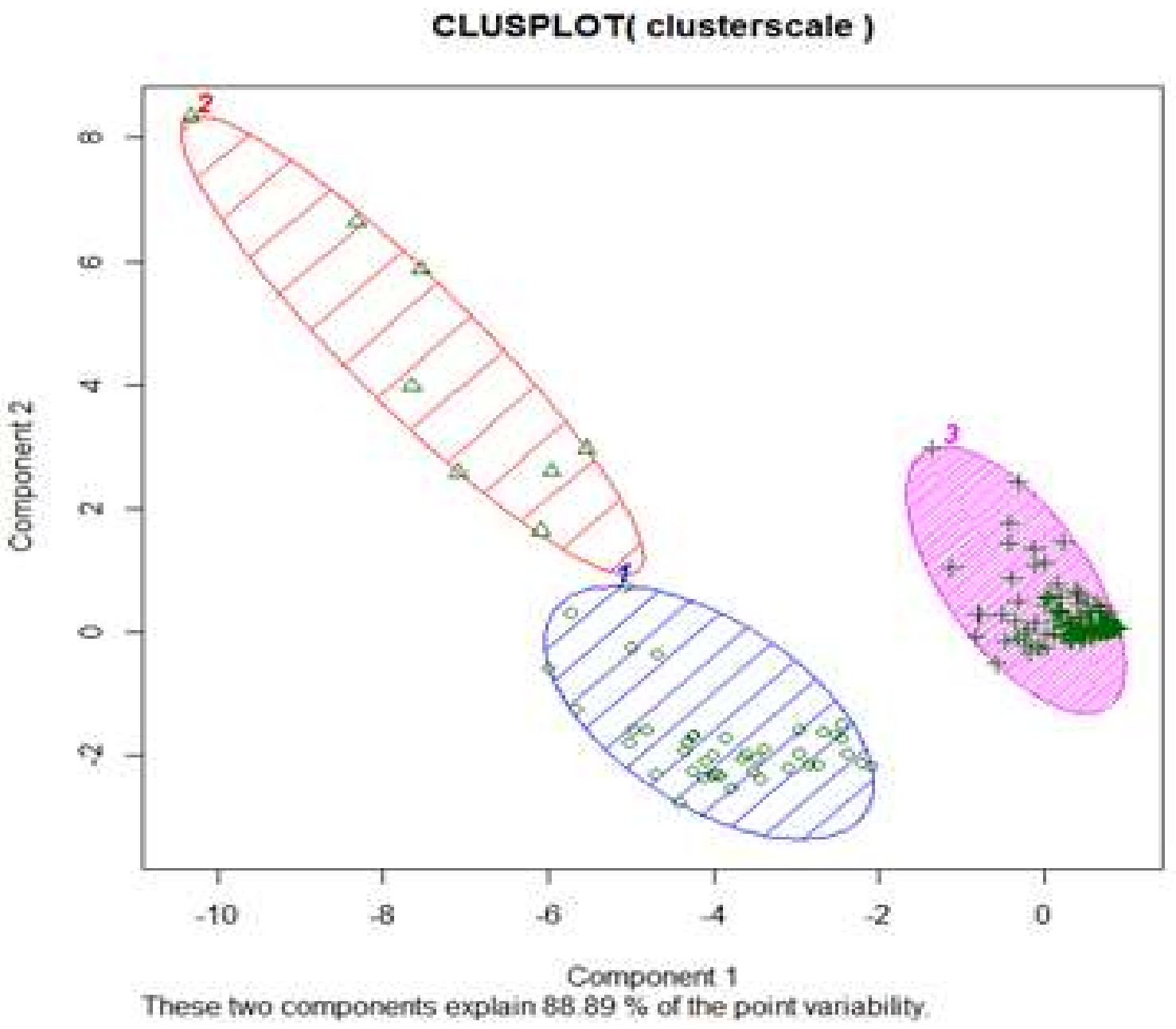

**Figure 74.** External students main clusters ($N$= 379, $k$= 3)

### 5.5.5 Case Study Results and Discussion

Within the previous clustering results, we studied the quantitative statistical values of each variable in every generated cluster. The next step was to compare these values within the same group. As a result, we classified a scale of ‘‘low,’’ ‘‘moderate’’ and ‘‘high’’ that describes the engagement level of each variable in every class. The scale was determined based on the mean value of activities. Table 14 describes the difference between MOOC indicators of each cluster.

The table shows that scaling values of variables vary between groups. Learner clusters are clearly distinguishable when observing the statistical dimensions. The ‘‘Dropout’’ cluster has a low level of engagement among all variables in both cases. Eminent values amid variables of the ‘‘Perfect Students’’ learner cluster are recorded in both use cases. Note that watching video lectures of the MOOC differs between both cases in the ‘‘Gaming the System’’ cluster. We believe that external students were more into learning than the university students. This is obvious when comparing the cluster size (44.88–11.08%). Additionally, university students were forced to get a certificate from the MOOC in order to pass the final university course which definitely made ‘‘Gaming the System’’ cluster the largest group in the first use case.

**Table 14.** Characteristics and comparison between engagement level of clusters

| Cluster | Reading Freq. Mean ± SD (scale) | Writing Freq. Mean ± SD (scale) | Video watches Mean ± SD (scale) | Quiz attempts Mean ± SD (scale) | Cluster size (%) | Certification ratio (%) |
|---|---|---|---|---|---|---|
| *University students (Undergraduates)* | | | | | | |
| Dropout | 6.25 ± 6.38 (L) | 0.01 ± 0.10 (L) | 2.44 ± 3.42 (L) | 2.76 ± 3.86 (L) | 95 (20.69 %) | 10.53 |
| Perfect students | 42.23 ± 23.23 (H) | 0.03 ± 0.19 (L) | 20.76 ± 6.01 (H) | 20.56 ± 3.84 (H) | 154 (33.55 %) | 96.10 |
| Gaming the system | 23.99 ± 11.19 (M) | 0.00 ± 0.07 (L) | 5.77 ± 4.01 (L) | 19.64 ± 3.84 (H) | 206 (44.88 %) | 94.36 |
| Social | 62.00 ± 53.68 (H) | 4.00 ± 1.41 (H) | 3.25 ± 4.72 (L) | 8.50 ± 9.61 (M) | 4 (<1 %) | 50 |
| *External Students* | | | | | | |
| Dropout | 6.03 ± 10.97 (L) | 0.23 ± 0.98 (L) | 1.24 ± 2.52 (L) | 0.68 ± 2.09 (L) | 329 (86.80 %) | <1 |
| Perfect students | 198.63 ± 63.05 (H) | 16.13 ± 9.42 (H) | 24.75 ± 6.34 (H) | 21.50 ± 3.82 (H) | 8 (02.11 %) | 100 |
| Gaming the system | 51.76 ± 43.22 (M) | 0.71 ± 1.88 (L) | 18.10 ± 8.36 (M) | 19.33 ± 6.06 (H) | 42 (11.08 %) | 76.19 |
| Scale *L* low, *M* moderate, *H* high | | | | | | |

The ‘‘Social’’ cluster appears to be a distinctive group in clustering. This cluster is the smallest class of the university students and could not be predicted in the external students group. A reasonable explanation behind this is due to the pessimistic beliefs that were recorded from the survey in Table 13. A design strategy could be implied in order to encourage university students into engaging in discussion forums. It seems that the high activity in the discussion forums guided the ‘‘Perfect Students’’ from the public to passing the course successfully. However, this is not always the case; for instance, a recent study found that not all top social contributors imply that they can pass a MOOC, but it might improve their performance (Alario-Hoyos et al., 2016).

Next, and after the clustering results, the opportunity to portray student engagement in the MOOC with the help of the previous cluster analysis becomes conceivable. Our next challenge was to check if there is a clustering experience from the traditional face-to-face classes that matches ours. The results directed us to a study by Elton (1996) which is based on Herzberg’s (1968) theory to motivate employees at work. The concept behind motivating a certain category of people leads to the main goals of this research study. We think it is quite engaging to see if an older framework like Elton’s might fit MOOC students. Figure 75 depicts the Elton’s clustering proposal, called the Cryer’s scheme.

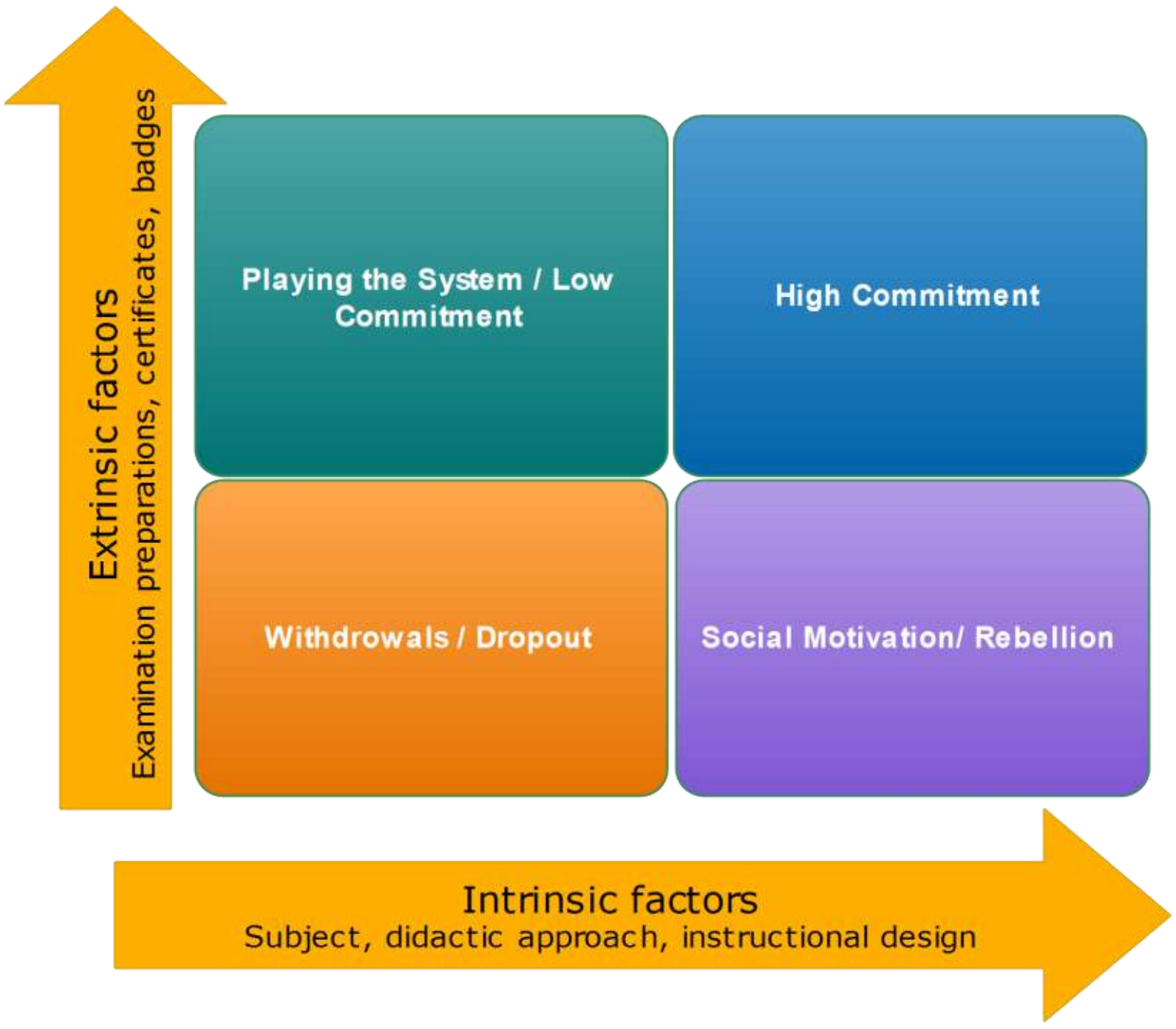

**Figure 75.** Cryer’s scheme of Elton based on levels of learner commitment in classes (Elton, 1996)

The figure shows a two-dimensional diagram that expresses student commitment in the class. The x-axis represents the intrinsic factors, including, but not limited to, achievements and subject. The y-axis records the examination preparation, which is named as the extrinsic factor.

The bottom left of the Cryer's scheme describes the students who are not interested in the course subject nor provide positive results. This class represents the rough description of students who disengaged or ‘‘Dropout.’’ This profile shares common patterns of being inactive among all the MOOC variables. The certification rate in this profile is low. The class at the top left describes learners who have a low commitment to the intrinsic factors. They were named ‘‘Gaming the System.’’ This term comes from a case when students are committed to doing specific tasks such as doing an assignment. The collected dataset analysis reveals many students who were watching learning videos with various skips. Interesting observations were recorded when some students started a quiz without watching the video lecturing material, which was also reported in our paper (Khalil & Ebner, 2015a). Such students meet the defined criteria of the ‘‘Gaming the System’’ cluster. The majority of this group could obtain the certificate at the end of the course.

The class at the bottom right is defined as the ‘‘social-motivated’’ category. This group of students shows sympathy towards the course but fails because of bad exam preparations or time shortage. In our cluster analysis, this group was identified in the undergraduate case as ‘‘Social.’’ Yet, it was difficult to detect them within the external participants group. This is because of the 100% completion rate of the ‘‘Perfect Student’’ group. We strongly believe that students from the ‘‘Perfect Student’’ class would be underrated in this category if they do not complete the course while retaining the high activity engagement in forums and vice versa. This category is characterized by holding active students in MOOC discussion forums. Participants of this group discern a conflict between their high intrinsic motivation of learning and extrinsic motivation. They may find themselves interested in the topic, but their commitment to finish the course faces obstacles. However, they might be involved in forums or watching videos from time to time.

The last class is the group which contains students whose commitment is high. ‘‘Perfect Students’’ reside in this class where their commitment is high. The data reveals that these students are satisfied intrinsically and extrinsically. They watch videos, discuss, do multiple quizzes for better performance, and their certification rate is quite spotless.

#### 5.5.5.1 Results Interpretation

The ideas presented in this article can be extended to other similar courses. The cluster analysis result relies mainly on quantitative data and recognizes the qualitative results from surveys and observations. While the Cryer's scheme was just a framework concept that could be applied to distance learning platforms such as MOOC environments, we see that it fairly fits our MOOC dataset. Stereotypes from traditional massive face-to-face classes might also occur in online courses. In Figure 76, we show the examined MOOC dataset being applied on the Cryer's scheme.

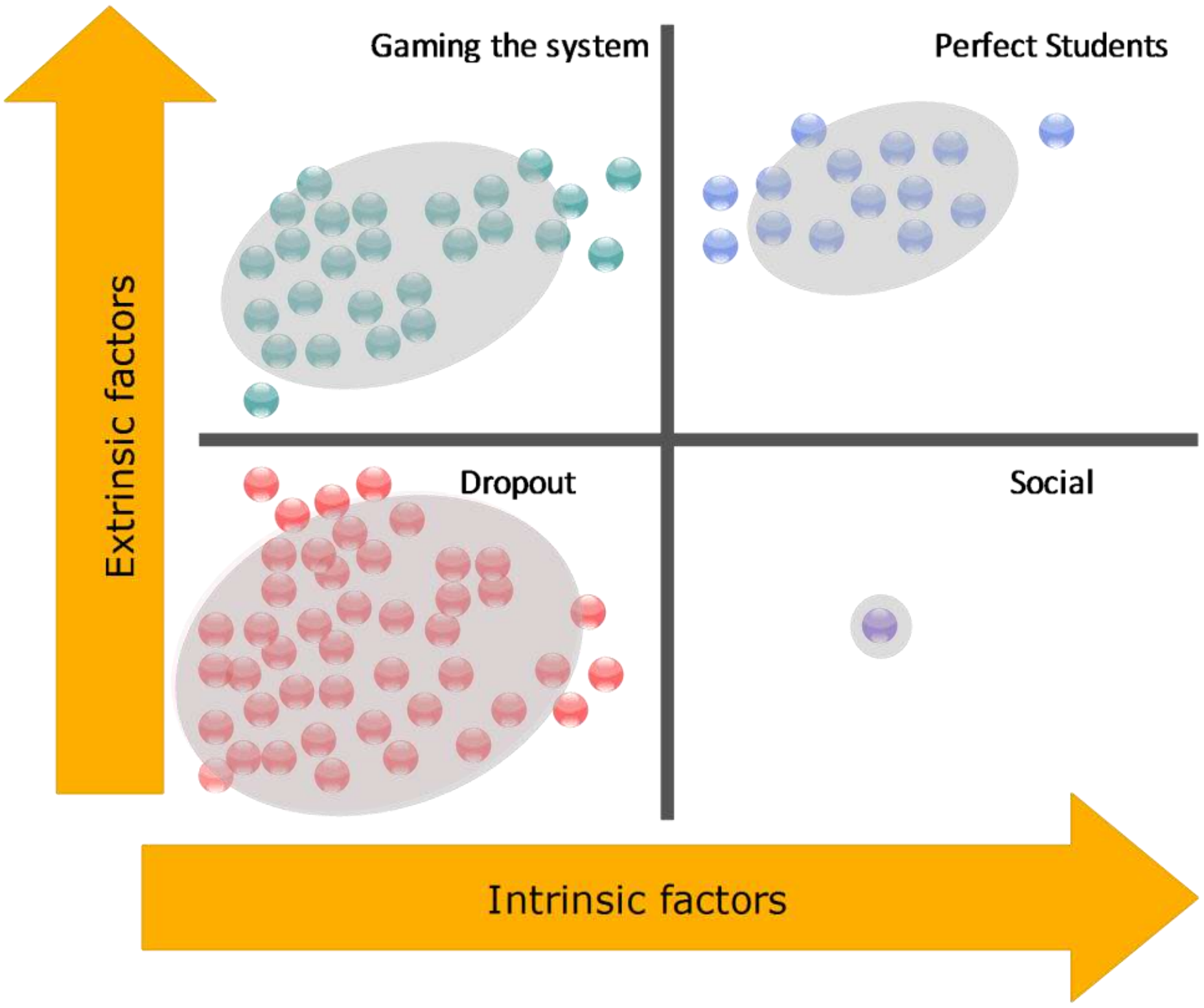

**Figure 76.** GADI MOOC data applied on Cryer's scheme

It must be stated that this scheme does not only include the specific shown profiles, but learners can also be ***distributed unequally*** among these four profiles. For instance, while the ''Social'' cluster does not appear in the external students' case, they could be represented somewhere near the ''Social'' profile in certain conditions. Observations from Figure 76 can be translated into the large quantity of MOOC students that are on the left side of the Cryer's scheme. This part represents either the ''Dropout'' or ''Gaming the System'' clusters.

Several interventions can be adapted thereafter: (a) It is better to consider reallocating students from the dropout class to the ‘‘Social’’ or ‘‘Gaming the System’’ class. The transfer to the social-motivated profile is feasible on one hand by concentrating on the improvement factors which by then increases the intrinsic motivation. The transmission to the ‘‘Gaming the System’’ is attainable on the other hand by focusing on the extrinsic factors. (b) Extrinsic motivation is not enough to lead to the high commitment cluster, therefore, appraising students or recognizing their efforts by providing badges, for instance (Wüster & Ebner, 2016), might not transfer students from ‘‘Gaming the System’’ to ‘‘Perfect Students’’ when they are not valuing an activity or expecting a desired learning outcome (Ryan & Deci, 2000). (c) Moving students from the ‘‘Social’’ to the ‘‘Perfect Student’’ is attainable; however, there are conditions that neither an optimal MOOC nor great didactical system could do without an initiative step by the learners themselves. (d) One of the steady lessons of this study is that paying a lot of attention to the extrinsic factors like shortening the MOOC, grades, certificates, or badges is not the only way to make positive student progress toward ‘‘Perfect’’ status. However, improving the intrinsic factors strategy such as investing in refining the instructional design and didactical approaches should transfer them to the fine edge of being motivated in the learning process.

### 5.5.6 Case Study Summary

This case study examined learner engagement in a mandatory MOOC offered by the iMooX platform. We studied patterns of the involved students and separated the main research into two use cases: university students (undergraduates) and external students (public). We performed a cluster analysis, which pointed out participants in MOOCs and whether they did the course on a voluntary basis or not. Furthermore, we found that the clusters can be applied to the Cryer’s scheme of Elton (1996). We also realized that the population in the GADI MOOC looks similar to a mass education scene happening in a large lecture hall. The experiment of this study leads to the assumption that tomorrow’s instructors have to think of increasing the intrinsic motivation for those students who are only ‘‘Gaming the system.’’ Finally, the intrinsic and extrinsic motivation should both be employed to liven up student engagement in MOOCs.

## 5.6 Case Study (6): Fostering Student Motivation in MOOCs[13]

MOOCs require student commitment and engagement to earn completion, certified or passing status. We saw in previous case studies how student engagement is strongly related to their activities. In this case study, we deeply thought about closing the loop of what we already discovered from online learners in MOOCs based on the available MOOC variables from iMooX, and the capabilities of iLAP log file system. Thus, we present an idea of enhancing students engagement and activities to spark their motivation in order to work more and do more activities, hoping for increasing the retention rate. The case study presents a conceptual learning analytics Activity-Motivation framework that looks into increasing student activity. The framework followed an empirical data analysis from MOOC variables using different short use cases. The framework strongly relies on a direct gamified feedback to drive their inner motivation of competency. After that, we implemented the framework on *Gratis Online Lernen 2016* MOOC to evaluate the results.

### 5.6.1 Background

MOOCs allow anyone to learn and interact through available learning variables such as video lectures, recommended articles, content downloads, discussion forums, assessment, etc. The characteristics of the open environment of MOOCs bring a distinct range of motivations and beliefs among students (Littlejohn et al., 2016). As long as the direct interaction between teachers and students does not reach the level of the traditional face-to-face lectures, students are forced to organize their ways of learning themselves. Also, online learning environments such as MOOCs differ from the traditional settings, which results in different student engagement. As a result, this raises students' load to self-regulate their learning, motivate themselves and monitor their own behavior (Zimmerman, 2000), as well as actively interact with online learning objects (Khalil, Kastl, Ebner, 2016).

Recent work on MOOCs revealed a high attrition scale in activities in the first two weeks (Balakrishnan & Coetzee, 2013). Balakrishnan and Coetzee reported a 50% dropout at the end of the second week. Some researchers, including us, suggested cutting course duration in half

---

[13] Parts of this section have been published in:

Khalil, M. & Ebner, M. (2017). Driving Student Motivation in MOOCs through a Conceptual Activity-Motivation Framework. *Journal for Development of Higher Education Institutions (ZFHE)*. In press.

(Lackner, Ebner & Khalil, 2015). Others pushed the concept of grabbing student attention by looking forward to boosting the extrinsic factors such as offering badges, certificates and honor awards (Wüster & Ebner, 2016). Researchers from the Northeastern University of China have noticed that the activities performed by students in the MOOC platform reflect their motivation (Xu & Yang, 2016). The authors found a strong relationship between someone's behavior and his/her evaluation of excitement. From there, they tried to find a relation between grade prediction and certification ratio along with their activities in the MOOC through a developed classification model. While such prediction might be hard to examine because of the nature of the predictive models and several MOOCs being difficult (Klüsener & Fortenbacher, 2015), others used online surveys and semi-structured interviews to identify learner motivation (Littlejohn et al., 2016). Further research about understanding the motivation of online learners in MOOCs was conducted by Stanford University researchers who listed 13 factors that could capture learner motivation (Kizilcec & Schneider, 2015). Despite their benefits, online surveys lack the proper target group and might provide inaccurate results if it takes more than a few minutes, and lack of the proper target group.

This case study was further influenced by a couple of learning analytics applications, which were considered in our proposed framework. One of these tools was the Course Signals (Arnold & Pistilli, 2012). It is an application that provides feedback according to the traffic light system. Whenever a green light is shown, it means that the student is on track, whereas the orange and the red lights imply at-risk situations and intervention(s) by either a teacher or an institution would be required.

The short background of this section as well as the literature review in the first chapters is strongly related to what this case study brings. By examining engagement and activity either in the discussion forums or video events, this research study leverages the data from MOOC variables to preserve student activities and motivate them to stay engaged. The existing literature, however, suffers from very little research in regards to direct learning analytics feedback for students on MOOC platforms. Despite the fact that showing statistics or gamification elements to students is usually obtainable in most online environments is a motivation factor, to the best of our knowledge, we could rarely find a module that describes direct feedback to students that looks into preserving a high activity level of students.

### 5.6.2 Methodology

Our methodology focused on obtaining data from the following MOOC variables: watching video lectures, login frequency, posts in forum, reading of forum posts, and quizzes, in order to identify a *competent activity level*. Henceforth, an analysis that includes finding patterns in visualizations and an examination using exploratory analysis on empirical data was conducted. Data collection was performed using the iMooX Learning Analytics Prototype (iLAP). We again followed the content analysis methodology by (Neuendorf, 2002), in which MOOC variables were measured and referenced. The study also employed Wang and Hannafin (2005) design-based research methodology that depends on identifying goals, collecting data during the whole design process, and refining according to the required goals.

The examined MOOC in this case study was the *Gratis Online Lernen* (GOL). We analyzed the offered courses of the 2014 and 2015 years. Students had to score at least 50% in the self-assessment quizzes to pass. The main content of the courses were video lectures with an average duration of 5 minutes per video. Students were rewarded with certificates whenever they successfully passed all the quizzes.

### 5.6.3 Case Study Analysis

In this section, we try to validate whether the certified students show more activity using MOOC variables (forums and videos) than the non-certified students. To our knowledge, we could not find a direct assurance that certified students participate in more activities with MOOC variables. As a result, we try to prove this hypothesis. For this purpose, we chose to analyze the following three MOOC variables: posts in forum, views in forum and videos. Similar to previous case studies, we split the students into two categories: *certified* and *non-certified*. The first group includes those who completed a MOOC and therefore received a certificate at the end of the course, while the second group includes the students who dropped out of the MOOC at any time during the course. The certified students in GOL-2014 and GOL-2015 were ($N$= 193, $N$= 117) respectively, while the non-certified students in GOL-2014 and GOL-2015 were ($N$= 810, $N$= 359). The analysis results in the following subsections proved that learning activities have quite a retention impact on students in a massive open online course.

### 5.6.3.1 Forum Reading Analysis

During the eight weeks of forum discussions, there were 22,565 views of forum threads in GOL-2014 and 8,214 views of forum threads in GOL-2015. Figure 77a and Figure 77b show the average number of thread views for both MOOCs. The difference between the reading activity of the two groups is quite obvious. Figure 77a depicts a maximum number of reads in week 1 for both groups, which rapidly drops until week 4. This follows the condition that attrition rate becomes more stable after the first four weeks of a MOOC (Lackner, Ebner & Khalil, 2015). However, in Figure 77b, we realized that certified student forum views escalated in week 5 and then dropped to around 4 views per user till the end of the MOOC. A study by Wong and his colleagues recorded similar student behavior (Wong et al., 2015). The authors showed that active users showed higher activity after the first weeks of the MOOC.

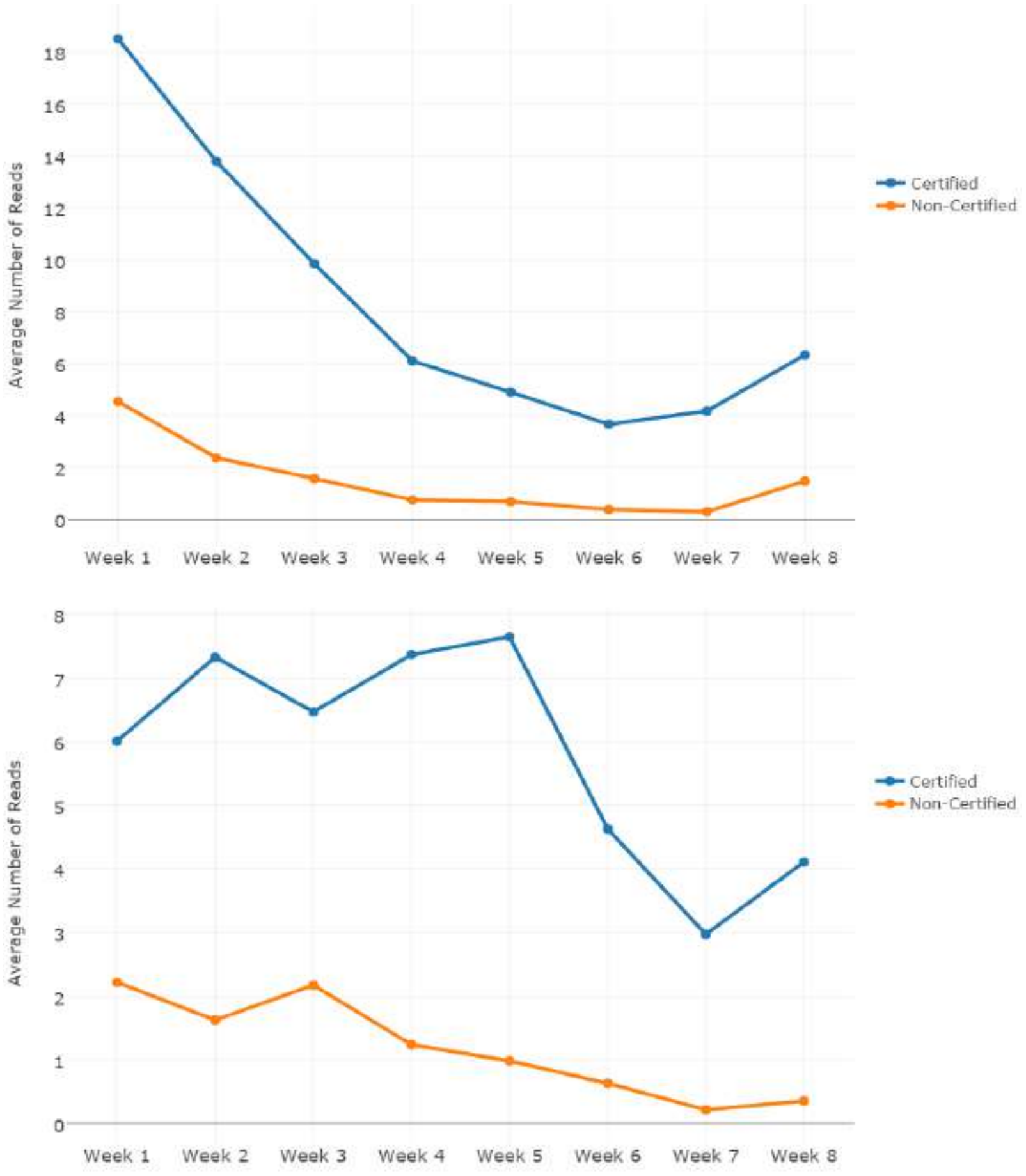

**Figure 77.** (a) Top: The average number of discussion forum views in GOL-2104 MOOC. (b) Bottom: The average number of discussion forum views in GOL-2015 MOOC

#### 5.6.3.2 Forum Posts Analysis

In total, there were 828 written posts in GOL-2014 and 408 written posts in GOL-2015. These posts took the forms of comments, threads, and replies. Figure 78a and Figure 78b illustrate the average number of written posts in both MOOCs forums.

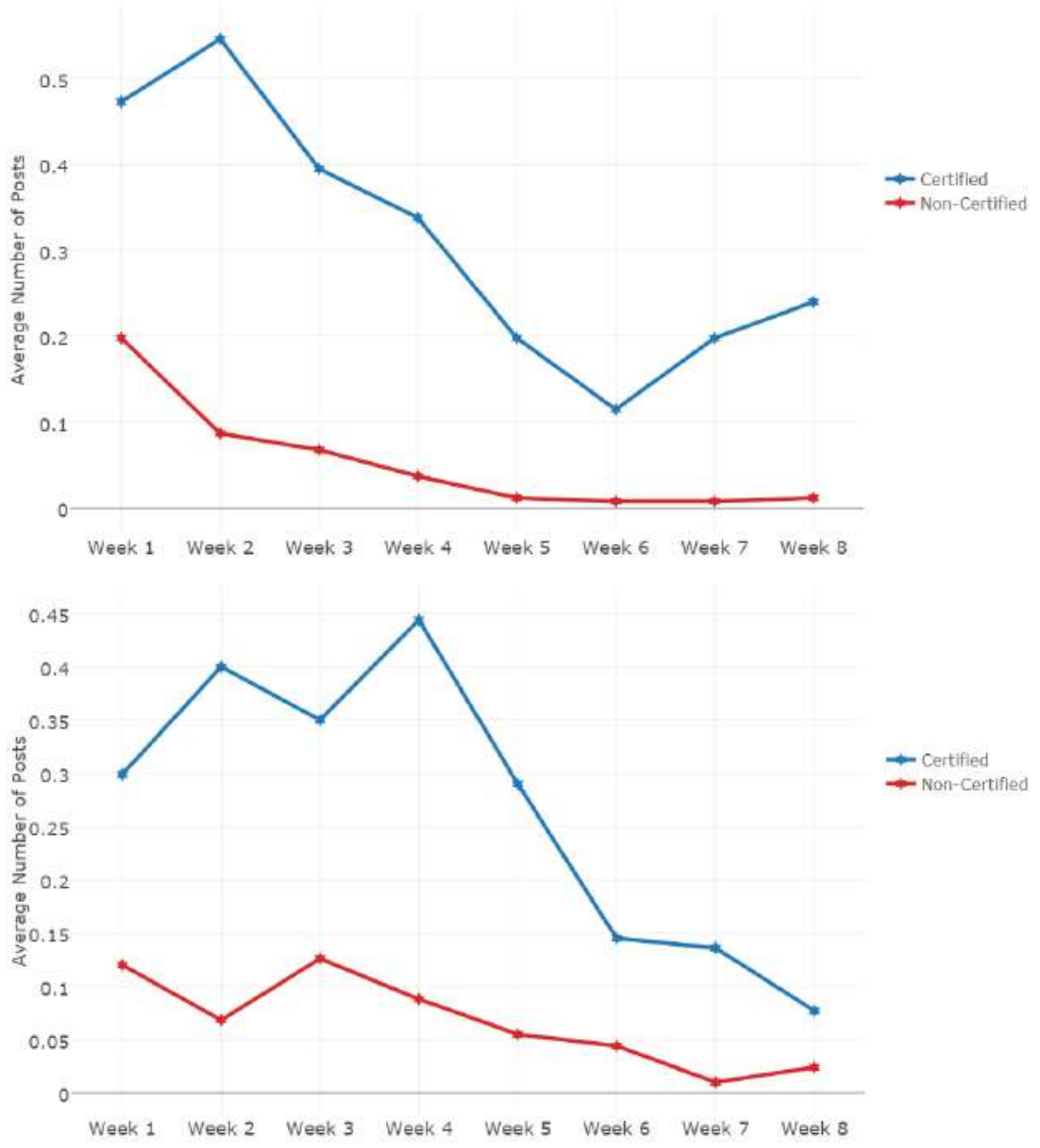

**Figure 78.** (a) Top: The average number of discussion forum posts in GOL-2104 MOOC. (b) Bottom: The average number of discussion forum posts in GOL-2015 MOOC

In fact, it is apparent that certified students are more active in posting and commenting in MOOC forums. In Figure 78a, the average number of contributions is very low after the fourth week. This is explained by different reasons like the steep dropout rate after the first weeks, or the low motivation to contribute and comment (Lackner, Khalil & Ebner, 2016).

### 5.6.3.3 Video Lectures Analysis

The third MOOC variable we analyzed was video lectures. Video contents were hosted on YouTube; however, the iLAP, as previously described, can only mine events of participants using play and pause/stop that happen on the iMooX platform. We summed up the total number of video interactions and showed the average number of events (play, pause, and full watch) per week. There were 17,384 video events in GOL-2014 and 8,102 video events in GOL-2015. Figure 79a and Figure 79b show a graph line of learner interactions in GOL-2014 and GOL-2015.

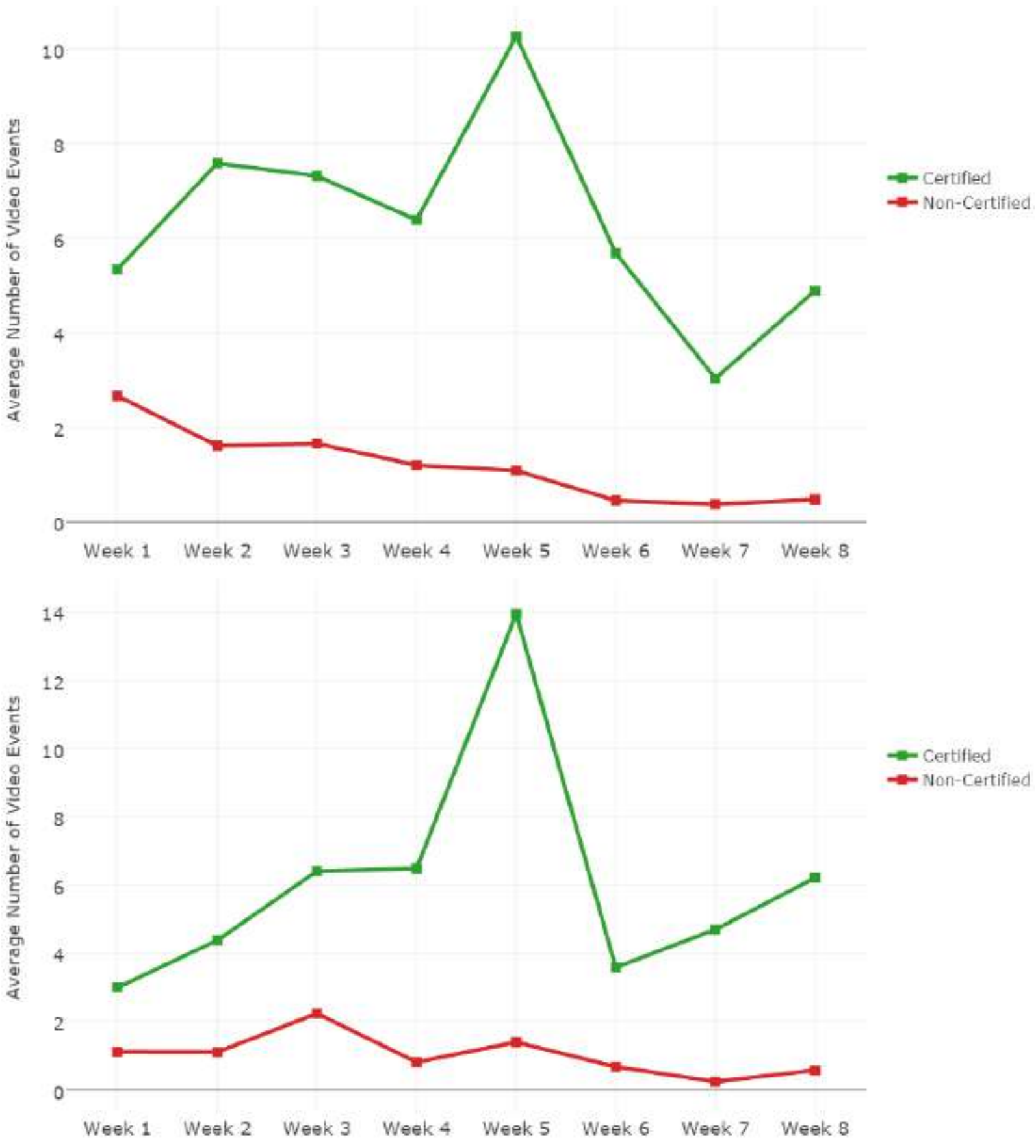

**Figure 79.** (a) Top: The average number of video events in GOL-2104 MOOC (b) Bottom: The average number of video events in GOL-2015 MOOC

The figures show that the average number of video events of certified students is undoubtedly higher than non-certified students. Non-certified students show weak video lectures activity.

### 5.6.4 Case Study Discussion and Evaluation

The six figures in the previous subsections showed that the active users demonstrated higher activity during the MOOC weeks. In fact, the earlier case studies about forum discussion, retention and activity revealed that students who complete MOOCs pursue more activity with MOOC variables. Nevertheless, we again like to prove our hypothesis by showing use cases on forums posts and reads as well as videos. With regards to student forum activities, we found there was an apparent gap between certified and non-certified students. Motivated students are more likely to engage in discussion forums (Lackner, Ebner & Khalil, 2015). Gilly Salmon identified four learner strategies in online discussions:

> (1) "Some do not try to read all messages," (2) "Some remove themselves from conferences of little or no interest to them, and save or download others," (3) "Others try to read everything and spend considerable time happily online, responding where appropriate," (4) "Yet others try to read everything but rarely respond" (Salmon, 2007).

The data presented in the forum posts and readings analysis correspond to Salmon's learner types 1, 2 and 4. Non-active students do not ask questions or comment in the forums. Presumably, certified students are more likely to post questions to ask a teacher or colleague for help which means they are more active in forums. The video analysis also showed the difference between certified and non-certified students. As MOOCs rely on videos, students need to watch them in order to pass quizzes. Thus, active students who want to pass quizzes need to watch videos, except for some unique cases where students game the MOOC system (Khalil & Ebner, 2016a).

#### 5.6.4.1 Activity-Motivation Framework

Per the previous analysis results and the impact of activities on student motivation to complete MOOC, we propose an Activity-Motivation framework to motivate learners to do more activities. In correspondence to the iMooX MOOC-platform potential of offering variables such as quiz attempts, logins, forum posts and views, and the tested empirical data, we designed this framework. The Activity-Motivation framework intends to assist in increasing student motivation and engagement.

The proposed model is shown in Figure 80 and consists of four main dimensions. Each of the dimensions contributes a portion to a gamification element.

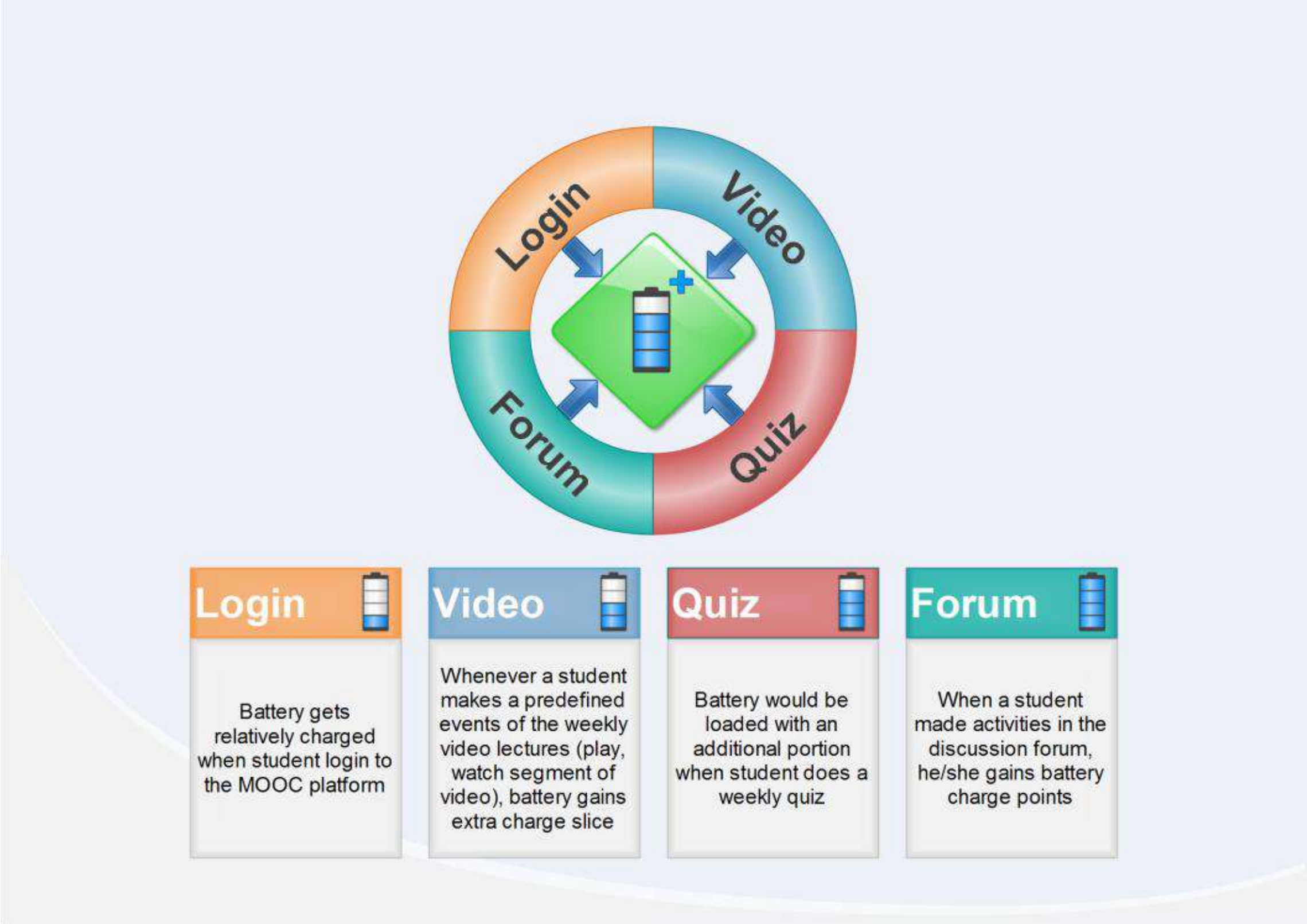

**Figure 80.** The Activity-Motivation framework that uses gamified element to increase the extrinsic motivation of MOOC students

Our choice of such items was a battery as it expresses a “filling up” animation. We thought that what happens to a battery is similar to what a student does with the MOOC activities. We aim to keep the students charged with activity, motivation, and incentive. Gamification elements have a positive impact on student motivation and learning (Gonzalez, Toledo, & Munoz, 2016).

The four dimensions of the proposed framework are: login, video, quiz, and forum. It is worth noting that these dimensions can be extended and are not exhaustive to the ones listed. For instance, extra dimensions involving readable content such as a downloadable article or assignments can be included when required. The gamification element was divided into four segments based on the number of the selected MOOC variables. The proposed Activity-Motivation model can be implemented as a plugin or as an independent tab on the MOOC page and would be updated on a weekly basis. In the following paragraph, we briefly elaborate on the model and describe its mechanism. Each element counts as 25% in the battery charged portions as follows:

- **Login**: When a student logs into the MOOC, he/she will reflect relatively on the gamification element (battery). The first segment of the battery will be 25% charged. Several logins will not increase the charged portion.
- **Video**: The second dimension is the video lecture. When a student interacts with the MOOC video lectures and completes a number of predefined events the battery is charged a bit further.
- **Quiz**: Battery will be filled with one additional portion when a student does a quiz. As previously described, iMooX MOOC-platform allows each student to try every weekly quiz up to five times. However, just one trial would be enough to indicate that the student is active. Identical to the previous dimensions, several attempts will not increase the battery's charged portion.
- **Forum**: The analysis of forums in the earlier case studies showed the relation of discussion forums and student activity. Being engaged in the forums either by writing or reading threads will increase the battery charging portion.

#### 5.6.4.2 Implementation of the Activity-Motivation Framework in iMooX

In this part of the case study, we show our results of implementing the activity motivation framework on *Gratis Online Lernen* MOOC of the 2016 year (GOL2016). The process of this implementation was done manually since we were looking for evaluation results at the first stage. A second stage of automatic implementation can be systemized on upcoming MOOCs if the results are impressive to the iMooX higher management.

We designed the gamification elements using open source software called Inkscape (http://www.inkscape.org, last accessed: December 2016). Figure 81 shows our design which consists of five gamification symbols.

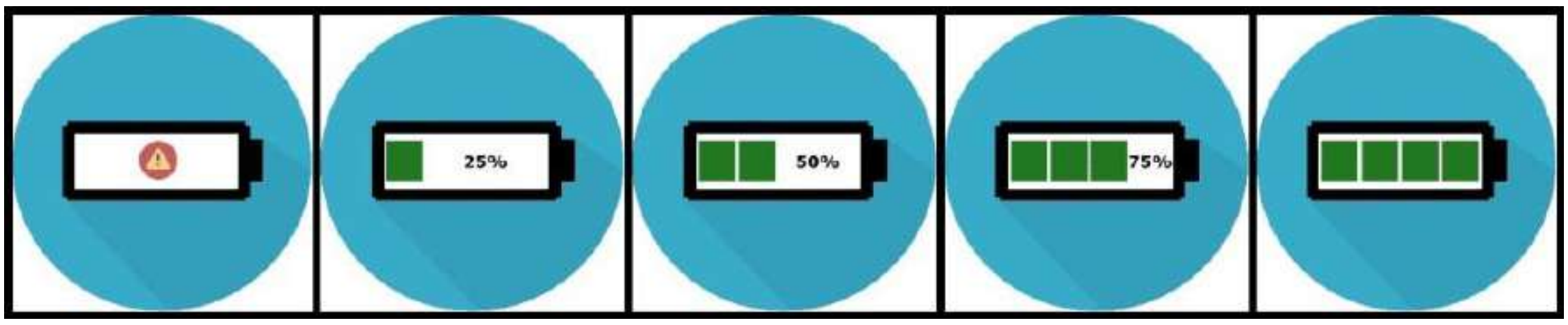

**Figure 81.** The gamification elements that were designed for the Activity-Motivation framework in GOL2016

Each item in Figure 81 was displayed on the user homepage of the GOL2016 MOOC. Every battery symbol represents the recorded MOOC activity of the user based on the learner's previous week's interaction. For example, by the end of the first MOOC week, we show the student his/her activity progress of that week on the first day of the next week and so on. The MOOC activities that were planned to be logged in our implementation were: a) logging into the MOOC homepage, b) doing a quiz, c) posting/commenting at least once or reading two threads in the MOOC's discussion forum, and d) watching a video. Nevertheless, we faced a major technical problem with logging user video activities. The problem was detected just a couple of days before the MOOC launch date (10th October 2016) when we discovered that the MOOC videos were embedded using *IFRAME* instead of *OBJECT* on the iMooX platform. Therefore, our simple and quick redress was that by assuming a user logs in, we show the "50%" battery symbol. The "25%" was not used at all because of the video logging problem. Therefore, the framework that we used was established by the following guidelines based on every week activities:

- Login activity: When a student logs into the GOL2016 MOOC, the activity will reflect relatively on the gamification element (battery). The first segment of the battery will be 50% charged. Several logins will not increase the charged portion.
- Quiz activity: The battery symbol will be filled with one extra portion (25%) when a student takes a quiz. As described before, the iMooX MOOC-platform allows each student to try the weekly quiz up to five times. However, just one trial would be enough to indicate that the student is active. Several attempts will not increase the battery's charged portion.
- Discussion forum activity: If a student is engaged in the forums either by writing one post or reading threads at least twice, then the battery charging portion will add another 25%.

We also arranged for each icon to have a *tooltip* or *mouseover* text so that the student can realize what these symbols imply. Table 15 shows the tooltips of each battery symbol.

**Table 15.** The gamification battery symbols with their tooltips.

| Battery symbol | Tooltip |
|---|---|
| 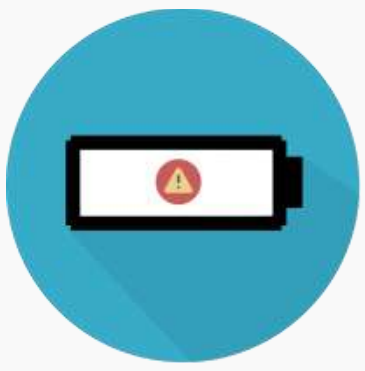 | No activity last week – we are looking forward to seeing you again this week! |
| 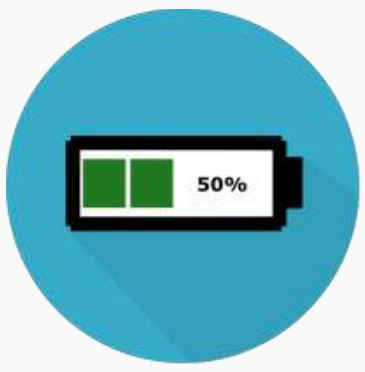 | Your activity last week is 50%. Good! Increase your activities to score better! |
| 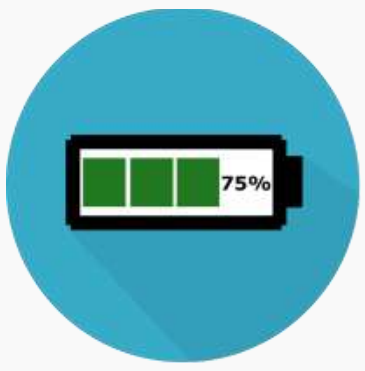 | Your activity last week is 75%. Great! Keep it up! |
| 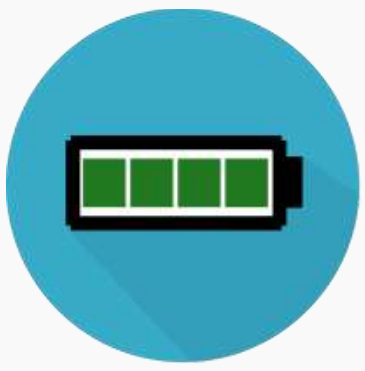 | Your activity in the previous week is 100%. Congratulations. Your commitment is excellent. Keep it up! |

Our further step was to evaluate the results of the Activity-Motivation framework. As stated previously, the examined MOOC was the *Gratis Online Lernen* 2016. The MOOC was finished on the 5$^{th}$ of December 2016. The descriptive analysis of the MOOC recorded that there were ($N$= 284) students who registered in the MOOC. The total number of students who were certified was ($N$= 51). Our implementation of this framework and its algorithm was concealed on purpose and students never knew how the battery symbol is charged/filled. The intention was to spark student motivation to do more activities and actions in the MOOC.

Next, we collected through our manual implementation the total number of students and their gamification statuses as shown in Figure 82.

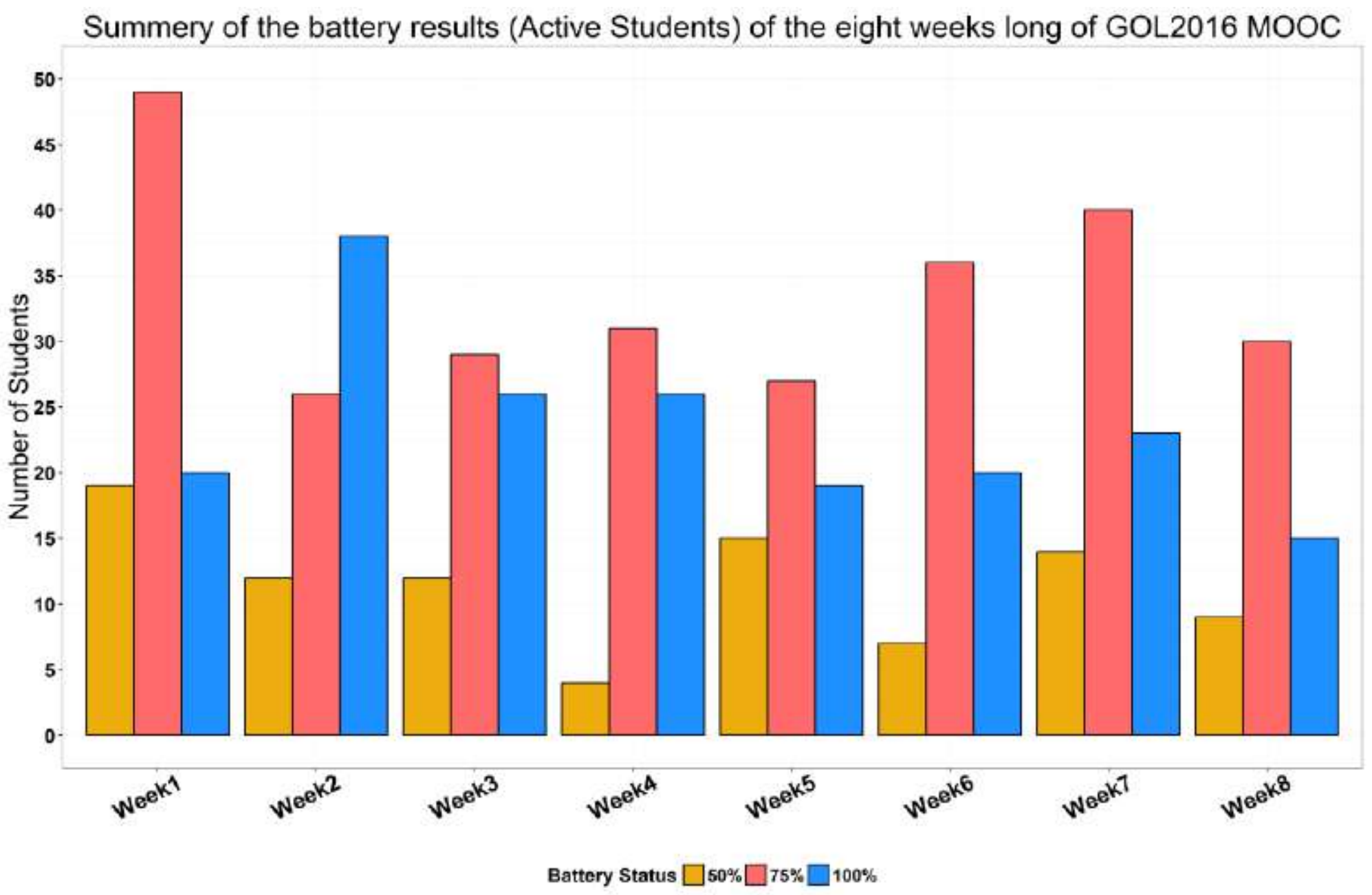

**Figure 82.** Summary of the battery results of active students in GOL2016 MOOC

The figure shows a noticeable elevation of the 75% battery status across all the MOOC weeks. Our explanation is that students who are committed to complete the MOOC intended to log in at least once a week and do the weekly quiz. Also, the figure shows that there is a slight decrement of the 50% battery status from week1 to week8. This was interesting, as students tried to push efforts to score higher than 50% activity or 0% status. Week4 showed the minimum score of 50% status. On the other hand, the full activity status 100% was obviously at its highest in the second week with around (*N*= 38) students. Our explanation of this behavior can be interpreted by the fact that students were pushing more efforts to improve their battery status. Likewise, the stability of the same situation is clear across all the weeks. It is worth pointing out that some students might do quizzes on different weeks. This can be complicated to track; however, we tracked students based on every week's quizzes, logins and forum activities.

The general status of the GOL2016 MOOC is shown in Figure 83. Only 284 students registered in the course with 51 certified students. Surprisingly, the number of active students (and those who at least did one quiz or wrote in the discussion forum, similar to the categorization in Figure 36) was 209 students, comprising 73.6% of the total number of registered students. This absolutely demonstrates that students did more activities to have a higher score with the gamification symbol. We compared GOL2016 with GOL2014 in respect to the active students and registrants ratio. In GOL2014, the ratio of active students to registrants was 47.33%, while the ratio in GOL2016 was 73.6%. The results were very impressive to us. Nevertheless, the certification ratio remained within the same level in GOL2014 and GOL2015, but the certification rate might increase in January 2017 since students need to complete a final questionnaire to get the certificate.

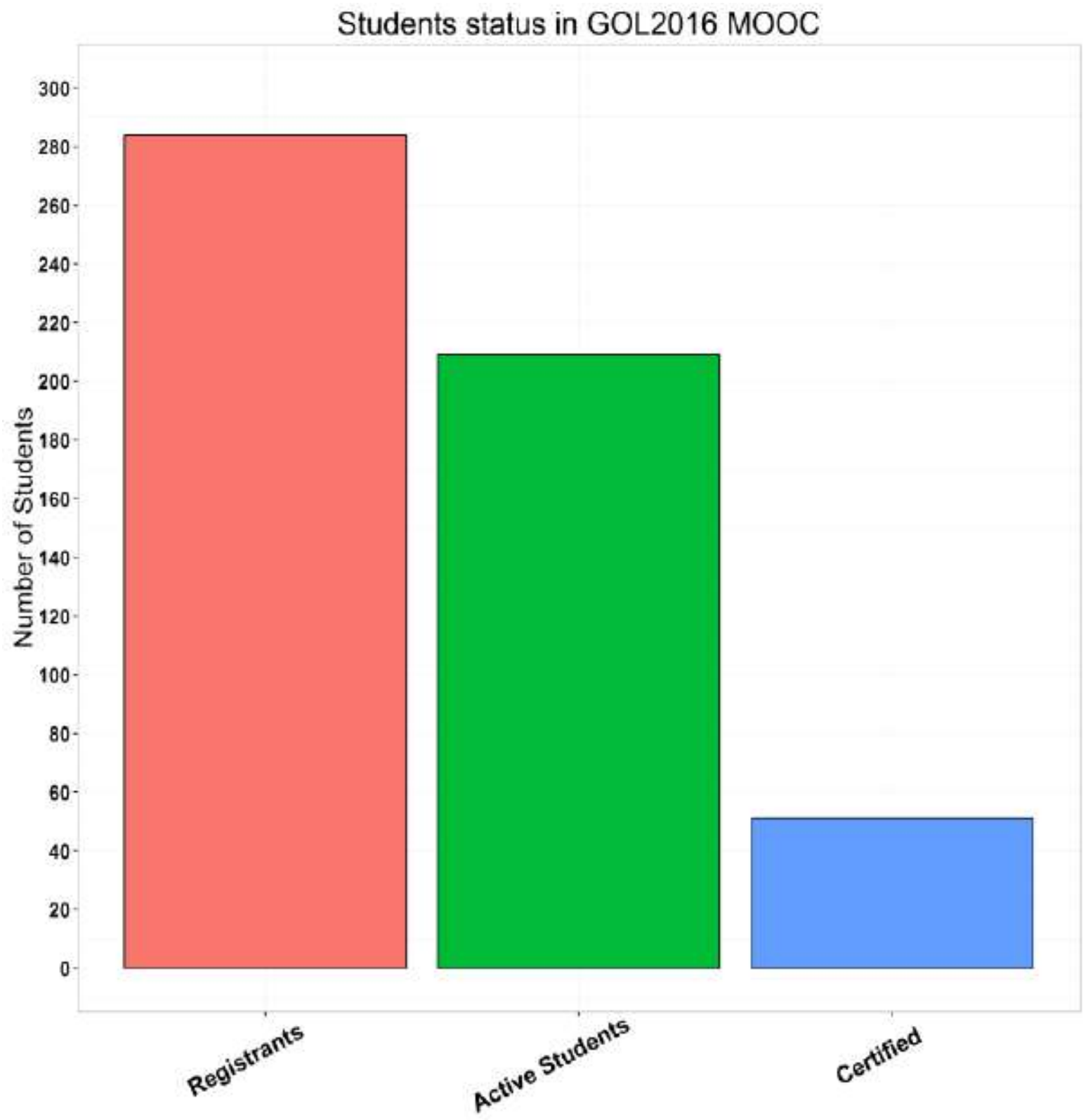

**Figure 83.** Summary of students' status after categorizing them to registrants, active students, and certified in GOL2016

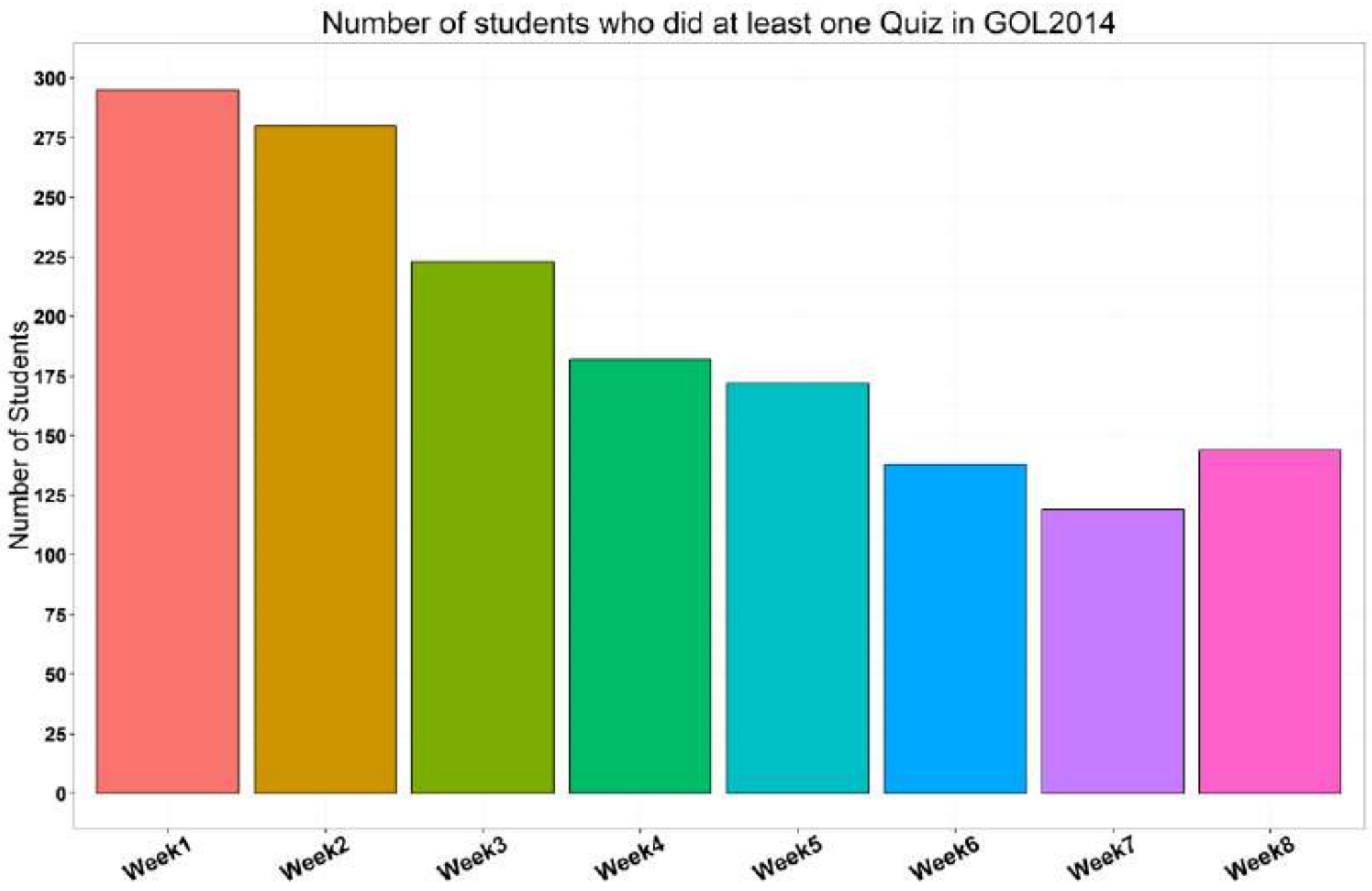
Number of students who did at least one Quiz in GOL2014
Number of Students
300
275
250
225
200
175
150
125
100
75
50
25
0
Week1
Week2
Week3
Week4
Week5
Week6
Week7
Week8

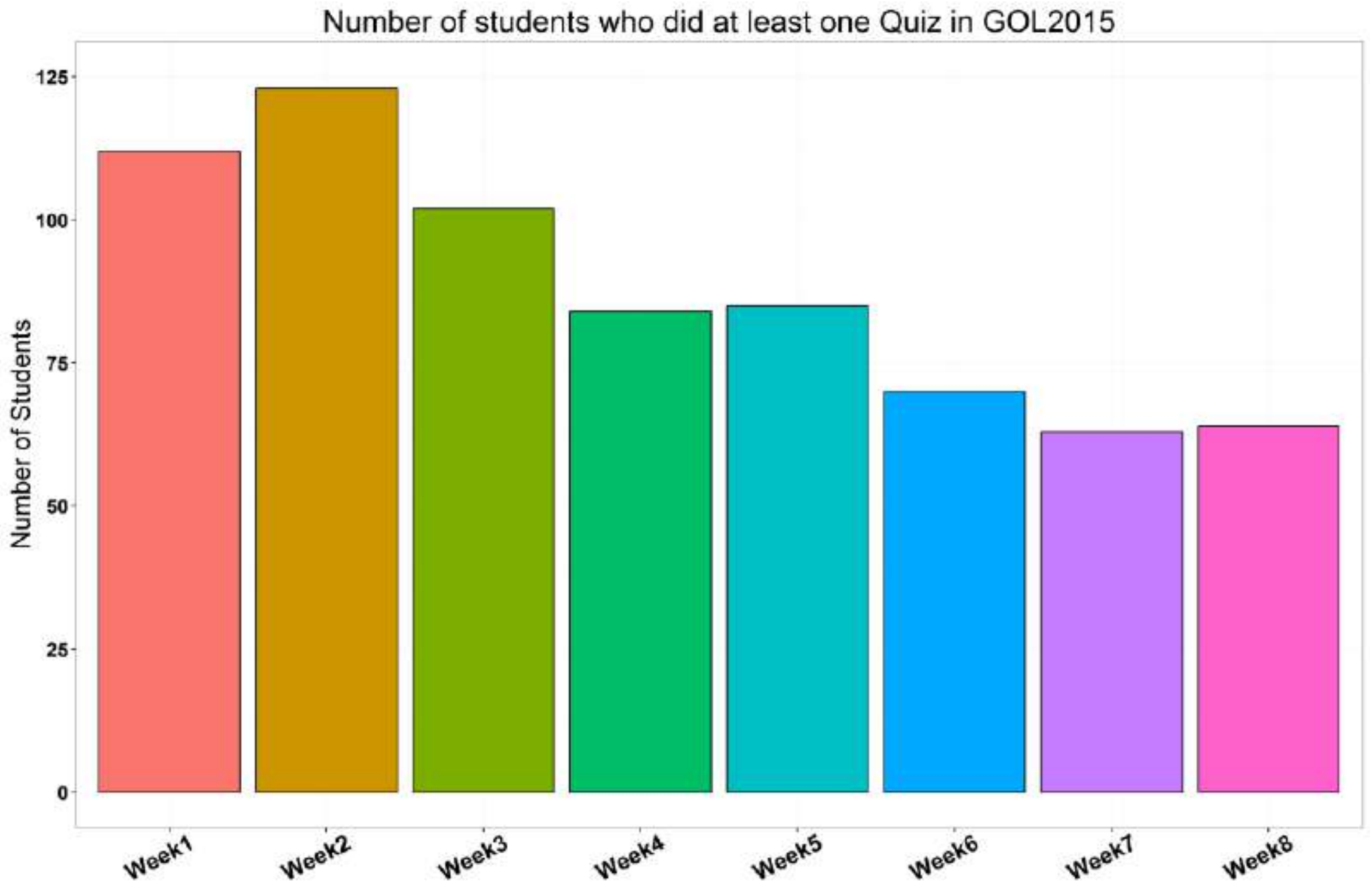
Number of students who did at least one Quiz in GOL2015
Number of Students
125
100
75
50
25
0
Week1
Week2
Week3
Week4
Week5
Week6
Week7
Week8

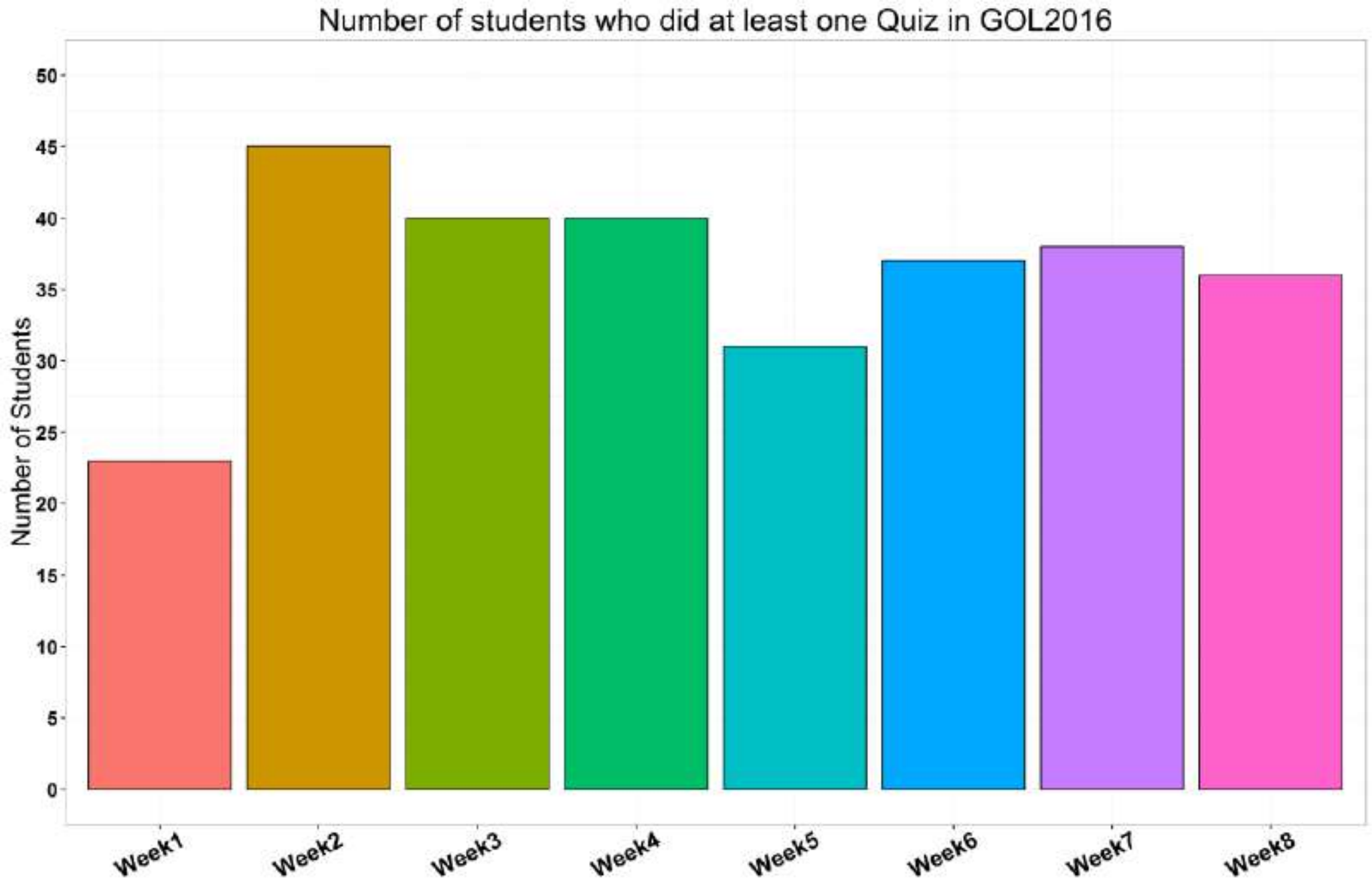

**Figure 84.** The number of students who did one quiz of each described week (a) Top: *Gratis Online Lernen* 2014 MOOC (b) Mid: *Gratis Online Lernen* 2015 MOOC (c) Bottom: *Gratis Online Lernen* 2016 MOOC

To check the validity of how active the students were within the MOOC variables, we inspected the quiz activities across every week of GOL2014, GOL2015 and GOL2016. Figure 84 depicts the examination of activities of the mentioned MOOCs across the weeks. The x-axis represents the MOOC weeks; the y-axis represents the number of students (*identical students not repetitive*). In Figure 84a, we can see that the number of students who did the GOL2014 quizzes every described week dropped except for a slight increase in the last week. Figure 84b shows nearly the same direction of GOL2014 behavior by which students were doing quizzes of GOL2015 actively in the first two weeks and then dropped till the last week. On the contrary, Figure 84c shows a very interesting behavior of the GOL2016 MOOC. The number of students who did the second week quiz nearly doubled when compared with the first week. Further, the figure shows that the number of students in week3 till week8 represents a stable participation rate in weekly quizzes.

Lastly, although the certification rate of the GOL2016 MOOC was nearly similar to the GOL2014 and GOL2015 MOOCs, the students in GOL2016 were obviously very active in comparison. Besides the apparent quiz activity in the GOL2016 MOOC, student actions were consistently registered in discussion forums and login frequency. We believe our Activity-Motivation framework was instrumental in increasing the student interactions with MOOC variables on one side and increasing the student motivation to complete quizzes on the other side, which leads to students wanting to complete the rest of the quizzes.

### 5.6.5 Case Study Summary

Since MOOCs are the new hype in the domain of Technology Enhanced Learning, higher education has come under pressure to adopt them in their educational systems. Although MOOCs are perfect for hosting a lot of students, the dropout, engagement, and motivation issues are still frustrating. In this case study, we utilized learning analytics on student data in order to investigate the hypothesis of the relationship between student activities and retention in MOOCs. We found that certified students who received certificates at the end of the course performed more activities than the non-certified students. Certified students engaged more in discussion forums; they viewed more forum posts and wrote more frequently than non-certified students. Additionally, they often interacted more with video lectures. In fact, our analysis on GOL2014 and GOL2015 showed that the active users demonstrated higher activity during the MOOC weeks. As a result, we became quite confident of the hypothesis that the more activities are done, the more likely the students are to complete the MOOC.

Based on the validation of this hypothesis, we proposed an Activity-Motivation model with the aid of learning analytics techniques and a gamification element. The framework was built on the previous analysis results in this study and earlier case studies using the inheritance of MOOC indicators. Our next step was to implement the framework and ttest it on GOL2016 MOOC to evaluate the results. The outcomes were stimulating to us since we found that we were successful in increasing the active student ratio when compared with earlier offered versions of the same MOOC.

# 6 LEARNING ANALYTICS CONSTRAINTS AND THE DE-IDENTIFICATION APPROACH[14]

We've already seen that learning analytics carries a lot of potential to the educational sector. However, the large-scale collection, processing, and analysis of data may steer the wheel beyond the borders to face an abundance of ethical breaches and constraints. Revealing learners' personal information and attitudes, as well as their activities, are major aspects that lead to identifying individuals personally. Yet, there are some techniques that can keep the process of learning analytics in progress while reducing the risk of inadvertent disclosure of learner identities. In this chapter, we will point out the learning analytics constraints and propose a conceptual approach that might offer a good solution to the ethical issues of learning analytics in MOOCs. Finally, we present Anonymizer, a prototype that is built on the de-identification framework which outputs anonymized files with good secure encryption level.

## 6.1 Learning Analytics Constraints

Learning analytics is an active area in the research field of online education and Technology Enhanced Learning. The applied analysis techniques as seen in previous case studies in the education data mainly aim to intervene and predict learner performance in pursuit of enhancing the learning context and its environment such as the MOOCs. However, ethical issues emerge while applying learning analytics in educational data sets (Greller & Drachsler, 2012). At the first International Conference on Learning Analytics and Knowledge (LAK '11), held in Banff, Alberta, Canada in 2011, participants agreed that learning analytics raises issues relevant to ethics and privacy and "it could be construed as eavesdropping" (Brown, 2011). The large-scale of data collection and analysis leads to ownership questions. After surveying the literature and in this

---

[14] Parts of this chapter have been published in:

Khalil, M., & Ebner, M. (2015b). Learning Analytics: Principles and Constraints. In *Proceedings of World Conference on Educational Multimedia, Hypermedia and Telecommunications* (pp. 1789-1799). AACE

Khalil, M., & Ebner, M. (2016b). De-Identification in Learning Analytics. *Journal of Learning Analytics*, 3(1), pp. 129-138. http://dx.doi.org/10.18608/jla.2016.31.8

subsection, we introduce eight-dimensional constraints that limit the beneficial application of learning analytics processes as shown in (Figure 85).

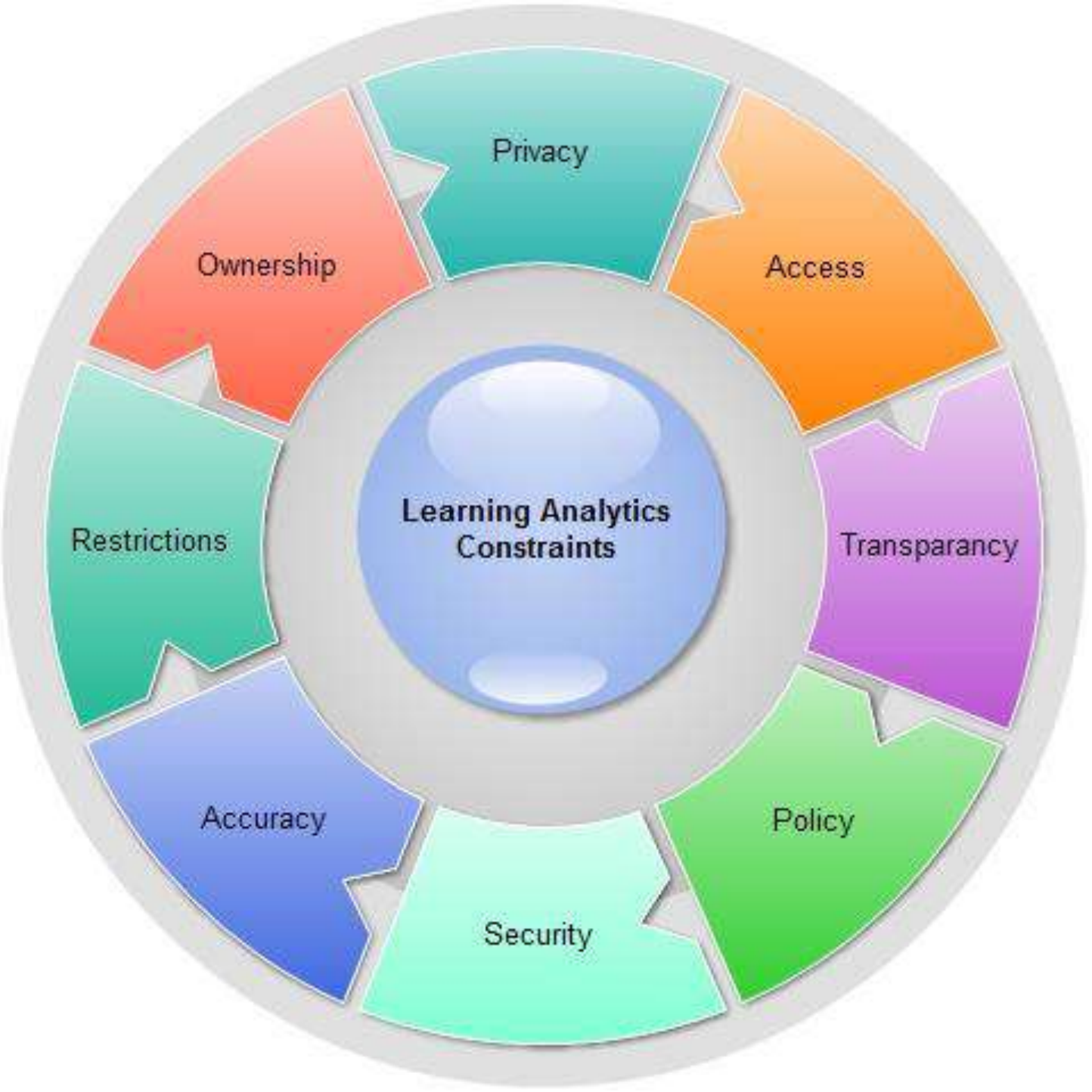

**Figure 85.** Learning analytics eight-dimensional constraints

- *Privacy*: Conforming to the main objective of learning analytics, namely predicting, a professor for instance, can point to a student who is at risk in his course. This could lead to the problem of labeling, wherein a learner is labeled as a “bad/good” student. The learning analytics committee needs to carefully consider potential privacy issues while analyzing student data. Data analysis and customization can reveal personal information, attitudes and activities about learners. Therefore, some educators claim that educational institutions are using software that collect sensitive data about students without sufficiently considering data privacy and how they eventually are used (Singer, 2014). Datasets may include sensitive information about learners. Thus, anonymization or de-identification may be required to preserve learner information. The student privacy law of Family Educational Rights and Privacy Act (FERPA) advocates the use of de-identification in higher education to preserve student records privacy. There are various cryptographic solutions, anonymization techniques and statistical methods that hide the owner's identity (Fung et al., 2010). The study of (Prinsloo & Slade, 2013) pointed to the requirement of de-identification of data before it becomes available for institutional use. This would serve the inclusion of learning analytics based on student activity and behavior while assuring the anonymity of their information. Regulations and laws are a good consideration to address the privacy issue. The Open University in England has made the first step to regulate laws specialized in learning analytics and privacy principles (OU, 2014). This is a good example that encourages other institutions to consider privacy as a fundamental element that should not be ignored.

- *Access*: Authentication assures that only legitimate users have permission to access and view specific data. Data access is relevant to policy regulations where these regulations must adhere to the authentication and authorization modules. The student should be allowed to update his/her information and provide additional information when needed. In order to achieve student privacy, there should be access levels for all learning analytics stakeholders. A student has the access level to view and update his information. A teacher is authorized to access student data without the possibility of viewing sensitive information such as ethnic origin or nationality. Decision makers can sustain the data in order to meet the institutional perspective. On the other hand, there are still unanswered questions about students, such as whether they have the right to access results of learning analytics, or do researchers have the morality to view and analyze students' data?
- *Transparency*: Disclosing information is a major challenge for information providers. Learning analytics methods should aim to be transparent and easily described to staff and students. The institution can assure transparency by providing information regarding data collection, usage and involvement of third parties in analyzing students' information.
- *Policy*: With the adoption of learning analytics in the educational field, institutions are required to adjust their policies with legislative framework. According to the study of (Prinsloo & Slade, 2013), many institutional policies failed to fully reflect the ethical and privacy implications of learning analytics. Learning analytics policies might describe regulations in data collection, usage of information, data security principles and description of the time period of keeping learners' data.
- *Security*: All learning analytics tools should follow expedient security principles in order to keep analysis results and student records safe from any threat. The widely-spread security model known to security experts is the CIA, which stands for Confidentiality, Integrity and Availability (Anciaux, Bouganim, & Pucheral, 2006). The confidentiality property guarantees that the data can never be accessed by an unauthorized user. Integrity property guarantees that the data cannot be altered, snooped or changed. The availability property means that the data should be available for authorized parties to access when needed. In the scope of learning analytics, student information and the analysis procedure should be kept safely and only accessible to authorized parties. A key component of protecting learner information is encrypting their data in order to achieve the confidentiality concept or anonymize information. Encryption guarantees that only authorized people can use the data. Moreover, assuring confidentiality can include invoking file permissions and granting a secure operating environment, while cryptographic hashing of datasets can assure the integrity property of students' records (Chen & Wang, 2008).

- *Accuracy:* Accuracy and validity of information is highly questionable. Making decisions, either by analysts or directors, based on a small subset of data could lead to fast judgments and hence trigger "false positives." Consequently, the accuracy of any forthcoming decision in a MOOC system will be influenced. For instance, if a group of students were "gaming the system" and an analyst builds a prediction model for all students based on MOOC indicators fulfillment, then a false positive action is triggered. As a matter of fact, learning analytics is not only based on numbers and statistics; judgments and opinions of researchers play a major role. We always see flounce on MOOC discussion forums activity and its correlation with performance. Some researchers find that more social activity in forums is reflected positively on performance while others go against this theory. In light of that, learning analytics is not always accurate.
- *Restrictions:* Data protection and copyright laws are legal restrictions that limit the beneficial use of learning analytics. Such legal restrictions are: limitations of keeping the data for longer than a specific period of time, which are regulated differently in each country; the data should be kept secure and safe from internal and external threats; data should be used for specific purposes and the results of any process should be as accurate as possible. The restrictions could be stronger when it relates to personal information.

- *Ownership:* Questions related to who owns the analyzed data of MOOCs can emerge anytime. Participants like to keep their information confidential, but at the same time, consent policy is essential to ensure transparency. Further, MOOC providers are encouraged to delete or de-identify personal information of their participants.

## 6.2 De-Identification Approach

Based on the previous section, learning analytics show that there are several drawbacks related to educational data breaches from the technical side and the ethical side. Data degradation (Anciaux et al., 2008), de-identification methods, or deletion of specific data records may offer a solution to preserve learners' information and increase the privacy level of learning analytics applications. In this case study, we will mainly focus on the de-identification process.

### 6.2.1 Background

Personal information is any information that can identify an individual. In fields such as the health sector, it is named Personal Health Information or PHI. In other fields, such as the education sector, this information is named Personal Identifiable Information or PII. The National Institute of Standards and Technology (NIST) defines PII as:

> Any information about an individual maintained by an agency, including 1) any information that can be used to distinguish or trace an individual's identity, such as name, social security number, date and place of birth, mother's maiden name, or biometric records; and 2) any other information that is linked or linkable to an individual, such as medical, educational, financial, and employment information. (McCallister, Grance, & Scarfone, 2010)

The personal information of learners can be categorized into details such as name, sex, photograph, date of birth, age, address, religion, marital status, e-mail address, insurance number, ethnicity, etc., or educational details such as qualifications, courses attended, degrees, and study records. As a criterion, a leak of an individual's personal information can induce misuse of data, embarrassment, and loss of reputation. However, organizations may be required to publish details extracted from personal information. For instance, some educational institutions are required to provide statistics about student progress; likewise, health organizations may need to report special cases from their patient records, such as communicable diseases. As a result, de-identification helps organizations to protect privacy while still informing the public. The de-identification process is used to prevent revealing individual identity and keeping the PII confidential.

In learning analytics, it is common for stakeholders to request additional information about the results extracted from educational data sets. Educational data mining and learning analytics mainly aim to enhance the learning environment and empower learners and instructors (Greller & Drachsler, 2012). Therefore, the analysis of these data may have interesting trends that could lead to further and deeper analysis by other institutions or researchers. Requests for more extensive analysis may involve the use of student-level data. Accordingly, ethical issues arise, such as privacy disclosure, and the need to de-identify the data becomes paramount.

Recently, Harvard and MIT universities released de-identified data from 16 courses offered in 2012– 2013 from their well-known edX Massive Open Online Course (MOOC) (MIT News, 2014). The Harvard and MIT edX ensure that the anonymity of the released data complies with the Family Educational Rights and Privacy Act (FERPA) (http://www2.ed.gov/policy/gen/guid/fpco/ferpa/index.html, last access January 2015). Furthermore, Prinsloo and Slade (2013) suggest different approaches that inform students in higher education of the implications of learning analytics on their private data.

De-identification of student records has been regulated in the United States and the European Union. The United States adopted FERPA regarding the privacy of student educational records. In the European Union, the Data Protection Directive (DPD; 95/46/EC) regulates the processing of personal data and the movement of such information. In respect to the European law, the most explicit citation of de-identification in is Article 26 on anonymization, in which "principles of data protection shall not apply to data rendered anonymous in such a way that the data subject is no longer identifiable." It is not obvious, however, what level of de-identification is required to anonymize education records under European law. However, the Article 29 Data Protection Working Party has an opinion on the identification of data: "Once a data set is truly anonymized and individuals are no longer identifiable, European data protection law no longer applies" (2014, p. 5).

### 6.2.2 De-Identification Drivers in Learning Analytics

A study by Peterson (2012) addressed the need to de-identify data used in academic analysis before making it available to institutions, to businesses, or for operational functions. Peterson (2012) pointed to the need to keep a unique identifier in case a researcher may need to study the behavior of a particular individual. Slade and Prinsloo (2013), however, drew attention to the ambiguity of data mining techniques in monitoring student behavior in educational settings. The authors linked de-identification with consent and privacy and stressed the need to guarantee student anonymity in their education records in order to achieve learning analytics objectives such as interventions based on student characteristics. An example of the link between consent and de-identification would be a questionnaire or survey that those filling it out are told will be used for research only. In that case, clearly the limitation of using their data will be just the one study. If the survey includes personal information, however, then assurances of anonymizing their data should be considered.

Ryan Baker (2013) discussed the demands of de-identifying educational data sets in his "Learning, Schooling, and Data Analytics." De-identification of these data sets means being able to share them among other researchers without violating FERPA regulations. Baker stressed that educational policies should include rules for anonymizing data in order to prevent identifiable information from being leaked without authorization. Furthermore, Drachsler and Greller covered the topic of anonymization in their DELICATE approach (Drachsler & Greller, 2016). De-identification techniques have been also reviewed as a right of access principle in learning analytics deployment (Pardo & Siemens, 2014). In addition, Pardo and Siemens (2014) further suggest that semantic analysis might be required to detect identifiable records in anonymized data sets.

### 6.2.3 De-Identification Approach

In this section, we propose a conceptual de-identification–learning analytics framework as shown in Figure 86. The framework begins with learners involved in learning environments. Currently, a large number of learning environments support online learning, such as LMS or MOOCs. These platforms offer environments with rich, vast amounts of data that can be quantitatively/qualitatively analyzed to benefit learners and enhance the learning context.

The next step is the de-identification process where techniques to convert personal and private information into anonymized data take place. De-identification techniques include such methods as anonymization, masking, blurring, and perturbation. The last step can include (but is not mandatory) the de-identified data linked with a unique descriptor that may be examined by learning analytics researchers and benefit stakeholders, but ultimately must be used only to the advantage of students.

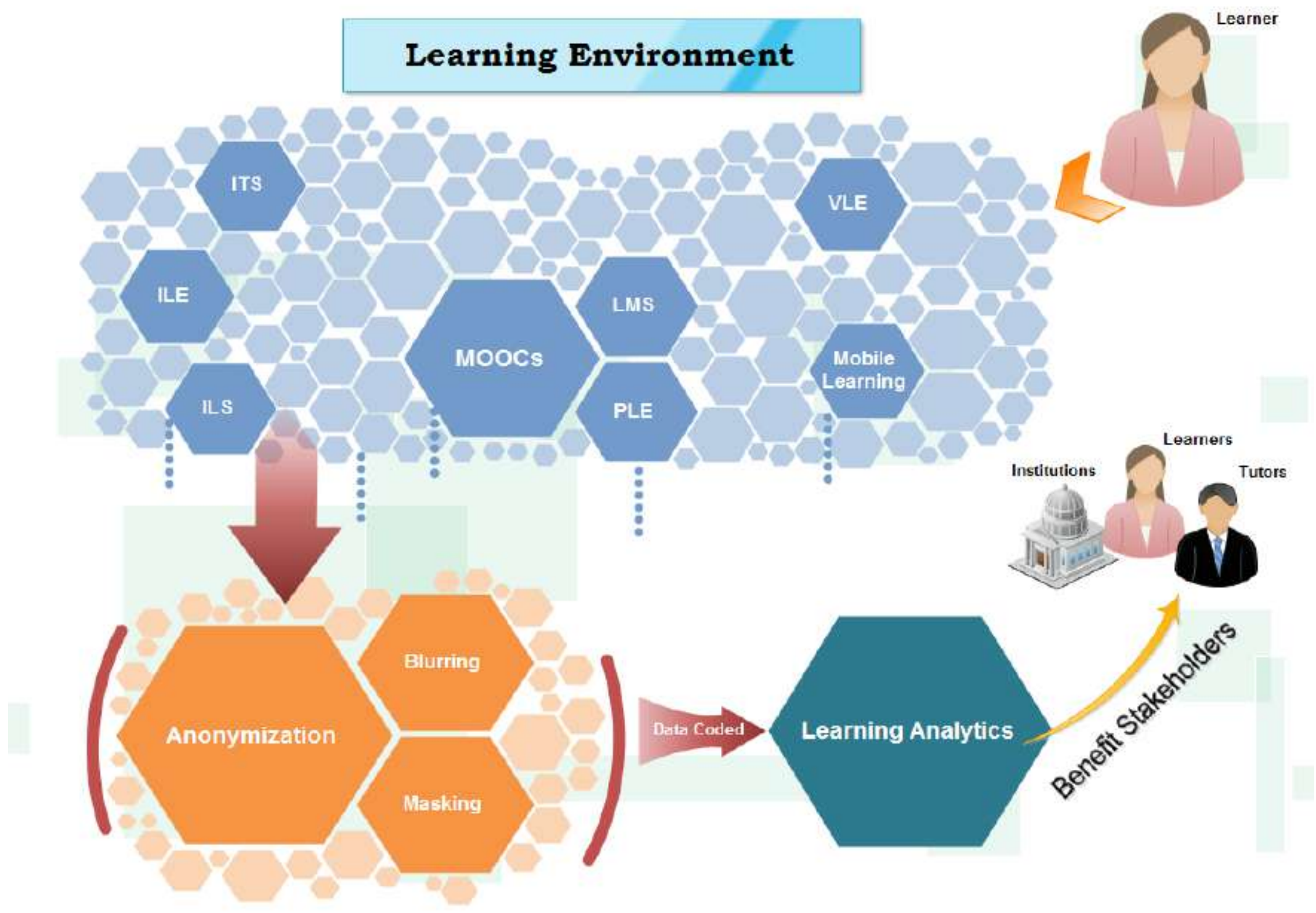

**Figure 86.** The proposed de-identification-learning analytics framework

### 6.2.4 De-Identification Techniques

In our proposed de-identification–learning analytics conceptual framework, there are several techniques available to de-identify student data records. Figure 87 lists several methods of de-identification and provides examples (based on Article 29 Data Protection Working Party, 2014; Cormode & Srivastava, 2009; Eurostat, 1996; Petersen, 2012).

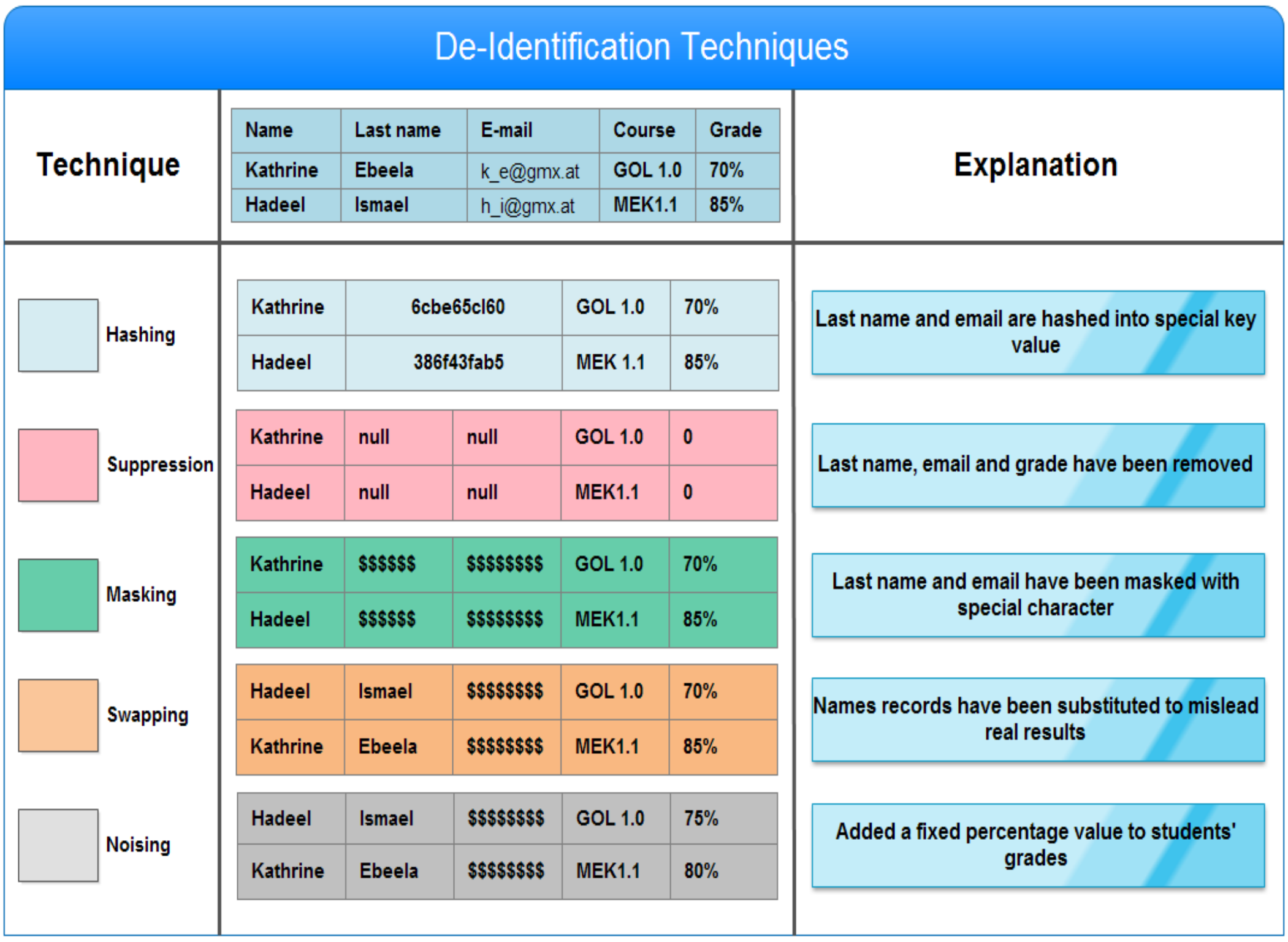

**Figure 87.** Examples of de-identification techniques

#### 6.2.4.1 Anonymization

Data anonymization techniques have recently been keenly researched in different structured data records with the goal of guaranteeing the privacy of sensitive information against unintended disclosure and a variety of attacks (Cormode & Srivastava, 2009). Ohm (2010) defined reasons behind anonymization when organizations want to release the data to the public, sell the information to third parties, or share the information within the same organization. The difference between anonymization and de-identification, however, is quite misunderstood. Anonymization principles are a subset of holistic de-identification methodologies. Data anonymization is the process of de-identifying data while preserving its original format (Raghunathan, 2013). In the educational context, anonymization refers to different procedures to de-identify student data in such a way that it cannot be re-identified (the opposite of de-identification) unless there is a record code. Anonymization is not reserved only for tabular data records, but can also be applied to other types of data — such as visualized data or graphs — where institutions intend to present their outcomes without revealing sensitive information.

On the other hand, in addition to anonymization, de-identification includes masking, randomization, blurring, and so on. For instance, replacing "Bernard" with "$$$$$$$" is a method of masking while altering "Bernard" to "Wolfgang" would be an example of anonymization. However, masking and blurring are not as well-known as anonymization. By any means, de-identification, pseudonymization, and anonymization are interchangeable topics under the information concealing umbrella. To clarify the differences in simple terms, pseudonymization means cloaking the original data with false information with the ability to track it back to its original formation; anonymization, conversely, cannot be reversed (Raghunathan, 2013).

As previously mentioned, educational data records may include private information, such as name or student ID, which singularly are called direct identifiers. Removing or hiding these identifiers does not assure a true data anonymization. Identifiers could be linked with other information that would allow identification of individuals (see Figure 88). However, quasi-identifiers can be used to ensure better de-identification of data. "Date of Birth + Sex + Name" is an example of a quasi-identifier. In 2006, AOL released the search records of 500,000 of its users. Several days after AOL's database release, New York Times journalists were able to reveal the identity of a 62-year-old widow using a similar process to that shown in Figure 88 (Soghoian, 2007). AOL admitted that the data release was a mistake and the research team responsible for sharing the data was fired.

| Course_ID | Course_Name | User_ID - deidentified | Grade |
|---|---|---|---|
| 001 | GOL | 0005 | 70% |
| 002 | MEK | 0009 | 90% |
| 001 | GOL | 0006 | 60% |

| Course_ID | Grade | Name | Country |
|---|---|---|---|
| 002 | 90% | Sabrina | Austria |
| 001 | 60% | Michael | Germany |
| 004 | 75% | Rebecca | Austria |

**Figure 88.** Linking data sources leads to name identification/labeling

Another example of identifying individuals was reported in 2000 when demographic information led to retrieving the names and contact information of patients whose medical data had been released in the United States (Sweeney, 2000).

Samarati and Sweeney (1998) provided a well-known anonymization technique, namely *k*-anonymization. This method addresses the problem of linking records to identify the individual's information when releasing data, thus safeguarding anonymity. The *k*-anonymity technique focuses on avoiding a data record from being identified with k individuals (Cormode & Srivastava, 2009).

#### 6.2.4.2 Masking and Blurring

Masking is a de-identification technique that replaces sensitive data with fictional data in order to disclose results outside the institution. Data masking can modify the data records so that they remain usable while keeping personal information confidential. For instance, character masking replaces a string with special characters.

On the other hand, blurring involves reducing precision to minimize the identification of data. There are several ways to achieve blurring, such as dividing the data into subcategories, randomizing the data fields, or adding noise to data records.

#### 6.2.4.3 Coding Data Records

In scientific research, data usually requires further investigation with researchers looking deeper into the details. Having de-identified data might be insufficient for these purposes; researchers may require additional information in order to do more analysis. The American federal Health Insurance Portability and Accountability Act (HIPAA), which is responsible for protecting the confidentiality of patient records, authorizes using an “assigned code” that can be appended to the records in order to permit the information to be re-identified for research purposes.[15] Based on that HIPAA rule, we found that FERPA 99.31(b) allows for using a unique descriptor for student data records in order to match an individual’s information for research and institutional use. Accordingly, we conclude that assigning a code to student records in our proposed framework can grant learning analytics researchers the ability to study behaviors of specific students and, therefore, can benefit learners. However, this step is not mandatory, as anonymized data can also be valuable in performing analysis if, for example, numerical values are swapped or noised.

### 6.2.5 De-Identification Limitations

Despite the fact that de-identification protects confidential information and privacy, the de-identified data still poses some privacy risks (Petersen, 2012). In many cases, some attributes are capable of identifying individuals; in other cases, attackers can link records together from different sources and therefore “code break” the de-identification. On the other hand, in their paper “Privacy, Anonymity, and Big Data in the Social Sciences,” Daries et al. (2014) assured that with de-identification, there is no guarantee of keeping the analysis process uncorrupted. Pardo and Siemens agree that “data can be either useful or perfectly anonymous, but never both” (2014, p. 447). The bottom line is that the stricter the de-identification guidelines, the greater the negative affect on the ultimate analysis.

---

[15] Rule 45 C.F.R. § 164.514(c).

## 6.3 Anonymizer

In this part of the chapter, we implemented a prototype using the proposed de-identification-learning analytics framework with three techniques: *k*-anonymity, masking and suppression. This mini-prototype, referred to as Anonymizer, is designed to anonymize the generated educational data files from the iLAP application. Anonymizer reads tabular structured files as CSV files and makes the researcher choose one of the above-mentioned de-identification techniques. The outcome will be an anonymized file that allows us to publish results to the public, or send the files to third-party researchers/teachers. The two primary goals of this mini-project are summarized in twofold: a) implement de-identification techniques on MOOC datasets, and b) anonymize the datasets in such a way that it is still possible to obtain useful knowledge out of them.

### 6.3.1 Anonymizer Architecture

Before designing the tool, we summarized our requirements of the prototype as the following:

1. The prototype needs to be a web-based application that reads and de-identifies data.
2. The application should handle the generated data files of iMooX and iLAP.
3. The tool should support usability of researcher needs.
4. The outcome should be useful information and secure.

After listing the requirements, we programmed the prototype backend. The programming was done using the Java Maven application using Servlets and a Servlet Container (Tomcat). We also intended to separate the client code and the server code so that it returned the anonymized files directly to the end user. The result can be either saved on the host server or can be downloaded. Figure 89 shows the architecture diagram of the Anonymizer prototype backend.

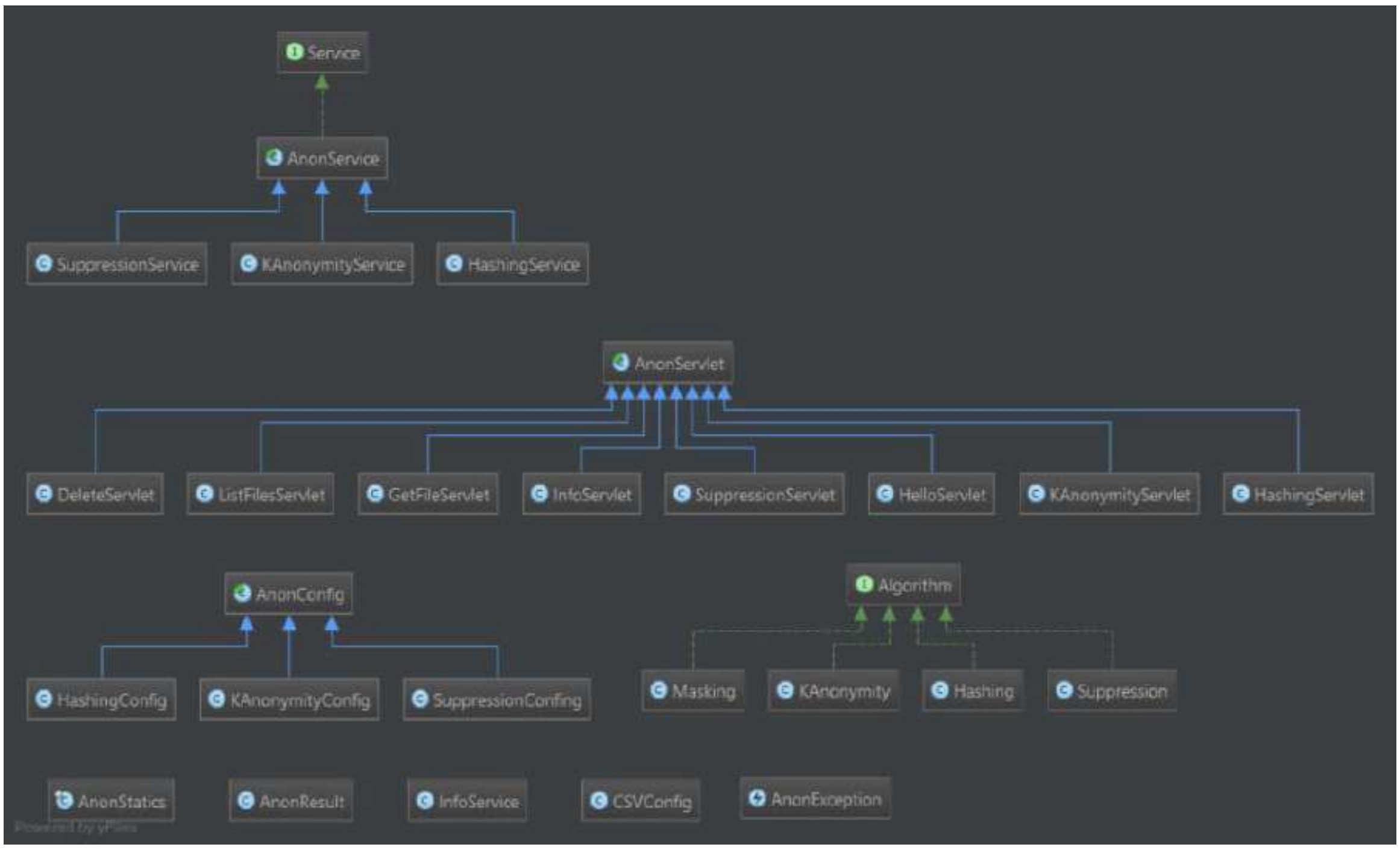

**Figure 89.** The architecture diagram of the de-identification prototype (Anonymizer)

The frontend interface was built using HTML5, CSS and JavaScript. The frontend procedure of the tool works as Figure 90 depicts.

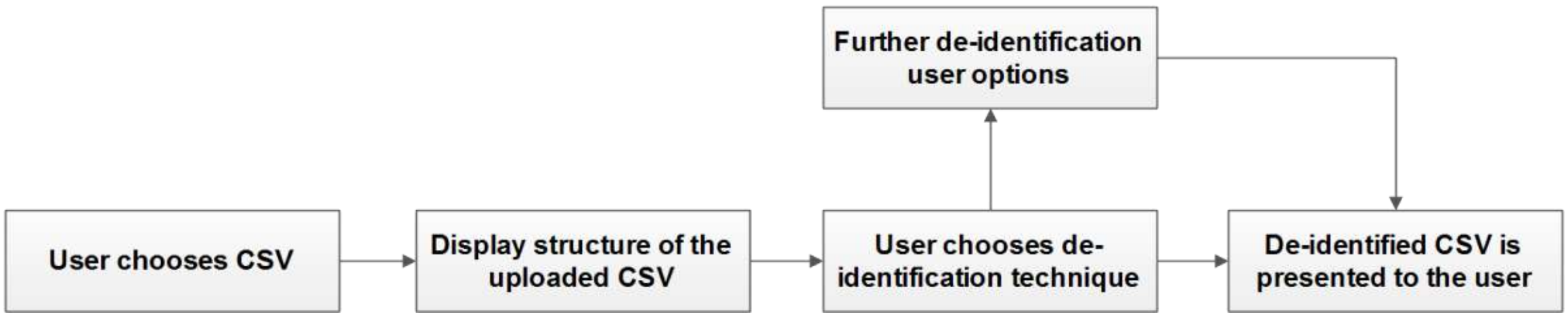

**Figure 90.** The frontend procedure of the de-identification prototype (Anonymizer)

The demonstration of the frontend interface of the prototype was simple and effective at the same time. Figure 91 shows the way in which the researcher can upload the CSV file of the educational data he/she has. After that, the user should define the delimiter as long as CSV supports colon and semi-colon, the operating system, and a new file name if needed. When these are defined, the application automatically shows the header titles and proposes an action to the next step. The researcher can then identify which de-identification algorithm he/she wants. Likewise, his/her action is required to select which columns need to be anonymized.

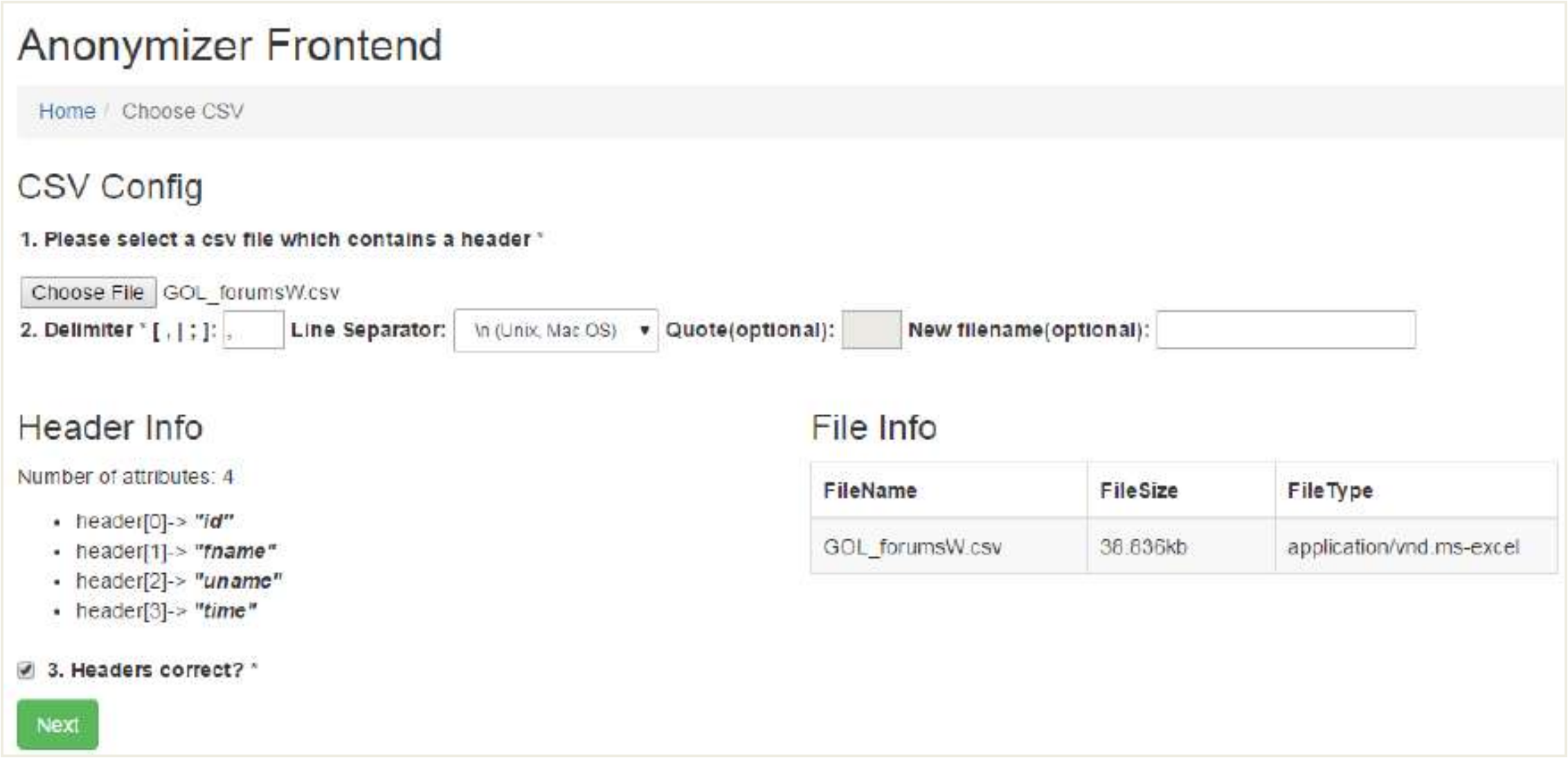

**Figure 91.** The Anonymizer frontend web page

The tool offers three types of anonymity. The hashing technique mixes up two columns' original information and gives a result of encrypted information. The suppression anonymization technique converts at least one column into a defined special character such as an asterisk. The third option is the *k*-anonymity which offers a strong anonymization algorithm that requests the user to define quasi-identifiers from another file and suppression at the same time. Figure 92 shows an example of a data file, a hierarchy for zip code anonymization of options up to (*k*= 5), and age hierarchy which anonymizes using a generalization method.

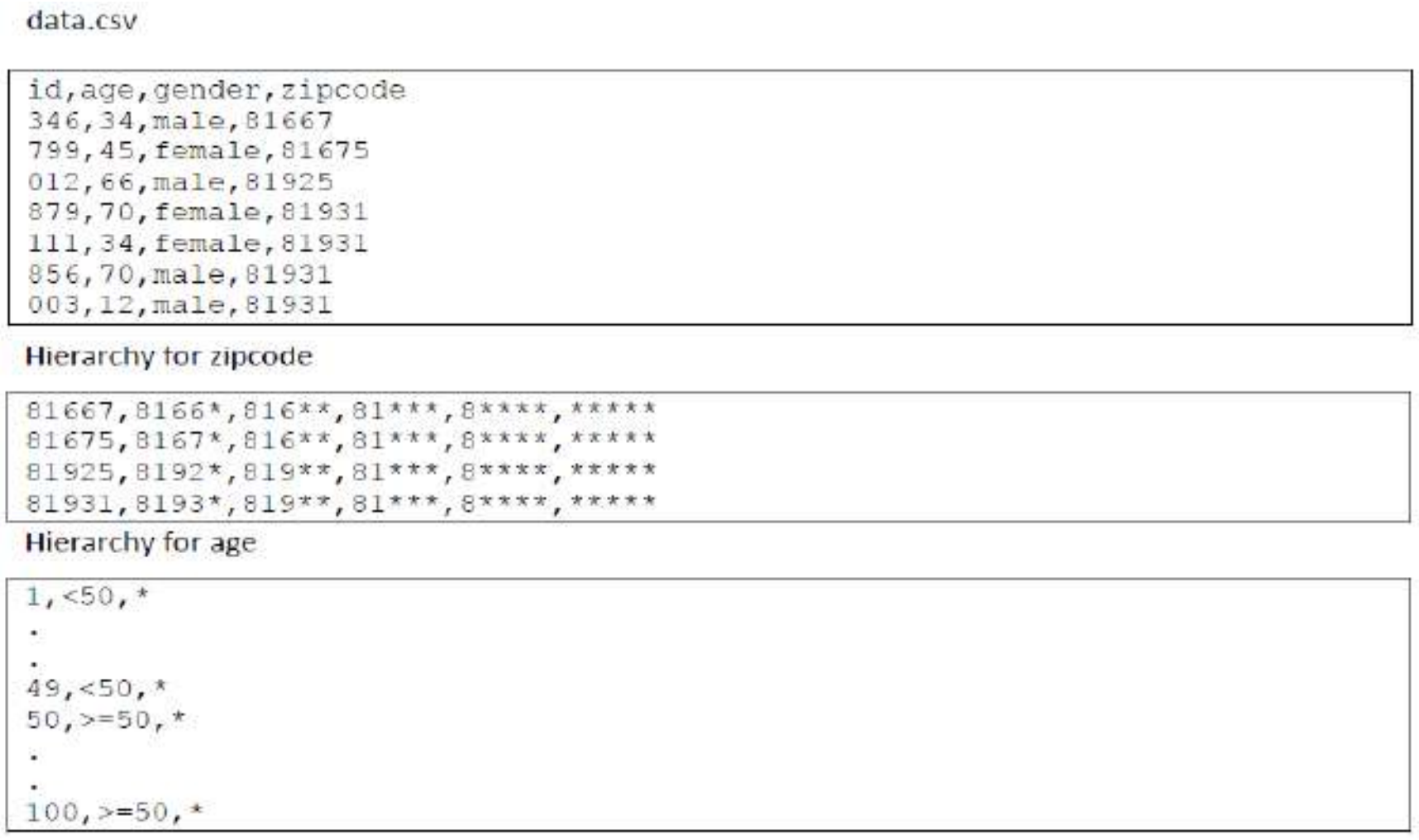
data.csv

```
id,age,gender,zipcode
346,34,male,81667
799,45,female,81675
012,66,male,81925
879,70,female,81931
111,34,female,81931
856,70,male,81931
003,12,male,81931
```

Hierarchy for zipcode

```
81667,8166*,816**,81***,8****,*****
81675,8167*,816**,81***,8****,*****
81925,8192*,819**,81***,8****,*****
81931,8193*,819**,81***,8****,*****
```

Hierarchy for age

```
1,<50,*
.
.
49,<50,*
50,>=50,*
.
.
100,>=50,*
```

**Figure 92.** An example from the Anonymizer prototype showing the k-anonymity de-identification technique

We believe the Anonymizer offered a good level of de-identification of our important records from iMooX. Nevertheless, caveats may occur if suppression or generalization data points are disproportionately used.

## 6.4 Chapter Summary

This chapter aims at introducing what security and ethical challenges can affect learning analytics in general and learning analytics in MOOCs in particular. We have summarized the challenges as eight-dimensional constraints. These constraints are strongly related to privacy, security, ethical issues, transparency and ownership. The very limited research in finding a solution to these problems compelled us to design a de-identification framework aimed at anonymizing educational data files. We shed light on this topic via US and EU regulations regarding data privacy. We proposed a conceptual approach with examples of de-identification techniques that assisted us with the iMooX platform and have the potential to help learning analytics specialists preserve confidential learner information. Finally, we implemented a mini-prototype that supported our framework and gave us the option to publish our results to other parties (researchers, teachers, and course designers).

Although de-identification is not a foolproof solution for protecting learner privacy, it is a valid consideration in the examination of the ethical dimensions of learning analytics.

# 7 CONCLUSION AND FUTURE RESEARCH

Due to the rhythm of modernization inherent in every aspect of our lives, educational technology has undergone a dramatic paradigm shift in the last decade. As technology prototypes are considered the main impetuses behind the creation of new online and distance learning models (Ajzen & Fishbein, 1977; Davis, Bagozzi & Warshaw, 1989), the umbrella of e-Learning currently covers a wide range of engaging online environments including Massive Open Online Courses (MOOCs). MOOCs have proven to scale education in different fields and subjects as well as support the movement toward a vision of lifelong learning (Kop, Fournier, H., & Mak, 2011). MOOCs have contributed drastic changes to higher education on one side and to elementary education on the other side. Furthermore, MOOCs have the ability to accommodate large numbers of students that would not be able to participate if in a conventional university or classroom face-to-face setting. This large volume aspect of MOOCs requires the hosting servers to handle massive amounts of data, which has led to the popular term “Big Data” in education. Additionally, there are two leading research communities oriented with respect to discovering new meanings of educational datasets activities: the educational data mining and learning analytics communities (Papamitsiou & Economides, 2014).

Even though MOOCs enjoy great popularity and bring many benefits to the educational community, some concerns arise with MOOC advancement. Alternately, learning analytics plays a significant role in scrutinizing and providing explanations of different views to MOOC issues (Clow, 2013; Knox, 2014). The motivation of this dissertation illustrates the need for learning analytics to unveil hidden information and patterns contained in the large educational data sets found in MOOCs. The main aim of this thesis was to investigate the potential of learning analytics in the Austrian iMooX-MOOC platform. Via this investigation, the purpose was then to provide deep analyses to open case studies and challenges that are strongly related to grouping and motivating students.

This thesis presents one step toward the interpretation of MOOC data using learning analytics to interrogate concerns, illuminate issues, and resolve problems. This final chapter proposes our vision for future directions and implementations as well as provides a summary of our primary findings to the posed research questions in Chapter 1. A solid background and a wide scan of related works in Chapter 3 is also provided. The methodology followed in this dissertation is presented in Chapter 2.

## 7.1 Conclusion

To better summarize our agenda regarding this dissertation, consider the idiom, “A picture is worth a thousand words” (see Figure 93), which presents a clear timeline of the thesis.

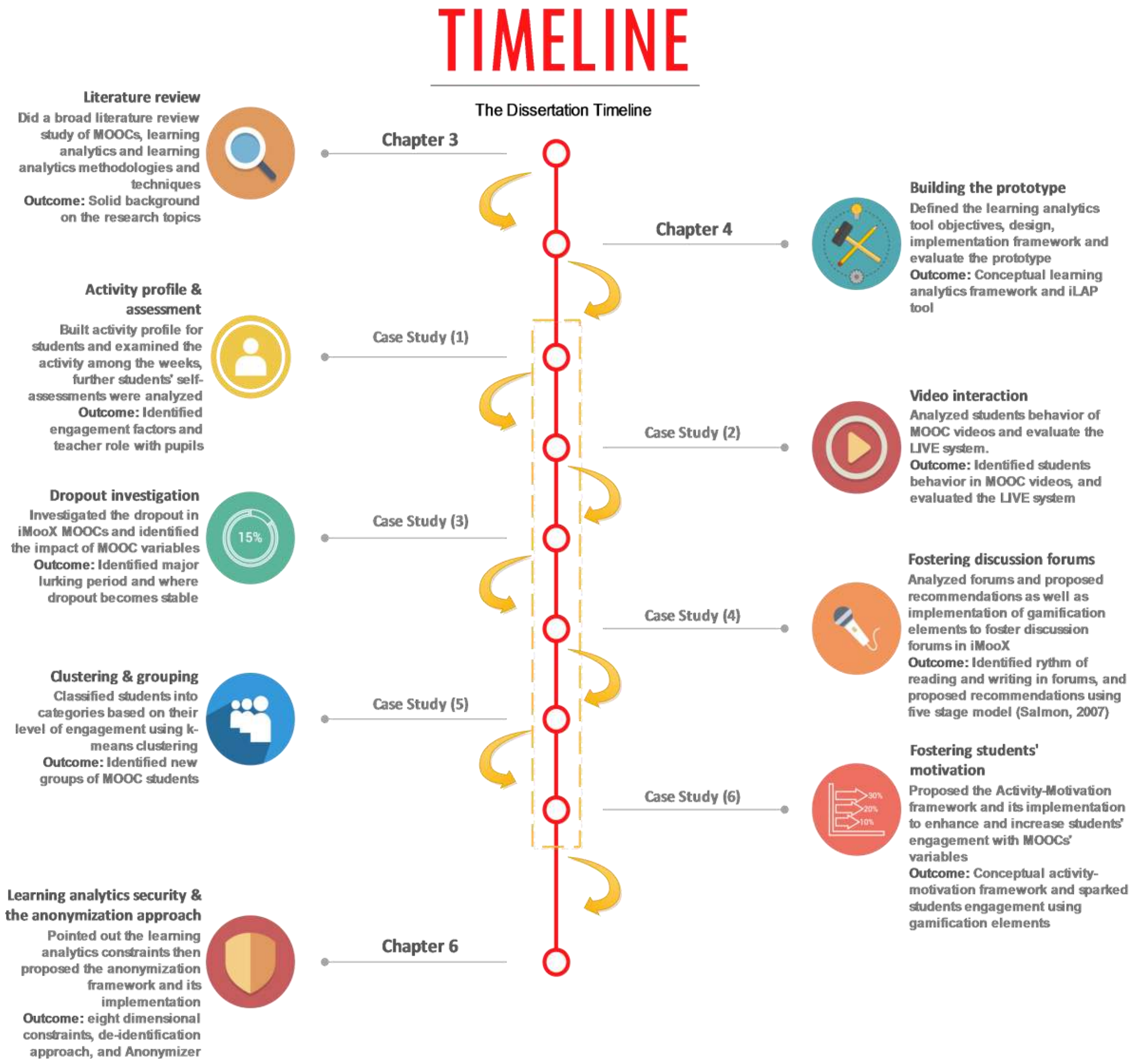

**Figure 93.** The dissertation timeline and procedure

Chapter 4 tackled the research question regarding the design of a learning analytics framework and its dimensions as well as prototyping of learning analytics in MOOCs. Our contribution in respect to the framework was "The learning analytics framework and lifecycle." The framework summarized the objectives, stakeholders, methodologies, and results as the body of a lifecycle design. Our purpose of such a draft came from the ideology related to "closing the loop," by which the learning analytics tool results would finally be optimized to the benefit of the main stakeholders (learners, teachers, environments). The second part of the question regarding the implementation of learning analytics in MOOCs was illustrated by the adoption of the framework mentioned above and the generation of two sub-architectures: first, the design architecture and second, the implementation architecture. The design architecture focused on collecting the data, processing them, and finding patterns within the data. The implementation architecture followed a system of logging, database management and visualization for research and administrative use. The contribution to the field is the iMooX Learning Analytics Prototype (iLAP).

Based on the iLAP, we gathered information about the offered courses and students. From here forward, Chapter 5 then introduced a total of six case studies that tackled open challenges and answered several research questions. The case studies were composed based on: 1) the nature of the offered MOOCs, 2) the release date of the MOOC, and 3) the number of enrollees.

Chapter 5.1 answered the question regarding the profiling of school children and examining their behavior in self-assessment quizzes in MOOCs. The built activity profiles divided student activities like quizzes and their video and social interactions on a weekly timeline. Further, the activity profile helped in the analysis of the student engagement factor and activity attrition as well as the discovery of certain types of students who could pass the MOOC without watching the MOOC's video lectures. The self-assessment behavior of school students revealed the significance of teacher interventions. Moreover, we discovered that guided learning undoubtedly helped students with their final performance. Chapter 5.2 lists our experience in answering the research question regarding student behavior in videos and the role of interactive videos in distinguishing between certified students and non-certified students. At present, MOOCs depend strongly on video lecturing. With the combination of iLAP and LIVE systems, we noticed that students watch video lectures not only to pass quizzes but also as a type of knowledge-building. Some students may visit the course just to watch certain segments of MOOC videos. Certified students slightly outperformed (with respect to time and outcomes) the non-certified students in answering the interactive LIVE's multiple-choice questions.

Is there a specific point in xMOOCs where learners drop the course or become lurkers? Through our analysis in Chapter 5.3, we can answer this question with the confidence that after the first four weeks, the dropout ratio becomes more stable. We relied on student activity in dealing with MOOC variables to answer this research question. We believe that shorter MOOCs can vigorously decrease the dropout ratio. In fact, during this dissertation, we found that dropout is particularly linked to student interaction with MOOC variables. In the iLAP evaluation subsection of Chapter 4, the dropout rate can be referenced to a new type of MOOC learners called "active students." Active students are those who complete at least some activities in MOOC. For example, if we imagine that our reference point for calculating the dropout rate is the number of students who only register but do not actually participate, then we can also imagine that it would look similar to a scenario of students who open a classroom door but do not enter to learn. In that case, controlling dropout by focusing on certain types of students becomes easier, according to Chapter 5.5's conclusion.

Chapter 5.5, "Clustering Patterns of Engagement to Reveal Student Categories," answered two of our research questions regarding the classification of students based on their engagement level and the research question concerning the distribution of students on the Cryer's scheme of Elton (1996). Students in MOOCs can be clustered and grouped according to their level of engagement. This section showed that depending on the level of engagement (high, medium, and low) in interactions with MOOC variables and the use of machine learning *k*-means clustering, we discovered some significant outcomes. External students (public) and undergraduate students (university) who are accredited with ECTS points share three student groups according to the defined *k*. One interesting group we found is called the "Gaming the System" cohort. This group involves students who try to trick the grading system. We found that this group comprised around two-fifths of the total number of undergraduate students. Chapter 5.5 closes with a distribution of students on Cryer's scheme of Elton (c.f. Figure 75) which was further discussed and explained in detail.

The research question related to the core behavior of students in the discussion forum and the fostering of their contribution therein was resolved in Chapter 5.4. Since interaction in forums was marked as an important component, we did an augmented study on this variable. Students' attitudes in forums sway between different points of view. Learners, for instance, actively read comments and posts in the first four weeks. Likewise, their demeanor was noted in discussions and writing during the same period. A moderate positive strength correlation between both elements (writing and reading) was recorded. Superposters as well as active instructor(s) can encourage activity in the discussions. With this in mind, we identified the rhythm of reading and writing in MOOC discussion forums. Learners may consider MOOCs as subsidiary material for knowledge ingestion since their interaction peaks in the afternoon. However, this remains a hypothesis that is open to debate since the data are collected based on a quantitative model (learning analytics). The second part of the research question was clarified based on Gilly Salmon's (2007) five stage model to foster student contributions in MOOC forums. In fact, we adopted stages wherein we tried to increase the usability as well as enhance the general design of the MOOC forums using gamification elements. Although the new design did not result in that much of a difference, key results obtained were improved usability, flexibility, and user satisfaction.

The last case study in Chapter 5 provided an answer to the research question on how to motivate students in MOOCs and increase their activity. As a matter of fact, all the case studies analysis of this dissertation have built the foundation to answer this particular question. Student engagement is strongly associated with their activity. With this in mind, we presented the idea of a battery icon gamification element to enhance student engagement and stimulate their motivation. The implementation of the battery gamification element revealed an obvious increase in the number of active students to 73% of the examined MOOC's student quota.

Student behavior in MOOCs was the center of attention in the previous case studies chapter. Since we employed learning analytics to answer those questions, we could not turn a blind eye to the negatives of learning analytics in general and learning analytics of MOOCs in particular. Having said that, the large-scale collection and processing of student information motivated us to tackle the question regarding security constraints and what solutions we can currently propose. Chapter 6 provided eight-dimensional constraints of the general learning analytics. We concluded that data ownership, personal information privacy, transparency, accuracy, and security are major ancillary issues. Flipping these constraints in the learning analytics of MOOCs will concern the above-mentioned issues in matters related to ethics and privacy. In order to face some of these issues, we proposed a de-identification approach by which we focused on anonymizing the private information on one side while preserving valuable and useful outcomes on the other side.

## 7.2 Future Directions

I have split the future research directions in two ways; the first is the future research directions of learning analytics. While the current learning analytics research encompasses extensive investigations in the computer science and machine learning disciplines, the field lacks research in psychology and pedagogy practice. Further, the field of learning analytics has been thoroughly researched with discourse that defines the essences and postulates of the field. Thus, focus should shift toward a reexamination of the position of theory in learning analytics research and the establishment of connections to other relevant domains such as cyber-learning, user modeling, technology enhanced learning, and collaborative learning. The prospective research in learning analytics endeavors to a greater focus on predictive analytics and providing feedback in learning environments such as LMS and MOOCs. Learners will be enthused with and encouraged by systems that forecast their next move. This can control, to some extent, the dropout issues as well as enhance final learning outcomes.

Another key concept worth mentioning is the fact that learning analytics can deliver many possibilities to elementary education. Learning analytics is not only dedicated to higher education but also to schools. Monitoring pupils either physically (using sensors) or digitally might be promising in the forthcoming future. Having said that, it should always be remembered that learning analytics is not only initiated to generate visuals and produce interventions but also to support learning and optimize teaching.

The second part of my future research directions is related to the learning analytics of MOOCs. I believe that MOOCs will continue to emerge in the following years. MOOCs have evolved from a simple idea in the education and technology domain to a large-scale area of business and marketing. In Chapter 3, we already found that the articles that combined learning analytics and MOOCs were the most cited papers according to the Google Scholar metric system. The research that impels learning analytics with MOOCs should strive for tools that improve motivation with prompt feedback. Such goals can be achieved using advanced visual analytics modules. Visual analytics can be employed with the use of dashboards with unique designs and models. Since research studies in the area show that the majority of MOOC enrollees belong to the age range of 20-30 years, we believe that the current delivery of analytics lacks a level of enjoyment and interactivity. In light of that, extrinsic factors like gamification elements might support learning in general and enhance student interaction with the system.

Furthermore, more research should be dedicated to particular categories of students in MOOCs. This will help to set up various strategies for learning improvement and the application of adequate interventions.

Tomorrow's instructors must think about increasing the intrinsic motivation of students by improving didactic approaches and the instructional design of their courses. The evidence collected from the case studies of this dissertation leads us to conclude that learning analytics might not offer that alone, however, course designers and instructors should consider analytics to improve student engagement in MOOCs. So, if we imagine the potential when course instructors can see the analytics interpretation and subsequently realize what is working and what is not in their courses, their next step will then be an absolute intervention that requires them to adjust their teaching style with respect to learning (Teaching Analytics).

Finally, the practical matters of ethical constraints and privacy remain. We already provided an anonymization approach to control personal information breaches. But is it secure enough or is it vulnerable to data leak? Every system has its own limitations. Thus, this matter is an important realm of interrogation in the future of learning analytics and learning analytics of MOOCs. In parallel with this direction, policy development in learning analytics practices is in strong demand in the area. Buckingham Shum (2015) stated that:

> The future of learning analytics depends to a large extent on the policy adopted by institutions and governments. Its practice will be greatly shaped by the regulatory framework which is established, the investment decisions made, the infrastructure and specifications which are promoted.

Therefore, how can policies control assumptions that lead to inaccurate judgments by learning analysts? And how can the regulation of learning analytics data ownership be illustrated in new policies? The answer to these questions require thorough investigation which hopefully will be enacted in the very near future.